# Photo-physics of Hybrid Metal Halide Perovskite Semiconductor Thin Films: Application in Photovoltaic

Submitted in partial fulfillment of the requirements

of the degree of

## Doctor of Philosophy

*by*

**Shivam Singh**

**(Roll No. 144120008)**

*Supervisor*

**Prof. Dinesh Kabra**

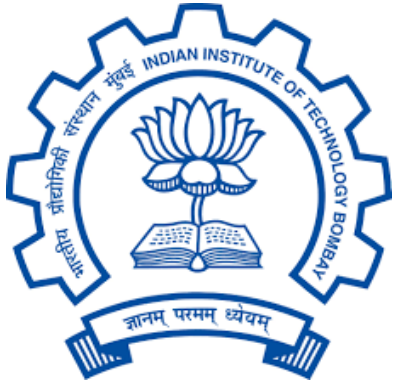

Department of Physics

INDIAN INSTITUTE OF TECHNOLOGY BOMBAY

**June 2020**

# Approval Sheet

This thesis/dissertation/report entitled **"Photo-physics of Hybrid Metal Halide Perovskite Semiconductor Thin Films: Application in Photovoltaic"** by Shivam Singh (144120008) is approved for the degree of Doctor of Philosophy.

Examiners

___________________________

___________________________

Supervisor (s)

___________________________

___________________________

Chairman

___________________________

**Date:** June 22, 2020

**Place:** IIT Bombay, Mumbai, Maharashtra, India.

# Declaration

I declare that this written submission represents my ideas in my own words and where others' ideas or words have been included; I have adequately cited and referenced the original sources. I also declare that I have adhered to all principles of academic honesty and integrity and have not misrepresented or fabricated or falsified any idea/data/fact/source in my submission. I understand that any violation of the above will be cause for disciplinary action by the Institute and can also evoke penal action from the sources which have thus not been properly cited or from whom proper permission has not been taken when needed.

_______________________________________
(Signature)

Shivam Singh

(Roll No. 144120008)

# Abstract

The perovskite photovoltaic (PV) technology is emerging as a prominent candidate in the PV community due to ease of fabrication, low cost, high efficiency (approaches to conventional Si based solar cells) and extra-ordinary opto-electronic properties. It is important to understand that what makes solution processable perovskite semiconductor so special, which is hard to achieve from others? This thesis demonstrates the photo-physics of hybrid metal halide perovskite ($CH_3NH_3PbI_3$) semiconductor thin films and its application to solar cells. This thesis includes the unprecedented study to reveal the mystery behind the positive temperature coefficient of the bandgap in halide perovskite, which is in contrast to other conventional semiconductors such as Si, GaAs etc. Temperature dependence of the band gap is due to combined effect of electron-phonon interaction and lattice dilation. It is conclude that the thermal expansion term dominates over the electron phonon interaction term with thermal expansion coefficient of 3.83 x $10^{-4}$ $K^{-1}$, which accounts for the positive temperature coefficient of the band gap. This thesis also demonstrates that structural properties do influence the optical properties of the perovskite semiconductor based thin films. The quality of perovskite film is highly dependent on fabrication condition, which ultimately decides the defect density in the final film. This study helps to understand the defects and disorder in the perovskite semiconductor even at room temperature and advised to apply some molecular or solvent additives engineering technique to reduce the defect states from the bulk as well as from the grain boundaries.

Thus, the role of additive engineering technique is investigated via solvent (DIO & phosphinic acid) and small molecule (Bathocuproine) additive to passivate the bulk and surface defects in order to suppress the non-radiative recombination on different class of Pb based perovskite semiconductors. The reduction in non-radiative channels is supported by enhanced time-resolved photoluminescence (PL) lifetime, higher electroluminescence quantum efficiency ($EL_{QE}$) and improved charge transport properties. This thesis also demonstrate the influence of iodine rich solvent additive on chemical (XPS) and electronic states (UPS) of perovskite thin film and can be applicable to other halide based solvent or molecular additives. The *p-i-n* configuration is used to fabricate the devices. The spin-coating of perovskite layer and electron extracting layers (EEL) are performed inside the glove box. This thesis also reports improved moisture stability via addition of Bathocuproine (BCP) which forms a physisorption kind of interaction between BCP and perovskite and it is validated by first principle DFT calculation. The BCP additive device shows a fill factor of 0.82 as a state of art.

This thesis also deals with a unique set of study where scanning photocurrent microscopy (SPM) and dielectric relaxation time constant of photo induced charge carriers are correlated. SPM helps to estimate the lateral transport length scale of hole and electron, separately, in a bipolar device and supports the enhancement in short circuit current density and fill factor of perovskite solar cells (PSCs) by solvent additive engineering. Although the Pb based perovskite PV technology is well promised, but toxicity of Pb is a big challenge for commercialization of PSCs. Sn emerges as an alternative of Pb in perovskite solar cells. However, oxidation, efficiency and stability are the main concern with Sn. This thesis demonstrate a comparative study on the influence of 'B' (Pb and/or Sn) cations on charge carrier recombination dynamics in Pb free based PSC ($MA_{0.20}FA_{0.75}Cs_{0.05}SnI_3$), pure Pb based PSC ($FA_{0.95}Cs_{0.05}PbI_3$) and Pb-Sn mixed PSC {$(MAPbI_3)_{0.4}$ $(FASnI_3)_{0.6}$} with the help of transient photovoltage (TPV) technique. Overall, this thesis provide a detail study of photo-physics and device physics of defect induced Pb halide based perovskite semiconductor and it will be useful to apply the fabrication and characterization techniques used in the thesis for Pb free PSCs.

# Acknowledgments

It is my privilege to work under the guidance of Prof. Dinesh Kabra. Not only has he led me towards the field of perovskite semiconductor based photo-physics & its application to the solar cells and let me experience the beauty of it. But, he also inspired me with his research spirit and passion, which would influence me throughout the rest of my life no matter which field I work in. I feel so fortunate to have a Ph.D. advisor like him and really appreciate his support and mentorship in those years. He always remains the source of motivation and encouragement for each and every challenge of my Ph.D. works that I have faced. He has always had a constant faith in my abilities from the very beginning of my Ph.D. and has always been there with all the ups and downs of my research career. He always gives me the freedom of choosing research problem and encourages my independent thinking. I am always grateful to him for giving me confidence in difficult situations and motivate to choose correct direction.

I am thankful to Prof. K. L. Narasimhan from Electrical Engineering Department, IIT Bombay for his constant helpful research feedback and scientific advice during my Ph.D. candidature. I am thankful to my research progress committee members, Prof. Pradeep Nair, Prof. Subhabrata Dhar and Prof. Aswani Yella for providing me valuable feedback on my research work. I am grateful to Prof. Parag Bhargava from Department of Metallurgical Engineering & Materials Science, IIT Bombay for his insightful feedback on my research works and he always allowed me to work freely in his laboratory. I am also thankful to Sudip Chakraborty from the Uppsala University, Sweden for DFT calculation on physisorption of Bathocuproine over $CH_3NH_3PbI_3$ based perovskite film. I am thankful to Prof. Sven Huettner and Dr. Cheng Li from University of Bayreuth, Germany for our collaborative work regarding temperature dependent band gap of $CH_3NH_3PbI_3$ based perovskite film. I am thankful to Prof. Soumitra Satapathi and Prof. Manojit Bag from IIT Roorkee for our collaborative work on the fabrication of perovskite solar cells in ambient air.

I would like to pay my profound gratitude to the IIT Bombay, Department of Science & Technology (Grant DST/TM/SERI/FR/186(G)), India, UKRI Global Challenge Research Fund project, SUNRISE (EP/P032591/1), Industrial Research and Consultancy Centre (IRCC), Centre of Excellence in Nano-electronics (CEN) and National Centre for Photovoltaic Research and Education (NCPRE).

I am thankful to Ms. Bosky Sharma for our many collaborative works on perovskite solar cells. I am also thankful to Dr. Ranjana Shourie for our collaborative work on interlayer engineering in perovskite solar cells. I am thankful to Dr. Amrita Dey and Ms. Santa Kolay for our collaborative works in the field of perovskite nano-crystals. I am also thankful to Dr. Dhanashree Moghe, Dr. Amrita Dey, and Mr. Gangadhar for helping me in learning of time-resolved PL measurement. I am thankful to Dr. Naresh Chandrasekaran for helping me in learning transient photo-voltage measurement. I am thankful to Mr. Rakshit Jain for helping me in the scanning photocurrent microscopy experimental set up. I am thankful to Ms. Urvashi Bothra for our collaborative work on perovskite-BHJ tandem solar cells. A special thanks to Ms. Laxmi for helping me in correcting this thesis.

I am grateful to Dr. Dushyant Kushwaha for encouraging me in the beginning of my research career. I have got three wonderful senior colleagues and friends; Amrita, Pravin and Naresh, who made me feel comfortable to work freely in the starting days of my Ph.D. Besides all the research activities, there are people who always remain beside me whether inside and outside of the lab environment. I have earned some lifelong friends here: Shiwani Pareek, Bosky Sharma, Laxmi and Kashimul Hossain. We have had together some wonderful memories to cherish throughout the life. I have also got some fabulous junior colleagues and friends: Bhupesh, Suraj, Gopa, Sumukh, Nrita, Rajib, Kalyani, Ayush and Neha. I am also thankful to my close friends Satyaveer Singh, Minita Surwade, Pooja Agarwal, Mayank Sharma, Deepika Rani, Jyoti Sharma, Jeetendra Sahani, Vinay Chaurasiya and Nidhi Tiwari, who have made my Mumbai chapter memorable. Finally, I am thankful to my parents (Mr. Surendra Singh & Mrs. Swarna Keshi Singh), siblings (Rana Sangram Singh, Sangeeta Singh & Jaya Singh), nephews Vatsal & Shashwat and niece Vartika for their immense support, love, and care without which my life would not have been flourished.

### *I dedicate this thesis to my late grandparents.*

## Publications included in this thesis:

## Miscellaneous Publications:

**16.** R. Sharma, B. Sharma, **S. Singh**, D. Kabra, P. Bhargava and S. Mallick "Nickel Oxide as a Potential Role Player for High Open Circuit Voltage in Perovskite Solar Cells" Chemistry Open **2020** (*Due for submission*).

## Presentations made in Conference:

1. **Title:** "Enhanced Photovoltaic Performance of Perovskite Solar Cells *via* Defect Passivation by Incorporating the Physisorbed Small Organic Molecule" Materials research society (MRS) Fall Meeting & Exhibit (Boston, USA) **2018** (Poster presentation).

2. **Title:** "Investigation on Iodine rich Lewis base solvent additive in perovskite film formation via Ostwald repining and ion exchange approach"10th International conference on Materials for Advanced Technology (Singapore) **2019** (Poster Presentation).

3. **Title:** "Influence of reducing agent on transport length scale in CH3NH3PbI3 based perovskite solar cells" 10th International conference on Materials for Advanced Technology (Singapore) **2019** (Oral Presentation).

4. **Title:** "Influence of Static vs Dynamic Disorder on the Electronic Properties of Hybrid Perovskite Semiconductor CH3NH3PbI3" International Conference on Hybrid & Perovskite Solar Cells (Delhi, India) **2019** (Poster Presentation).

5. **Title:** "Charge carrier recombination dynamics in Pb *vs.* Sn based Perovskite solar cells" SUNRISE Symposium (Bangalore, India) **2020** (Poster Presentation).

6. **Title:** "Correlation between Charge Transport Length Scales and Dielectric Relaxation Time Constant in Hybrid Halide Perovskite Semiconductor." SUNRISE Symposium (Bangalore, India) **2020** (Oral Presentation); **Won Best Oral Presentation Award.**

# Table of Contents

# List of Figures

## Chapter 3: Methodology

**Chapter 5: Solvent additive engineering for improved photovoltaic performance via controlling the crystal growth kinetics and defect states.**

**Chapter 6: Investigation on Organic Molecule Additive for Moisture Stability and Defect Passivation via Physisorption in CH$_3$NH$_3$PbI$_3$ Based Perovskite.**

**Chapter 7: Correlation between charge transport length scales and dielectric relaxation time constant in hybrid halide perovskite semiconductor**

**Chapter 8: Charge carrier recombination dynamics in Pb vs Sn based halide perovskite semiconductors**

**Chapter 9: Summary and future outlooks**

# List of Tables

**Chapter 6: Investigation on Organic Molecule Additive for Moisture Stability and Defect Passivation via Physisorption in CH₃NH₃PbI₃ Based Perovskite.**

**Chapter 7: Correlation between charge transport length scales and dielectric relaxation time constant in hybrid halide perovskite semiconductor**

**Chapter 8: Charge carrier recombination dynamics in Pb vs Sn based halide perovskite semiconductors**

# Chapter 1

# Introduction to Perovskite Solar Cells

# CHAPTER 1

# Introduction to Perovskite Solar Cells

## 1.1    Background

Energy sources are essential to meet the daily need in our society such as lighting, communication, cooking, transportation, building, electricity generation etc. But, the conventional resources such as coal, fossil fuels, oils, natural gas etc. are finite. As per recent report on *31$^{st}$ October 2019 by 'Ministry of Power, Government of India'*, the total installed electricity generation capacity in India is around 364.9 GW. However, 63% contribution to the total installed power is from non-renewable sources such as coal, lignite, diesel, petroleum and gas. The dependency on energy sources are going to increase day by day as the India's population is increasing with time. As per report, the growth in energy consumption is almost linear with time.[1] Hence, there is requirement of energy security. Renewable energy sources are standing out as a prominent candidate for the energy dependency and security purpose.[2] It also helps in maintaining the balance between sustainability and environment such as dependency on resources coming from foreign countries (eg. petroleum) along with reduction in carbon emission (burning coal). Solar, wind, biomass, hydro and geothermal energy are few examples of renewable energy sources.[3] India projects to reduce the use of coal to 50% in 2030 by promoting the technique of renewable energy sources (figure **1**).[4]

Among all these, this thesis will focus on solar energy considering the geographical location of the country. There are many ways to harvest the solar energy such as solar water heater,[5] solar cooker,[6] concentrated solar power (steam engine)[7] and photovoltaic (PV) or solar cell.[8] Among all, solar cell technology plays a pivotal role in energy harvesting and become more popular.[9] Even in solar energy technology, there are numerous PV techniques emerging with time such as crystalline Si cells,[10] Single junction GaAs,[11] CIGS,[12] CdTe,[13] Amorphous Si:H,[14] multi-junction cells[15] and thin films based solar cells.[16] In present time,

Si based solar cells are commercialised and there are continuous efforts towards reduction in their manufacturing cost. However, there is not plenty of room for reduction in manufacturing cost of Si solar cells and the cost reduction rate is saturating.[17] This suggests that there is necessity of some different class of semiconductors whose material cost is cheap and is easy to fabricate. Thin film based PV technology stands out as one of the alternative. Currently, dye-sensitized solar cells,[18] organic cells,[19] inorganic cells,[20] quantum dot based cells[21] and perovskite solar cells[22] are emerging in thin film based PV community. Out of all thin film based PV technology, perovskite solar cells (PSCs) show a huge jump in efficiency (3.9 % in 2009 to 25.2% in 2019) in such a short span of time, which is approximately close to single junction crystalline Si solar cells in terms of efficiency (figure **2**).[23]

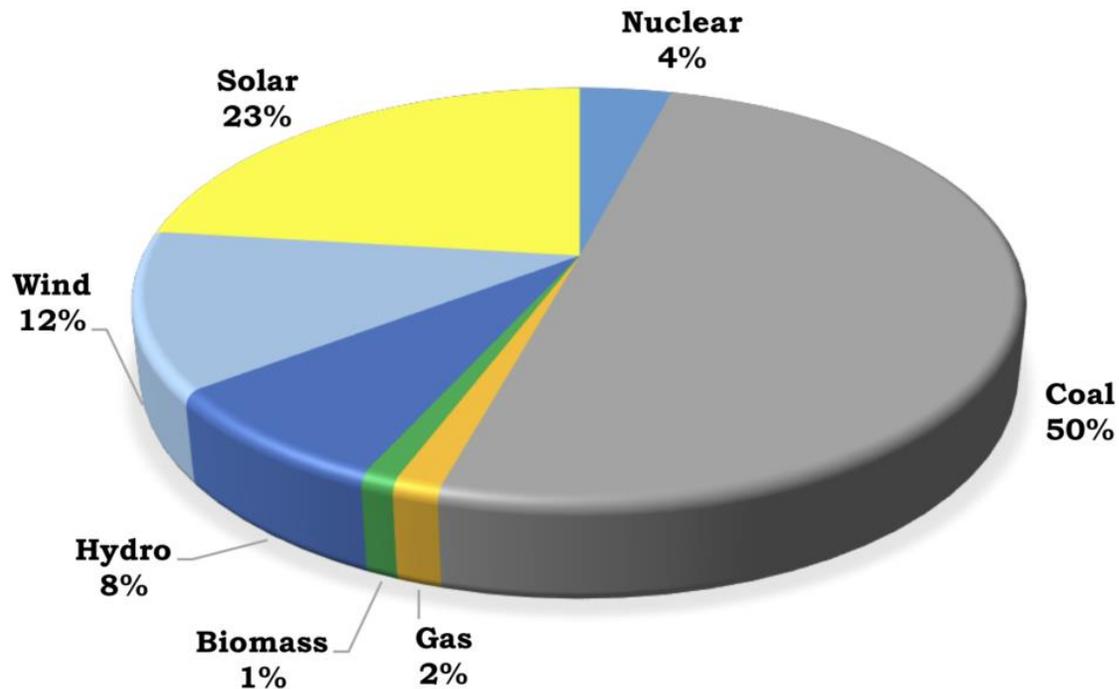

***Figure 1:*** *Pie chart of likely gross generation (831 GW) in 2029-2030. (Source: Central Electricity Authority, Government of India.[4]*

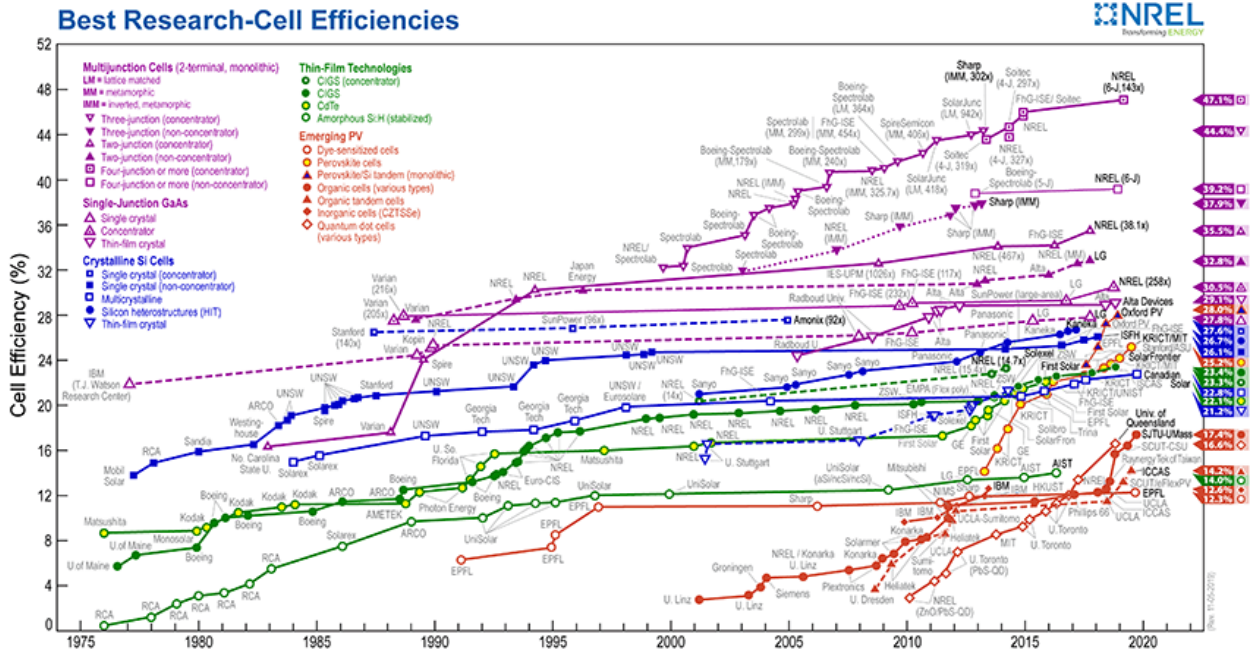

***Figure 2:*** *Best research solar cell efficiency chart from NREL.*[24]

## 1.2    Why Perovskite?

Perovskite is a kind of material which has a similar structure to calcium titanium oxide (CaTiO₃).[25] The general chemical formula of perovskite is ABX₃ (A & B = cations with different sizes and X = anion) and is shown in figure **3**.[26] For an ideal cubic perovskite structure, 'A' cation should be larger than the 'B' cation. 'B' cation is surrounded by octahedron of anions and is in 6 fold coordination. 'A' cation is in 12 fold coordination. In order to decide the perovskite crystal structure, the ionic radii of the two cations 'A' & 'B' and the anion 'X' should follow the tolerance factor and octahedral rules:[27]

$$Tolerence\ Factor: 0.81 < t = \frac{R_A + R_X}{\sqrt{2(R_B + R_X)}} < 1.11$$

$$Octahedral\ Factor: 0.44 < \mu = \frac{R_B}{R_X} < 0.9$$

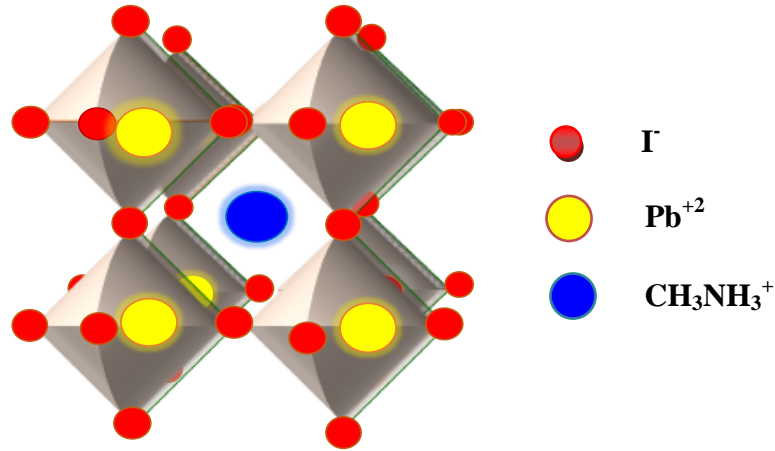

I⁻

Pb⁺²

CH₃NH₃⁺

*Figure 3:* *Crystal structure of Perovskite (ABX₃, more specifically CH₃NH₃PbI₃).*

The high optical absorption and direct band gap property of perovskite semiconductor makes it superior than the first (Si) and second generation (GaAs) solar cells. The absorber layers of first and second generation solar cell are around 300 μm and 2 μm respectively. However, perovskite based absorber layers needs even less than 500 nm thickness to produce absorption coefficient of the order of $10^5$ cm-1. The factors that govern the optical absorption of semiconductors are:[28]

(a) Transition matrix element between the conduction band and the valance band

(b) Joint density of states

Hence, the optical absorption coefficient of a semiconductor is highly dependent on the electronic structure. The optical absorption process for first, second generation and perovskite based solar absorber are shown in figure **4**. In first generation solar cell (Si), the optical absorption takes place in between *p* orbital of valance band to the *p* and *s* orbital of the conduction band. Since, Si is an indirect band gap material; the probability of optical transition is lower than that of direct band gap semiconductor. This makes the absorber layer of Si based solar cells two order more thicker than the direct band gap semiconductors, which increases the total cost of the material. However, the GaAs (second generation) and hybrid halide perovskite based solar absorber are direct band gap semiconductors but the absorption coefficient of perovskite absorber is higher than that of GaAs due to different electronic structure. The valance band of GaAs is made up of *p* orbital of As and the conduction band is composed of *s* orbital of Ga and *s* orbital of As. However, the conduction band of perovskite is made up of degenerated *p* orbital of Pb and valance band is composed of *s* orbital of Pb and *p* orbital of X (I, Br, Cl). The *s* orbital is dispersive or delocalised in comparison to the *p* orbital. Hence, the density of states in lower states of conduction band of perovskite is higher than that of GaAs. In addition, the intra-atomic transition is occurring in between valance band (Pb s) to conduction band (Pb p) which makes the probability of transition higher in case of perovskite. Hence, the optical absorption strength is higher in perovskite than GaAS. Apart from high absorption coefficient, the other properties such as higher diffusion length (> 1μm), lower binding energy (< 10 meV), higher charge carrier mobility (> 10 cm$^2$V$^{-1}$S$^{-1}$), band gap tunability (UV-VIS-NIR) and low temperature solution processable technique promises the future of perovskite solar cell in PV community.

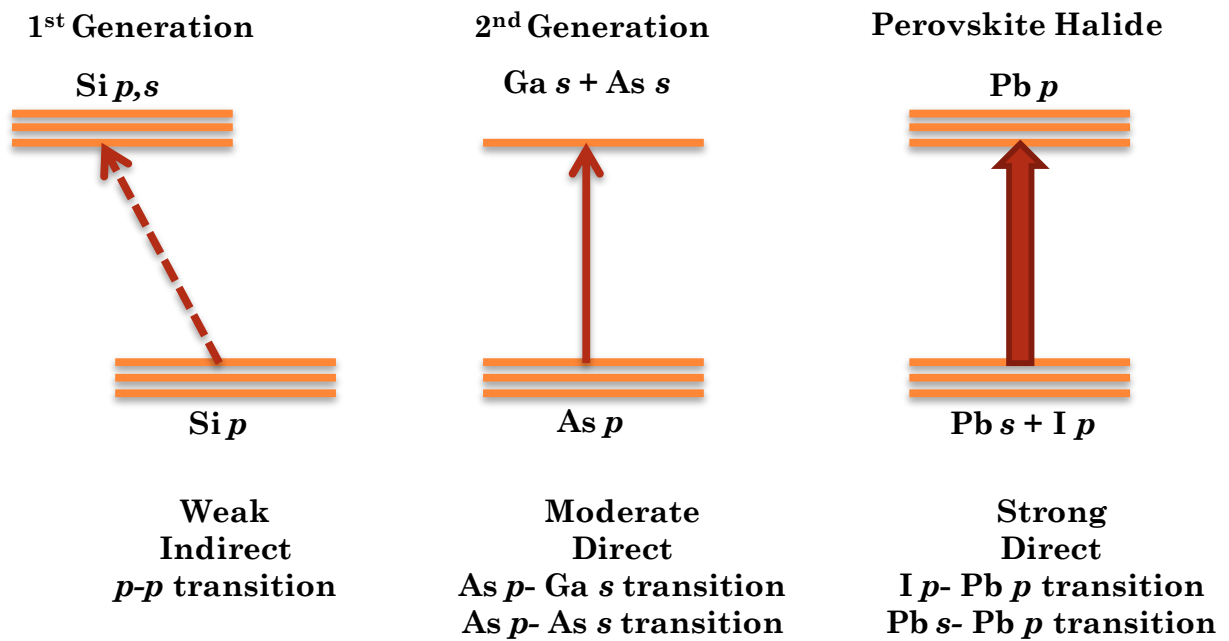

***Figure 4:*** *Mechanism of optical absorption in first generation, second generation and organic inorganic hybrid halide perovskite based absorber layer.*

## 1.3    Motivation

Although the perovskite PV technology is well promised, but scalability and stability are the two major issues that need to be addressed for commercialization of PSCs. Mostly all the high efficiency PSCs are reported on small area devices (~0.1 cm$^2$) and anti-solvent approach is applied to control the rate of crystallization in order to get pin hole free and smooth perovskite thin film. However, it is difficult to transfer the anti-solvent approach to the large area devices due to the non-uniformity in morphology of the perovskite thin films and thus it causes irreproducibility in terms of efficiency number. Hence, additive engineering, printing, doctor blade, spray coating, carbon based devices etc. are the alternative approaches which can be transferred for the fabrication of large area devices.[29,30] The efficiency number for an active area of 1cm$^2$ PSC had been reached close to that of 0.1 cm$^2$ device.[31] But, the efficiency (10%) is highly reduced for large area PSC modules (169 cm$^2$).[32] Greatcell Solar reported an efficiency of 12% on 100 cm$^2$ PSC module.[33] The other important challenge is the stability of PSCs. Quasi 2D-3D based devices exhibit higher moisture stability than the 3D PSCs due to the large organic ligands in low dimensional materials.[34] Small molecule additives such as pyridine, BCP, urea, PC$_{61}$BM, thiophene etc. are also used to passivate the grain boundaries and provide moisture stability.[35,36,37] There are various ways to check the stability of PSCs using AM1.5 xenon lamp with or without UV filter, metal halide lamp, UV free LED based lamp over encapsulated or unencapsulated devices measured under inert atmosphere or inside the glove box. Transparency in reporting the stability of PSCs is needed in the Perovskite PV community.[38] Apart from the scalability and stability, photo-physics behind the perovskite materials also surprised every day. Few of them are; double emission peak of perovskite materials at low temperature,[39] contrast behaviour of enhanced steady state PL intensity *vs* decreased lifetime at low temperature with respect to room temperature in perovskite nano-crystals,[40] origin of non-radiative recombination: photon recycling *vs* defect mediated recombination,[41] role of ions *vs* ferroelectricity, etc.[42] Perovskite semiconductors are considered as full of defects, disorders (static *vs* dynamic), dislocation, distortion, ion migration etc.[43] Ionic motions in the octahedral cage of PbX$_6$, Rashba splitting, slow carrier relaxation, dynamic disorder are the probable reasons behind the exotic physics of this novel class of semiconductor.[44]    Thus, it is necessary to have a better understanding of these imperfections (defects and disorders) in order to further improve the quality of perovskite materials by some molecular or solvent additive engineering and thus to understand the device physics of the PSCs. However, complete removal of toxic atom (Pb) from the perovskite is necessary to commercialise the PSCs. Replacement of Pb by Sn open a new door in this direction but there is still requirement of a novel engineering approach to fabricate Pb free PSC without compromising on its efficiency and stability.[45]

## 1.4    Outline of the thesis

This thesis is mainly focused on developing an understanding about the physics of organic – inorganic hybrid metal halide based perovskite based thin films and solar cells. Chapter **2** demonstrate the device physics of solar cells which includes generation, recombination & transportation process and charge collection in good details. Chapter **3** describes the device fabrication procedure and various experimental techniques that help us to support our experimental results and final outcome. Chapter **4** deals with the temperature dependent absorption and emission spectroscopy in order to understand the dynamic *vs* static disorder present in the perovskite semiconductor. Here, we also reveal the mystery behind the positive temperature coefficient of band gap in perovskite semiconductor, which is in contrast to conventional semiconductors such as Si, GaAs etc.[43] Along with that, this chapter also provide a novel method for tuning the band gap of a particular perovskite semiconductor by embedding the perovskite nano-crystal in a low dielectric constant organic matrix.[46] This technique can open new door for many optoelectronic devices such as wavelength tunable light emitting diodes. Chapter **4** helps to understand the defects and disorder in the perovskite semiconductor even at room temperature and advised to apply some molecular or solvent additives engineering in order to reduce the defect states from the bulk as well as from the grain boundaries. Chapter **5** demonstrates the role of solvent additives (1,8 diiodooctane & phosphinic acid) on structural, morphological, optical, chemical and electronic state properties of perovskite semiconductors and relates it to defect passivation in the perovskite thin films which improves the overall PCE of PSCs.[47] Apart from efficiency number, moisture and thermal stability are other big issues in the field of PSCs. Hence, there is need for such kind of additives which can improve the efficiency and stability both at the same time. Chapter **6** investigates the role of small organic molecule, Bathocuproine, on the moisture stability and defect passivation via physisorption in $CH_3NH_3PbI_3$ based perovskite semiconductor.[48] This chapter also deals with the thermal stability of PSCs along with moisture stability by using double hole and electron extraction layers. PEDOT:PSS is mostly known for its acidic & hydrophilic nature and is one of the main reason for degradation at the interfaces in *p-i-n* configuration based PSCs. $MoO_3$ is used as a double hole extraction layer along with PEDOT:PSS to improve the charge extraction as well as thermal stability of the perovskite solar cell.[49] However, it is important to understand the mechanism behind defect passivation by molecular or solvent additive engineering, which facilitate the charge transport and thus improve the PCE of PSCs. Chapter **7** demonstrate a novel technique of scanning photocurrent microscopy (SPM) which helps to estimate the lateral transport length scale of hole and electron, separately, in a bipolar device and supports the enhancement in short circuit current density and fill factor of PSCs by solvent additive engineering.[50] Pb based perovskites are emerging rapidly in the field of PV due to its unique opto-electronic properties. But, toxicity of Pb is another challenge for commercialization of perovskite solar cells. Sn emerges as an alternative of Pb for B cation ($ABX_3$) in perovskite solar cells. However, oxidation, efficiency and stability are the main concern with Sn. Chapter **8** presents a comparative study on influence of B (Pb and/or Sn) cations on charge carrier recombination dynamics in Pb free based PSC ($MA_{0.20}FA_{0.75}Cs_{0.05}SnI_3$), pure Pb based PSC ($FA_{0.95}Cs_{0.05}PbI_3$) and Pb-Sn mixed PSC $\{(MAPbI_3)_{0.4}\,(FASnI_3)_{0.6}\}$ with the help of transient photovoltage (TPV) technique. Finally, chapter **9** concludes all the research work included in the thesis and it also demonstrates the future plan related to the thesis.

# Chapter 2

# Physics of Solar Cell and Literature Survey

# CHAPTER 2

# Physics of Solar Cell and Literature Survey

## 2.1 Introduction

Chapter **1** describe the motivation of this thesis to understand the defect physics in the PSCs and the concept of additive engineering to reduce it along with understanding of charge transport properties. The working principle and charge transport properties of perovskite solar cells will be discuss in chapter **2**. The literature review of imperfection in perovskite and passivation techniques used to reduce those imperfections will be also discussed in this chapter. A solar cell is an electronic device that converts light into electricity by photovoltaic effect.[1] The electrical characteristics such as voltage, current and resistance of a solar cell change, when it is exposed to the light. The combination of individual solar cells can form module and commonly it is known as solar panels.

## 2.2 Working principle

The basic structure of a solar cell is P-N junction as shown in figure **1**. In general, when light is incident on the solar cells, photons with energy equal to or higher than the band gap of the semiconductor material gets absorbed and then the electrons and holes are generated inside the semiconductor material either via direct inter band transition or via dissociation of exciton at room temperature. The photo-generated electrons and holes get collected by their respective electrodes by drift and diffusion process.[2] The working principle of solar cells depends upon the following three things:

(a) Absorption & generation process: For an efficient solar cell, the generation rate of electron and holes should be as high as possible. The photo-generated electrons and holes are collected to the electrodes which lead to provide photocurrent in the solar cells.[3]

(b) Recombination process: The lifetime of photo-generated electrons in the excited state is of the order of few nanoseconds and after that electrons get annihilated with holes via recombination process. The rate of recombination inside the semiconductor decides the amount of charge carriers collected at the electrodes. Hence, it is important to understand the recombination mechanism which can help to minimize the non-radiative recombination rate in order to get higher $V_{OC}$ in the solar cells.[4]

(c) Charge transport process: The free charge carriers which are left after recombination process get collected at the electrodes and are responsible for the photocurrent produced in the solar cells. The process of charge transport to the electrode takes place via drift and/or diffusion processes. The carrier concentration gradient decides the diffusion process, however, drift based charge transport is driven by the electric field inside the material. Since the solar cells are operating at maximum power point (MPP), where the electric field is smaller than that at the short circuit position. Hence, the charge transport process in solar cell is mainly dominated by diffusion process. Therefore, for better performance of the solar cell, semiconductor material should have higher diffusion length of charge carriers.2[5]

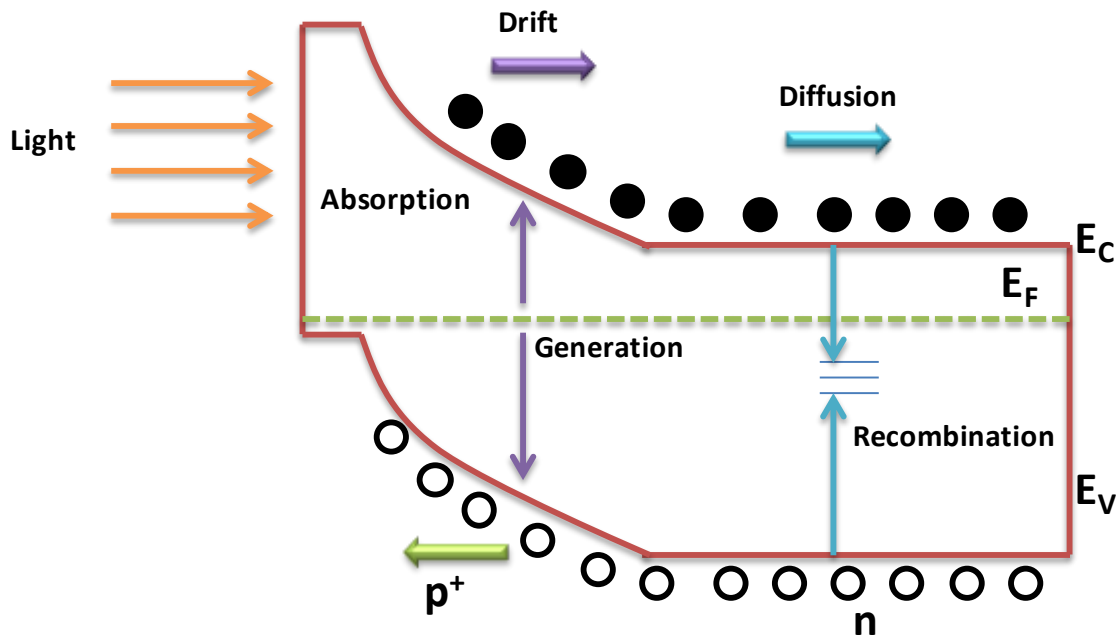

**Figure 1:** *Typical band diagram of $p^+n$ junction solar cell. Charge carrier generation, recombination and transport in solar cells.*

### 2.2.1   Absorption/Generation

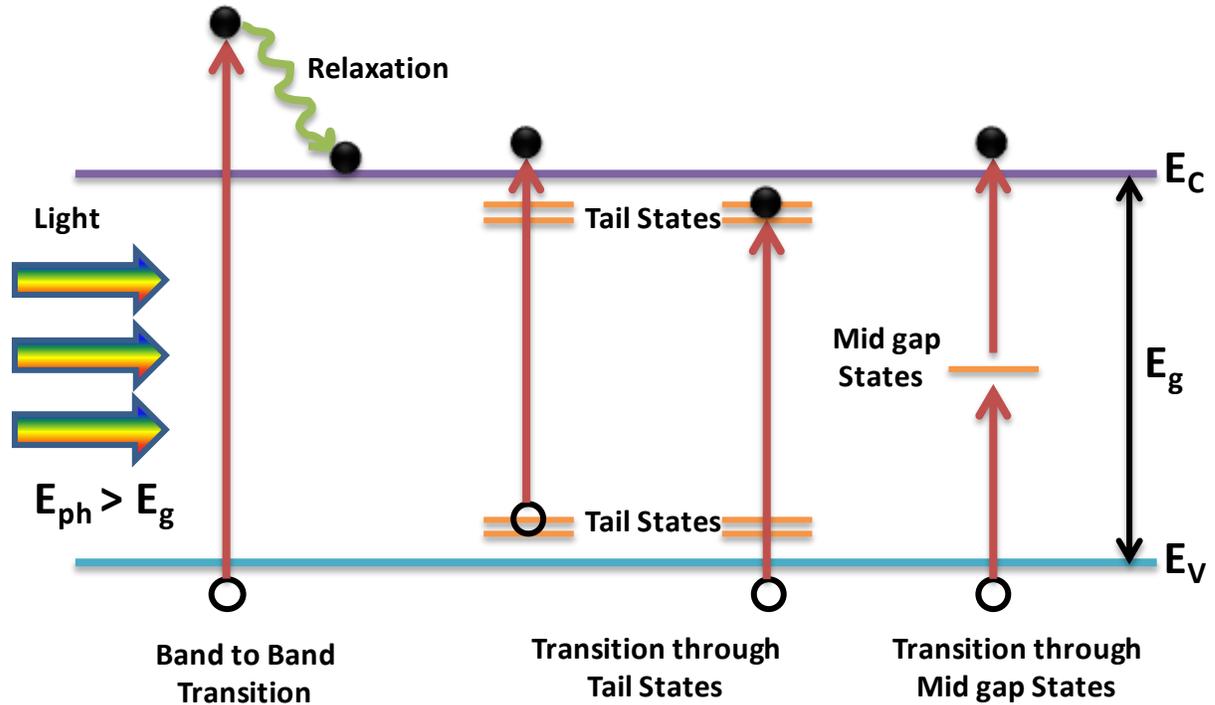

**Figure 2:** *Absorption process through band to band transition, tail states and mid gap states transition.*

In order to produce photo-current, the generation of free holes and electrons are essential. The absorption of photon can take place via three processes: (i) band to band transition (which is generally refereed as intrinsic optical transition), (ii) transition through tail states and (iii) mid states (extrinsic absorption). The contribution of extrinsic absorption in producing the photovoltaic current is almost negligible due to many order lower absorption coefficient than that of intrinsic absorption. However, the extrinsic absorption via tail states and mid gap states influence the optical and charge transport properties of the semiconductor. The absorption processes through band to band, tail states and mid gap states are shown in figure **2**. When a photon incident on a semiconductor material with energy higher than the band gap. Then, it can liberate one electron from the valance band to the conduction band. The excess energy i.e. $hv$-$E_g$ dissipated as thermal loss in the lattice. The excited electrons collide with the lattice (electron-phonon interaction) and relax to the minima of the conduction band. This phenomenon is attributed to band to band transition in semiconductors. When a photon having energy less than the band gap of the material, absorption process can take place through the tail states, deep or localized states. This kind of optical transitions helps in understanding of band structure of the semiconductors.[6]

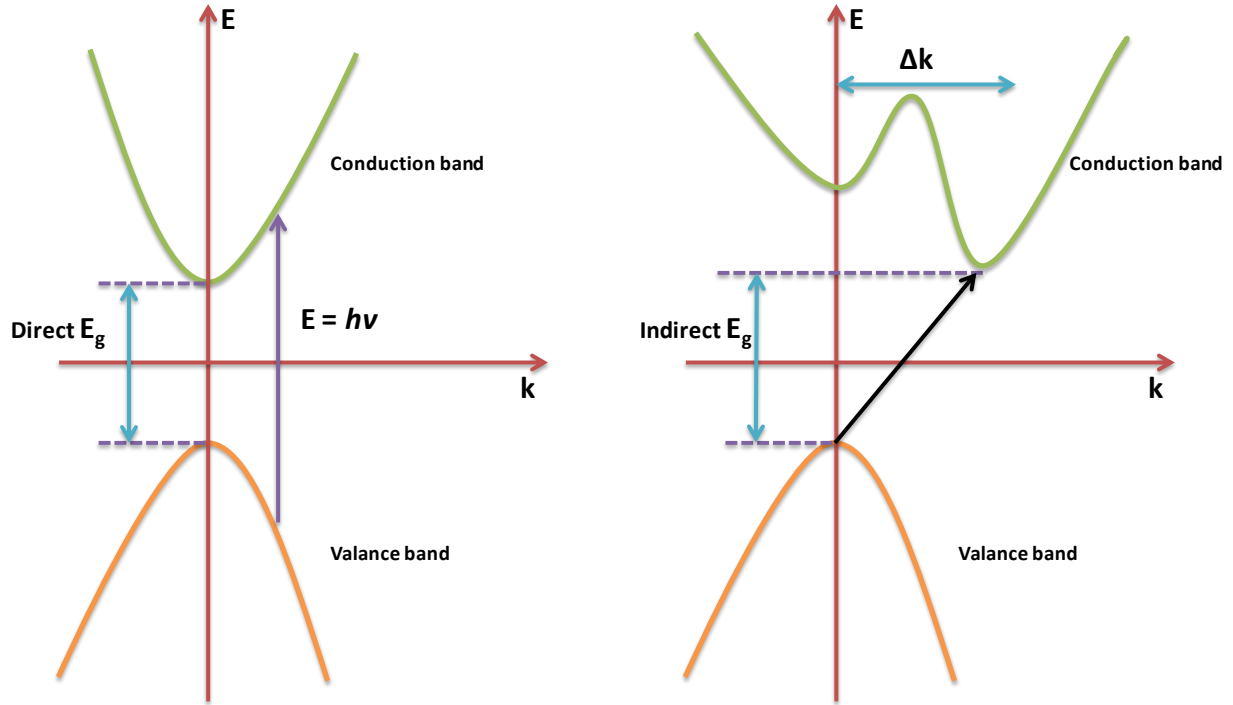

*Figure 3: Band diagram of typical direct and indirect optical transition.*

Band to band transitions are basically of two types: direct and indirect band transition and is shown in figure **3**. A direct transition is momentum conserving transition in which only photon is involved. However, in indirect transition, the momentum is changed for state transition with involvement of phonon.

For direct transition, the absorption coefficient (α) of a semiconductor can be expressed as[7]

$$\alpha = A(h\nu - E_g)^{\frac{1}{2}} \qquad (1)$$

Where, A is a constant and correlated with mass of electrons and holes, $\nu$ is frequency and $h$ is the Planck's constant.

For indirect transition, the absorption coefficient depends upon the phonon absorption and phonon emission process;

If phonon absorption takes place with energy $E_P$, the absorption coefficient will be[8]

$$\alpha = A \frac{(h\nu - E_g + E_P)^2}{e^{\frac{E_P}{K_B T}} - 1} \qquad (2)$$

However, if the transition involves emission of phonons

$$\alpha = A \frac{(h\nu - E_g - E_P)^2}{1 - e^{\frac{-E_P}{K_B T}}} \qquad (3)$$

Absorption coefficient of direct and indirect band gap semiconductors is shown in figure **4**.[9] Sub band gap transition do not contribute significantly on the absorption coefficient but it provide useful information about the distribution of sub band gap states such as Urbach energy ($E_U$), which is a measure of disorder in the semiconductor.[10]

$$\alpha\ (E) = \alpha_0 \exp\left[\left(h\nu - E_g\right)/E_u\right] \qquad (4)$$

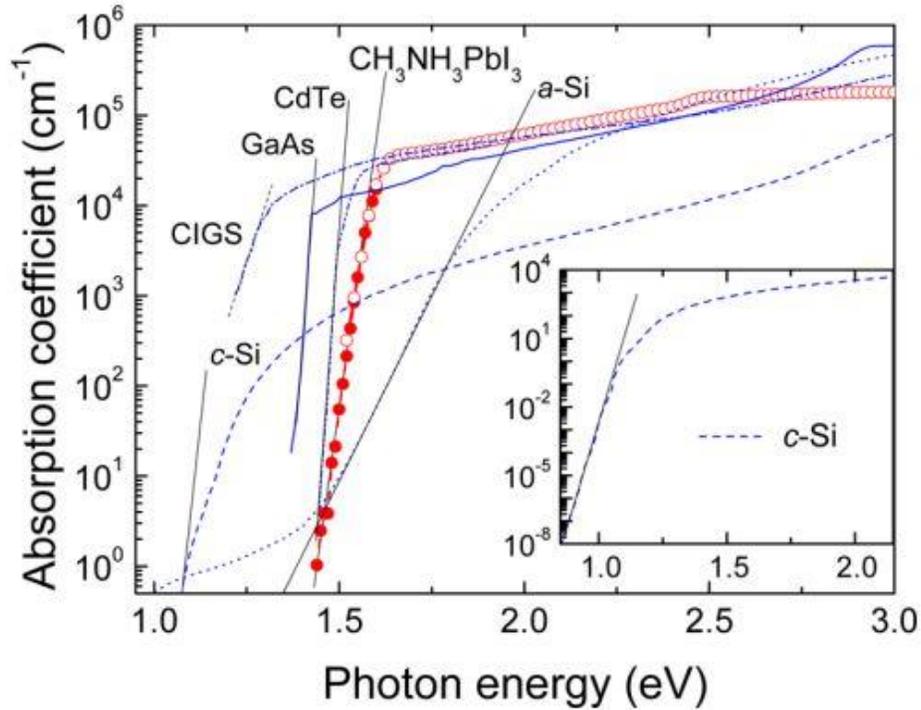

**Figure 4:** *Absorption coefficient of different direct and in direct band gap photovoltaic materials.*[9]

### 2.2.2   Recombination

Once the free electrons and holes are generated in the bulk of the semiconductor through optical absorption process, the free charge carriers tries to collect at their respective electrodes before recombination of electrons and holes. The recombination process decreases the carrier density from the equilibrium concentration. The electrons into the conduction band will recombine with the holes into the valance band either by radiative or non-radiative recombination. The typical kind of recombination processes in the semiconductor material is shown in figure **5**. The band-to-band recombination (radiative recombination) leads to emission of photons whose energy is equal to the band gap of the semiconductor material. However, non-radiative recombination (trap assisted recombination) process involves phonons. In the case of Auger recombination, the energy is transferred to another electron in the form of kinetic energy. We will discuss each kind of recombination in details below:

**(a) Band-to-band recombination**

The band-to-band recombination occurs only in direct band gap semiconductors, where an electron in the conduction band falls directly into an empty state which is associated with hole in the valance band and recombination of free electron and hole occurs. This is also known as bimolecular recombination.[11] It depends upon the electron and hole density in the conduction and valance band, respectively. Hence, the recombination rate is proportional to the available density of electrons (n) and holes (p). As there is no net generation or recombination in the thermal equilibrium condition, the rate of recombination must be equal to the rate of generation. Thus, $. p = n_i^2$ , where $n_i$ is the intrinsic charge carrier density in thermal equilibrium condition. Hence, the rate of recombination can be expressed as[12,13]

$$R_{b-b} = B_r(n.p - n_i^2) \qquad (5)$$

Here, $B_r$ is the band-to-band recombination coefficient.

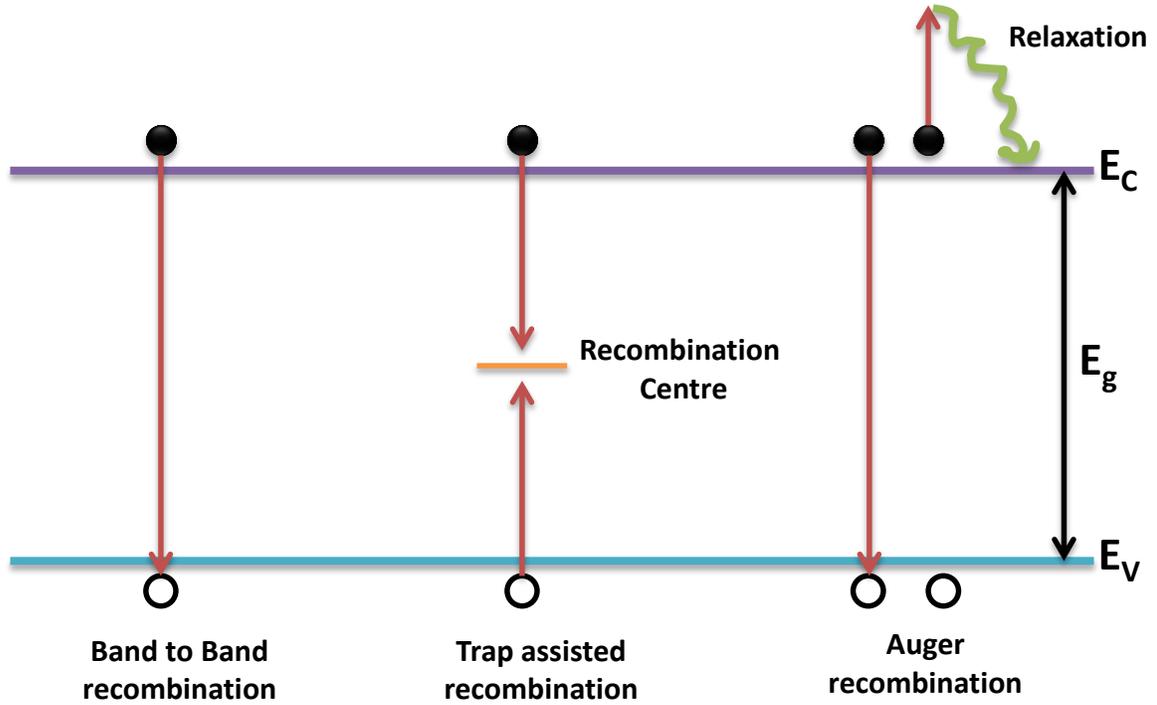

*Figure 5: Recombination processes in the semiconductor: band to band, trap assisted and Auger recombination.*

For an n-type semiconductor (n >> p and n>>$n_i$) in which doping level is low i.e. $\Delta n$, $\Delta p$ << n , the recombination rate can be written as

$$R_{b-b} = B_r n. \Delta p = \frac{\Delta p}{1/nB_r} = \frac{\Delta p}{\tau_r} \qquad (6)$$

Where, $\tau_r$ is the lifetime of minority charge carriers.

When the doping level is high ($\Delta n$, $\Delta p$ >> $n_0$, $p_0$), Equation 5 can be written as

$$R_{b-b} = B_r \Delta n. \Delta p = B_r n. p = B_r n^2 = B_r p^2 \qquad (7)$$

**(b) Trap assisted recombination**

The trap assisted recombination occurs through a sub band gap state which can be formed due to inclusion of a foreign atom or structural defects. The trap assisted recombination is a two steps recombination process; in first step the electron fall into a trap state in the mid gap state and then it fall into the unoccupied state of valance band to complete the recombination process. The trap assisted recombination is also known as Shockley Read Hall (SRH) recombination.[12] There are four possibilities of SRH generation or recombination process:[14]

1. An electron can be captured into a trap state with energy $E_T$ and lies in between the conduction band ($E_C$) and valance band ($E_V$). It leads to transfer the excess energy $E_C$-$E_T$ to the crystal lattice in the form of phonon emission. The process is call electron capture.

2. Then trapped electron can fall into the valance band and recombine with the hole or the hole is captured by the electron occupied in the trap state. This process also leads to emission of phonon with energy $E_T$-$E_V$ and is called hole capture.

3. An electron can be captured in the trap state from the valance band when a photon with energy lower than the band gap of the semiconductor is incident. This process creates a hole in the valance band. It means a hole is moved from the unoccupied trap state to the valance band. The energy required for completion of this process is $E_T$-$E_V$ and the process is called hole emission.

4. An electron trapped into the trap state can move to the conduction band and the process is called electron emission. The energy required for completion of this process is $E_C$-$E_T$.

Process 3 and 4 are for SRH generation of electron hole pairs which requires energy from the lattice. The recombination rate for SRH recombination can be written as

$$R_{SRH} = \frac{n.p - n_i^2}{\tau_p(n + n') + \tau_n(p + p')} \qquad (8)$$

Where,

$$n' = n_i \exp\{(E_T - E_i)/kT\}$$

$$p' = p_i \exp\{(-E_T - E_i)/kT\}$$

$$\tau_n = \frac{1}{B_n N_T} \quad and \quad \tau_p = \frac{1}{B_p N_T}$$

$\tau_n$ and $\tau_p$ are the lifetime of minority charge carriers of electrons and holes, respectively. $N_T$ is the defect density; $B_n$ and $B_p$ are the SRH recombination coefficient for minority electrons nd holes, respectively.

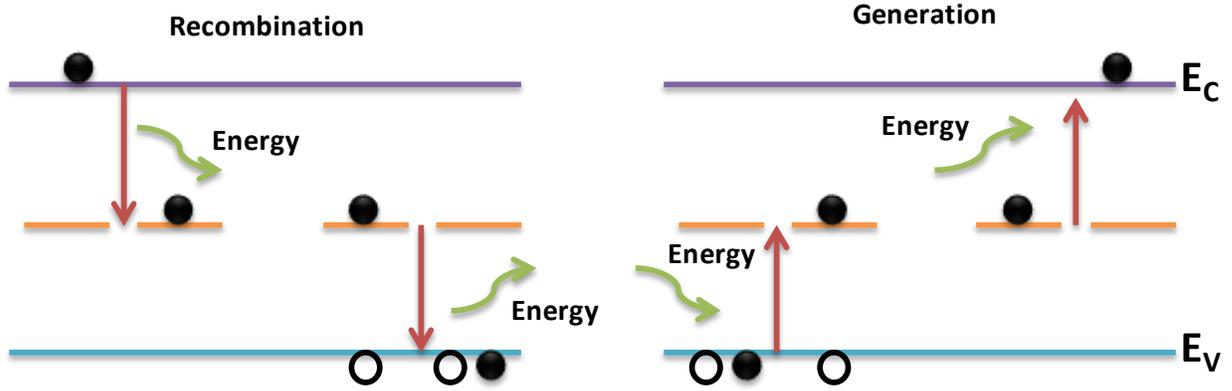

**Figure 6:** *Four possible sub processes in the SRH recombination.*

### (c) Auger recombination

The energy released during the band to band transition or trap assisted transition can transfer to another electron in the conduction band, which finally collide with the crystal lattice and relaxes to the minima of the conduction band. This process is known as Auger recombination.[15] In order to Auger recombination takes place, the semiconductor should be highly doped i.e. the concentration of free charge carrier should be high. The Auger recombination rate can be written as following:

$$R_A = B_A n(np - n_i^2) \qquad (9) \qquad \text{(for n type semiconductor)}$$

$$R_A = B_A p(np - n_i^2) \qquad (10) \qquad \text{(for p-type semiconductor)}$$

The recombination rate equation for Auger recombination is similar to the band-to-band recombination in addition to the density of holes or electrons, which gain energy during the electron hole recombination in band-to-band transition.

In general, all the three recombination (band to band, trap assisted and Auger recombination) occurs in the semiconductor at the same time. Hence, the total lifetime of the minority charge carrier is given by

$$\frac{1}{\tau} = \frac{1}{\tau_r} + \frac{1}{\tau_T} + \frac{1}{\tau_A} \qquad (11)$$

where, $\tau_r$, $\tau_T$ and $\tau_A$ are the lifetime of the minority charge carriers due to band-to-band, trap assisted and auger recombination.

The surface and interface of a semiconductor contains large number of electrically active states due to sudden termination of crystal growth at the surface. These active states act as recombination center at the surface and interface, which will finally affect the performance of the device. However, the recombination rate equation for surface recombination is similar to the trap assisted recombination in the bulk (SRH).

### 2.2.3   Charge transport and collection

The absorption of photons leads to the generation of electron-hole pair in the bulk of the semiconductor which causes an excess minority charge carriers. The final current in the device depends upon the charge recombination and collection mechanism. The charge collection is determined by the contribution of both drift and diffusion based components. The drift based transport is depends upon the electric field in the semiconductor, however, the charge carrier concentration in the semiconductor will define the diffusion based transport.[16] One dimensional charge transport equation due to contribution of both drift and diffusion component can be written as:

$$\frac{\partial n}{\partial t} = \frac{D_n \partial^2 n}{\partial x^2} + \mu_n \mathcal{E} \frac{\partial n}{\partial x} + \mu_n n \frac{\partial \mathcal{E}}{\partial x} - \frac{n - n_0}{\tau_n} + G_n(x) \qquad (12)$$

$$\frac{\partial p}{\partial t} = \frac{D_p \partial^2 p}{\partial x^2} + \mu_p \mathcal{E} \frac{\partial p}{\partial x} + \mu_{pn} p \frac{\partial \mathcal{E}}{\partial x} - \frac{p - p_0}{\tau_p} + G_p(x) \qquad (13)$$

Where, $D_n$ & $D_p$ are the diffusion coefficient of electron and holes, $\partial n$ and $\partial p$ are the excess electron and hole concentration, $\mu_e$ & $\mu_p$ are the electron and hole mobility, $\mathcal{E}$ is the electric field $\tau_n$ & $\tau_p$ are the minority carrier lifetime of electron and hole, respectively. G is the generation rate.

The one dimensional charge transport equations (12 and 13) can be solved by Poisson's equation. In solar cells, rate of generation of holes and electrons are equal in steady state condition. Hence, a single transport equation is sufficient to describe the excess in minority charge carrier density

$$\frac{\partial \Delta n}{\partial t} = D' \frac{\partial^2 n}{\partial x^2} + \mu_n \mathcal{E} \frac{\partial n}{\partial x} + \mu' n \frac{\partial \mathcal{E}}{\partial x} - \frac{\Delta n}{\tau_n} + G \qquad (14)$$

where, $\Delta n = n - n_o$; The diffusion coefficient and mobility can be written as

$$D' = \frac{n \mu_n D_p + p \mu_p D_e}{n \mu_n + p \mu_p} \qquad (15)$$

$$\mu' = \frac{\mu_n \mu_p (p - n)}{n \mu_n + p \mu_p} \qquad (16)$$

From equation 12 and 13, we observed that charge transport in solar cells is dominated by minority charge carriers.

**(a) Diffusion based component (Lateral photo-voltage regime)**

The first term in equation 14 represents the diffusion component of charge transport. Consider a system in which the electric field in the material is zero and there is no charge generation, then equation 14 becomes

$$\frac{\partial \Delta n}{\partial t} = D' \frac{\partial^2 n}{\partial x^2} - \frac{\Delta n}{\tau_n} = 0 \qquad (17)$$

The solution of equation 17 is

$$\Delta n = \Delta n_0 \exp\left(-\frac{x}{L_n}\right) \qquad (18)$$

Where $L_n$ is the diffusion length of minority charge carriers; $L_n = \sqrt{D_n \tau_n}$

From equation 18, it can be seen that the minority charge carrier density decreases exponentially with distance.

**(b) Drift based component (High field regime)**

The second term in equation 14 represents the drift component of charge transport. In general, the charge transport process in semiconductor with high diffusion length of charge carriers is mainly governed by the diffusion process. But, for organic semiconductor the charge carrier diffusion length is small and thus, an electric field is required to improve the charge transport and collection property.

$$\frac{\partial \Delta n}{\partial t} = \mu_n \mathcal{E} \frac{\partial n}{\partial x} + \mu' n \frac{\partial \mathcal{E}}{\partial x} - \frac{\Delta n}{\tau_n} = 0 \qquad (19)$$

The solution of equation (19) is given by

$$\Delta n = \Delta n_0 \exp\left(-\frac{x}{R_n}\right) \qquad\qquad (20)$$

Where, $R_n = \mu_n \tau_n \varepsilon$ is the drift range of minority electrons due to electric field.[17] Equation (20) is similar to the equation (18), the difference is equation (20) represents a drift range $R_n$ instead of diffusion length $L_n$.

**(c) Drift-diffusion based component (Low field regime)**

In Low field regime, both drift based component and diffusion based component exist in the charge transport, a critical electric filed is defined as[18]

$$\mathcal{E}_c = \frac{k_B T}{q L_n} \qquad (21)$$

Which is the electric filed when $L_n = R_n$

So, when the electric filed $\varepsilon > \varepsilon_c$, carrier transport is dominated by drift, and when

$\varepsilon < \varepsilon_c$, carrier transport is dominated by diffusion.

Drift component in charge transportation is very important for organic based solar cells due to their smaller diffusion length in comparison to the Si based solar cells. Hence, the active layer is sandwiched between the p and n type layer and the device configuration is either *p-i-n* or *n-i-p*. In *p-i-n* device configuration, the direction of electric field is from *n* to *p*. Hence, the photo-generated electrons and holes are swept towards n and p type layer before recombining with each other, respectively.

## 2.3 Shockley-Queisser (SQ) limit

In 1961, William Shockley and Hans Queisser calculated the maximum theoretical power conversion efficiency of a solar cell is 33.7 % with a band gap of 1.34 eV using an AM 1.5G solar spectrum with an assumption that loss mechanism is only due to radiative recombination.[19] The modeling is done by assuming the temperature of sun and the device as 6000 K and 300 K, respectively. The maximum efficiency of a single junction solar cell was modeled as a function of band gap and is shown in figure **7a**. When the band gap is below 1.34 eV, the most of the energy losses is due to relaxation process and when the band gap is higher than the 1.34 eV, the losses are attributed to absorption loss.

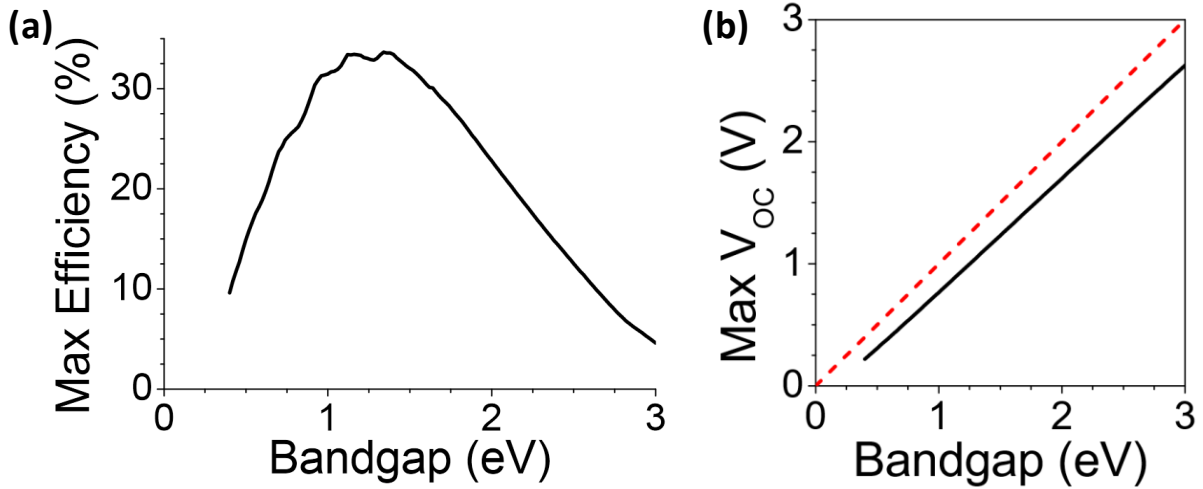

***Figure 7:*** *Plot between (a) Maximum theoretical efficiency vs band gap and (b) maximum V_OC vs band gap of a single junction solar cell.*[20]

The primary considerations in Shockley Queisser limit are following

(a) Blackbody radiation: Any material which is kept above the zero Kelvin, it will emit electromagnetic radiations. This effect is known as black body radiation. However, the solar cell kept at room temperature (300 K), it will emit 7% of the energy that is incident on the solar cells. Any energy which is not contributing in absorption will provide heat to the solar cells. Hence, the temperature of the solar cells increases and thus, black body radiation effect also increases until equilibrium is reached. For Si solar cell, the equilibrium temperature is ~360 K. Hence, the efficiency of a solar cell is always less than its rating at room temperature.

(b) Recombination: Recombination process again limits the efficiency by approximately 10%. The effect of recombination is directly affecting the $V_{OC}$ of the solar cell and it is shown in figure 6b. the red dotted line represent the $V_{OC}$ of the solar cell when recombination losses are absent i.e. $V_{OC}$ is equal to the band gap of the semiconductor. Solid black line represents the $V_{OC}$ limited by recombination process and it is always less than the band gap.

(c) Spectrum loss: The band gap of Si is 1.1 eV which means that any solar photon having energy higher than 1.1 eV leads to generation of electron hole pairs. In other words, we can say that up to absorption process takes place in Si up to 1100 nm wavelength. Red, yellow, green and blue photons are responsible for charge generation. However, most of the infra red, micro waves and radio waves are the losses in this case and they cannot contribute to charge generation which limits the short circuit current density and is shown in figure **8a**. Figure **8b** represents the plot of different loss mechanism vs band gap of single junction solar cell.

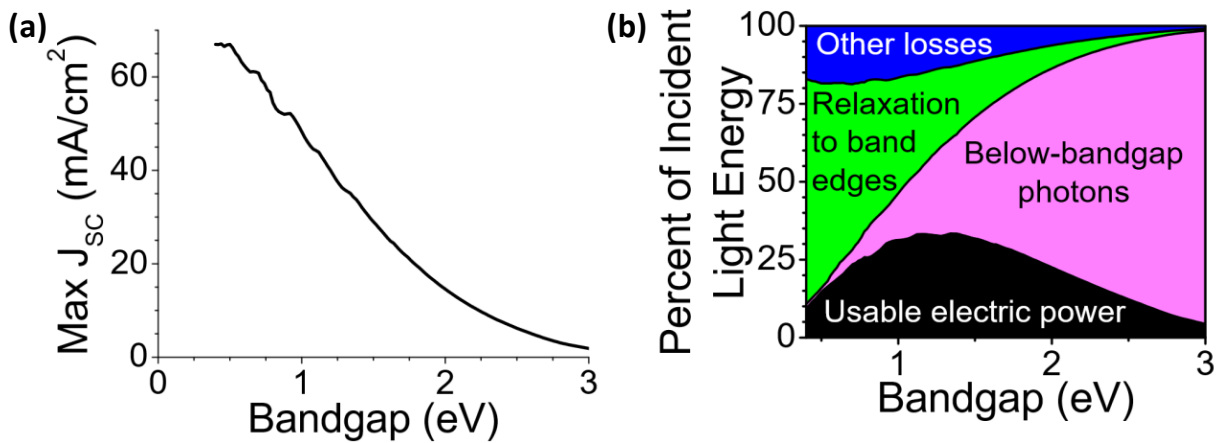

**Figure 8:** *Plot between (a) Maximum JSC vs band gap of a single junction solar cell. (b) Different loss mechanism vs band gap of single junction solar cells.*[20]

### 2.4    *I-V* characteristics

The single diode equation is

$$I = -I_L + I_0 \left[ exp\left(\frac{qV}{\eta kT}\right) - 1 \right] \qquad (20)$$

Where, $I_L$ is the current generated due to illumination of light, $I_0$ is the dark saturation current, η is the ideality factor, q is elementary charge and T is the temperature of the cell.

Under illumination, the diode equation becomes

$$I = -I_L + I_0 exp\left(\frac{qV}{\eta kT}\right) \qquad (21)$$

The term '-1' in the equation 20 can be neglected because exponential term is usually much larger than 1 except at lower voltages (below 100 mV). Along with that, at low voltage, $I_L$ is dominating over $I_0$, and thus equation 21 is valid under illumination. Typical *I-V* curve for a solar cell is shown in figure **9**. Here, the maximum power point ($P_{mp}$ = V X I) is used to operate the solar cell in order to provide maximum output power.

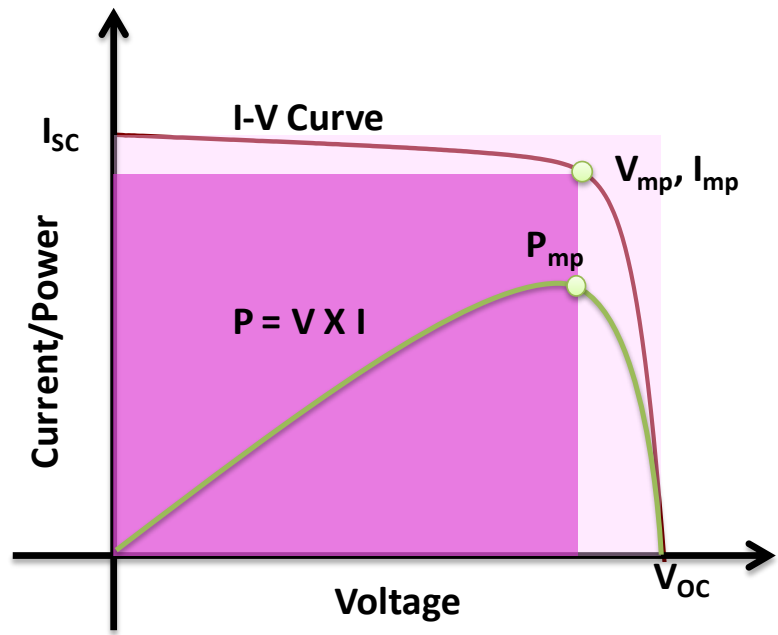

**Figure 9:** *A typical current-voltage (I-V) curve for a solar cell under illumination.*

The open circuit voltage ($V_{OC}$) is defined as the maximum voltage a solar cell can acquire at zero current. At $V = V_{OC}$, equation (2) becomes

$$V_{OC} = \frac{\eta KT}{q} ln\left(\frac{I_L}{I_0}\right) \qquad (22)$$

The short circuit current ($I_{SC}$) is the amount of current passing through the cell at zero applied bias. It is dependent upon the generation and collection of charge carriers with in the solar cell. JSC is the maximum current that a solar cell can draw. $I_{SC}$ depends upon the area of the cell, number of absorbed photons, optical properties and collection properties. The ISC and VOC are the maximum voltage and current in the solar cell, but, the power at both the points is zero. The fill factor (FF) determines the maximum power from a solar cell in conjunction with ISC and VOC. The FF is defined as

$$FF = \frac{P_{mp}}{V_{OC} \, X \, I_{SC}} \qquad (23)$$

The power conversion efficiency of a solar cell can be expressed as following

$$PCE = \frac{P_{max}}{P_{irradiation}} = \frac{\max(V \, X \, I)}{0.1 \, W/cm^2} = \frac{V_{OC} X \, I_{SC} \, X \, FF}{0.1 \, W/cm^2} \qquad (24)$$

The series and shunt resistances are the two important parameters which affect the performance of the solar cell. Series resistance ($R_S$) depends upon the contact resistance between the active layer and the inter-layers along with sheet resistance of top and bottom contacts. The main effect of $R_S$ is on the FF and if the value of $R_S$ is quite high, it may also affect the current as well. However, shunt resistance ($R_{Sh}$) are typically decided by the manufacturing defects rather than the poor design of the solar cell. The main impact of $R_{Sh}$ is on the current of the solar cell. For better performance of the solar cell, $R_{Sh}$ should be high and $R_S$ should be low. The $R_S$ and $R_{Sh}$ can be extracted from the illuminated I-V curve.

$$R_S = \frac{\partial V}{\partial I} \, at \, V = V_{OC} \quad \& \quad R_{Sh} = \frac{\partial V}{\partial I} \, at \, V = 0$$

The $R_S$ and $R_{Sh}$ can be also calculated from the dark *I-V* curve and it is more accurate than that extracted from the illuminated *I-V* curve because illuminated *I-V* curve can be affected through several factors such as slight fluctuation in the light intensity, low diffusion length and poor collection properties etc. The diode equation in the presence of $R_S$ and $R_{Sh}$ can be written as[21]

$$I = -I_L + I_0 \left[ exp\left( \frac{V - IR_S}{\eta \frac{KT}{q}} \right) - 1 \right] + \frac{V - IR_S}{R_{Sh}} \qquad (25)$$

At low voltage, the current is flowing mainly through shunt path and the dark current is dominated by shunt current (region **1** in the figure **10**). The third term in the equation 25 represents the shunt current and this portion is used to extract the $R_{Sh}$ by calculating the inverse of the slope of region 1. When the applied bias increases, the current increases exponentially in region 2. This represents the recombination current and is shown by the exponential term in the equation 25 and is used to calculate the $\eta$. Further increase in the applied bias, the current will be limited by the series resistance and it becomes constant after a certain voltage (region 3). The inverse of slope of region 3 will provide the value of $R_S$.

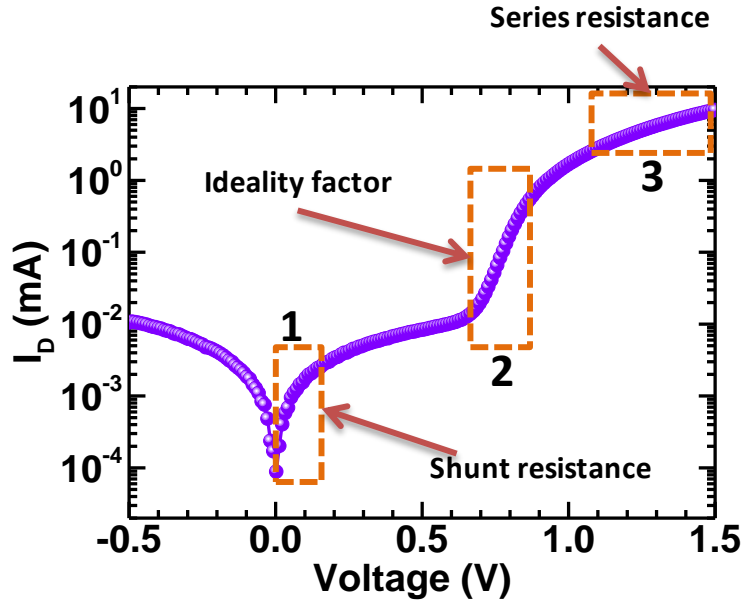

***Figure 10:*** *Typical dark I-V curve of a perovskite solar cell to extract $R_S$, $R_{Sh}$ and $\eta$.*

## 2.5    Defects in perovskite thin films

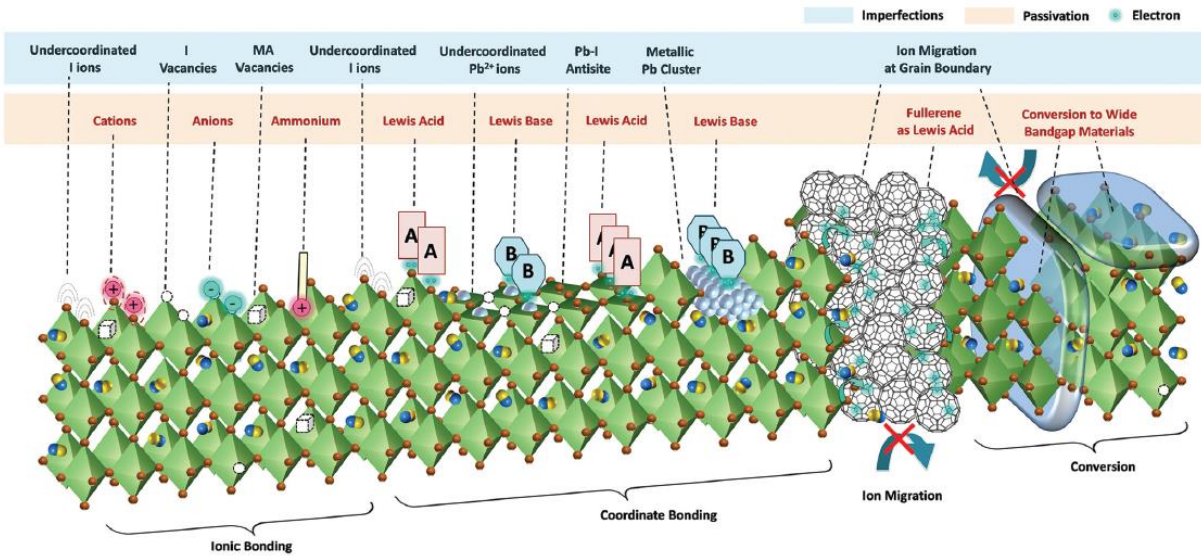

**Figure 11:** *Imperfection in organic inorganic lead halide perovskite films and the ways to passivate the imperfection.*[22]

In the starting, methyl ammonium lead triiodide ($CH_3NH_3PbI_3$) which is commonly known as $MAPbI_3$ is the most common perovskite material used as an active layer in perovskite solar cells (PSCs). But it is highly sensitive to the moisture and degradation occurs very rapidly. In addition, the movement of ionic ions are responsible for the hysteresis in the device and it again leads to degradation of PSC.[23] However, in recent years, there are developments in the processing techniques to form compact and smooth perovskite films, understanding of photo-physics of perovskite semiconductor, and especially the defect passivation of the perovskite material which is associated with the grain boundaries and within grains.[24] The low temperature solution processable technique for fabrication of perovskite films creates impurities and defects in the bulk, at the grain boundaries and on the surface of the perovskite film due to low formation energy. These defects acts as recombination centres and thus it affects the charge transport and opto-electronic properties of perovskite film, which is ultimately detrimental to the performance of PSC.[25] The types of possible imperfections in the perovskite thin films during fabrication process are shown in figure **11**. Now, we will discuss various ways for passivation of such defects in the next section.

## 2.6    Defect passivation in perovskite thin films

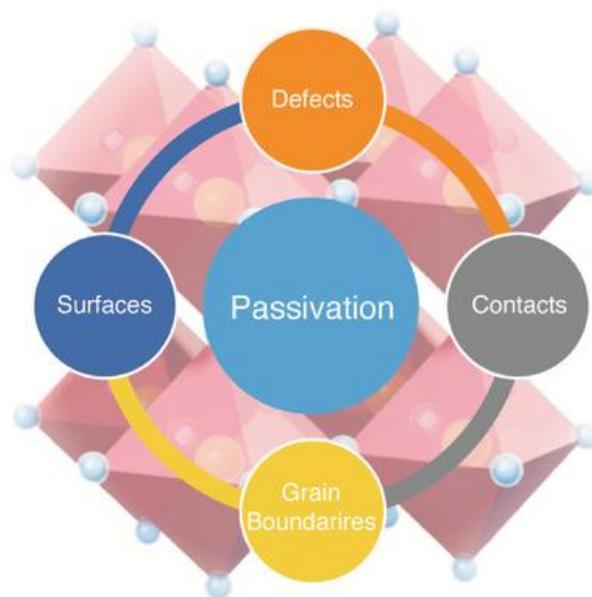

***Figure 12:*** *Defect and contact passivation for perovskite solar cells.*[26]

The imbalance in the stoichiometry of perovskite components during the fabrication process is responsible for formation of such defects (figure **11**). The uncoordinated lead ($Pb^{2+}$) or iodine ($I^-$) acts as Lewis acid and base, respectively. Thus they undergoes reduction and oxidation processes. Reduction of $Pb^{2+}$ leads to the formation of $Pb^0$ which is metallic lead and it is widely explored as a recombination centre in the perovskite community via X-ray photoelectron spectroscopy (XPS). In a similar manner, $I^-$ undergoes oxidation process and form $I_2$ molecule, which is volatile in nature. The release of $I_2$ molecule results into decomposition of perovskite. Most of the uncoordinated $Pb^{2+}$ and $I^-$ are on the surface of the perovskite films which again leads to create interfacial defects in PSCs. In literature, there are several efforts to reduce the defect states and are discussed below:

### 2.6.1   Bulk defect passivation

It is widely known that the point defects such as ion vacancies, antisite and interstitial site substitution are the possible reasons for the bulk defects in the perovskite film. The trap states associated with such kind of point defects are usually close to the conduction band minima or valance band maxima and they can trap the charge carriers which ultimately affect the device performance of PSCs. In order to passivate point defects with shallow energy levels, doping of specific ions in a small proportion to the perovskite helps quite a lot. To passivate the I interstitial and Pb-I antisite substitution, positively charged metal cations are used. $Cs^+$ ion is widely used metal cation in the perovskite community due to its reasonable size. However, addition of small amount of alkali metal iodide such as KI, NAI, RbI, LiI or CsI has been explored in past few years. Doping of KI results into reduced degree of hysteresis in *J-V* curve in comparison to the other alkali metal iodide due to the lower formation energy of K-site in the $(FAPbI_3)_{0.875}(CsPbBr_3)_{0.125}$ based perovskite (figure **13**).[27] Later on, it is reported that $K^+$ can passivate the uncoordinated halides ($I^-$ and/or $Br^-$) via accumulating at the surface and the grain boundaries of the perovskite film (figure **14**).[28] There are reports on the divalent and trivalent dopant such as $Ni^{2+}$, $Eu^{3+}$, $Y^{3+}$, and $Fe^{3+}$ in the perovskite precursor solution to passivate the bulk defects.[29] Wang et al. shown that $Ni^{2+}$ suppresses the formation Pb interstitial and Pb-I antisite substitution via distribution among MA cations or $[PbI_6]^{4-}$ octahedral. Among all the three trivalent cations, $Eu^{3+}$ doping results into better performance and stability of PSCs. $Eu^{3+}$ acts as a 'redox shuttle' which oxidized Pb interstitial and reduced I interstitial.

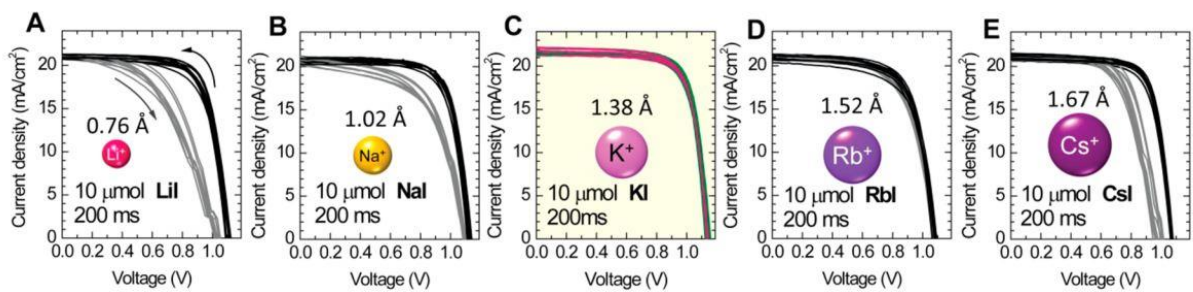

***Figure 13:*** *J-V characteristics of perovskite solar cells with doping of different alkali metal iodide. KI doped PSC shows reduced degree of hysteresis in comparison of LiI, NaI, RbI and CsI doped PSCs.*[27]

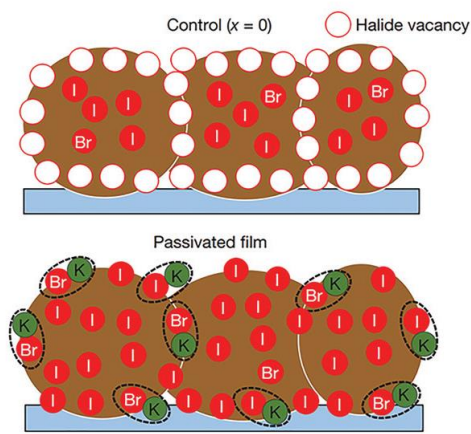

***Figure 14:*** *Schematic representation of halide vacancy in the perovskite film which is passivated by coordination with K atom at the grain boundaries and perovskite surface.*[28]

Similarly, anions are used to passivate the Pb interstitial and halide vacancy. Chlorine ion doping via $PbCl_2$, MACl and FACl facilitate the crystal growth of perovskite along with passivation of grain boundaries. During thermal annealing, Cl volatilized and its residual passivate the grains and surface of perovskite. Kim et al. Used MACl in $FAPbI_3$ based PSC and achieved PCE of 23.48%.[30] Another report by Kim et al. revealed that electrons and holes separation is facilitated by $Br^-$ doping.[31] Being the most electronegative atom among all the halides, $F^-$ have tendency to make strong hydrogen bond with organic cation and robust ionic bond with Pb and shows a strong surface passivation via accumulation.[32] Other anions such as $SCN^-$ and ionic liquids ($BMIMBF_4$) are also used to passivate the point defects in perovskite.[33,34]

### 2.6.2 Surface and grain boundaries defect passivation

The residual of $PbI_2$ in the perovskite film after annealing can be identified by looking the scanning electron microscopy (SEM) images (figure **15a**), where perovskite domains are surround by white $PbI_2$ phase. Since the band gap of $PbI_2$ (2.3 eV) is higher than that of perovskite (1.6 eV), the conduction and valance band of $PbI_2$ is at higher and lower energy than that of perovskite, respectively. Hence, the charge transport through perovskite grains is highly affected by the surrounded $PbI_2$ rich grains. By native passivation of $PbI_2$, You et al. achieved PCE beyond the 21%.[35] Alkyl ammonium halogen salts are widely used as both dopant and by layer engineering technique to passivate the surface and grain boundaries.

Guanidinium (Gua[+]),[36,37] *n*-butylammonium (BA[+]),[38] ethylammonium (EA[+]),[39] and isobutylammonium (iso-BA[+]) are the alkaneamine functionalities which is used to provide a low dimensional layer on the surface of the 3 dimensional (3D) perovskite to passivate the surface defects and improve the moisture stability. Alharbi et al. reported that addition of EA[+] into the perovskite precursor will modify the perovskite surface by forming a mixed EA/FA phase and it forms a 1D passivation layer (figure **15c**).[36]

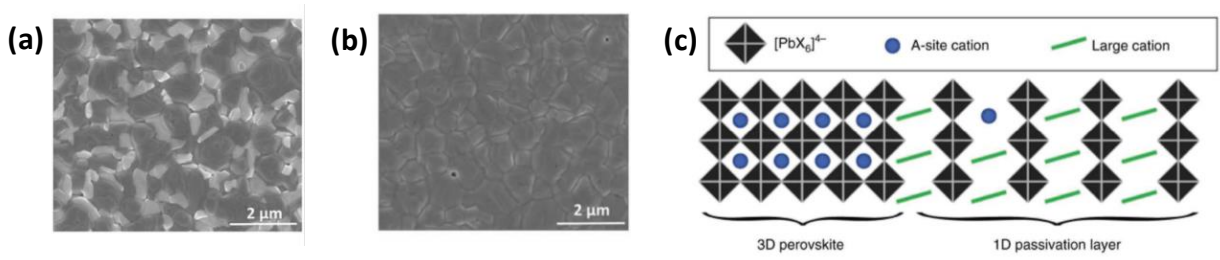

***Figure 15:*** *SEM images of perovskite films with different PbI2/FAI combinations, (a) observable white phase of PbI2 formed on the perovskite surface and (b) no PbI2 was observed on the surface.[35] Schematic representation of the 1D/3D heterostructure evidenced by solid-state NMR proximity measurements.[36]*

Hu et al. reported that alkylammonium halogen group such as PEAI (phenethylammmonium iodide) can form perovskite bilayer which improves the charge transport by hindering the electron transport and reducing the recombination at the interface.[40] Additionally, You et al. pointed out that PEAI can passivate the perovskite surface without annealing from photoluminescence measurement.[41] 4-fluroaniline (FAL) was used by Zhao et al. as a passivation layer by vapor post treatment method to facilitate the charge transport and defect passivation.[42]

### 2.6.3   Defect passivation by Lewis acid and base

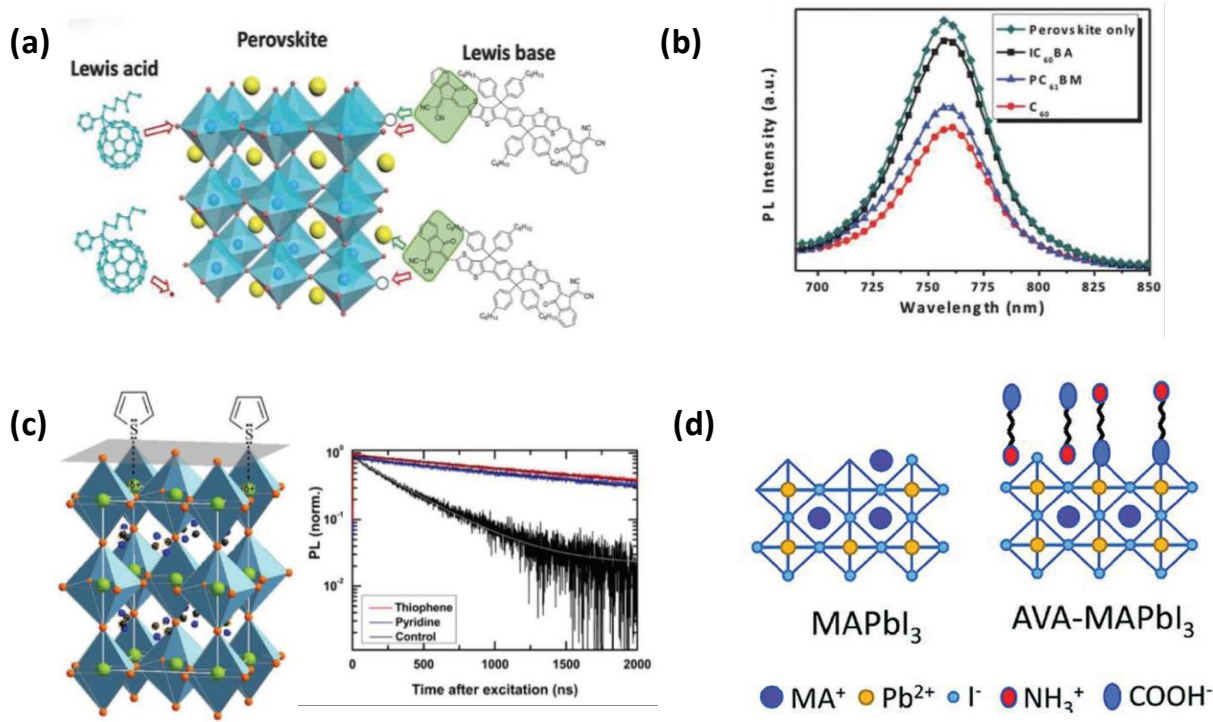

***Figure 16:*** *(a) Interaction of Lewis acid and base with the perovskite crystal.[43] (b) Steady state PL spectra of perovskite film with different fullerene based quenchers.[44] (c) Passivation of defect states and enhanced time resolved PL life time of perovskite film via surface treatment of pyridine and thiophene group.[45] (d) Schematic diagram of defect passivation through AVA at the lattice termination point of MAPbI3.[48]*

A Lewis acid is a molecule or ion that accepts an electron pair however, a Lewis base donate a pair of electrons. A Lewis acid can passivate defects which have free electrons such as Pb-I antisite or free iodide ions. On the other hand, a Lewis base can passivate electron deficient defects such as $Pb^{2+}$ interstitial (figure **16a**).[43] Such molecules can form acid –base complex without electron transfer and are able to eliminate deep level defects associated with anion or cations. Fullerene ($C_{60}$) and its derivatives are the commonly used Lewis acids for passivation. $C_{60}$, PCBM and ICBA are investigated by Ling et al. to compare their passivation effect and it is found that $C_{60}$ shows better passivation effect than other two fullerene derivatives. The superiority of $C_{60}$ is shown through improved charge transport via steady state PL measurement (figure **16b**).[44] Nitrogen, oxygen, sulphur and phosphorous atoms carry the lone pair of electrons and molecules having such atoms can be used as Lewis bases for passivation. Pyridine and thiophene derivatives are examples of Lewis bases. Time resolved PL lifetime shows an enhancement by thiophene and pyridine treatment on the perovskite surface (figure **16c**) and this enhance the efficiency.[45] Trioctylphosphine oxide (TOPO) is used by deQuilettes et al. for passivation.[46] Zheng et al. proposed a novel idea of addition of Lewis acid and base together to form zwitterions molecule which can passivate both positive and negative charge defects.[47] The molecule with both carboxyl and amine group will have multiple electron pairs and have better ability to passivate the defects. Aminovaleric acid (AVA) is an example which proves its ability to passivate the defects and enhance the efficiency (figure **16d**).[48]

### 2.6.4   Hydrophobic materials for passivation

Moisture stability is a big challenge in commercialization of the PSCs. Perovskite degraded very rapidly in presence of $H_2O$ and $O_2$, hence there is need of some hydrophobic materials which can resist moisture degradation as well as passivate the defect and promote the charge transport. There are various polymer materials such as polyvinylpurrolidone (PVP),[49] poly (methyl methacrylate) (PMMA),[50] and poly (ethylene-co-vinyl acetate) (EVA)[51] prevent perovskite from $H_2O$ molecule due their macromolecular structure. A range of insulating materials (ultrathin layer) such as Teflon, fluoro-silane etc. are also used for passivation effect and selective carrier tunneling.[52] The surface contact angle of perovskite increased tremendously with aid of these hydrophobic materials.[53] Organic small molecules such as pentafluoro-phenylethylammonium (FEA),[54] 2,2',2''-(1,3,5-Benzinetriyl)-tris(1-phenyl-1-H-benzimidazole) (TPBI)[55] and Bathocuproine (BCP)[56]  are also used to passivate the defect states and reduce the non-radiative losses in the perovskite. There are only few reports on small organic molecule additive in the perovskite due to limited solubility of these materials into the host solvent (dimethylsulfoxide, N,N dimethylformamide).

## 2.7    Towards eco-friendly perovskite solar cells and tandem devices

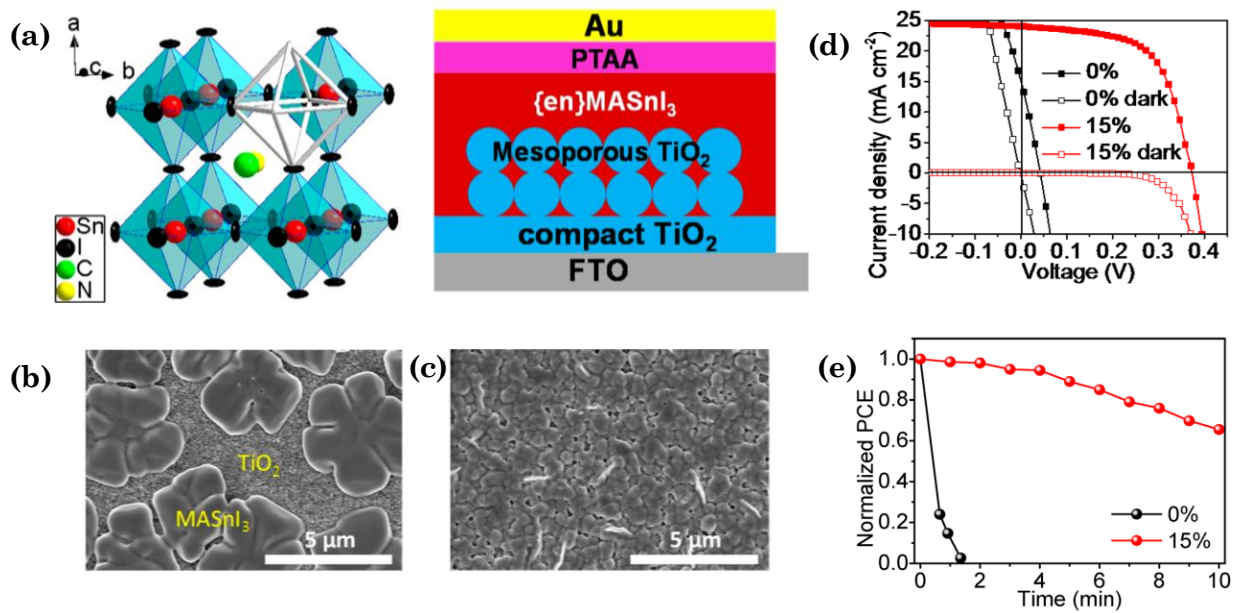

***Figure 17:*** *(a) crystal structure of MASnI₃ and device structure (n-i-p) of Sn based perovskite solar cells. Scanning electron microscopy images of (b) pristine MASnI₃ and (c) with (en) cation additive MASnI₃ based perovskite film. (d) J-V curves and (e) thermal stability of pristine and (en) additive MASnI₃ based PSCs.[58]*

In present time, Pb based PSCs stands out in photovoltaic community due to its outperformance than other conventional solar cells. However, commercialization of Pb based PSCs is still a challenge due to two major issues: toxicity and stability. In this section, we will discuss about efforts to reduce the toxicity of PSCs by replacing Pb with tin (Sn) either completely or partially. However, there are some other non-toxic compounds exists as well which can replace Pb from the perovskite structure such as Sn-Ga based halides, Bi-Sb based halide perovskites etc. Among all, Sn based perovskites are emerging fast due to their interesting optical and electronic properties which is in good agreement with Pb based perovskites. Hao et al. reported 6% PCE of the first Pb free or Sn based perovskite solar cell in 2014.[57] After that there are so many reports which focus on enhancing the PCE and stability of Sn based PSCs. Ethylenediammonium (*en*) cation is used in Sn based PSC to enhance the PCE and stability by forming a new kind of 3D hollow perovskite (figure 17).[58,59] MASnI₃ suffers from poor morphology, so the mixture of FA and MA provides improved morphology and thus enhanced efficiency.[60] However, thermal stability is very poor in these Sn based PSCs. Other cations such as butylammonium (BA) and phenylethylammonium (PEA) are used in 3D Sn based perovskite to reduce the dimensionality in order to fabricate more stable PSCs.[61] The highest PCE for Sn based PSCs has been reported to achieve 9.6% by Jokar *et al.* in which 20% guanidinium (GA) cation is used along with 1% *en* cation in Sn based perovskite.[62] However, after so many efforts, achieving the stability and high performance simultaneously for Sn based PSCs is still a big challenge for the community. Hence, Pb-Sn mixed based perovskite open a new door to reduce the toxicity of Pb based PSCs and enhance the stability of Sn based PSCs. Recently; Prasanna *et al.* demonstrate a new design for Pb-Sn mixed based PSC by improving the morphology of the perovskite film and capping with sputtered indium zinc oxide (IZO) electrode to achieve high stability.[63] Lin et al. reported a novel method of introducing metallic Sn into the perovskite precursor solution to reduce the Sn vacancy in Pb-Sn mixed based PSC (figure **18**).[64] Further they used Pb-Sn mixed low band gap perovskite along with wide band gap material to fabricate all perovskite tandem solar cells. The device structure of all perovskite tandem solar cell is shown in figure **18a**. This study opens a novel approach towards low band gap and less toxic PSC with thermal stability of 463 hours. A PSC as a top cell combined with bottom cell such as Si and CIGS as a tandem design is the talk of the town.[65] Sahli et al. reported a 25.2% efficient fully textured monolithic tandem cell by using a p-i-n perovskite top cell.[66] The company Oxford Photovoltaics demonstrated a tandem cell with 27.3% efficiency in June 2018.[67] However, the current record efficiency stands at 28% as of December 2018 (NREL efficiency chart).

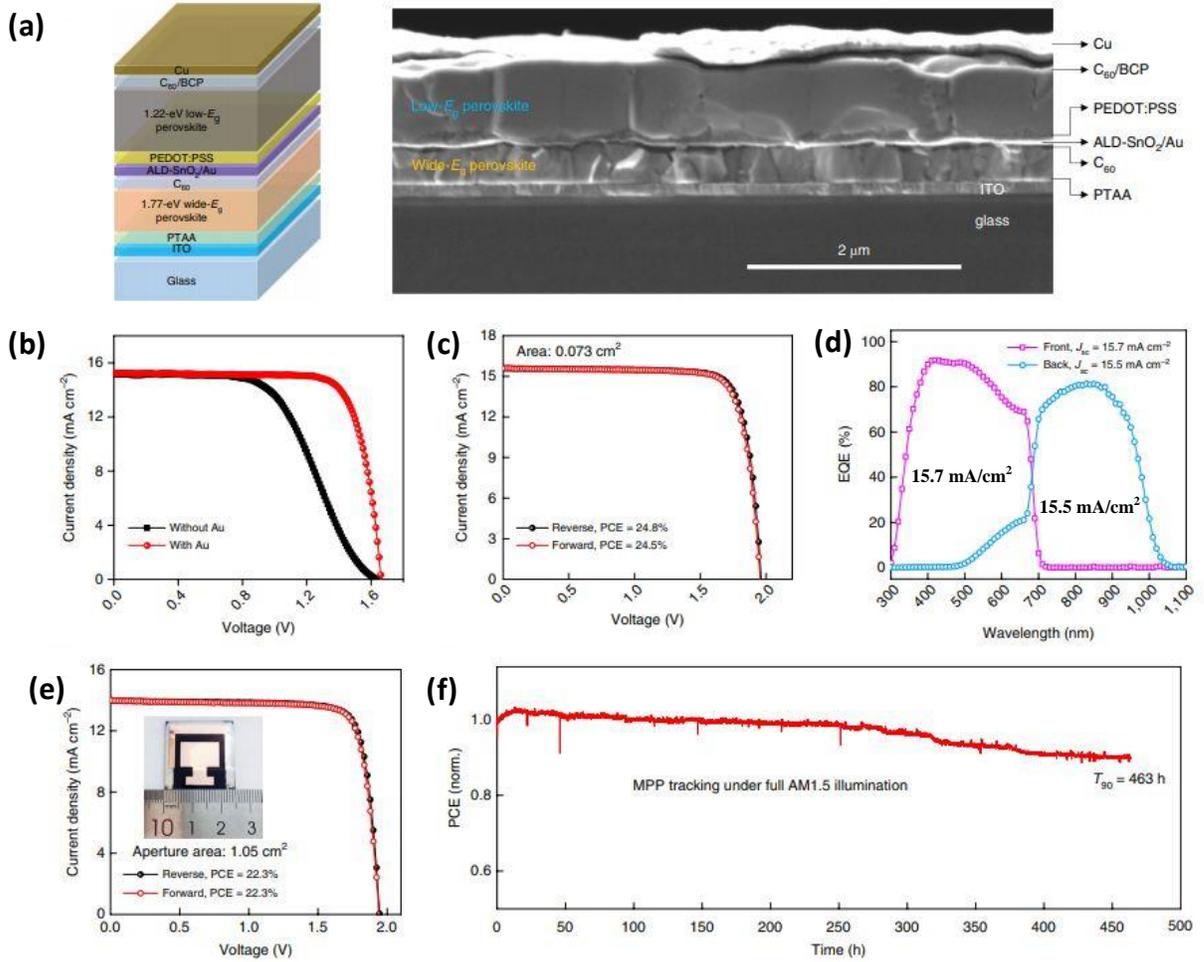

***Figure 18:*** *(a) Device structure and corresponding cross-sectional SEM image of a tandem solar cell. (b) Illuminated J–V curves of tandem solar cells without and with an ultrathin gold (Au) layer. (c) J–V curves and (d) corresponding EQE of the best performing small-area tandem solar cell, respectively. (e) J–V curves of a large-area tandem solar cell. The inset of figure 18(e) shows the digital photo of the large-area device. (f) Maximum power point tracking of non-encapsulated small area tandem solar cell for 463 h in a glovebox under AM1.5 solar illumination (100 mW cm$^{-2}$) without UV filter.*[64]

# Chapter 3

# Methodology

# CHAPTER 3

# Methodology

The characterization techniques employed in this thesis to characterize both the perovskite thin films & devices will be discussed in this chapter. In order to understand the photo-physics and device physics of perovskite semiconductor, we have used different experimental techniques such as steady state photoluminescence, absorption, temperature dependent photoluminescence, time resolved photoluminescence spectroscopy, field emission scanning electron microscopy, atomic force microscopy, X-ray photoelectron spectroscopy, ultra-violet photoelectron spectroscopy, X-ray diffraction, contact angle measurement, dektak profilometer, device fabrication, dark & illuminated current density-voltage (J-V) measurement, external/internal quantum efficiency measurement, steady state electroluminescence, current density - voltage - light (J-V-L) measurement, transient photovoltage and photocurrent measurement and scanning photocurrent microscopy.

## 3.1 Thin film and device fabrication

### 3.1.1   Substrate preparation

Indium tin oxide (ITO) sputtered glass substrates (10 $\Omega$ sq$^{-1}$) with dimension of 50 mm X 50 mm are purchased from Lumtech Ltd. The patterning of the ITO coated substrate for device fabrication is done by following steps:

(a) Clean the 50 mm X 50 mm ITO coated glass substrate with soap solution, deionized (DI) water, acetone and isopropanol for 10 minutes each in ultra-sonicator.

(b) Bake the clean ITO coated glass substrate on the hot plate at 95$^0$ C for 3 minutes for the complete removal of the solvent.

(c) Spin-coat the positive photo-resist (PPR) over cleaned ITO substrate at 500 rpm and 3000 rpm for 10 sec and 30 sec, respectively. The PPR coated substrates are baked at 95$^0$ C for 3 minutes.

(d) Keep the baked PPR coated ITO substrate under the UV curing light (250 W) for 20 sec. The substrates are covered with the lithography mask having four strips of dimension 8 mm X 50 mm and the distance between two strips is 4 mm.

(e) Pour the UV cured substrate into the developer (MF 319) for 30 sec.

(f) Rinse the substrate immediately with DI water after taking out the substrate from the developer.

(g) Dry the substrate with nitrogen gun and bake it on the hot plate at $95^0$ C for 3 minutes.

(h) Prepare an ITO etchant, which consist of 200 mL of DI water, 60 mL of hydrochloric acid (HCl) and 15 ml of nitric acid ($HNO_3$), in a beaker of capacity 500 mL.

(i) Keep the etchant on the hot plate and maintain its temperature ~$60^0$ C. The temperature of the solution is measured by thermometer.

(j) Pour the prepared substrate into the etchant for 2-3 minutes and then pour it into DI water after taking out from the etchant.

(k) Remove the PPR using acetone.

(l) The 50 mm X 50 mm ITO coated glass are further cut into 16 pieces with dimension of 12.5 mm X 12.5 mm. Finally, 12.5 mm X 12.5 mm ITO coated substrate is prepared by patterning the substrate as shown in figure **1**.

(m) The substrates are cleaned subsequently with soap solution, DI water, acetone and isopropanol (IPA) for 10 minutes each in ultra-sonicator.

(n) The cleaned substrates are dried with nitrogen gun and keep on hotplate at $110^0$ C for 10 minutes under laminar air flow to remove the residual solvents.

(o) Finally the substrates are cleaned with oxygen plasma for 10 minutes.

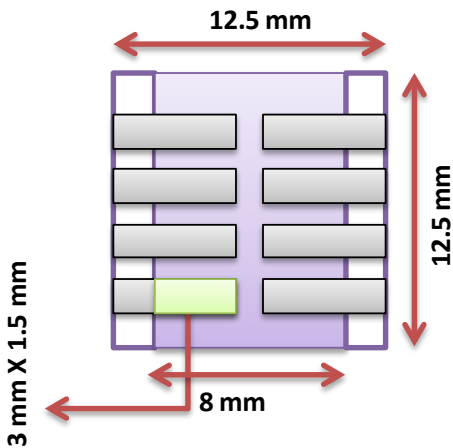

**_Figure 1:_** _Schematic representation of patterned ITO substrate for perovskite solar cell fabrication. The green portion represents the device area which will be decided by the metal evaporation mask._

### 3.1.2 Perovskite solar cell fabrication

The *p-i-n* (inverted) based configuration is used to fabricate the perovskite solar cells (PSCs) in this thesis. A schematic diagram of the *p-i-n* configuration based PSC is shown in figure **2**. The typical structure of *p-i-n* configuration used in this thesis is ITO/PEDOT:PSS/perovskite/PCBM$_{60}$/BCP/Ag.[1,2] In *p-i-n* configuration, the photo-generated electrons and holes are transferred through electron (PCBM$_{60}$ & BCP) and hole (PEDOT:PSS) transporting layers, respectively. Finally, the electrons and holes are collected at Ag and ITO, respectively. For hole transporting layer, PEDOT:PSS (Clevios, PH 4083) is spin-coated over plasma treated ITO substrate and annealed on hot plate under nitrogen atmosphere. Then, PEDOT:PSS spin-coated substrates were transferred into nitrogen filled glove box and perovskite precursor solution was spin-coated over PEDOT:PSS/ITO and annealed on hot plate at $100^0$ C. After that, PC$_{60}$BM was spin-coated at 1000 rpm for 60 sec and then BCP was spin-coated at 5000 rpm for 20 sec. Finally, 100 nm Ag was deposited under a vacuum of $1 \times 10^{-6}$ mbar with a rate of 1 Å/sec. For all studied devices, the device area was defined as 4.5 mm$^2$ by metal shadow mask. The specific details for different perovskite layers such as annealing temperature and time are discussed in each chapter.

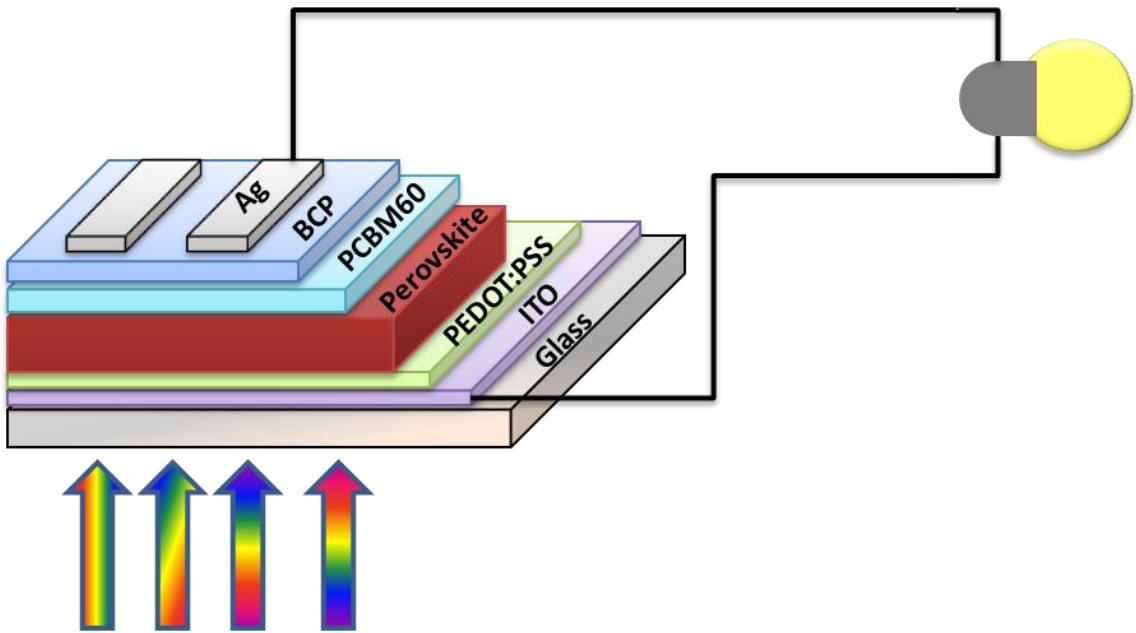

**Figure 2:** *Schematic diagram of p-i-n configuration based perovskite solar cells.*

### 3.1.3   Device encapsulation

After the metal evaporation, the devices are attached to the edge clips to take connection for ITO and Ag (figure **3**). Then the devices are encapsulated with Devcon epoxy resin (ITW 2T) and glass. The encapsulated devices are kept in dark inside glove box for minimum 6 hours for the epoxy to cure. The devices are encapsulated to enhance the moisture stability of PSCs.

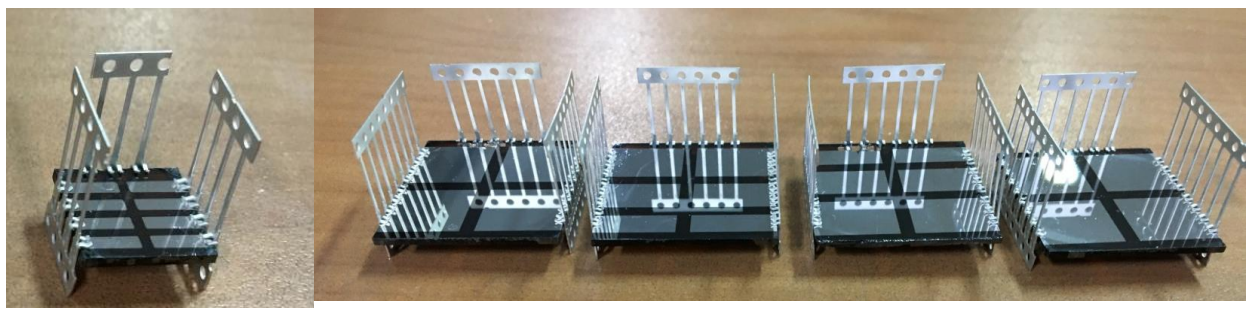

***Figure 3:*** *Images of perovskite solar cells with edge clips to take the connection for ITO and Ag for different active area pixel of devices.*

### 3.2 Thin Film Characterizations

### 3.2.1   Steady state absorption

Perkin Elmer Lambda 950 spectrophotometer with integrated sphere is used to collect the steady state absorption/ transmission spectra of all the solution processed perovskite thin films. It consists of two light sources: deuterium and tungsten halogen lamp. The samples are illuminated from the front side for absorption & transmission spectra and are placed at position '1' in the figure **4**. For reflection spectrum, a reflecting metal have been deposited over the perovskite thin film and are placed at position 2 in the figure **4**. Samples are illuminated from the glass side such that light reflects through the reflecting metal and enters into the integrating sphere. Lambda 950 provides full range of UV-Vis-NIR spectrum (175 nm- 3300 nm). The photo-detector used in the Lambda 950 are InGaS & PbS and are placed at the bottom of the integrating sphere as shown in the figure **4**.

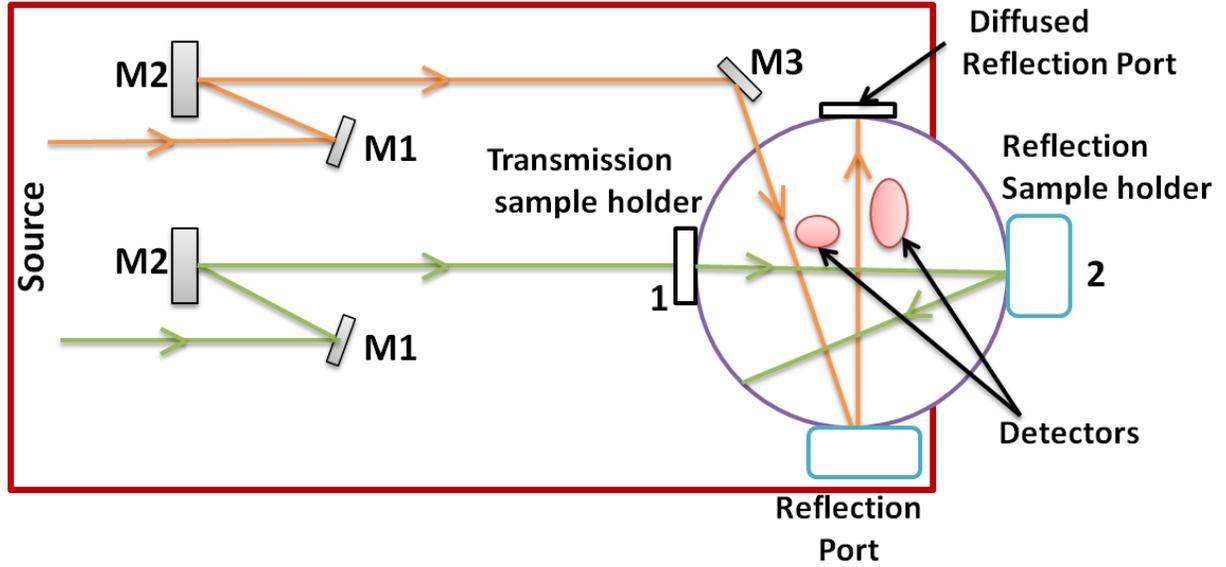

**Figure 4:** *Schematic representation of absorption spectroscopy through Perkin Elmer Lambda 950 spectrophotometer.*

The absorbance of the thin films can be expressed as[3]

$$A\,(\lambda) = -log\,\frac{I}{I_0} \qquad (1)$$

where, I and $I_0$ are the intensity of transmitted and incident light, respectively.

The absorption coefficient can be expressed as[4]

$$\alpha\,(\lambda) = -\frac{2.303}{d}\log\frac{I}{I_0} \qquad (2)$$

here, d is the thickness of the perovskite active layer.

The urbach energy ($E_u$), which is a measure of disorder in the bulk of the perovskite semiconductor material, can be calculated from absorption spectrum by the following relation[5]

$$\alpha\,(E) = \alpha_0 \exp\left[\left(E - E_g\right)/E_u\right] \qquad (3)$$

where, E and $E_g$ are the photon energy and band gap of the semiconductor, respectively.

### 3.2.2 Steady state and time resolved photoluminescence

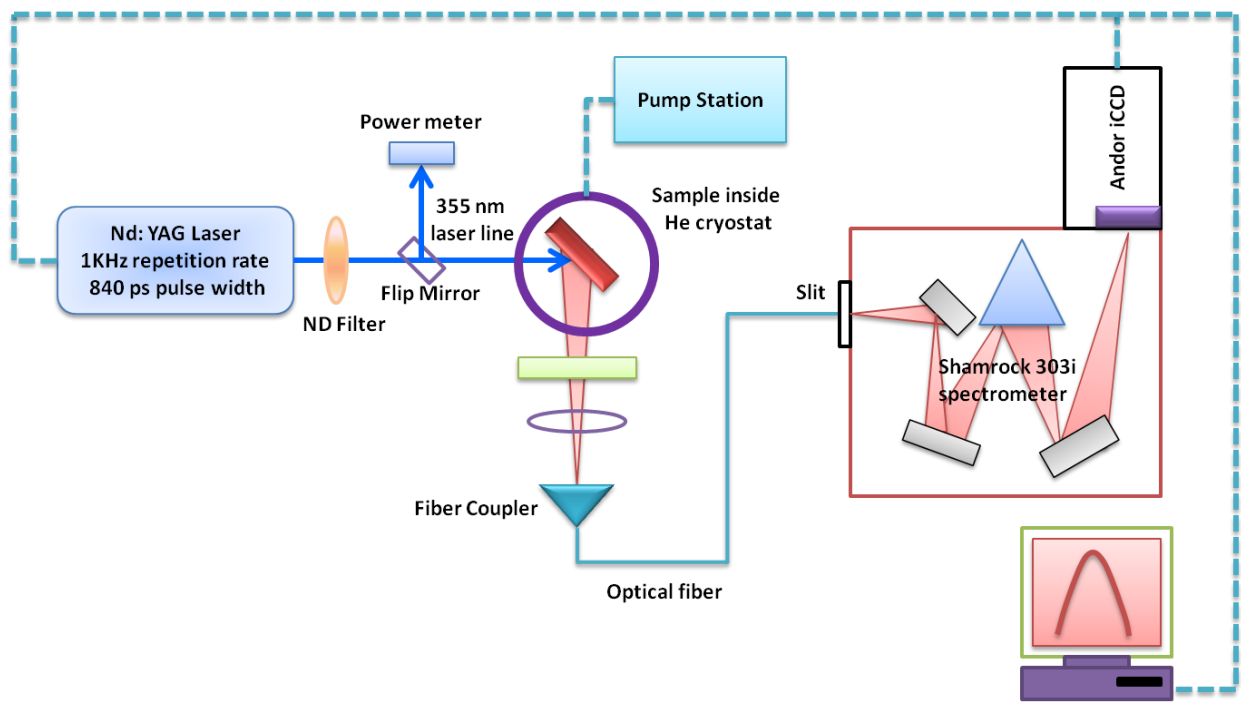

**Figure 5:** *Schematic diagram of steady state and time resolved photoluminescence spectroscopy using a pulse laser and gated i-CCD coupled with spectrometer.*

In the photoluminescence (PL) experiment, photons coming from a light source (laser) with energy equal to or higher than the band gap of the semiconductor material are absorbed by the material and electrons & holes are generated in the valance band & conduction band, respectively. The absorption of incident light takes place on the time scale of femtoseconds. After excitation, the electrons undergo a fast relaxation process and reach the conduction band minima within the time scale of picoseconds. Then the transition of an electron from the minima of conduction band to valence band maxima gives reemission of the photon which is characteristics to the material band-gap. The time scale of the reemission of photons varies from picoseconds to few nanoseconds. . Steady-state PL (SSPL) measurement is performed on thin-films in vacuum at a pressure of $10^{-5}$ mbar in a custom made chamber. Since SSPL gives information about emission peak and PL intensity. Hence, to understand the excited-state charge carriers decay dynamics in the perovskite semiconductor, time-resolved PL (TRPL) is a useful technique. For TRPL measurement, a gated intensified charge coupled device (i-CCD, Andor iStar) is used for detection with excitation laser (3rd harmonic at 355nm of Nd:YAG solid-state laser with the fundamental wavelength of 1064 nm) of a pulse width of 840 ps pulse with 1 kHz repetition rate. The pulse width of the laser should be lower than the expected lifetime of the charge carriers. The photo-excitation of perovskite thin film leads to generate electrons in the conduction band with a particular concentration depends upon the power of the incident laser. The power of incident laser is measured using the Si based power meter as shown in figure **5**. The maximum energy provided by Nd:YAG laser is 22 μJ and it can be vary with the help of neutral density filters. The electrons in the excited state recombine with the holes in the valance band *via* radiative and non-radiative recombination. The decay process of charge carriers from excited state to ground state is characterized by the lifetime of the material (τ). In order to capture the decay in PL, the gate time of the i-CCD is directly triggered by the external circuit of the laser. Since i-CCD does not have any predefine zero time. The time taken by the optical signal to reach the i-CCD can be taken as zero time. However, it is triggered with electronic signal and the actual input signal is optical so there is generally a delay of ~78ns between gate opening time of i-CCD and optical signal arrival time. Hence we put 78 ns as initial delay time to make it the zero time. The PL emission enters into the Shamrock 303i spectrometer, which is a grating spectrometer with 500 nm blaze and 600 lines/mm ruling. Andor i-CCD is integrated with the spectrometer. The minimum gate time is 780 ps. For temperature-dependent steady state and time resolved PL measurements, He cryostat (Oxford Instruments: Janis CCS-300S/204) is used. The sample is mounted on the cold finger of the cryostat under the vacuum of $10^{-5}$ mbar. The temperature range is 300K to 7K.

### 3.2.3   Field emission scanning electron microscopy (FESEM)

FESEM is widely used to study the surface morphology (crystallite size, grain boundaries), cross section view (compactness and thickness of individual layer of multi-layered device) and composition of material (each constituting atom percentage). FESEM is superior over the optical microscope because of small wavelength of electrons in the range of 0.859-0.037 Å, it gives high resolution up to magnification of 300,000. High energy electron beam (1-30 KeV) generated from electron gun in ultrahigh vacuum chamber ($10^{-9}$ mbar) is used as a source.

These primary electrons are accelerated using anode then different optical lenses are used to converge the beam size. First condenser lens is used to reduce the beam size and then objective lens is used to further confine it into a spot size of around 100nm on the sample. Scanning coils are used to deflect the electron beam over the sample area in zigzag motion to scan the sample. Samples are inserted into main chamber via airlock chamber using magnetic rods to maintain the high vacuum ($10^{-6}$ mbar). When primary electron beam hits the atomic electron of samples then new different kind of electrons (secondary electron, backscattered electron, x-rays etc) are generated. Different detectors are used to detect generated different kind of electrons. Then signal from detectors sent to the computer to see the digital images of samples.[6]

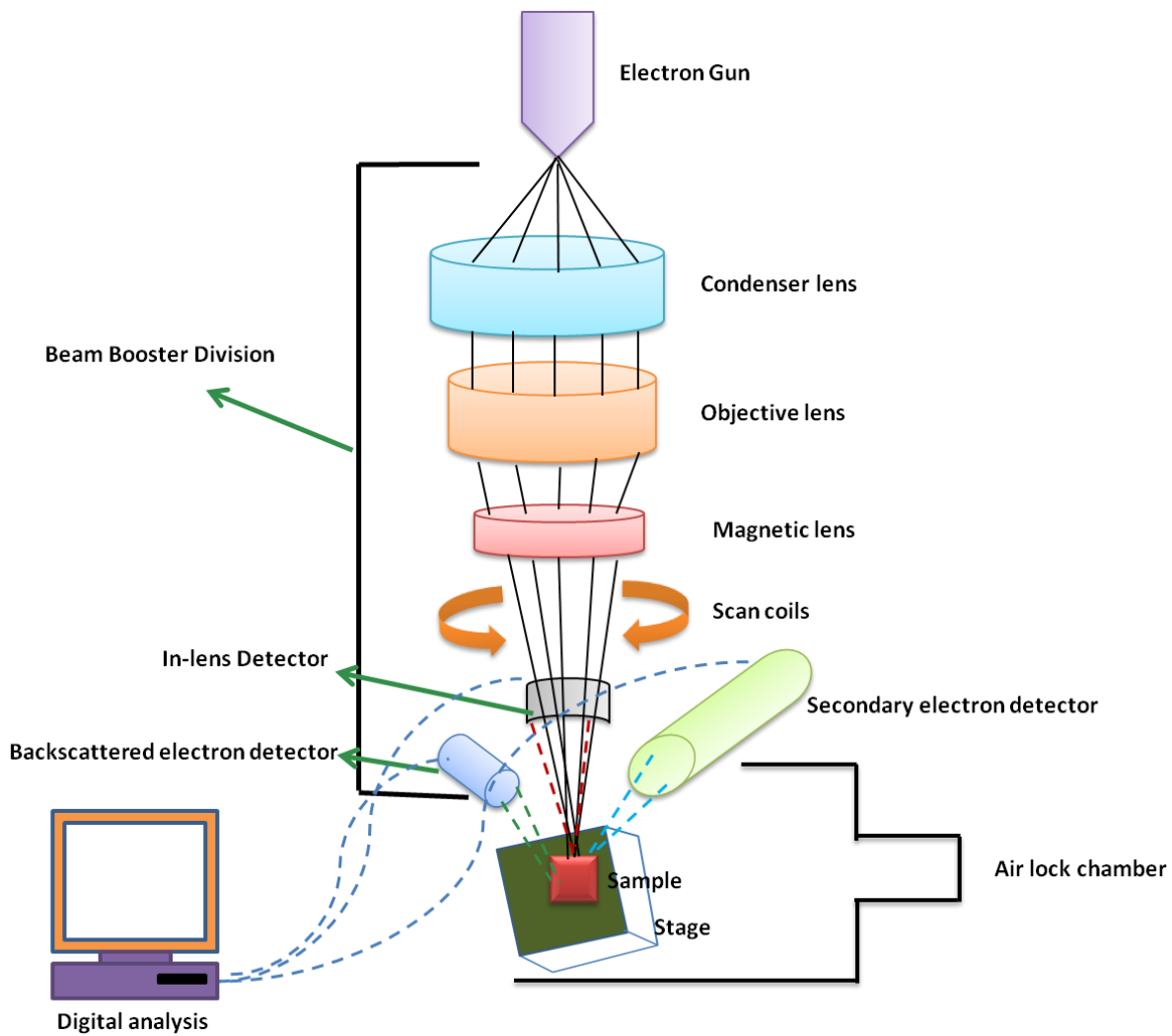

***Figure 6:*** *Schematic diagram of field emission scanning electron microscopy.*

### 3.2.4    Atomic force microscopy (AFM)

AFM characterization technique is used to study the surface topography of the perovskite thin films. AFM provides the information about the surface roughness, grain size, topography or morphology and domain structures of the perovskite films. We have used digital instruments multimode Nanoscope IV AFM instrument to carry out the topographical study of the perovskite films. Principle of AFM is based on beam deflection detection. The AFM consist a cantilever with a tip attached to one end and the diameter of the tip is in few nm. The cantilever with tip is made up of silicon or silicon nitride material. The AFM is used to scan the surface of the perovskite film in either contact mode or tapping mode. The cantilever deflects accordingly the roughness of the perovskite surface. At the same time the reflection of the laser coming from the cantilever also changes its position over the surface of the photo-detector.

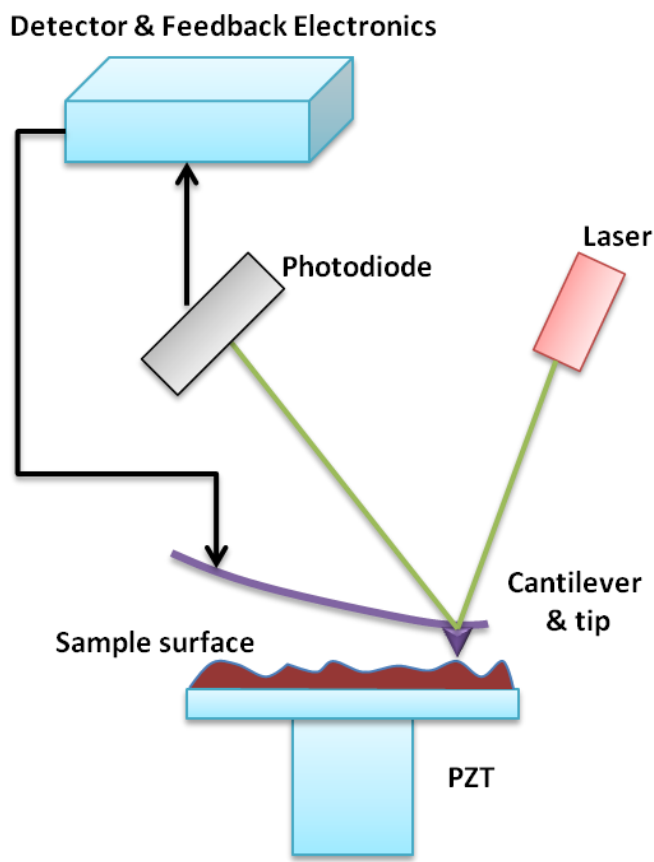

***Figure 7:*** *Schematic diagram of atomic force microscopy (AFM).*

### 3.2.5  Photoelectron spectroscopy

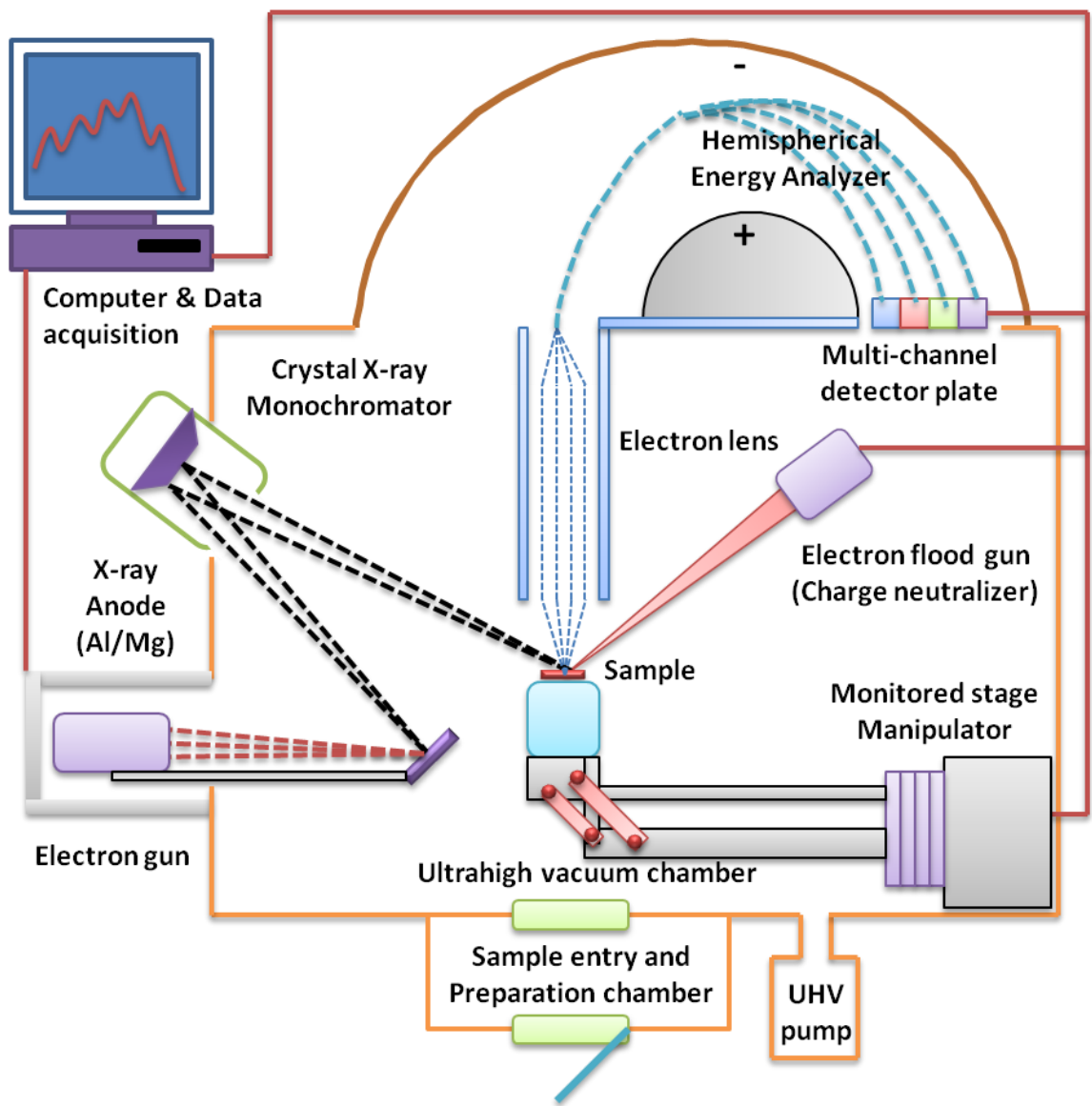

*Figure 8:* Schematic representation of X-ray photoelectron spectroscopy.

The basics of X-ray photoelectron spectroscopy (XPS) is lies in photoelectric effect, for which Einstein got Nobel Prize in 1921 and it was first discovered in 1905. Kai and his co-workers applied this effect in developing the XPS tool to study the elemental composition, chemical state, electronic state and empirical formula of the material. Later on, in 1981, Kai received Noble Prize for XPS in Physics.[7] When a photon is incident on the surface of a material, it gets absorb and emits an electron by transferring its energy to the atomic orbital electron. The emitted electron is called photoelectron and the phenomenon is known as photoelectric effect.

A typical XPS tool consist of an ultra high vacuum pump (turbo molecular pump, which can create vacuum ~$10^{-9}$ Torr), monitored stage manipulator (to locate the position of samples), X-ray Anode (Mg or Al Kα with energy 1253.6 eV or 1486.6 eV, respectively), charge neutralizer (electron flood gun), crystal X-ray monochromator, electron collection lens, electron energy analyzer, detector and data acquisition system. The sample size should be smaller (5 mm X 5 mm) and thickness should be below 5 mm. The prepared samples are transferred into the XPS tool through the flexi-lock chamber (preparation chamber) and then pump started. When the vacuum of the flexi lock chamber becomes comparable to the vacuum of analysis chamber (ultrahigh vacuum chamber), then the door between flexi lock and analysis chamber opened and sample transferred to the analysis chamber through commands given by the software. Stage is calibrated with respect to the X-ray beam in the X-Y direction. When the pressure inside the analysis chamber is ~$10^{9}$ Torr, X-ray anode source turns ON. When X-ray photon is incident on the sample, electron absorbs its energy and liberates one electron from the atomic orbital depending upon the binding energy of that particular emitted electron associated with an atom. At the same time, charge neutralizer is ON during the whole experiment in order to prevent the formation of holes on the surface of the sample, which can create screening effect for other electrons. The emitted electron flow through the electron collection lens in a particular direction (as shown in figure **8**) and enters the hemispherical electron energy analyzer. The incident photon can transfer some part of its energy to liberate the electron from the atomic orbital and rest will provide kinetic energy to the emitted electron. Depending upon the amount of energy contributed for kinetic part for each electron, the path of each electron is different in the hemspherical energy analyzer. The electron analyzer detects the kinetic energy of the emitted electrons. Binding energy of the emitted electrons are calculated by the following relation

$B.E. = h\nu - K.E. - \phi$     (4)

Where, $h\nu$ and $\phi$ are the energy of the incident photon and work function of the XPS spectrometer, respectively. The Fermi level ($E_F$) of the sample and the spectrometer are at the same energy position so that the binding energy of the emitted electron will be referred as the $E_F$. Finally, the electrons are collected by multichannel detector plate and data acquisition is done by the software. Ultra violet photoelectron spectroscopy (UPS) is used to obtained the electronic states of the materials (mostly conductors or semiconductors). For UPS measurement, the sample should be prepared on conducting substrate and the charge neutralizer should be OFF during the measurement. In general, He I source is used in UPS measurement with energy 21.22 eV.[8] A typical full scan XPS and UPS spectra for $CH_3NH_3PbI_3$ based perovskite thin film is shown in figure **9**.

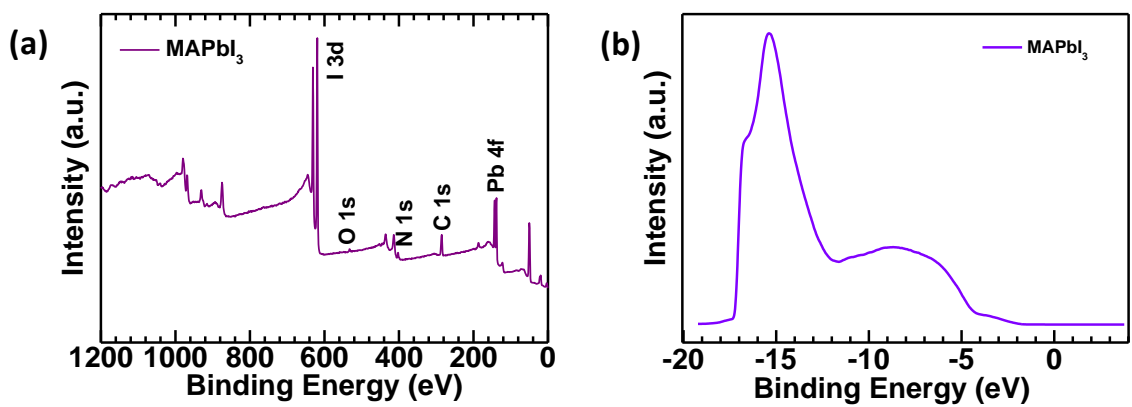

*Figure 9:* *Full scan (a) XPS and (b) UPS spectra of MAPbI₃ based perovskite thin film.*

### 3.2.6 X-ray diffraction (XRD)

The phenomenon of elastic scattering of X-ray photons through the atoms of a periodic lattice is known as X-ray diffraction (XRD). XRD occurs for crystalline materials only, in which the components of the material (such as ion, molecule or atom) are arranged in a periodic order. The X-rays photons diffracted from the atoms of the periodic lattice can interfere either destructively or constructively with each other. However, the detector can read only those X-ray photons which interfere constructively with each other. The schematic representation of XRD is shown in figure **10**. As it is shown in figure **10** that the incident X-ray photons are scattered through atoms of different planes of the crystalline material and the path length is also different for different X-ray photons. The type of interference (either constructive or destructive) depends upon the distance between the two lattices and the path length travelled by two X-ray photons. This constructive interference occurred in XRD by crystal lattice is explained by Bragg's law:[9]

$$n\lambda = 2dSin\theta \quad (5)$$

The directions of the diffracted X-ray photons are highly dependent on the shape and size of the unit cell. Most of the crystals are not single crystal and made up of different small unit cells aligned in different planes. If a polycrystalline material is placed below monochromatic X-rays, the incident photons can interact with all possible orientation of the planes and it can provide a different and specific diffractogramm for that particular material which is a fingerprint of that material. We have used Rigaku lab spectrometer XRD machine using Cu K$\alpha$ radiation ($\lambda$=1.54 Å). XRD pattern is also used to calculate the crystal size (D) of the spherical domains of the perovskite films by using Debye Scherrer formula:[10]

$$D = \frac{k\lambda}{\beta Cos\theta} \quad (6)$$

Where, k, $\lambda$, $\beta$ and $\theta$ are the shape factor (0.9), wavelength of X-rays, broadening at the full width half maxima (FWHM) and Bragg angle, respectively.

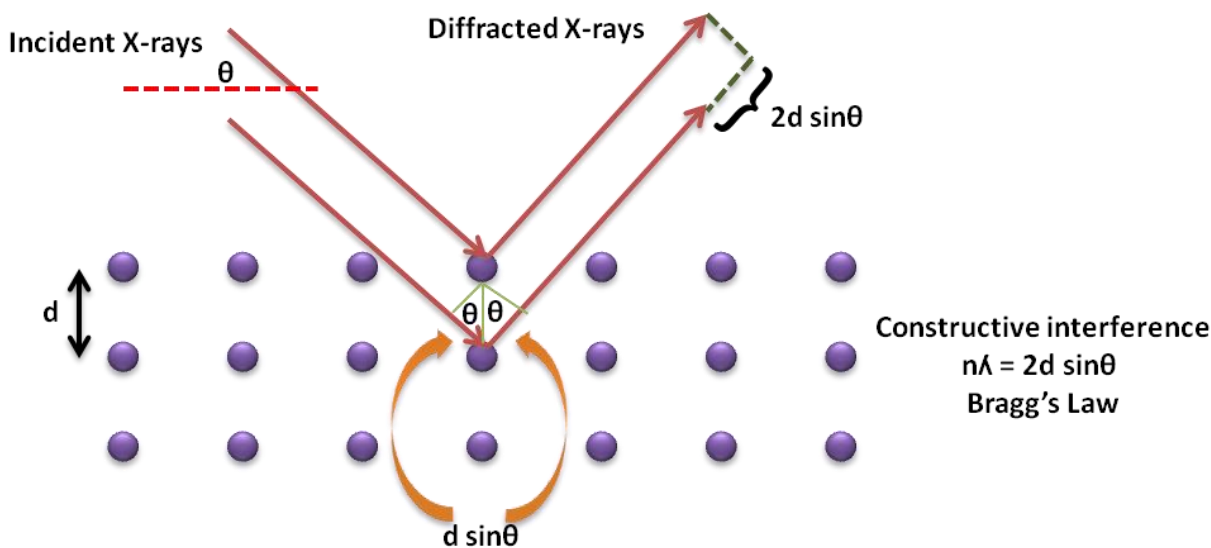

**Figure 10:** *X-ray diffraction by atoms of the periodic crystal lattice.*

### 3.2.7    Contact angle measurement

The contact angle measurement is used to study the moisture resistive behavior of the perovskite films via different chemical engineering method which will be discussed in chapter **6**. The schematic representation of contact angle measurement is shown in figure **11**. The water droplet (2 µL) is used to check the wettability of the perovskite surface. The angle between the water droplet and the surface of the perovskite film will decide the hydrophobic or hydrophilic nature of the perovskite film. The volume of the water droplet is controlled via micro-pipette. Once the water droplet touches the perovskite thin film surface and spread with respect to time, it has been recorded by the camera. The surface tension of the sample (material) plays an important role in deciding the contact angle. Moreover, adding the organic molecule as additive makes the additive based film more hydrophobic due to low surface energy of organic molecules

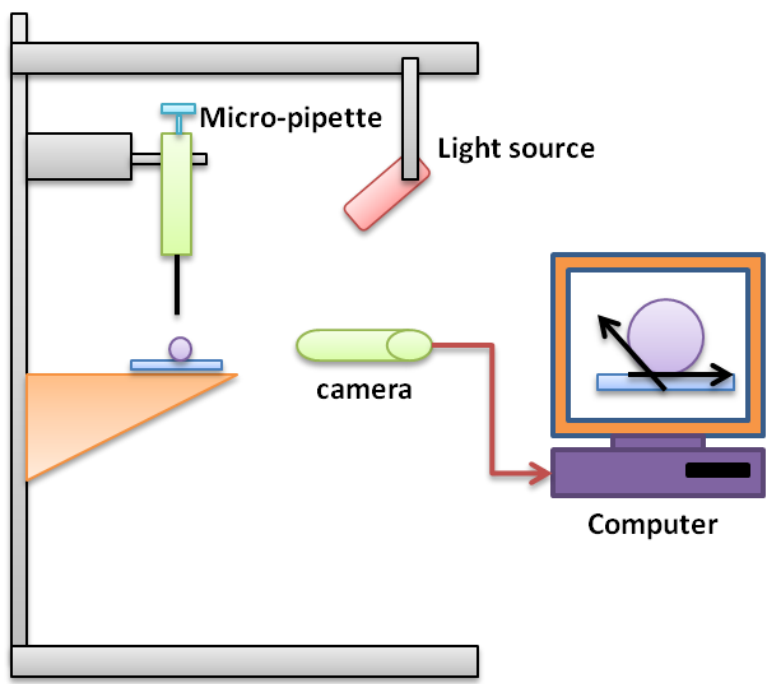

**Figure 11:** *Schematic representation of contact angle measurement between water droplet and surface of perovskite film.*

### 3.2.8    Dektak profilometer

The thickness of the solution processed perovskite thin films are measured with the help of Dektak stylus surface profilometer (model : Dektak-XT). The stylus has radius of 50 nm and force of 1 mg to ensure that stylus do not make scratches on the perovskite sample. The stylus is used to scan the perovskite surface along with low inertial sensor with a minimum resolution of 1 Å in the vertical direction. We have made four sharp scratches on the surface of the perovskite thin film with the help of a sharp pin in order to get average thickness of the film at four different locations. Thickness measurement of the $CH_3NH_3PbI_3$ based perovskite thin film coated on the indium tin oxide (ITO) substrate is shown in figure **12**.

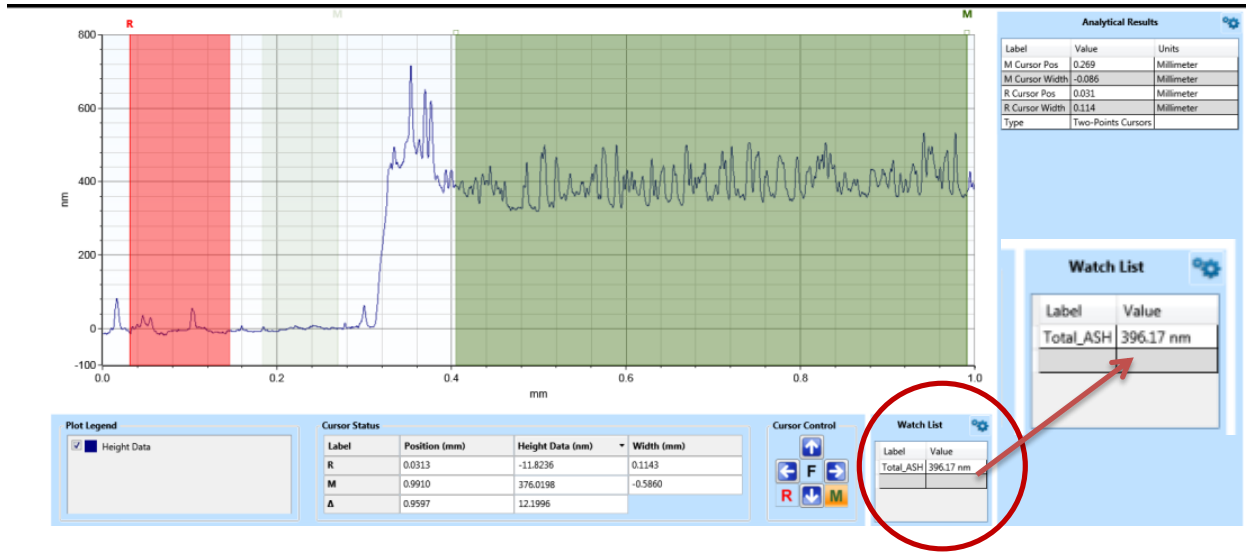

***Figure 12:*** *Thickness measurement of the $CH_3NH_3PbI_3$ based perovskite thin film coated on the indium tin oxide (ITO) substrate and the thickness of the film is ~ 400 nm.*

### 3.3 Device Characterization

### 3.3.1    J-V characterization

Photocurrent density-voltage (J-V) measurement is the most important characterization technique of the PSCs to determine the quality of the device in the term of power conversion efficiency (PCE). In this measurement, we measured the current flowing through the diode by varying the voltage across the device. Figure **13** shows the equivalent circuit diagram of the solar cell including shunt and series resistance under illumination with single diode model. In this equivalent circuit, $I_L$, $R_{Sh}$, and $R_S$ are the photo-generated current, shunt and series resistance, respectively. The current flowing through the device after applying a bias voltage is given by:[11]

$$I = -I_L + I_0 \left[ exp \left( \frac{V - IR_S}{\eta \frac{KT}{q}} \right) - 1 \right] + \frac{V - IR_S}{R_{Sh}} \qquad (7)$$

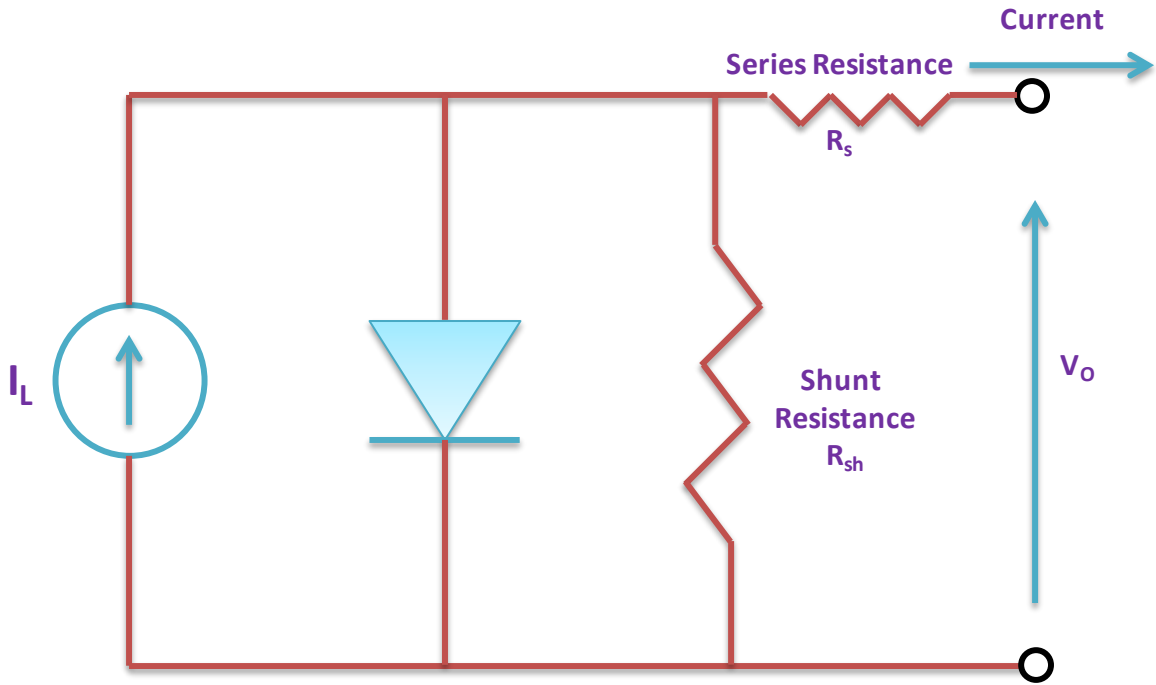

**Figure 13:** *Circuit diagram of the solar cell including shunt and series resistance under illumination with single diode model.*

J-V measurements are carried out using a Keithley 4200 semiconductor characterization system and LED solar simulator (ORIEL LSH-7320 ABA) after calibrating through a reference solar cell provided from ABET. All the J-V measurements were performed using scan speed of 40 mV/s. Light-intensity dependent measurements were carried out under the same solar simulator using neutral density (ND) filters.

### 3.3.2    External/Internal quantum efficiency

External quantum efficiency (EQE) is an experimental technique which is used to investigate the absorption of incident photons and charge collection properties of the solar cells. EQE measurement is carried out to measure the photo-response as a function of wavelength using Bentham quantum efficiency system (Bentham/PVE300). The EQE of the solar cells is defined as the ratio of number of charge carriers collected to the number of incident photons at each wavelength.

$$EQE = \frac{number\ of\ charge\ collected\ (\lambda)}{number\ of\ incident\ photons\ (\lambda)} \quad (8)$$

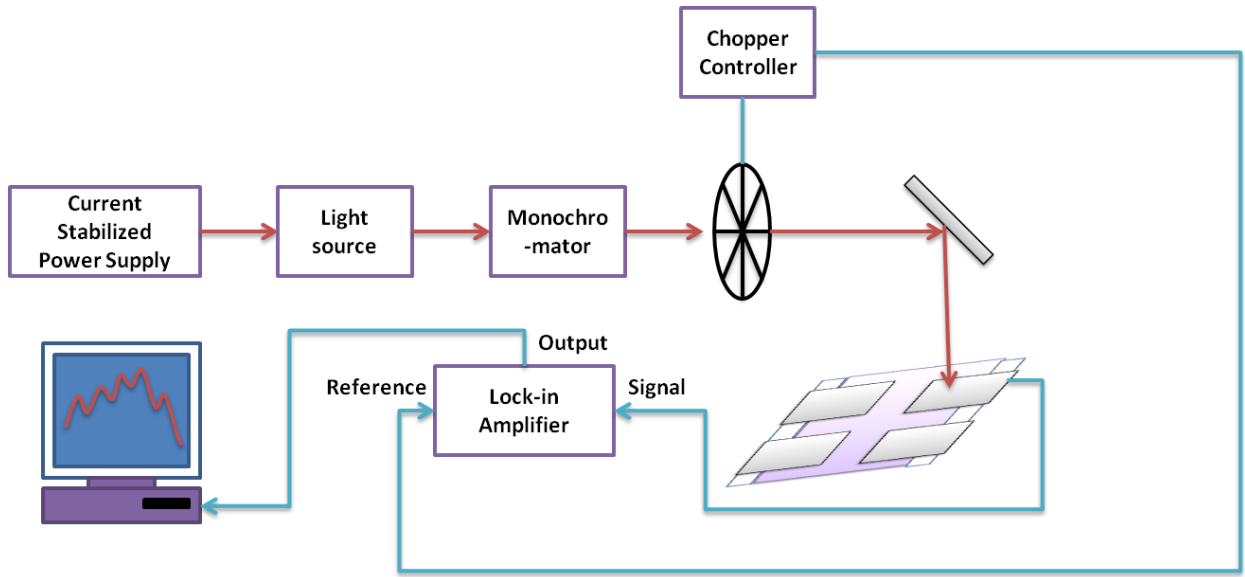

**Figure 14:** *Block diagram of external quantum efficiency measurement set up.*

In EQE, the xenon and quartz halogen lamps (Newport 250 W QTH) are used as white light source to produce photons. The intensity of the light used to measure the EQE response of the solar cells is less than 1mW/cm$^2$. Then the light passes through the monochromator (Oriel Cornerstone 130) in order to get a desire wavelength. The light coming from Monochromator is chopped through optical chopper with frequency of 490 Hz. The chopper converts the DC signal into AC signal and chopper frequency is considered as reference signal for the lock-in amplifier. The light coming from the optical chopper passes through optical lenses and converges into a focus light spot over the device. The light spot should be lower than the area of the device and it should fall inside the active area of the device. The device is connected to the input port of the lock-in amplifier. By matching the phase between the reference signal and the device signal, lock-in amplifier provides the device response for each wavelength. However, the intensity of the light source is calibrated through a calibrated Si detector (Thorlabs FDS100CAL). The internal quantum efficiency (IQE) of the sample is calculated from EQE and reflective absorption spectrum and is defined as:

$$IQE = \frac{EQE}{1 - Reflective\ absoption} \qquad (9)$$

### 3.3.3 Electroluminescence measurement

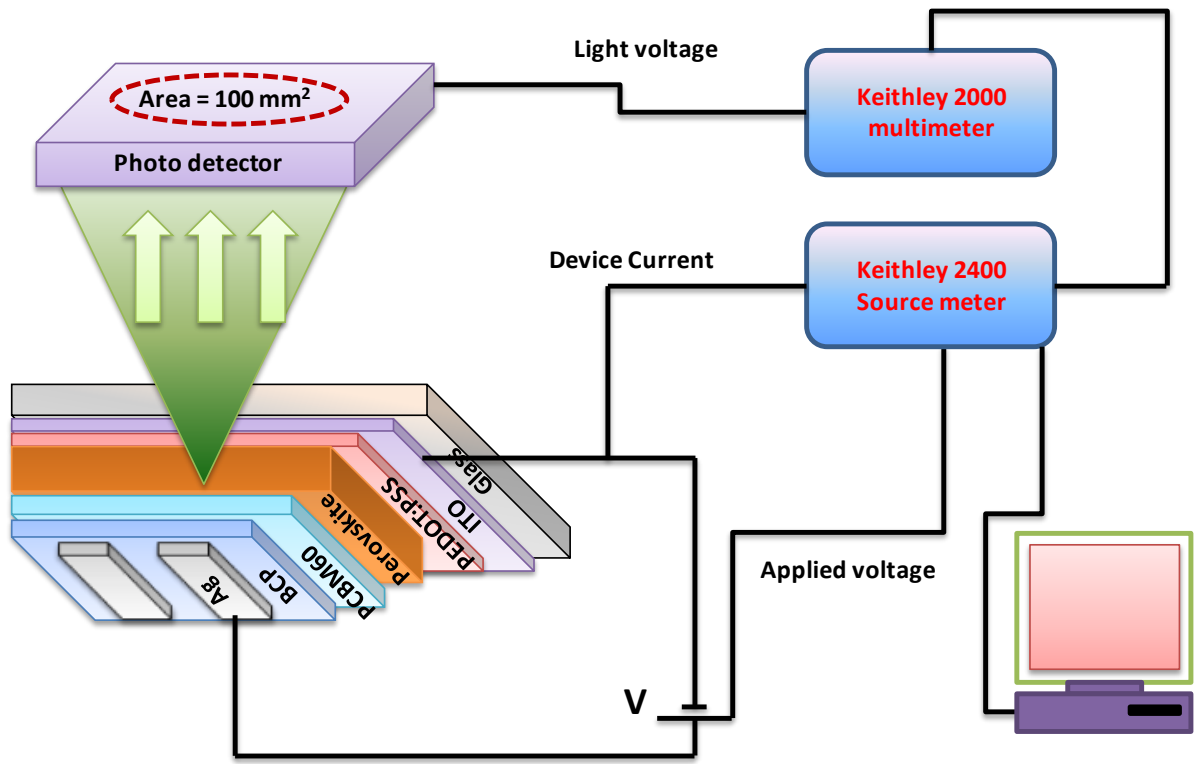

**Figure 15:** *Schematic representation of current-voltage-light measurement.*

The steady state electroluminescence (EL) is measured by biasing the device through Keithley 2400 source meter in constant current mode. The steady state EL emission is collected through optical fiber and measured by USB 4000 ocean optics spectrometer. Steady state current–voltage–light characteristics are measured using a Keithley 2400 source meter, Keithley 2000 multimeter, and calibrated Si photodiode (RS components). Keithley 2000 multimeter is used to measure the voltage out of the detector. The source meter and multimeter are connected in series with each other using National Instruments (NI) GPIB to GPIB cable. Both are also interfaced with the computer through a Labview program using GPIB to USB connector (NI).

### 3.3.4 Transient photovoltage (TPV) measurement

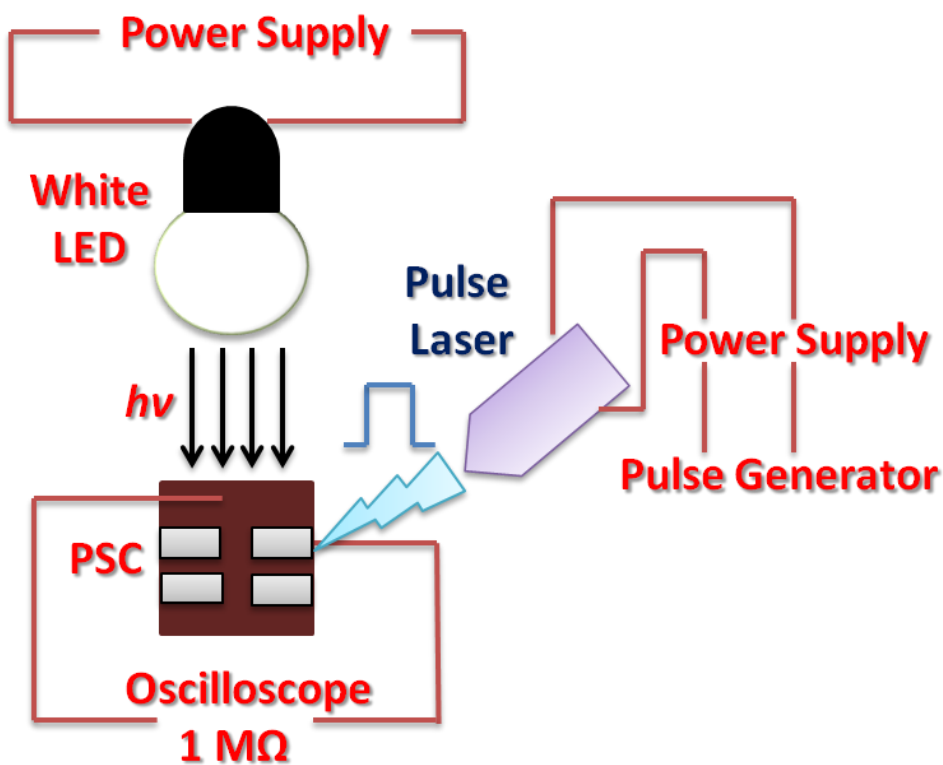

*Figure 16:* Schematic diagram of transient photovoltage set up.

TPV measurement is a well established technique and widely used in organic & dye-sensitize solar cells to study the charge carrier recombination dynamics under open circuit condition. The schematic representation of TPV experimental set up is shown in figure **16**. Different background light intensity condition for the devices was created by a constant DC LED light source and the devices were held at open circuit condition (1 MΩ termination). A small charge perturbation *Δn* was added to the system with the help of an AC pulse laser (490 nm and pulse width of 500 ns with a repetition rate of 1 ms). The decay profile of the extra *Δn* charges was fitted using a mono-exponential curve to get the lifetime $\tau_{\Delta n}$ of the transient charges.[12,13] The TPV technique is discussed in chapter **8** with good details.

### 3.3.5   Scanning photocurrent microscopy (SPM)

*Figure 17:* Scanning photocurrent microscopy.

Figure **17** represents the schematic diagram of SPM technique. A 490 nm TOPTICA class 3B laser diode was mechanically chopped (1000 Hz) and focused on the sample through 60X (0.7 NA) objective in an inverted Olympus IX73 microscope. The spot size of 490 nm diode laser is ~450 nm. The intensity of the laser was decided in such a way that the laser spot will not form any circular fringes around the central spot. The power of laser used in this experiment is ~100 nW and it was measured through Si based photo-detector and lock-in amplifier. When the laser spot is scanned in extended ITO direction (figure **18**), holes are easily collected by the ITO. But electrons have to travel in the lateral direction in order to reach the electron collecting electrode (Ag). Hence, as the laser spot moves away from the Ag electrode, the efficiency of electrons to reach the electrode reduces. As a result of this, we observed an exponential decay in the photocurrent profile as a function of distance from the Ag electrode towards the film, which is used to calculate the lateral diffusion length of electrons in PSCs.[14,15] In a similar way, one can calculate the diffusion length for holes by scanning the laser spot in extended Ag direction. SR830 lock-in amplifier coupled with the same mechanical chopper mentioned previously is used to measure the electron and hole photocurrent. SR830 lock-in amplifier measures R and θ simultaneously for an incoming signal. Here, R represents the photocurrent and θ represents phase delay between the chopped reference signal and device signal. The lock-in amplifiers and scanning stage are interfaced via LabView VI. LabView software was customized to collect the photocurrent and phase delay simultaneously with respect to (x or y) position of the device. SPM technique is discussed in good details in chapter **7**.

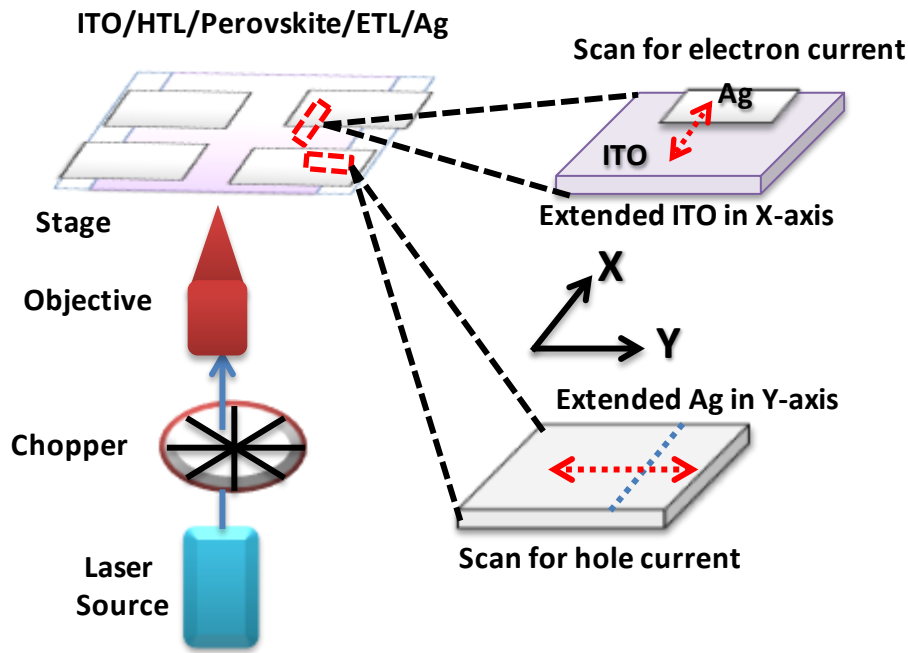

**Figure 18:** *Schematic of SPM to estimate the charge transport length of photo-generated electron and hole, separately.*

# Chapter 4

# Photo-Physics of Hybrid Metal Halide Perovskite based Semiconductors

# CHAPTER 4

# Photo-Physics of Hybrid Metal Halide Perovskite based Semiconductors

***Abstract:*** In this chapter, the effect of thermal and structural disorder on the electronic band gap of hybrid perovskite semiconductors is investigated by measuring the exciton line width and Urbach energy as a function of temperature. A detailed analysis of the exciton line width is shown to distinguish between static and dynamic disorder. The static disorder component, manifested in the exciton line width at low temperature, is small (~11 meV). Above 60 K, thermal (dynamic) disorder increases the exciton line width. Temperature dependence of the band gap is due to combined effect of electron-phonon interaction and lattice dilation. It is found that the thermal expansion term dominates over the electron phonon interaction term with thermal expansion coefficient is of the order of $10^{-4}$ $K^{-1}$, which accounts for the positive temperature coefficient of the band gap. This study provides important insights into the electronic and structural properties of $CH_3NH_3PbI_3$ based perovskites which should be transferable to many related organic metal-halide perovskites. Along with that, this chapter also present the modulation of electronic states of $CH_3NH_3PbX_3$ perovskite nano-crystals (PNCs) (X = Br$^-$ and I$^-$) as a function of crystallite size in organic semiconductor matrix [(4,4'-bis[9-dicarbazolyl]-2,2'-biphenyl (CBP) & bathocuproine (BCP)] forming a type-I hetero-structures with the bulk perovskites. This results into an easy growth of $CH_3NH_3PbX_3$ PNCs of tunable sizes from ~110 nm to ~10 nm in organic semiconductor matrix. A blue shift is observed in the photoluminescence peak ($PL_{max}$) energy of $CH_3NH_3PbX_3$ PNCs embedded in organic matrix. Results about blue shift in PL peak could be explained using the classic particle in a box *vs* excitonic Bohr radius model under weak confinement regime.

**Part I: Influence of static *vs* dynamic disorder on the electronic structure of hybrid perovskite semiconductor CH₃NH₃PbI₃.**

## 4.1 Introduction

In recent times, lead based hybrid perovskite materials have found a lot of attention, exhibiting solar cell efficiencies >25%,[1,2,3,4] high efficiency light emitting diodes (EQE > 20%)[5,6,7] and other optoelectronic applications.[8,9] The band gap of these materials can be tuned from NIR to UV by changing the halide ion. Along with these applied research, there are fundamental research aspects also being studied. As it has been shown that the temperature dependence of the band gap is anomalous i.e. the band gap increases with temperature.[10] However, a clear understanding on the origin of this temperature dependence was not been established. In this chapter, a detailed study on the temperature dependence of the optical absorption and emission of high quality $CH_3NH_3PbI_3$ ($MAPbI_3$) films prepared on quartz substrate is presented. The temperature dependence of the optical band gap, the exciton line width ($\Gamma_{EX}$), and the sub-band gap absorption in $MAPbI_3$ based perovskite thin film are studied. Because the material undergoes a structural transition at low temperature, this chapter encompasses the room and low temperature phases of the $MAPbI_3$ perovskite semiconductor.[11]

## 4.2 Experimental section

***CH₃NH₃I synthesis:*** All materials were purchased from Sigma-Aldrich and used as received. Methylammonium iodide (MAI) was synthesized as discussed elsewhere.[12,13] MAI was synthesized by reacting 24 mL of methylamine (33 wt. % in absolute ethanol) and 10 mL of hydroiodic acid (57 wt% in water) in a round-bottom flask at 0 °C for 2 h with stirring. The raw precipitate was recovered by removing the solvent in a rotary evaporator at 50 °C. The raw product was washed with dry ether, dried in vacuum at 60 °C and re-dissolved in simmering absolute ethanol. The pure MAI recrystallized on cooling is filtered and dried at 60 °C in a vacuum oven for 24 h.

***CH₃NH₃PbI₃ film preparation:*** For perovskite film formation, a published procedure was adapted.[13,14] The quartz substrates were cleaned with detergent diluted in deionized water, rinsed with deionized water, acetone and ethanol, and dried with clean dry air. Clean substrates were transferred in a glovebox filled with nitrogen. For perovskite formation, $PbI_2$ (1M) was dissolved in N,N-dimethyl formamide (DMF) overnight under stirring conditions at 100 °C and 80 µl solution was spin coated on the quartz substrates at 2000 rpm for 50 s, and dried at 100 °C for 5 min. Powder of MAI (100 mg) was spread out around the $PbI_2$ coated substrates with a petridish covering on the top and heated at 165 °C for 13 h. To protect the samples from air and humidity, 40 mg/ml poly(methylmethacrylate) (PMMA; Aldrich) in butyl acetate was spin-coated on top of the perovskite at 2000 rpm for 30 sec. All steps were carried out in nitrogen atmosphere inside glove box.

***Optical characterization:*** The temperature dependent optical absorption spectra of MAPI based perovskite film was done in the laboratory of Prof. Sven Huettner, University of Bayreuth, Germany. The films were measured in a quartz windows fitted helium flow cryostat unit. The absorption spectra were recorded using an Ocean Optics DS3000 Halogen-Deuterium light source and an Ocean Optics QE Pro spectrometer coupled with fiber optics and lenses to the cryostat system. A temperature-step profile was applied to obtain the UV-Vis spectra at the respective temperatures, allowing 10min equilibration time.

The temperature dependent photoluminescence spectra were measured in our laboratory (IIT Bombay, India). To perform the temperature-dependent photoluminescence measurement, the sample ($MAPbI_3$ film) was placed on the cold finger of the helium cryostat (Janis CCS-300S/204) under a dynamic vacuum of $\approx 10^{-5}$ mbar. The sample was excited at 355 nm, the third harmonic of Nd:YAG laser at 1 kHz repetition rate and 840 ps pulse width, and collected by Andor gated ICCD (istar model no. 05560) integrated with Shamrock spectrometer (model no: SR303i-B) with 10 ms integration time.

***XRD characterization:*** The x-ray diffraction of perovskite films were done in the laboratory of Prof. Sven Huettner, University of Bayreuth, Germany. Perovskite films were prepared as described, but then scratched off in the glovebox and transferred into a Kapton sample holder (Aluminium discs sealed with Kapton tape, serving as windows). The samples were transported in inert atmosphere to the Australian Synchrotron, where they were measured at the SAXS beamline in the liquid nitrogen cooled temperature stage (Linkam) which was purged with nitrogen. The X-ray energy was at 10keV using a 4k Pilatus detector for the WAXS signal collection.

## 4.3    Results

### 4.3.1    Structure and lattice expansion:

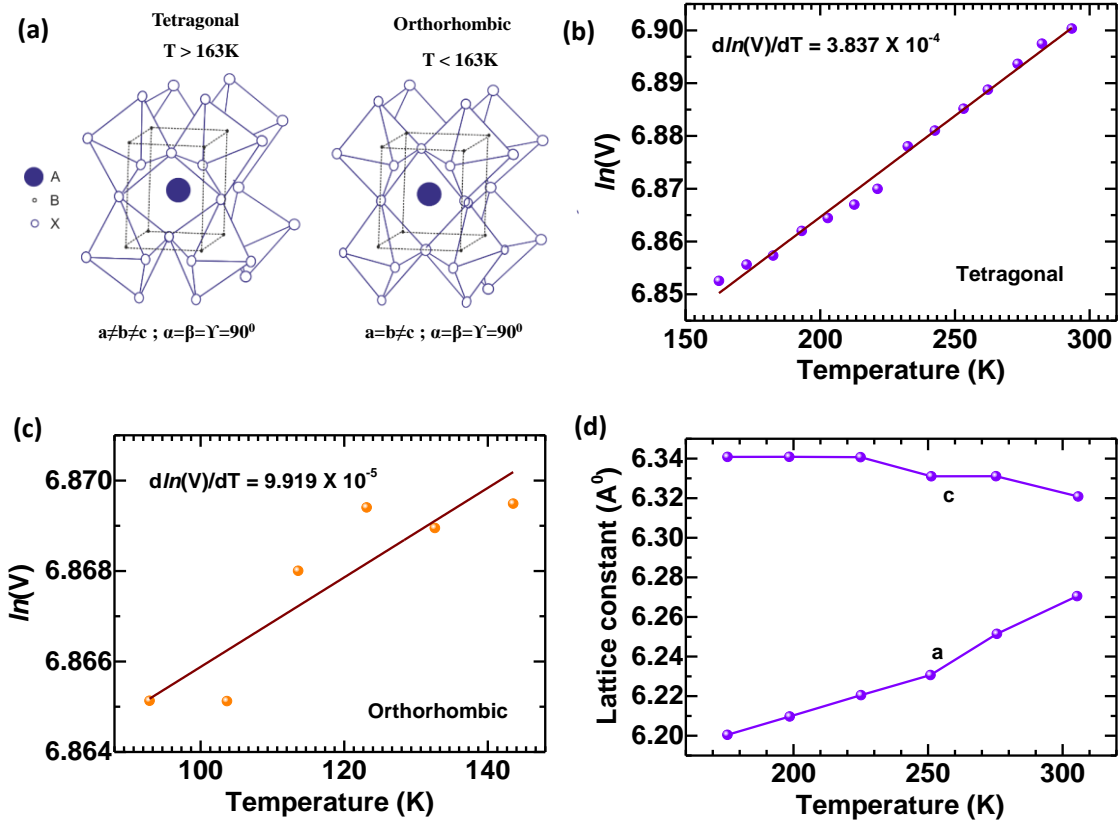

***Figure 1:*** *(a) Crystal structure of MAPbI₃ in tetragonal and orthorhombic phase. ln of volume of MAPbI₃ in (b) tetragonal and (c) orthorhombic phase as a function of temperature. Solid line (wine color) is a linear fit. (d) Lattice constant of MAPbI₃ in tetragonal phase is obtained from temperature dependent XRD data.*

Hybrid perovskite MAPbI₃ is known to have temperature dependent structural phase transitions. As depicted in figure **1a**, for temperatures T < 163 K it is in the orthorhombic phase ($a\neq b\neq c$; $\alpha=\beta=\gamma=90^0$) and for temperatures 163 K $<$T$<$ 327.3K it remains in the tetragonal phase $a=b\neq c$; $\alpha=\beta=\gamma=90^0$. Beyond 327.3 K, MAPbI₃ forms a cubic phase ($a=b=c$; $\alpha=\beta=\gamma=90^0$). With the temperature dependent X-ray measurements, we are able to determine the lattice expansion coefficient $\left(\frac{dln(V)}{dT}\right)_P$. In particular, we have estimated this coefficient for the tetragonal (figure **1b**) and orthorhombic phase (figure **1c**) using temperature dependent X-ray scattering to be in the order of $10^{-4}$ K⁻¹ which is in agreement with previously reported data by Kawamura et.al.[15] who found $\left(\frac{d\ln(V)}{dT}\right)_P$ = (1.35±0.014) x  $10^{-4}$ K$^{-1}$. We note that this value is almost 50 times higher than conventional Si. Using this value, we estimate $\left(\frac{dE_g}{d\ln(V)}\right)_T$ to be 1.26 eV which is in good agreement with theoretical values estimated for the perovskite.[16] V corresponds to the volume of the unit cell calculated as depicted in figure **1b** and **1c**. The lattice constant 'a' and 'c' for tetragonal phase is shown in figure **1d**. The direct relation of the lattice expansion on the optical properties will be discussed in the next section.

### 4.3.2   Absorption

Figure **2a** represents the optical absorption spectra of MAPbI$_3$ film on quartz substrate for a range of temperature i.e. from 6 K to 305 K. It is observed that the onset of the optical absorption moves to higher energies with increase in temperature from 6 K to 150 K. After 150 K, a discontinuity in the trend of increase in the optical absorption edge is observed. This is because of change in structural phase of MAPbI$_3$ from orthorhombic to tetragonal. A similar trend of increase in absorption edge is observed in the tetragonal phase (175 K to 305 K) as well. Figure **2b** and **2c** show the optical absorption spectra of a MAPbI$_3$ film on quartz substrate in its orthorhombic and tetragonal phase at three different temperatures for clear visibility, respectively. Strong excitonic absorption features dominate the optical absorption edge[10] and become even more pronounced at lower temperatures. In order to separate the band-to-band absorption *vs* excitonic absorption from the UV-Vis spectra (figure **2d**), Elliot theory [6,17,18] is used to find the contribution of excitonic and inter-band absorption, i.e.,

$$\alpha_{UV-Vis}(\varepsilon) = \alpha_{excitonic} + \alpha_{band-to-band} \tag{1}$$

In order to determine the electronic bandgap and excitonic properties of these materials, the experimental results are modeled using Elliot's theory of Wannier exciton in 3D semiconductors. The following equation is used to fit measured UV-Vis spectrums of MAPbI$_3$ films:

$$\alpha(E) \propto \frac{\mu^2}{E}\left[\sum_n \frac{2E_x}{n^3} sech\left(\frac{E-E_n^X}{\Gamma}\right) + \int_{E_g}^{\infty}\left[sech\left(\frac{E-E_n^X}{\Gamma}\right)\frac{1}{1-e^{-2\pi\sqrt{E_x/E_x-E_g}}}\frac{1}{1-\frac{128\pi\mu b}{\hbar}(E_1-E_g)}dE_1\right]\right] \tag{2}$$

where E$_x$, $\mu$, $\Gamma$, E and b are exciton binding energy, transition dipole moment, FWHM of excitonic peak, photon energy and non-parabolic contribution, respectively. This equation is valid for bulk semiconductors with E$_x$ much smaller than the E$_g$ (Wannier excitons) and was used to describe optical transitions to bound and/or ionized excitonic states in model inorganic semiconductors. In above equation there are two terms, first term represents excitonic levels below the conduction band of various perovskite semiconductors and second term represents the continuum of states beyond the energy of $E_g$ as band-to-band transition contribution in the overall optical absorption. The absorption in the continuum spectrum does not simply follow the square root dependence of the density of states on energy $\alpha(h\nu)^2 = A(E-E_g)$ as expected for bare band-to-band transitions between uncorrelated electron and hole particles. The excitonic enhancement of the optical density of states at band-edge depends on the strength of the Coulomb interaction, through the exciton binding energy, and provides a measure of the degree of overlap of electron-hole wave-functions in bulk of semiconductor.

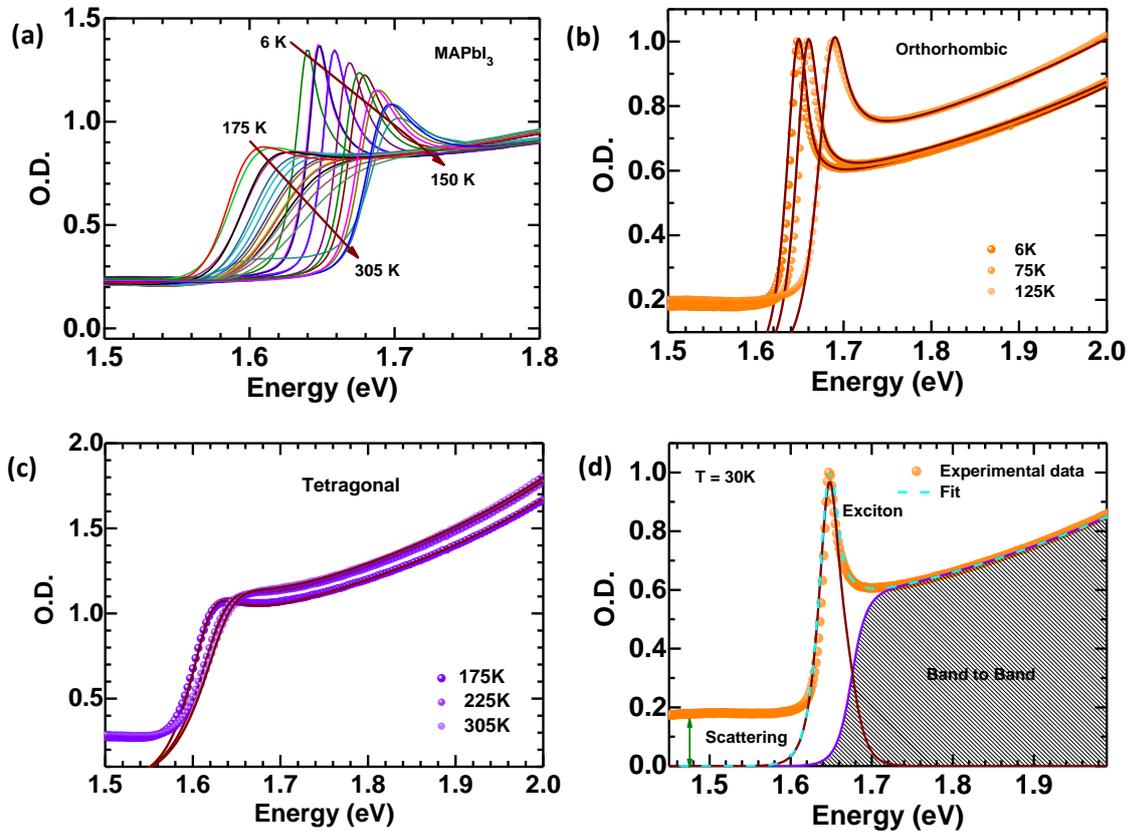

***Figure 2:*** *(a) Optical Absorption spectra of MAPbI₃ at different temperature range (6 K to 305 K). UV-Vis absorption spectra of MAPbI₃ thin film at three different temperatures in the (b) orthorhombic and (c) tetragonal phase. (d) Illustration of the fitting of the absorption using Elliot's theory with the excitonic and band to band contribution.*

Fitting the optical absorption using this model provides the exciton binding energy ($E_x$), FWHM ($\mathbf{\Gamma_{EX}}$) of the exciton peak and the electronic bandgap ($E_g$)[6]. The excitonic contribution in the absorption spectra for different temperatures is separated out from band to band transition and shown in figure **3a** and **3b** for orthorhombic and tetragonal phase, respectively. Table **1** shows the respective fitting parameters and figure **3c** shows the variation of the band gap with temperature. The band gap increases with temperature in contrast to what is normally observed in crystalline semiconductors where it decreases.[19,20] In the vicinity of 160K, the band gap exhibits a discontinuity, due to the structural phase transition from the orthogonal to the tetragonal phase.[11] The respective temperature coefficient of bandgap ($dE_g$ / dT) is larger in the orthorhombic phase and in general the temperature dependence of the band gap $E_g$, can be expressed as [20,2122]

$$\left(\frac{dE_g}{dT}\right)_P = \left(\frac{dE_g}{dT}\right)_V + \left(\frac{dE_g}{dln(V)}\right)_T \times \left(\frac{dln(V)}{dT}\right)_P \qquad (3)$$

The first term explicitly represents the electron phonon coupling through the deformation potential which results in a decrease of the band gap with temperature. The second term implicitly represents the lattice dilatation term which causes the band gap to increase with temperature.

For most inorganic semiconductors the lattice dilatation term is much smaller than the electron phonon interaction term. Hence the band gap decreases with temperature for many inorganic semiconductors.[21] This, however, is different in organo-metal halide perovskites such as MAPbI$_3$, which can be seen when we express the experimental data using equation **3**. The second term, i.e. the lattice dilatation term, dominates the temperature dependence of the band gap for the lead perovskite reported here. From figure **3c** we estimate $\left(\frac{dE_g}{dT}\right)_P$ to be $(2.50 \pm 0.11)$ x $10^{-4}$ eV/K for the tetragonal phase and $(4.85 \pm 0.28)$ x $10^{-4}$ eV/K for the orthorhombic phase. The volume expansion coefficient (vide supra) is estimated for the tetragonal phase to be $\left(\frac{dln(V)}{dT}\right)_P = (3.837 \pm 0.018)$ x $10^{-4}$ K$^{-1}$. Using this value, we estimate $\left(\frac{dE_g}{dln(V)}\right)_T$ to be 1.26 eV which is in good agreement with theoretical values estimated for the perovskite.[16] We would like to point out that this value is a lower limit as we have ignored the (negative) contribution to $\left(\frac{dE_g}{dT}\right)_P$ from the electron phonon term in equation **3**. The thermal expansion coefficient of the perovskite samples is almost 120 times larger than similar data for Si (3 x $10^{-6}$ K$^{-1}$).[22] We mention in passing that other lead compounds- like lead chalcogenides also have a positive temperature coefficient of the band gap, where volume expansion coefficient is higher than MAPbI$_3$ resulting in an even higher $\left(\frac{dE_g}{dT}\right)_P$ than MAPbI$_3$.[23,24]

We will now present the temperature dependence of the excitonic absorption. The exciton binding energy $E_x$ accounts for the energy difference between the electronic band gap and the excitonic peak position. The line width of the excitonic peak is a signature of the disorder present in the semiconductor. The disorder can be of static (structural) and/or dynamic (thermal) nature - both will broaden the excitonic line width.[25,26] Figure **2a** shows the FWHM line width **Γ$_{EX}$** as a function of temperature. **Γ$_{EX}$** decreases linearly with decrease in temperature till about 60K and begins to flatten out at lower temperature and decreases continuously through the phase transition. At very low temperature, broadening due to static disorder dominates **Γ$_{EX}$** and is about 11.5 meV, which can be seen as a lower-bound of the inherent disorder in the film. For all higher temperatures, the **Γ$_{EX}$** is influenced by dynamic disorder (which will be analyzed in more detail further below). The instrument broadened line width with about 3 meV is much smaller than measured **Γ$_{EX}$** at 60K.[25]

A very typical approach well-known for inorganic solids attributes the dynamic disorder to the interaction with phonons. Then, the broadening $\Gamma_{EX}$, can be fitted using a Bose-Einstein type expression.[27,28]

$$\Gamma_{EX}(T) = \Gamma_0 + \frac{\Gamma_{ep}}{\exp\left(\frac{E_{LO}}{k_B T}\right) - 1} \qquad (4)$$

where, $\Gamma_0$= intrinsic **Γ$_{EX}$** width at T=0K, $\Gamma_{ep}$= coupling constant and E$_{LO}$ = phonon energy. Figure **3c** shows the fit for two phases separately over the whole temperature range. The respective fitting parameters are summarized in Table **1**. We note that $\Gamma_0$= 11.64 meV (orthorhombic) and $\Gamma_0$= 15.05 meV (tetragonal) is small for both phases, which is an indicative of high crystal quality of the film. $\Gamma_0$ is marginally smaller for the orthorhombic phase than tetragonal phase, which suggests that the orthorhombic phase is relatively more ordered than tetragonal phase in agreement with literature.[29] Considering that E$_{LO}$ (using equation 2) is ~ 210 cm$^{-1}$ first principle calculations have shown that such low energy phonon modes relate to the coupled phonon mode between the inorganic cage and the MAI$^+$ motion.[29,30] In MAPbI$_3$, the orthorhombic phase restricts the molecular ion motion relatively more compared to the tetragonal phase resulting in a lower dynamic disorder.

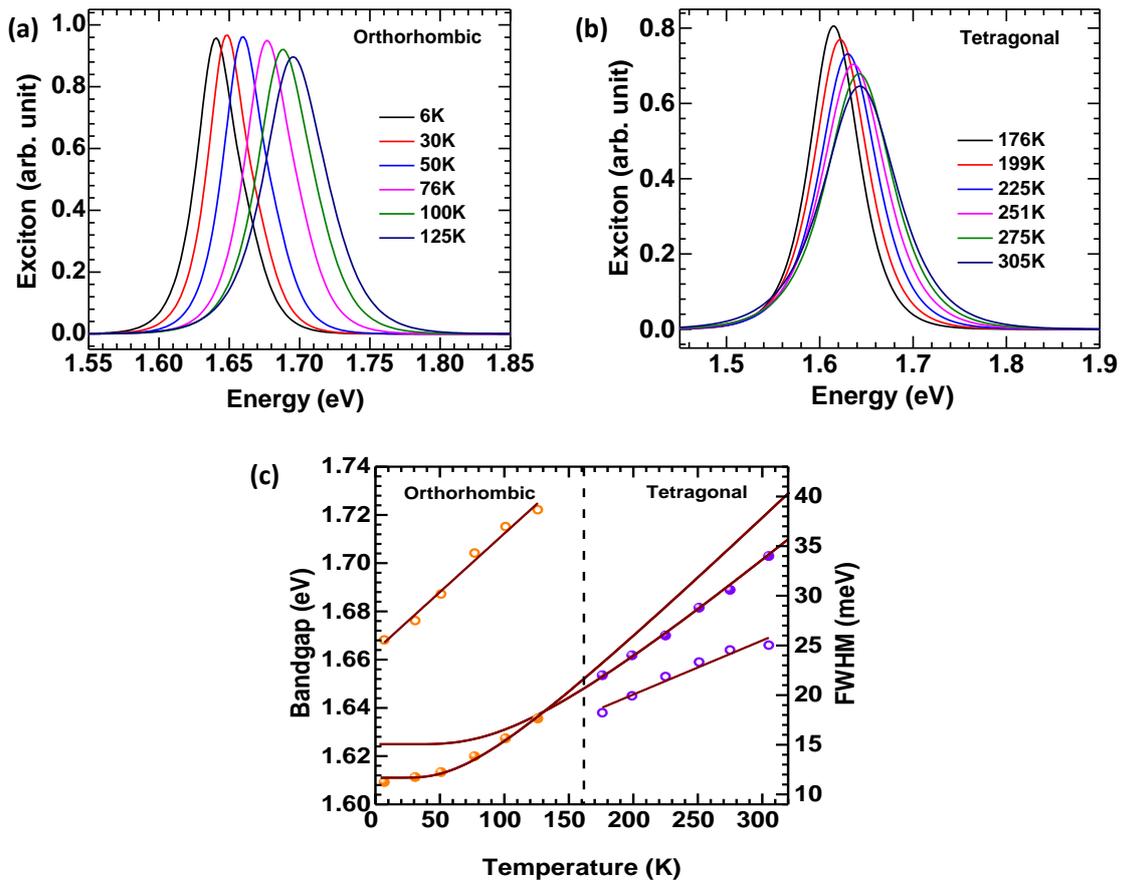

**Figure 3:** *Excitonic contribution in (a) orthorhombic phase and (b) in tetragonal phase, visualizing the broadening with higher temperature. (c) Variation of band gap and FWHM of excitonic peak with temperature in orthorhombic and tetragonal phase of MAPbI₃. Orange hollow circles (○) and violet hollow circles (○) represent the band gap at different temperatures in the orthorhombic and tetragonal phases, respectively. Similarly, orange and violet solid spheres represent the FWHM of excitonic peak in orthorhombic and tetragonal phase, respectively. Wine solid lines are fits to the experimental data in the orthorhombic and tetragonal phase.*

***Table 1:*** *Slope of $E_g$ versus T, $E_g$ versus $E_u$, volume versus T, and extracted parameters $\Gamma_0$, $\Gamma_{ep}$, $E_{LO}$, $\Theta$, P, and $E_u$ (T = 0) in Orthorhombic and Tetragonal Phases of MAPbI$_3$ Film.*

| MAPbI$_3$ | Orthorhombic | Tetragonal |
|---|---|---|
| dE$_g$/dT ($10^{-4}$ eV/K) | $4.85 \pm 0.28$ | $2.50 \pm 0.11$ |
| $\Gamma_0$ (meV) | $11.64 \pm 0.17$ | $15.05 \pm 0.51$ |
| $\Gamma_{ep}$ (meV) | $26.11 \pm 1.47$ | $36.35 \pm 1.42$ |
| E$_{LO}$ (meV) | 17.81 | 27.98 |
| $\Theta$ (K) | $197 \pm 25$ | $324 \pm 33$ |
| P | $8.92 \times 10^{-4}$ | 0.002 |
| $\sigma_0$ | $0.403 \pm 0.017$ | $0.464 \pm 0.005$ |
| h$_{vp}$ (meV) | $17.81 \pm 1.4$ | $27.98 \pm 1.19$ |
| E$_u$ (at T = 0 K) (meV) | 21.47 | 32.37 |
| d(ln V)/dT (K$^{-1}$) | $9.91 \times 10^{-5}$ | $1.35 \times 10^{-4}$ |
| dE$_g$/dE$_u$ | $4.91 \pm 0.49$ | $1.57 \pm 0.11$ |

Figure **4a** and **4b** show the plots of log α vs hν at different temperatures for the low energy side of the absorption of exciton peak for the two phases. The low energy edge of the exciton peak satisfies the following empirical relation[26,31]

$$\alpha = \alpha_0 \exp[\frac{\sigma(h\nu - E_0)}{K_B T}] \qquad (5)$$

Where α is the absorption coefficient, σ the steepness parameter (vide infra) and k$_B$ the Boltzmann constant. This empirical relation is valid for many semiconductor materials and is known as the Urbach rule.[31] The exponential absorption is indicative of tail states which are a consequence of disorder.[21,25,26,32] In a logarithmic plot the fitting of the band edge at different temperatures results in a common focus E$_0$. This Urbach focus is one of the characteristic features of Urbach absorption. As shown in figure **4**, the lines come to a separate common focus (E$_0$) for each of the phases. A single E$_0$ cannot fit both phases, reflecting the prevalence of different disorder of each of the phases. In most semiconductors the Urbach focus E$_0$ is located at energies which are greater than the optical band gap providing theoretical maximum bandgap estimation.[21,25,26] However, in the present case, as T decreases, the bandgap decreases and the slope of optical absorption becomes steeper. Since the temperature dependence of the band gap is positive, the Urbach focus (E$_0$) now takes place at energies less than the band gap. The applicability of the Urbach rule in this case provides an elegant way in describing the minimum possible bandgap which can be achieved in MAPbI$_3$ at zero disorder. We note that due to the high absorption cross section and due to the primary relevance of higher tail states a fit accounting of even less than one order of optical density suffices this formalism. In particular, the excellent convergence substantiates the applicability of the Urbach rule within this rather small optical density range. As can be seen from photothermal deflection measurements[33] the exponential tail may extend another 2-3 magnitudes yet this is partially obscured here by scattering effects.

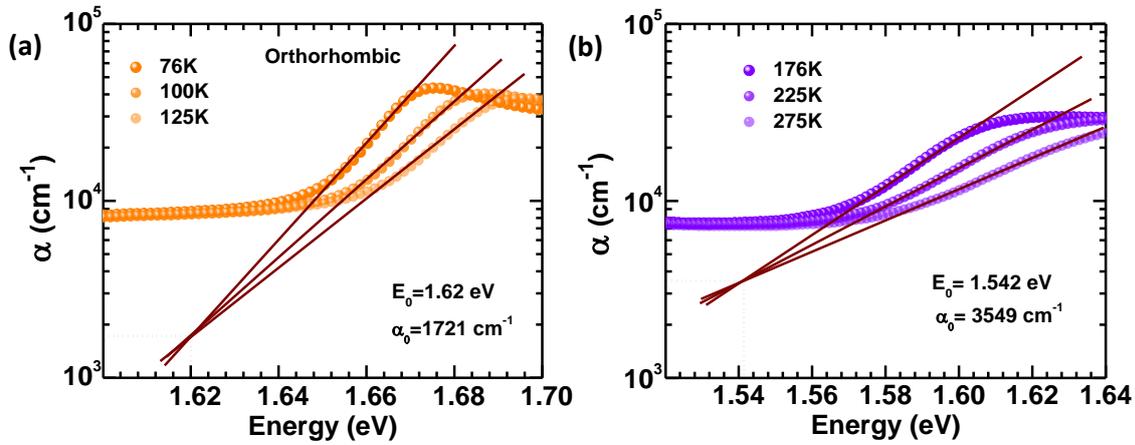

***Figure 4:*** *Logarithmic variation of absorption coefficient (α) with photon energy at different temperatures in (a) tetragonal and (b) orthorhombic phase of MAPbI$_3$.*

Furthermore, the prevalence of the common focus shows that the Urbach rule is compatible with the above explained band gap shift due to lattice dilatation. The temperature dependence is accounted by the steepness factor $\sigma$ which is described in the following section. Figure **5a** shows the steepness parameter σ as a function of temperature obtained by using equation 4. The steepness parameter is temperature dependent and following the Urbach formalism, is approximated by[26]

$$\sigma = \sigma_0\left(2kT/h\nu_p\right)tanh(h\nu_p/2kT) \qquad (6)$$

$\sigma_0$ is a constant which is characteristic of the excitation and $h\nu_p$ quantifies the energy of involved phonons. Table **1** summarizes the values of $\sigma_0$ and $h\nu_p$ obtained for the two phases. Using the values of $\sigma_0$ we calculate the Urbach energy. The energy $kT/\sigma = E_u$ is known as the Urbach energy and is related to the degree of disorder. Following Cody, the total disorder can be thought of as the sum of two terms - thermal disorder and static disorder.[21] The thermal disorder arises from excitations of phonon modes and the static disorder is due to inherent structural disorder. Figure **5b** shows the Urbach energy as a function of temperature. This can be written quantitatively as a sum of thermal and structural disorder:[21]

$E_u$ $(T, X)$ =K ($<U^2>_T + <U^2>_x$)

Here, $<U^2>_T$ is related to the mean square displacement of atoms (similar to the Debye Waller factor) and $<U^2>_x$ is the quenched inherent structural disorder. The temperature dependence of $E_u$ can be estimated by describing the phonon spectrum to be an assembly of Einstein oscillators[21]. Following the procedure of Cody et al we write[25]

$$E_u(T,P) = K \left(\frac{\theta}{\sigma_0}\right) \left[\frac{1+P}{2} + \{\exp\left(\frac{\theta}{T}\right) - 1\}^{-1}\right] \qquad (7)$$

where, $\Theta$ is the Einstein characteristic temperature which corresponds to the mean frequency of lattice phonon excitation; $P$ is the structural disorder and is defined as $P = <U^2>_x / <U^2>_0$; the suffix 0 denotes the zero point vibrational mode. In a perfectly ordered semiconductor film $P$ =0. Equation **5** is used to fit the data of $E_u$ vs T for the two phases and the results are summarized in table **1**.

The small value of $P$ reflects again the high crystal quality in these films and we again note that $P$ is larger for tetragonal phase than orthorhombic phase confirming that the orthorhombic phase is more ordered. For comparison, structural disorder value $P$ is found to be approximately 2.2 for a:Si.[21] It is worth mentioning that the phonon modes, responsible for thermal disorder in the two phases match the ones responsible for exciton $\mathbf{\Gamma_{EX}}$ broadening (see Table 1).The Einstein characteristic temperature $\Theta$ is lower for the orthorhombic phase confirming that the orthorhombic phase is more ordered than the tetragonal phase.

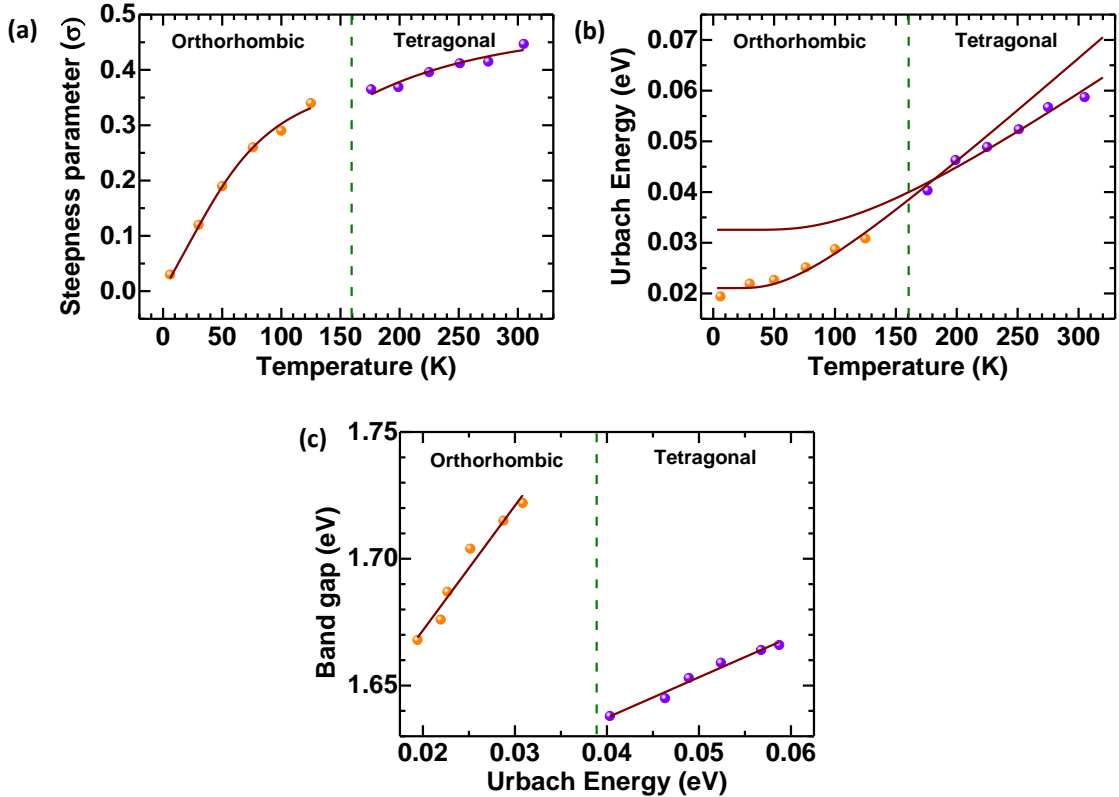

***Figure 5:*** *(a) Steepness parameter and (b) Urbach energy as a function of temperature for MAPbI3 together with corresponding fit (wine solid lines) of the experimental data using eq 4. (c) Band gap as a function of Urbach energy. Orange and violet solid spheres represent the experimental data in orthorhombic and tetragonal phase, respectively.*

In order to discuss how much is the extrinsic disorder ($E_u(T, P = 0)$, which can be thermal or structural due to different processing conditions)[32] of the material affects the band gap with respect to the intrinsic disorder ($E_u(T = 0, P)$), we analyse the change in bandgap and $E_u$ at a particular temperature. Figure **5c** shows $E_g$ vs $E_u$ for the MAPbI$_3$ film for both phases. As $E_u$ increases, $E_g$ also increases in both phases. The slope of $E_g$ vs $E_u$ is almost three times higher for the orthorhombic phase ($dE_g/dE_u = 4.91\pm0.49$) than for the tetragonal phase ($dE_g/dE_u = 1.57\pm 0.11$). This analysis establishes a relationship between the bandgap and the width of the absorption tail, i.e., $E_u$ for each phase, which suggests that the bandgap of this material is determined by the degree of the thermal disorder in the film at given T. Since the orthorhombic phase is the relatively more ordered phase, a small change in $E_u$ results in a larger change in $E_g$ as compared to the tetragonal phase of MAPbI$_3$.

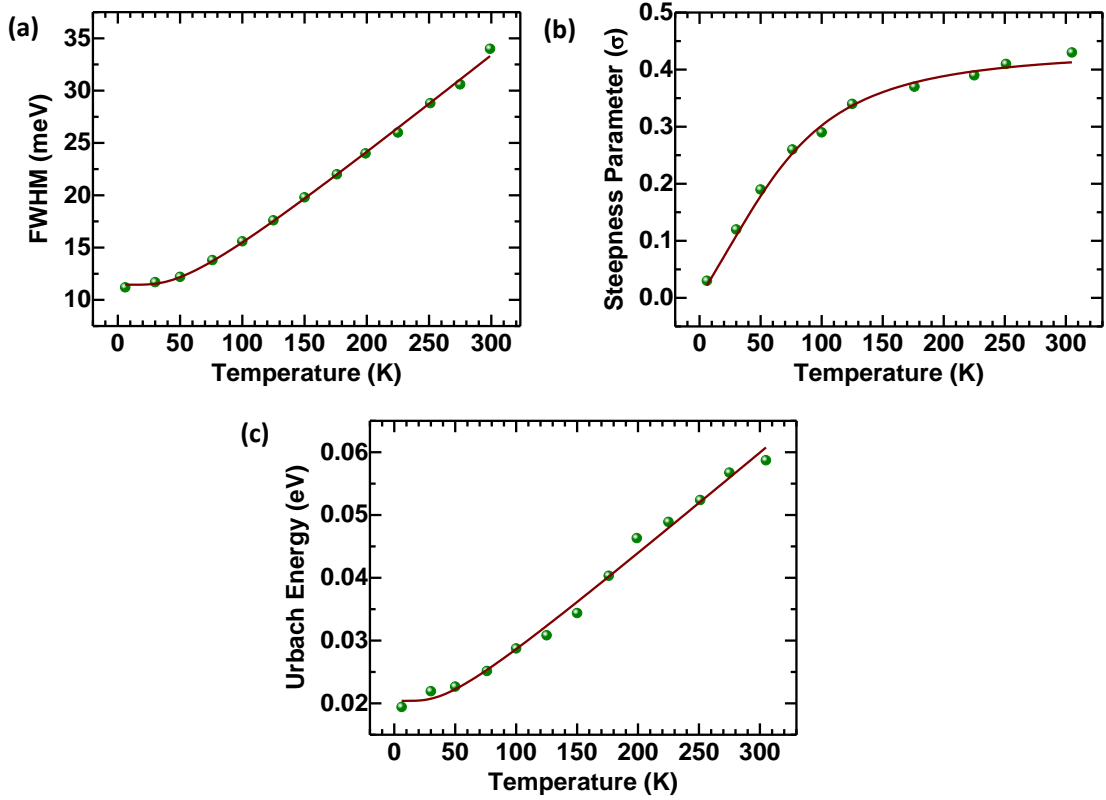

**Figure 6:** *(a) Variation of FWHM of excitonic peak with temperature in MAPbI₃. (b) Steepness parameter as a function of temperature for MAPbI₃. (c) Urbach energy as a function of temperature for MAPbI₃. Wine solid line is fit to the experimental data. Fittings are done without separating phases of MAPbI₃.*

**Table 2:** *Fitting parameters without separating the phases of MAPbI₃ film.*

| MAPbI₃ | Without separating phases |
|---|---|
| $\Gamma_0$ (meV) | $(11.45 \pm 0.21)$ |
| $\Gamma_{ep}$ (meV) | $(14.55 \pm 1.67)$ |
| $E_{LO}$ (meV) | $(13.14 \pm 1.21)$ |
| $\Theta$ (K) | $(223 \pm 19)$ |
| P | $(0.017 \pm 0.0.05)$ |
| $\sigma_0$ | $(0.433 \pm 0.009)$ |
| $h_{vp}$ (meV) | $(20.58 \pm 1.13)$ |
| $E_u$ (at T = 0 K) (meV) | 21.47 |
| $E_u$ (at T = 305 K) (meV) | 58.72 |

Figure **6** represents the FWHM of excitonic peak, steepness parameter and urbach energy as a function of temperature and the fitting of experimental data are done without separating the orthorhombic and tetragonal phases of MAPbI₃. The fitting parameters are listed in table **2** and it is found in good agreement with the fitted parameters listed in table **1**.

### 4.3.3   Photoluminescence

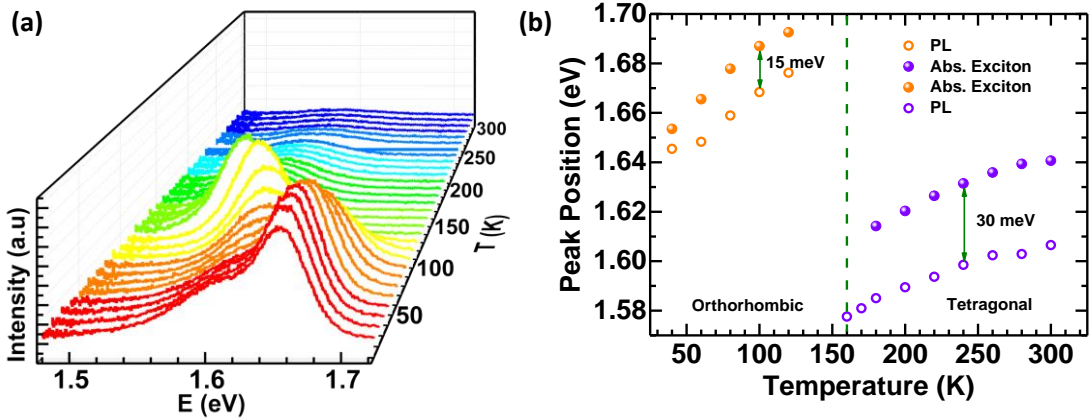

***Figure 7:*** *Photoluminescence (PL) of MAPbI₃ film on glass substrate as a function of temperature. (b) Stokes shift in PL peak versus excitonic absorption peak at different temperatures.*

Temperature dependent photoluminescence (PL) studies are carried out on MAPbI₃ films. Figure **7a** shows the PL spectra at different temperatures for the orthorhombic, (T< 167K) and tetragonal (T>167K) phase, respectively. The PL peak moves towards lower energy as temperature decreases in agreement with the band gap shift as seen in absorption measurements (Figure **7b**). For the PL spectra of the orthorhombic phase, we observe a remaining PL feature from the tetragonal phase around 1.6 eV in the vicinity of the transition temperature.[34,35,36] The temperature dependent energetic positions of the PL peak and the already determined excitonic peak from the absorption data is shown in Figure **7b**. The energy difference between absorption and emission is about twice as large for the tetragonal phase than for the orthorhombic phase. We consider this difference arises from the broadening of the density of states induced by thermal disorder and is beautifully consistent with the results obtained from the absorption spectra analysis. This conclusion can be further substantiated by considering the evolution of the photoluminescence spectra with temperature.

For the analysis of photoluminescence spectra we focus on the high energy edge where the spectrum is not affected by lower energy emission features such as remaining tetragonal incorporations or possible bound exctions.[8,34,36,37] We fit the blue edge by a superposition of a Gaussian peak – attributed to the inhomogeneously broadened DOS due to static disorder and exponential tails that reflect the additional broadening caused by thermal disorder (Figure **8**).[26] Considering the relative contributions of the Gaussian part (A) and the exponential contribution (B), we find that a Gaussian peak is fully sufficient to describe the emission below 60 K, yet from 60 K onwards there is an increasing exponential contribution (Figure **8c** and **8d**). This is nicely consistent with the increase in FWHM observed in the absorption spectra (Figure **3c**) and thus further supports our overall approach.

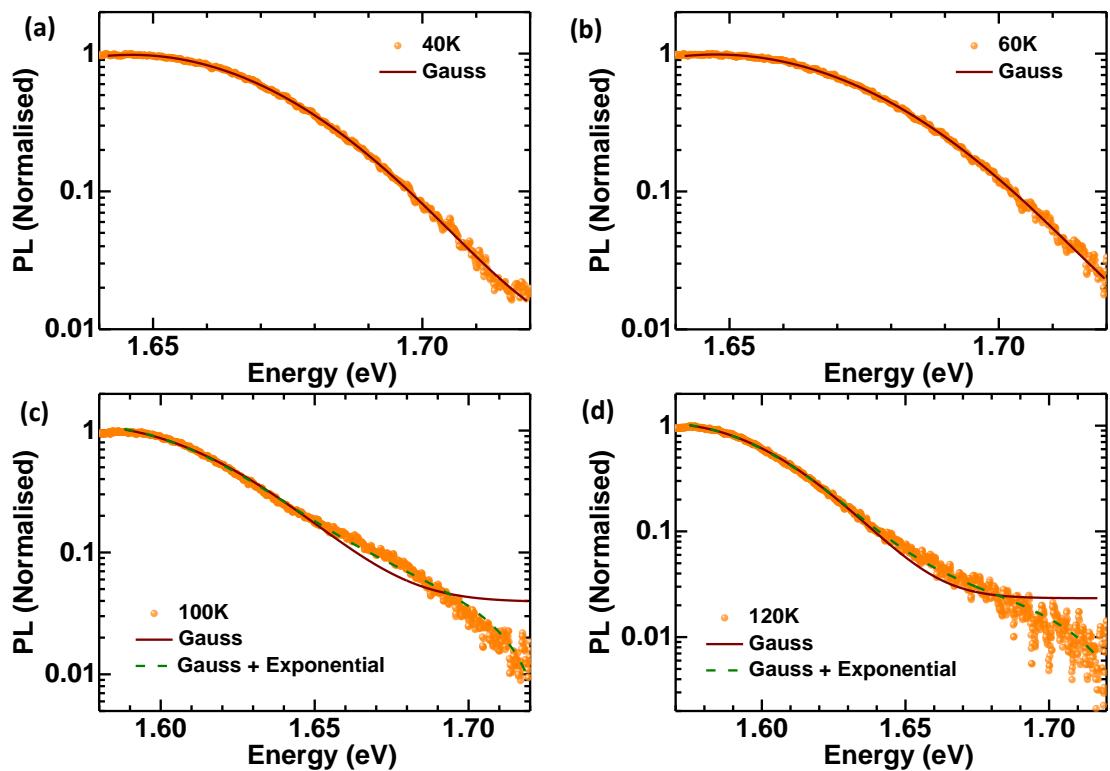

**Figure 8:** *PL spectra at (a) 40K (b) 60K, (c) 100 K and (d) 120K together with best fits of the high energy side of the spectra using either a superposition of a Gaussian and an exponential (dash green line) or only a Gaussian (wine line).*

## 4.4 Discussion

We have studied the temperature dependence of the optical properties of MAPbI$_3$ perovskite films. The onset of optical absorption is dominated by exciton absorption. The band gap increases with temperature which is in strong contrast with the decrease in band gap with temperature seen in most crystalline inorganic semiconductors. This is confirmed from both absorption and photoluminescence measurements. We show that the positive temperature coefficient of the bandgap relates to the large temperature coefficient of lattice expansion in these materials. Lattice dilatation plays a much more significant role than electron–phonon interactions. The volume coefficient of MAPbI$_3$ is (3.837±0.018) x $10^{-4}$ K$^{-1}$, which is ~ 120 times larger than that for crystalline Si. A model, using Einstein oscillators fits to the exciton linewidth **$\Gamma_{EX}$** indicates that the orthorhombic phase is slightly more ordered than the tetragonal phase. The low energy dependence of the exciton absorption is given by an exponential absorption tail reminiscent and consistent of classical Urbach absorption. From the temperature dependence of the Urbach energy, we estimate the disorder and show that it is surprisingly small in these samples, remarkable for solution processed semiconductors. The analysis on temperature dependent PL data consistently supports this picture. It allows to analyse the impact of static and dynamic disorder on the spectra by differentiating between Gaussian and exponential contributions to the shape of the PL. Analysis of temperature dependent PL broadening has been used widely to understand the physics of electron phonon coupling in perovskite semiconductors. The evolution of FWHM of the PL can be fitted by taking in account of temperature independent inhomogeneous broadening ($\Gamma_0$) and interaction between charge carrier (electron) and LO-phonon coupling which is described by Frouhlich Hamiltonian[38]

$$\Gamma(T) = \Gamma_0 + \frac{\Gamma_{LO}}{exp\left(\frac{E_{LO}}{k_B T} - 1\right)} \qquad (8)$$

Where, $\Gamma_{LO}$, $E_{LO}$ and $k_B$ are electron-phonon coupling strength, energy of LO phonon and Boltzmann constant, respectively. Equation 8 is similar to equation 4. The comparison of the electron phonon coupling constant and phonon energy for the MAPI based thin films and single crystal (SC) are listed in table 3 using PL technique.

***Table 3:*** *Reported linewidth (FWHM) parameters for MAPI based thin films and SC.*[38,39,40,41,42,43] *{PL = Photoluminescence, PC = Photocurrent, Abs = Absorption}*

| Techniques | $\Gamma_0$ (meV) | $\Gamma_{LO}$ (meV) | $E_{LO}$ (meV) | Reference |
|:---:|:---:|:---:|:---:|:---:|
| PL (SC) | - | 62±1.6 | 18.7±0.8 | **39** |
| PL (SC) | 4 | 37.3 | 16.3 | **40** |
| PC (SC) | 4.1±0.4 | 57±22 | 16.1±3.4 | **41** |
| PL (Thin Film) | 39 | 40 | 15 | **38** |
| PL (Thin Film) | 26±2 | 40±2 | 11.5 | **42** |
| Abs (Thin Film) | 11.5 ± 0.2 | 14.6 ± 1.7 | 13.2 ± 1.2 | **43*** |
| | | | | ***present work** |

It is known that perovskite SC shows lower density of defect states in comparison to the perovskite thin films. This is quite evident from such a small value of $\Gamma_0$ in the case of perovskite SC. In general, the electron-phonon coupling strength is higher in the case of perovskite SC than the thin films (**table 3**) due to low defect density in the former one. In thin films, charge carrier easily trapped in the defect states which results into the screening for interaction between charge carriers and phonons.[42] Thus, the electron-phonon coupling strength weakens in the case of thin films. However, the defect density in the case of perovskite SC can also vary depending upon the fast or slow crystallization and thus it results into different coupling strength.[39,40]

## Part II: Modulation of electronic states of hybrid lead halide perovskite embedded in organic matrix

### 4.5 Introduction

Second part of this chapter deals with the average crystal size effect of the perovskite nanocrystals (PNCs). The band gap of the perovskite materials can be easily tuned over the entire visible and near IR wavelength, by changing the composition of the halide part in it.[6,44,45] Recently colloidal nanocrystals based on halide perovskite are becoming popular in the field of perovskite based optoelectronics due to their superior photoluminescence (PL) properties and easy band gap tunability.[46] Butkus *et al.* have used broadband ultrafast transient absorption spectroscopy to examine the onset of quantum confinement in size-tuned cubic CsPbBr$_3$ NCs compared with the bulk material.[47] Sapori *et al.* described the quantum confinement in nano-platelets of CsPbX$_3$ and hybrid organic–inorganic perovskites in 2D and 3D structures.[48] In an alternative approach, Di *et al.* have shown the size of the perovskite crystals can be controlled by blending MAPbBr$_3$ with 4,4'-bis[9-dicarbazolyl]-2,2'-biphenyl (CBP).[49] Later on, Li *et al.* have successfully demonstrated efficient perovskite based light emitting diodes based on perovskite nanocrystals prepared by similar method where MAPbBr$_3$ is blended with polyimide precursor dielectric (PIP).[50] Masi *et al.* show that the blending of MEH-PPV (poly[2-methoxy-5-(2-ethylhexyloxy)-1,4-phenylenevinylene]) with MAPbI$_3$ helps to produce a smooth layer of perovskite by controlling the crystalline domain formation of perovskite in polymer matrix.[51] In many of such nanocomposite studies, either PL position did not change or even when it got blue shifted could be hypothesized using Mie modes or formation of 2D perovskites, which can complicate the understanding for these experimentally observed results.[52] Hence, it is important to study this using wider variety of test samples to understand these optical properties of organometallic halide perovskites in good details.

Herein, an easy alternative way of growing MAPbI$_3$ and MAPbBr$_3$ based PNCs by blending it with small molecular organic semiconductor (BCP and CBP, separately) is demonstrated and thus, tuning of the photoluminescence (PL) peak position over a wide range of electromagnetic spectrum. MAPbI$_3$ PNCs embedded in BCP organic matrix shows a blue shift of ~200 meV at volumetric ratio of 1:3 (Pe:BCP) with crystallite size of ~11 nm. This smaller perovskite crystallite size suggest that PL tuning can be considered to fall into weak quantum confinement regime and a classical particle in a box model can explain experimental findings with good level of accuracy with pretty reasonable fitting parameters related to MAPbI$_3$. However, MAPbBr$_3$ being relatively low dielectric constant material has much smaller exciton Bohr radius and also poor electronic confinement with used organic semiconductor host matrix which provides relatively lesser blue shift of PL peak with respect to PNCs size.[46] Along with this, we also observed higher tuning of PL peak position in case of MAPbI$_3$:BCP blend than in MAPbI$_3$:CBP blend for a given volumetric concentration of the organic component in the fixed concentration of perovskite precursor solution. Out of all four combinations, it is found that MAPbI$_3$ PNCs embedded in BCP organic matrix have highest PL tuning of ~200 meV.

**4.6 Experimental section**

***Preparation of MAPbI₃ and MAPbBr₃ precursor solution****:* $CH_3NH_3I$, $CH_3NH_3Br$ and $Pb(Ac)_2.3H_2O$ is purchased from Greatcell and sigma Aldrich, respectively. A 20 wt.% $MAPbI_3$ precursor has been prepared with 3:1 molar ratio of $CH_3NH_3I$ and $Pb(Ac)_2.3H_2O$ in anhydrous DMF. Similarly, a 20 wt.% $MAPbBr_3$ precursor has been prepared with 3:1 molar ratio of $CH_3NH_3Br$ and $Pb(Ac)_2.3H_2O$ in anhydrous DMF. Both the solutions are stirred at $70^0C$ for 30 minutes before use.

***Preparation of CBP and BCP solution:*** CBP *and* BCP (sigma Aldrich) has been dissolved in anhydrous DMF with concentration 15mg/ml and kept for stirring at $70^0C$ for 30 minute to get a clear solution.

***Composite synthesis and film preparation:*** $MAPbI_3$ and each of the organic (CBP and BCP) precursors are mixed together in desired volumetric ratio to get different proportion of $MAPbI_3$: CBP/BCP blends. The perovskite and the composite solutions are spin coated on alumina coated glass substrate at 2000 rpm for 45 seconds and subsequently annealed at $100^0C$ for 5 minutes.

***XRD measurement:*** The XRD measurement are carried out in Rigaku smartlab diffractometer with Cu $K_\alpha$ radiation (λ=1.54Å). 2θ scan has been carried out from $10^0$-$50^0$ with step size $0.01^0$.

***Steady state PL and absorption measurements:*** The steady state PL spectra are obtained by exciting the samples by 355 nm pico- second pulsed laser with repetition rate 1kHz at the fluence of 50μJ/$cm^2$. The UV-Vis absorption spectra have been recorded using Perkin Elmer lambda 950 UV-Vis spectrometer in the integrating sphere mode.

**4.7 Results**

**Figure 9a** shows the typical crystal structure of perovskite material ($MAPbX_3$), where A = $CH_3NH_3^+$, B = $Pb^{2+}$ and X is the anion from halogen group (X= $Br^-$ and/or $I^-$). The chemical structures of CBP (4,4'-bis[9-dicarbazolyl]-2,2'-biphenyl) and BCP (bathocuproine) molecules are also shown in **figure 9a**. **Figure 9b** shows the energy levels of the $MAPbI_3$/ $MAPbBr_3$ perovskite along with CBP and BCP organic semiconductors, which forms a type-I hetero-structure with perovskite. Energy level values for perovskites and organic molecules are taken from literature.[53,54] **Figure 9c and 9d** shows the first order (110) X-ray diffraction (XRD) peak of MAPbI$_3$ in the $2\theta$ range between $13.5^0$ to $14.7^0$ for the pristine MAPbI$_3$ and CBP & BCP matrix based blend thin films on glass substrate for different volumetric concentration of organic components, respectively. Whereas, **Figure 9e and 9f** shows the first order XRD peak of MAPbBr$_3$:CPB and BCP matrix based blend films in the $2\theta$ range between $14.5^0$ to $15.5^0$, respectively.

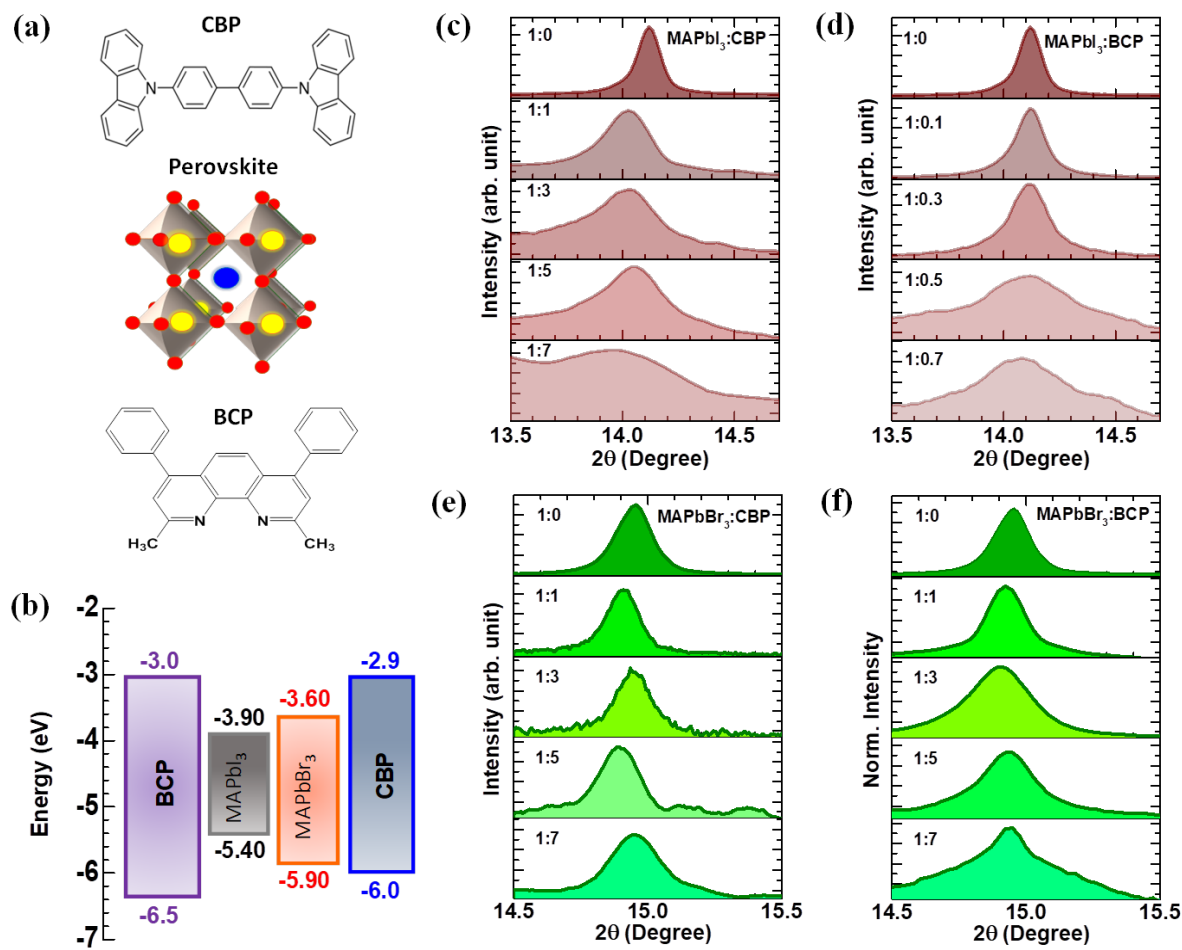

*Figure 9:* (a) Typical crystal structure of perovskite and chemical structure of CBP and BCP molecules. (b) Energy levels alignment of MAPbI$_3$/MAPbBr$_3$ with BCP and CBP. First order (110) XRD peak for (c) MAPbI$_3$ : CBP, (d) MAPbI$_3$ : BCP, (e) MAPbBr$_3$ : CBP, and (f) MAPbBr$_3$ : BCP based perovskite-organic blend thin films.

The XRD peak positions with error bar of MAPbI$_3$ and MAPbBr$_3$ with CBP and/or BCP blend films are listed in **table 4** and **table 5,** respectively. It is observed that the (110) XRD peak of MAPbI$_3$ based film gets broader (**figure 9c** and **figure 9d**) with increase in the CBP and the BCP concentrations. However, the peak broadening effect is stronger in the case of the BCP based blend films. The corresponding full width half maxima (FWHM $=\beta_{FWHM}$) of the first order diffraction peak can be correlated with the crystallite size ($D_{avg}$) of the pristine MAPbI$_3$ and the blend films.

**Table 4:** *Average crystallite sizes and 1$^{st}$ order XRD peak position for the pure MAPbI$_3$ and CBP/BCP based composite films.*

| MAPbI$_3$:CBP | 2θ (Degree) | Average Crystallite Size (nm) | MAPbI$_3$:BCP | 2θ (Degree) | Average Crystallite Size (nm) |
|---|---|---|---|---|---|
| 1:0.0 | (14.12±0.01) | 104 | 1:0.0 | (14.12±0.01) | 104 |
| 1:0.1 | (14.12±0.01) | 95 | 1:0.1 | (14.12±0.01) | 77 |
| 1:0.3 | (14.12±0.01) | 88 | 1:0.3 | (14.11±0.01) | 59 |
| 1:0.5 | (14.11±0.01) | 81 | 1:0.5 | (14.10±0.01) | 19 |
| 1:0.7 | (14.10±0.02) | 62 | 1:0.7 | (14.05±0.02) | 17 |
| 1:1.0 | (14.03±0.02) | 50 | 1:1.0 | (14.04±0.02) | 15 |
| 1:3.0 | (14.03±0.01) | 45 | 1:3.0 | (14.01±0.02) | 11 |
| 1:5.0 | (14.05±0.01) | 43 | 1:5.0 | - | - |
| 1:7.0 | (13.97±0.02) | 22 | 1:7.0 | - | - |
| 1:10 | (13.83±0.02) | 20 | 1:10 | - | - |

**Table 5:** *Average crystallite sizes and 1^st order XRD peak position for the MAPbBr₃ and CBP/BCP based composite films.*

| MAPbBr$_3$:CBP | 2θ (Degree) | Average Crystallite Size (nm) | MAPbBr$_3$:BCP | 2θ (Degree) | Average Crystallite Size (nm) |
|---|---|---|---|---|---|
| 1:0.0 | (14.94±0.01) | 110 | 1:0.0 | (14.94±0.01) | 110 |
| 1:0.1 | (14.94±0.01) | 108 | 1:0.1 | (14.94±0.01) | 106 |
| 1:0.3 | (14.94±0.01) | 106 | 1:0.3 | (14.93±0.01) | 102 |
| 1:0.5 | (14.94±0.01) | 104 | 1:0.5 | (14.93±0.01) | 97 |
| 1:0.7 | (14.94±0.02) | 101 | 1:0.7 | (14.93±0.02) | 91 |
| 1:1.0 | (14.91±0.02) | 97 | 1:1.0 | (14.92±0.01) | 79 |
| 1:3.0 | (14.94±0.01) | 89 | 1:3.0 | (14.90±0.01) | 46 |
| 1:5.0 | (14.89±0.01) | 76 | 1:5.0 | (14.94±0.02) | 57 |
| 1:7.0 | (14.95±0.02) | 72 | 1:7.0 | (14.94±0.02) | 25 |
| 1:10 | (14.93±0.02) | 60 | 1:10 | (14.93±0.02) | 23 |

We use Debye-Scherrer formula ($D_{avg}=k\lambda/\beta\cos\theta$, k is the shape factor, λ is the X-ray wavelength, β is the FWHM of XRD peak, θ is the Bragg angle) to determine the average crystallite size of the PNCs in the organic matrix using the first order (110) diffraction peak (**table 4 and 5**).

   **Figure 10a and 10b** shows the steady state PL spectra of the pure MAPbI₃ and corresponding CBP and BCP based blend thin films. The PL peak shift towards the blue side of the electromagnetic spectrum with increase in CBP and BCP concentrations. However, the amount of blue shift is more prominent in the case of BCP blend films in comparison to CBP blend films. **Figure 10c and 10d** represents the steady state PL spectra of MAPbBr₃ films blended with different concentration of CBP and BCP. We notice that the PL peak is tuned over ~200 meV (1.63 eV to 1.83 eV) for MAPbI$_3$:BCP blend films whereas it is only ~ 40 meV (2.37 eV to 2.41 eV) in case of MAPbBr$_3$:BCP blend films for the similar vol% of BCP.

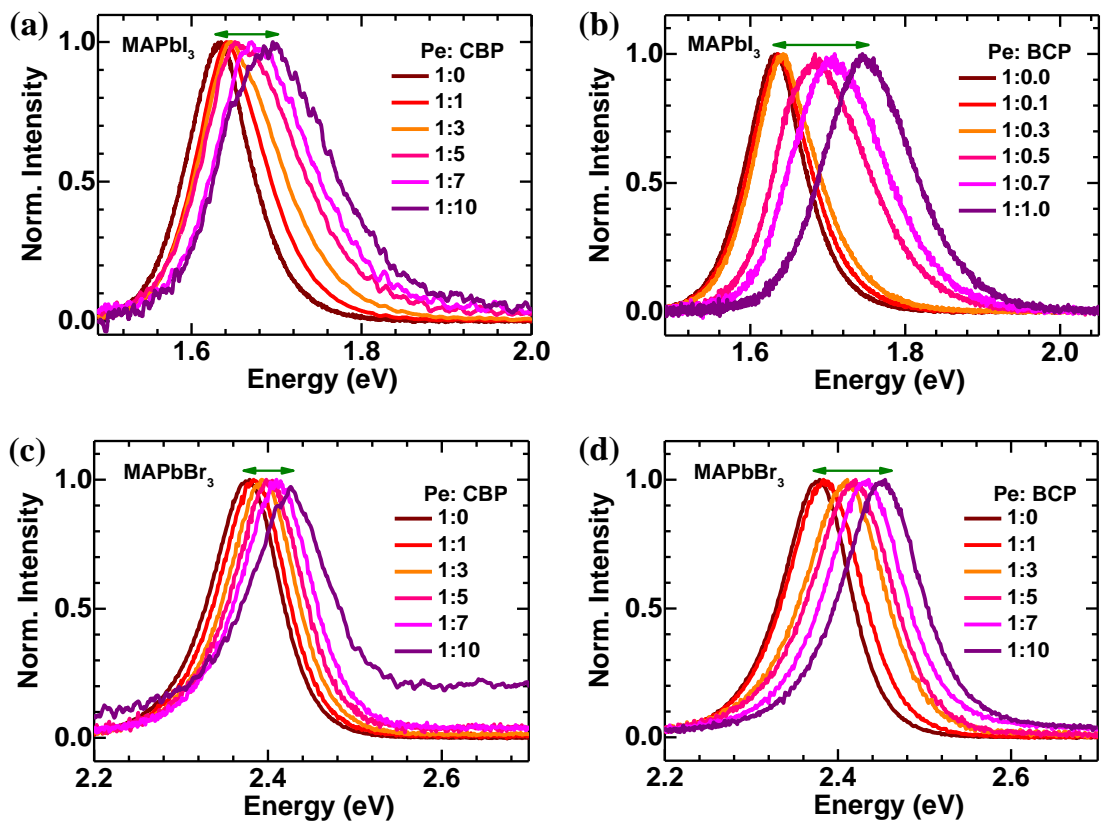

***Figure 10:*** *Photoluminescence (PL) spectra of the (a) MAPbI$_3$ :CBP, (b) MAPbI$_3$ :BCP, (c) MAPbBr$_3$ :CBP and (d) MAPbBr$_3$ :BCP, blend films with different concentration of organic compound. The PL peaks get blue shifted with increase in organic concentration (both for CBP and BCP blending). However, the effect is much stronger for BCP additive films.*

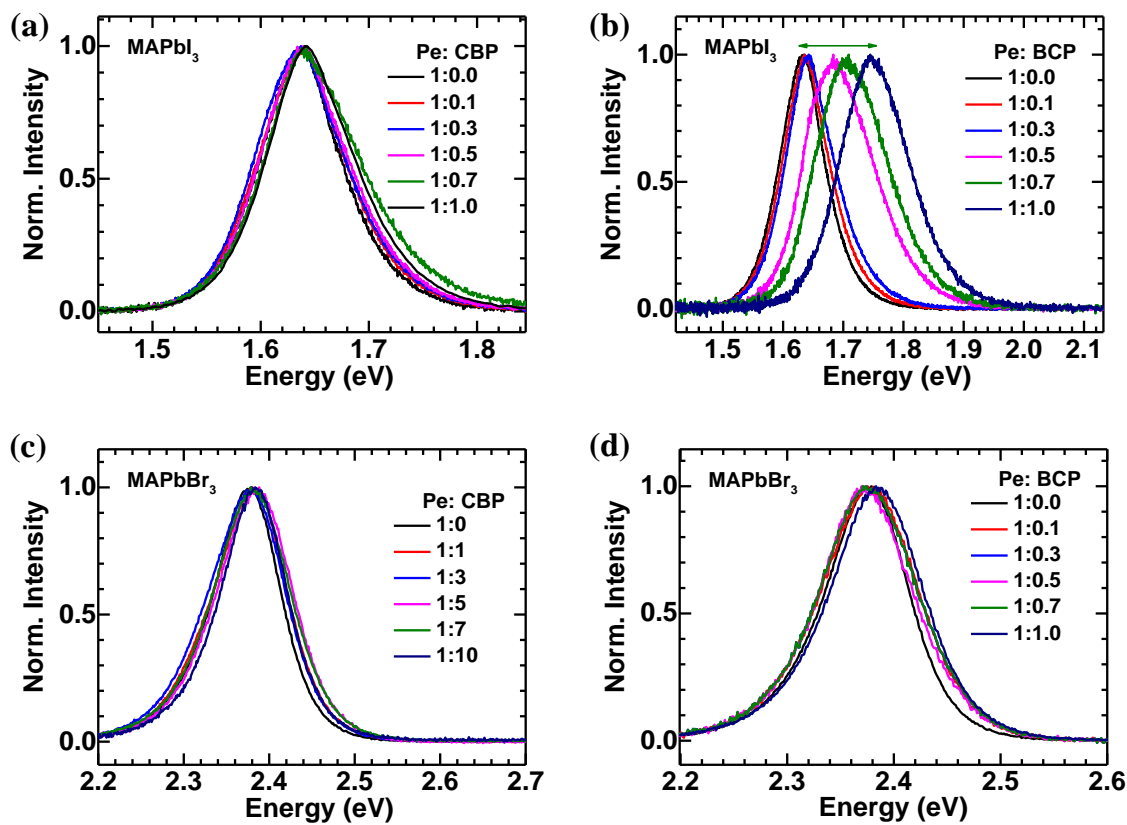

***Figure 11****: Photoluminescence (PL) spectra of the (a) MAPbI₃ :CBP, (b) MAPbI₃ :BCP, (c) MAPbBr₃ :CBP and (d) MAPbBr₃ :BCP, blend films with different concentration of organic compound.*

PL tuning of perovskite:organic composite based films at lower concentration of organic semiconductor is also shown in **figure 11**. UV-Vis absorption spectra of MAPbI₃:BCP based films also shows blue-shift in the band edge (**figure 12a**) with increase of BCP in MAPbI₃, in good agreement with steady state PL peak position. However, the PL peak energy does not show significant blue-shift in the case of MAPbI₃:CBP, MAPbBr₃:CBP and MAPbBr₃:BCP based films for blending ratio upto 1:0.7 (Perovskite: organic), which is evident in the UV-Vis absorption spectra as well (**figure 12b and 13**). However, for volumetric ratio of 1:1, these films show a slight blue-shift in the PL peak energy in agreement with absorption excitonic peak.

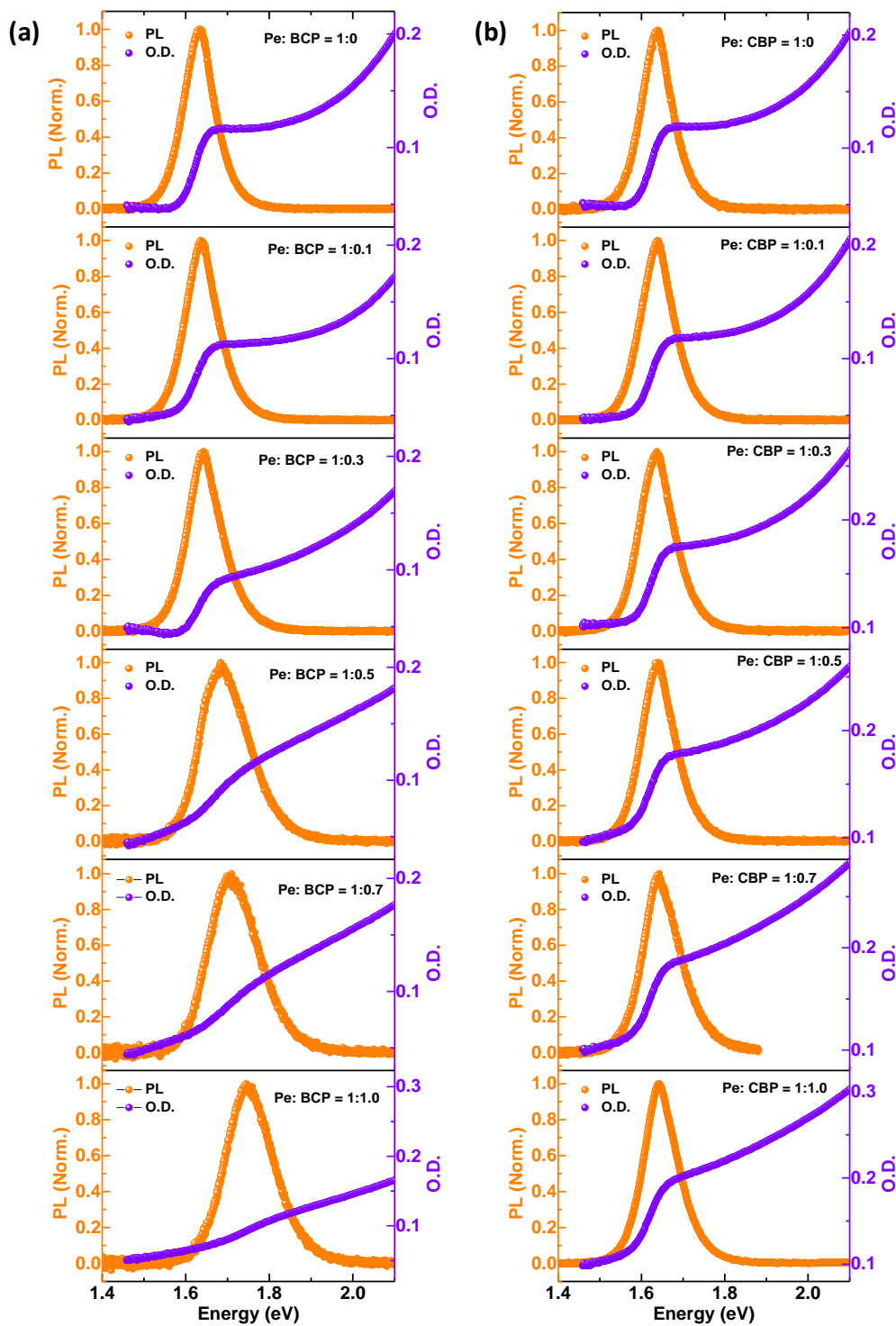

**Figure 12:** *PL and UV-Visible absorption spectra for MAPbI₃ based perovskite thin films blended with different volumetric ratio of (a) BCP and (b) CBP.*

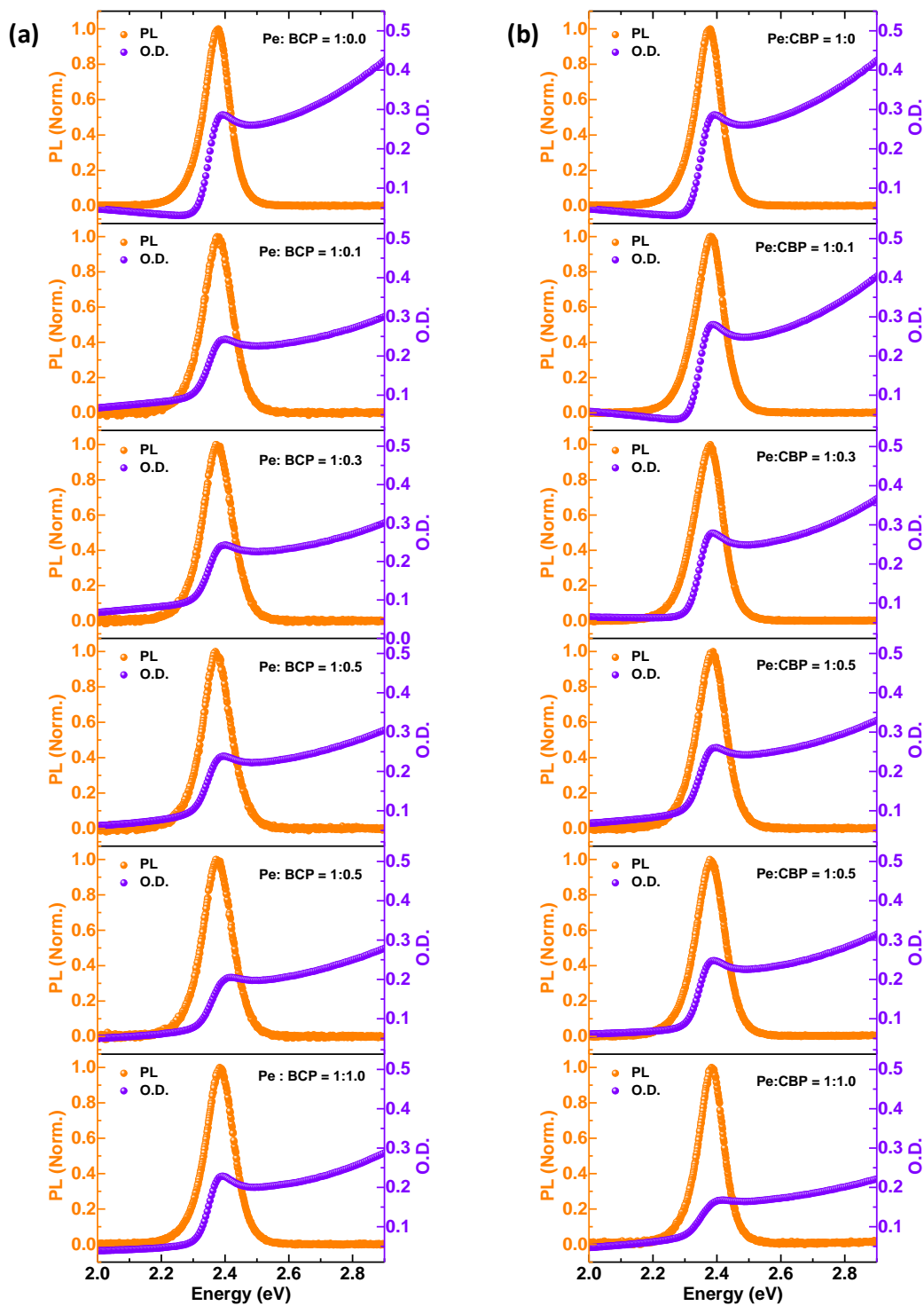

**Figure 13:** *PL and UV-Visible absorption spectra for MAPbBr₃ based perovskite thin films blended with different volumetric ratio of (a) BCP and (b) CBP.*

We observe that PL does get broader for higher concentration of organic matrix (**figure 12**). Higher organic concentration in the perovskite precursor solution increases the disorder in the final perovskite film, which can be understood by higher urbach energy (1/slope of absorption) of the 1:10 (MAPbI$_3$:BCP) in comparison to the pristine one (1:0).[55] In our earlier publication, we have done first principle calculations by considering the size of BCP molecule that it does not enter into perovskite structure and stays only at grain boundaries via physisorption at a distance of 0.29 nm.[56] For CBP also, it was verified by small angle x-ray diffraction studies to confirm that it does not form 2D structure (unpublished work). This indicates the blue-shifted PL in case of blend films originates from the perovskite crystals only. The blue shift in the PL spectra are explained by quantum confinement effect where the total band gap consists of the band gap of the bulk material as well as bang gap enhancement factor due to quantum confinement.[57] However, in the present case, the PNCs size is larger than the exciton Bohr radius which is generally below 5 nm in such hybrid perovskite semiconductor.[58] Hence this confinement effect can be considered as a weak confinement effect and attributed to lighter effective reduced mass of carriers in these semiconductor systems.[59] The optical band gap (E$_g$) *vs* crystallite size has been fitted with particle in box approach using following relation:[49]

$$E_g = E_{g,bulk} + \frac{3\hbar^2\pi^2}{2m^*D^2}$$

Where, $E_{g,bulk}$ is the band gap of bulk component, $D$ is the average size of the PNCs and $m^*$ is the effective reduced mass of the carriers. The fitted curve gives the effective reduced mass of ~0.093 $m_e$ and 0.097$m_e$ (**figure 14a and 14b**) for BCP and CBP blended MAPbI$_3$ based films, respectively, which is in good agreement with literature.[60,61,49] However, the optical band gap *vs* crystallite size of MAPbBr$_3$:BCP/CBP blend films cannot be fitted with equation (1) due to higher crystallite size and thus, it will not come into quantum confinement regime (figure **14c** and **14d**).

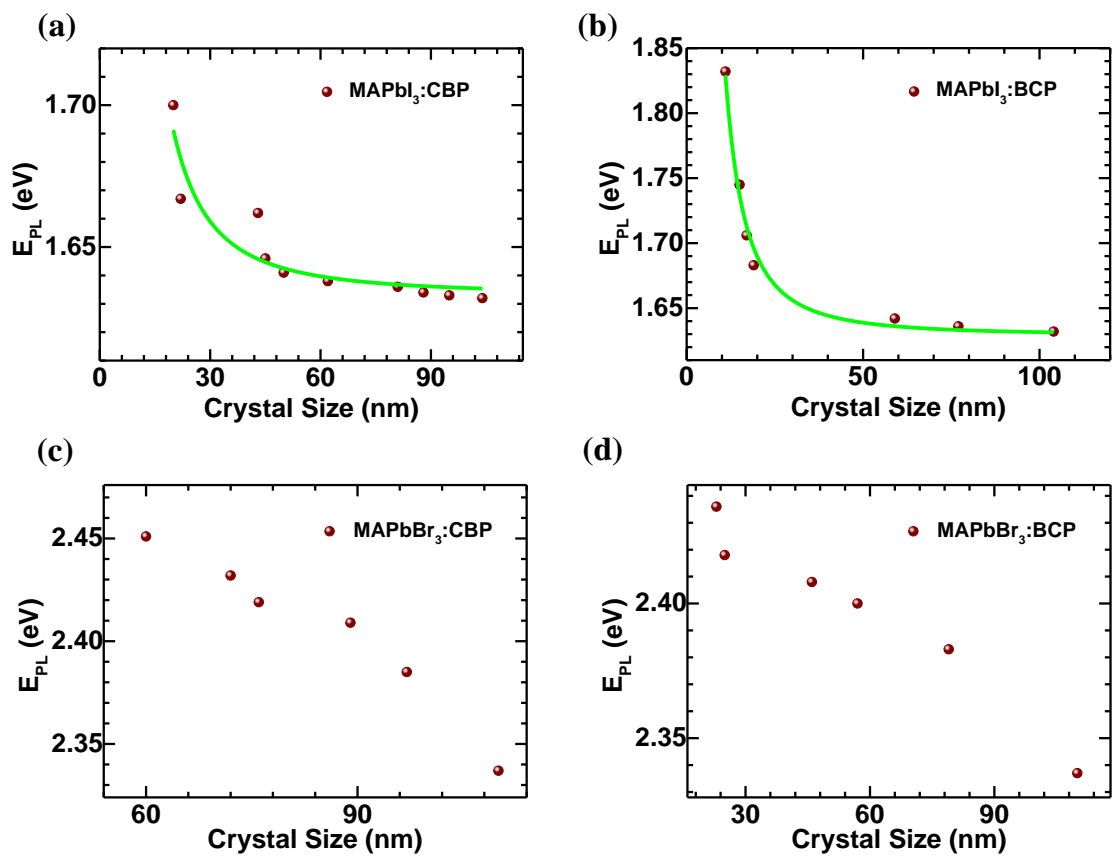

**Figure 14:** *The energy of the PL emission peak as a function of the average PNC size for (a) MAPbI₃:CBP, (b) MAPbI₃:BCP, (c) MAPbBr₃:CBP and (d) MAPbBr₃:BCP based blend perovskite films.*

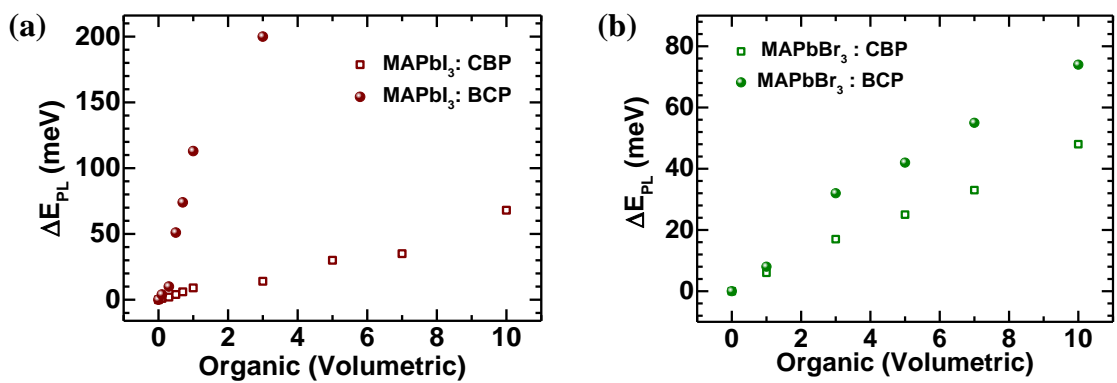

**Figure 15:** *Shift in PL peak position on energy scale of (a) MAPbI₃ and (b) MAPbBr₃ with different volumetric concentration of organic molecules (BCP or CBP).*

### 4.8    Discussion

Our results indicate that the band gap of the perovskite material can be tuned (from ~110 nm to ~11 nm) effectively within the organic matrix of CBP and BCP,[54,62] which forms type-1 hetero structure with the bulk perovskite material. Therefore, it reduces the bulk crystal sizes by forming a quantum well type structure with organic semiconductors. However, we observed variation in the crystallite size of $MAPbI_3$ *vs* $MAPbBr_3$ for a particular volume concentration of the organic semiconductor (either CBP or BCP). We correlate this discrepancy with the charge carrier confinement in the perovskite crystal surrounded by organic molecular matrix. There are energy offset ($\Delta E_{offset}$) for the valance band maxima (VBM) and conduction band minima (CBM) in this hetero-structure in the range of ~ 1000 meV and ~900 meV, respectively, in case of $MAPbI_3$:BCP hetero-structure. However, the $\Delta E_{offset}$ for VBM and CBM are ~600 meV and ~1000 meV in case of $MAPbI_3$:CBP hetero-structure, respectively. $MAPbBr_3$ have deeper VBM (5.9 eV) than $MAPbI_3$ (5.4 eV)[45,53] and thus the $\Delta E_{offset}$ between VBM of $MAPbBr_3$ and organic molecules (BCP or CBP) further reduces. In case of $MAPbBr_3$:BCP blend, the $\Delta E_{offset}$ is ~600 meV and it reduces to ~ 100 meV for $MAPbBr_3$:CBP blend. Hence, as per basic quantum mechanics, the hole wave function will be partially leaked in case of $MAPbBr_3$:CBP blend due to low $\Delta E_{offset}$ (<$4K_BT$) and thus, the shift in the PL peak is only ~48 meV with a blend volume ratio of 1:10 ($MAPbBr_3$:CBP). However, even a lower vol% blending of BCP in $MAPbI_3$ (1:3) provides a PL peak shift of ~200 meV due to more effective hole confinement ($\Delta E_{offset}$ ~1010 meV). As shown in **figure 15**, PL peak position of PNCs in BCP matrix increases more rapidly in comparison to CBP matrix as we increase the organic concentration. Therefore, a larger offset in the VBM levels can cause a better quantum confinement resulting in higher amount of PL energy tuning (~200 meV) in the case of $MAPbI_3$ PNCs embedded in BCP matrix due to successful energy transfer at the type - I  quantum well hetero-structures.

### 4.9 Conclusion

In summary, our study in chapter 4 establishes a methodology using optical techniques for the concise characterization of disorder in the perovskite materials. It correlates fundamental aspects such as lattice dilatation and static & dynamic disorder to the optical properties of perovskite semiconductors. This provides important insights into the electronic and structural properties of $MAPbI_3$ based perovskites which should be transferable to many related organic metal-halide perovskites. In addition, this chapter also shows an easy way of processing PNCs by blending the bulk perovskite with wide band gap organic semiconductor (BCP and/or CBP) forming type-1 hetero-structure. This, in turn, largely tunes the band gap over a wider range of electromagnetic spectrum due to weak confinement. We also demonstrate that wide range of band gap tuning can be explained broadly by Bohr radius vs average size of crystallites in efficient Type-1 hetero-structure, as it is clear from the CBP or BCP blending in $MAPbI_3$ vs $MAPbBr_3$. Considering the relatively higher crystallite size (<50 nm), our results suggest that along with weak confinement, the $\Delta E_{offset}$ in the valance band (VB) of both the semiconductors might also play a role in deciding the amount of PL shift. This room temperature solution processable perovskite organic blending with wide band gap tunability in the organic matrix opens up an easy alternative fabrication process of perovskite-organic composite for application in the field of color tunable LEDs and other optoelectronic devices. We present a generalized study on two organic matrix *vs* two halide perovskite which brings a clear picture of size dependence as strong mechanism to explain these complex results in simpler manner. Moreover, this study could also provide some of the semiconductor parameters with much simpler experimental studies with ease of sample preparation and ease of characterization studies.

## 4.10    Post Script

Overall, this chapter demonstrates the photo-physics of organic-inorganic hybrid metal halide based perovskite semiconductors. This chapter provides important insights into the electronic and structural properties of $MAPbI_3$ based perovskites which should be transferable to many related organic metal-halide perovskites. Combine study of temperature dependent absorption and emission spectroscopy reveals some surprising facts about two different phases of perovskite semiconductor. It also helps to understand the defects and disorder in the perovskite semiconductor even at room temperature and advised to apply some molecular or solvent additives engineering in order to reduce the defect states from the bulk as well as from the grain boundaries. In addition, an easy technique for modulation of electronic states of $MAPbX_3$ PNCs embedded in organic matrix is demonstrated and could open new doors in the field of optoelectronic devices such as wavelength tunable LEDs. However, further improvement in the size and distribution of PNCs is required, which can extend the range of emission tunability.

# Chapter 5

# Solvent Additive Engineering for Improved Photovoltaic Performance via Controlling the Crystal Growth Kinetics and Defect States

# CHAPTER 5

## Solvent Additive Engineering for Improved Photovoltaic Performance via Controlling the Crystal Growth Kinetics and Defect States

**Abstract:** This chapter deals with the role of iodine rich solvent additive 1,8-diiodooctane (DIO) in wide band gap $CH_3NH_3PbBr_3$ ($MAPbBr_3$) perovskite films and phosphinic acid in $CH_3NH_3PbI_3$ ($MAPbI_3$) based perovskite films, which provides a favorable chemical environment for film formation and reduces the defect states. In case of DIO, we experimentally verify the proposed model where Iodine being a soft Lewis base has a tendency to chelate with soft Lewis acid (lead, Pb). The I-Pb chelation is pretty evident in XRD and can be understood by red shift in absorption edge & photoluminescence (PL) peak position. Ultra-violet photoelectron spectroscopy (UPS) studies suggest that electronic states of perovskite thin films modulate with the fractional concentration of DIO in perovskite precursor solution. Furthermore, Chemical analysis of perovskite films shows that lower binding energies of Bromine (Br 3d) and lead (Pb 4f) for solvent additive films, supports the chelation of $I^-$ with Pb by replacing $Br^-$, which is in good agreement with optoelectronic and structural studies. Hence, incorporation of DIO into the perovskite precursor solution alters the kinetics of crystal growth and nucleation *via* Ostwald ripening and ion exchange approach, which results in the improved performance of perovskite solar cells and light emitting diodes. However, in case of phosphinic acid, we observed an improved morphology of the perovskite film with lower number of grain boundaries. Addition of phosphinic acid results in suppressed non-radiative recombination by reducing the number of defect states at grain boundaries and at surface, which is pretty evident by enhanced steady state and time resolve photoluminescence lifetime in solvent additive films. The reduction in defect state can be also understood by independent behavior of $EL_{QE}$ with respect to applied bias in phosphinic acid additive based perovskite solar cells. The suppressed non-radiative recombination enhances the overall power conversion efficiency (PCE) of MAPbI$_3$ based solar cells up to 16.2 % .

## 5.1    Introduction:

In literature, there are various processing techniques including one-step solution deposition[1,2], two-step sequential deposition[3,4], vapor assisted solution deposition[5], solvent additive,[6] anti-solvent treatment,[7] solvent-solvent extraction,[8] vacuum deposition and atomic layer deposition[2] used to improve the morphology and crystallinity of perovskite thin film. The perovskite film morphology and crystallinity can be affected by nucleation and growth rate of the deposited thin film. As per literature, partial substitution of Cl$^-$ with I$^-$ prolong the crystallization process. The evidence of this prolonged crystallization is the requirement of 2 to 3 hours annealing time in mixed halide perovskites films.[9] However, pure iodine perovskites require only less than 30 minute annealing.[10] Since perovskite is known to have defects and disorder in the bulk as well as on the grain boundaries, there is a need of some chemical engineering or technique to reduce the number of defect states in order to improve the photovoltaic performance. This chapter will deal with two different kind of solvent additives; one is iodine based additive 1,8-diiodooctane (DIO) and other is a strong reducing agent (phosphinic acid). Among all the above mentioned processing techniques, solvent additive method is commonly used in the perovskite community to get a compact and smooth perovskite thin film. The choice of solvent additive depends mainly on the two factors: (1) solvent additive should have higher/lower boiling point/vapor pressure than the host solvent (DMF or DMSO); (2) Solvent additive should increase the solubility of perovskite in the host solvent.[11] The role of these additives in altering the kinetics of crystallization and thus effect on the structural, morphological and optoelectronic properties of perovskite thin films will be discussed. Firstly, we present the insight on the perovskite film formation with and without (w/o) DIO additive, with different weight (wt.) % of DIO into the perovskite precursor, supported by detailed chemical and electronic states analysis (using photoelectron spectroscopy). We show that there is a favorable chemical environment to improve the film morphology by addition of DIO in perovskite solution, which prolong the time for film formation. We will also analyze the effect of DIO additive on MAPbI$_3$ and MAPbI$_{3-x}$Cl$_x$ based perovskite systems to compliment the study done on MAPbBr$_3$. The focus of this study is to investigate the role of DIO on optoelectronic properties and chemical analysis of MAPbBr$_3$ films and connect it to device results in terms of efficiency. However, another solvent additive phosphinic acid being a reducing agent maintain the iodine (I) *vs* lead (Pb) stoichiometry in the final perovskite film by reducing the oxidation of I$_2^-$ into I$^-$. Reduction in non-radiative recombination channels can be understood by enhanced steady state photoluminescence (PL) and time resolved PL lifetime. Addition of phosphinic acid into the perovskite precursor solution provides an improved morphology with increased domain size, which helps in increasing the charge transport rate in the perovskite solar cells (PSCs). As a results of reduction in defect states due to increased domain size, the phosphinic acid additive based solar cells shows an improved power conversion efficiency of 16.2%.

## Part I: DIO as iodine rich Lewis base solvent additive

## 5.2    Introduction

Halide based additive are used in the perovskite precursor solution to balance the Pb:I stoichiometry in order to enhance the PCE of perovskite solar cell with improved morphology . Leung *et al.* used hydroiodic acid (HI) and hydrochloric acid (HCl) as an additive to alter the crystal growth kinetics and improve the perovskite thin film quality.[12] Song *et al.* demonstrate that perovskite crystallinity could be improved by incorporating 1-chloronaphthalene (CN) additive in perovskite solution.[13] Heo *et al.* retarded the nucleation time during the spin-coating process by inclusion of HBr aqueous solution in MAPbBr$_3$/DMF solution [14]. Jen and his co-workers developed a method to enhance the crystallization of MAPbI$_{3-x}$Cl$_x$ based perovskite by using DIO as an additive to the perovskite precursor solution[15]. In another report, they show the role of alkyl chain length by incorporating DIO, 1,4-diiodobutane (DIB) and 1,10-diiododecance (DID) in order to improve the film quality[16]. Most of their work was confined to device performance related studies, without providing good details on influence of DIO on perovskite itself. The role of DIO to affect the optical, structural, charge transport property, chemical and electronic states of the MAPbI$_{3-x}$Cl$_x$ based perovskite is not explained in good detail so far in literature. Interestingly, the use of DIO is not only limited to MAPbI$_{3-x}$Cl$_x$ based perovskite but it can be also extended to other Pb based perovskites such as MAPbBr$_3$ and MAPbBr$_{3-x}$I$_x$ , which contains another halide having high electronegativity than I⁻ (coming from DIO) and hard Lewis base behavior. This approach is extremely useful for tuning of band gap in desired region of electromagnetic spectrum for display and tandem solar cell application.[17]

## 5.3    Experimental details

***Materials and synthesis:***  Lead chloride ($PbCl_2$), lead iodide ($PbI_2$), lead bromide ($PbBr_2$), bathocuproine (BCP) and 1-8 diiodooctane (DIO) were purchased from Sigma Aldrich and used as received. Poly(3,4-ethylenedioxythiophene)-poly(styrenesulfonate) (PEDOT: PSS) and [6,6]-Phenyl $C_{61}$ butyric acid methyl ester ($PC_{61}BM$) are purchased from Clevios and Solenne BV, respectively. Methylammonium iodide (MAI) was synthesized by reacting 24 mL of methylamine ($CH_3NH_2$, 33 wt. % in absolute ethanol) and 10 mL of hydroiodic acid (HI, 57 wt% in water) in a round-bottom flask at 0 °C for 2 hours with stirring. The raw precipitate was recovered by removing the solvent in a rotary evaporator at 50 °C. The raw product was washed with diethyl ether, dried in vacuum at 50 °C and dissolved in minimum possible absolute ethanol. The pure MAI recrystallized by adding diethyl ether. The recrystallized precipitate is filtered and dried in vacuum for 24 h.[18] Similarly, methylammonium bromide (MABr) was synthesized by reacting methylamine 33 wt% in absolute ethanol (Sigma Aldrich) solution and hydrobromic acid (HBr) 47wt% in aqueous solution (Merck) at equal (1:1) molar ratio. Rest of the process is same as for the synthesis of MAI.

***Solution preparation:***  40 wt.% perovskite precursor solution of $MAPbI_{3-x}Cl_x$ was prepared by mixing MAI and $PbCl_2$ in 3:1 ratio in anhydrous DMF and stirred for overnight at room temperature. 40 wt.% of $MAPbI_3$ solution was prepared by mixing MAI and $PbI_2$ in 1:1 ratio in anhydrous dimethylformamide (DMF) and stirred for overnight at room temperature. The $MAPbBr_3$ solution was prepared by mixing 112 mg of MABr, 367 mg of PbBr2, and 78 mg of dimethyl sulfoxide (DMSO) in 0.6 mL DMF and stirred for overnight at room temperature. The Different weight percentage (wt.%)  of DIO is added to the perovskite precursor solution for DIO additive perovskite thin films and kept for stirring at room temperature for overnight in case of $MAPbBr_3$ solution. However, for $MAPbI_{3-x}Cl_x$ and $MAPbI_3$ solutions were kept at 70°C for 15 min before use. 15 mg $PC_{61}BM$ is dissolved in 1 mL of 1,2-dichlorobenzene and kept at room temperature for 12 hours stirring. 1 mg BCP is dissolved in 1 mL of isopropanol and kept at $70^oC$ for 15 min before use.

***Device Fabrication:*** Indiums tin oxide (ITO) coated glasses were cleaned with soap, deionized (DI) water, acetone, methanol, and isopropanol. Oxygen plasma treatment was carried out on the cleaned ITO coated glasses for 10 min. PEDOT: PSS was spin-coated on the ITO surface under 5000 rpm for 45 sec and then annealed at $120^0C$ for 20 min under nitrogen atmosphere. The perovskite precursor solution of $MAPbI_{3-x}Cl_x$ was spin-coated above the PEDOT: PSS layer at 2000 rpm for 45 sec and then annealed at $100^0C$ for 2 hours in nitrogen filled glove box. The $MAPbBr_3$ precursor solution was spin-coated in two steps: 1000 rpm for 10 sec and 5000 rpm for 20 sec. During the second step, 100 µL of chlorobenzene is dropped on the film. After that, films are annealed at $100^0C$ for 30 min in nitrogen filled glove box. $MAPbI_3$ based perovskite thin films were fabricated in the same way as it was used for $MAPbBr_3$ based thin films. $PC_{61}BM$ in 1,2-dichlorobenzene solution was spin-coated onto the perovskite layer at 1000 rpm for 60 sec and leave the samples for 5 min at working bench. After that BCP solution was spin-coated over the perovskite/$PC_{61}BM$ layer. Finally, the devices were transferred to a vacuum chamber for Silver (Ag) electrode evaporation of thickness 100 nm. The device area is 4.5 $mm^2$. Different wt.% of DIO was added to the perovskite precursor solution to study the effect of the solvent additive on $MAPbI_{3-x}Cl_x$ and $MAPbBr_3$ based PSCs and light emitting diodes.

***Thin films and Device Characterization:*** All the photovoltaic measurements were carried out in ambient atmosphere. Photocurrent density-voltage (J-V) measurements were carried out using a Keithley 2400 source meter unit and AM1.5G solar simulator after calibrating through a reference solar cell provided by Accreditation Board for Engineering and Technology (ABET). Light-intensity dependent measurements were carried out under the same solar simulator using neutral density (ND) filters. All the J-V measurements were performed using scan speed of 40 mV per sec. External quantum efficiency (EQE) and Internal quantum efficiency (IQE) measurements were carried out to measure the photo-response as a function of wavelength using Bentham quantum efficiency system (Bentham/PVE300).

The X-ray diffraction (XRD) measurements were carried out in Rigaku smart lab diffractometer with Cu $K_\alpha$ radiation ($\lambda$=1.54Å). $\theta$-2$\theta$ scan has been carried out from 5$^o$-50$^o$ with resolution limit up to 0.001$^o$. The structural and morphological analysis was done using a Rigaku lab spectrometer X-ray diffraction machine using Cu $K\alpha$ = 1.54 Å and field emission scanning electron microscopy (FESEM), respectively. Optical studies were carried out using a spectrometer (PerkinElmer LAMBDA 950) and photoluminescence (PL) spectrometer (Horiba FluoroMax 4). Steady-state PL measurement was done on thin-films in vacuum at a pressure of $10^{-5}$ mbar in a custom-made chamber. Excitation energy and wavelength were 60 nJ and 355 nm, respectively. For transient PL, a gated i-CCD (i-CCD, Andor iStar) coupled with Shemrok spectormeter was used for detection with excitation laser (3rd harmonic Nd: YAG with an emission wavelength of 355 nm) pulse width of 840 ps pulse with 1 kHz repetition rate. Steady-state current-voltage–light characteristics were measured using a Keithley 2400 source meter unit, Keithley 2000 digital multimeter and a calibrated Si photodiode (OSI semiconductors, RS components). Surface chemical element analysis were carried out using high-resolution X-ray photoelectron spectroscopy (XPS, KRATOS) with an Al $K\alpha$ (1486.6 eV) X-ray source. Electronic states are derived from ultra-violet photoelectron spectroscopy (UPS, KRATOS) with He (I) source (21.22 eV). Chamber pressure was below 5.0E-9 Torr for XPS and UPS measurement.

## 5.4    Results

### 5.4.1    Morphological Studies

Morphology of the perovskite thin film is the key factor in deciding the performance of the solar cell. In order to understand the importance of DIO concentration in the perovskite film formation, we examined the effect of different wt.% of DIO on the morphological properties of perovskite thin films *via* FESEM and atomic force microscopy (AFM). Figure **1** represents the microstructures of MAPbBr$_3$ with different wt.% of DIO deposited over PEDOT: PSS/ITO coated glass substrates. We deposited perovskite film over PEDOT:PSS coated substrate considering the *p-i-n* configuration used in this work.

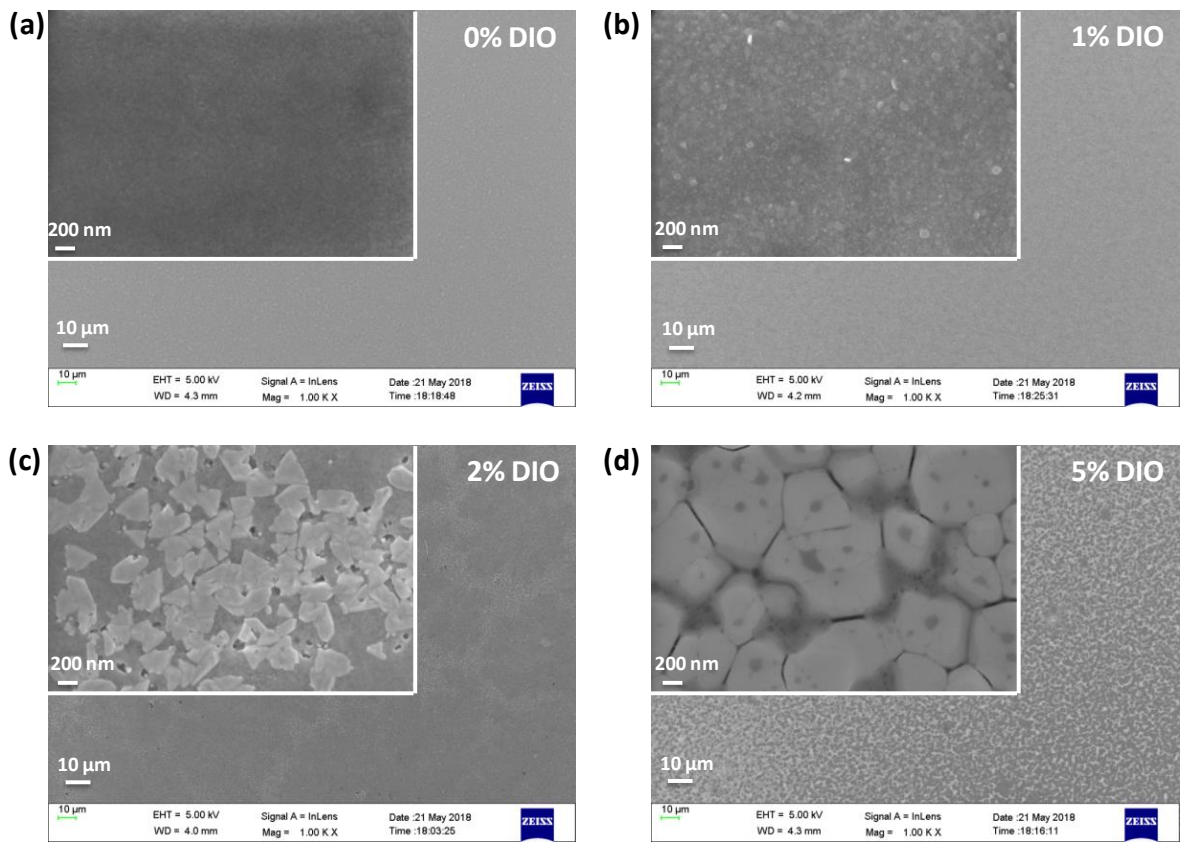

**Figure 1:** *Top-view SEM images of (a) 0 wt%, (b) 1 wt%, (c) 2 wt% and (d) 5 wt% DIO additive MAPbBr3 based perovskite films. Scale bars, 10 mm and 200 nm (inset).*

We observe that the morphology of w/o DIO additive MAPbBr$_3$ based perovskite thin film is compact and smooth but domains are very small. However, the morphology of 1 wt.% DIO additive based perovskite film is also compact and smooth but having slightly increased domain size (figure **1b**). With 2 wt.% DIO additive, the crystal size of MAPbBr$_3$ film becomes larger than the w/o DIO and 1 wt.% DIO additive films along with good number of small domains  (figure **1c**). Further increase in the concentration of DIO to 5 wt.%, results in even more larger domains with some small perovskite nanocrystals (figure **1d**). The crystal grain size growth observed with increasing wt.% of DIO is consistent with a typical Ostwald ripening process. The concentration of dissolved mass is always higher for smaller domains than the bigger domains. Hence, there will be mass transfer occurs from smaller to bigger domains, which leads to continuous growth of bigger domains until the inverse flux of the dissolved components from the small domain stops[19].  However, there is clear evidence of relatively poor morphology for the 2 wt.% and 5 wt.% DIO additive perovskite films (figure **1c** and **1d**).

DIO contains iodine ($I^-$), which dissociates during thermal annealing and generates extra Iodine. As a soft Lewis base, $I^-$ interact with $Pb^{+2}$ (soft Lewis acid) by replacing another borderline/Hard Lewis base (Bromine ($Br^-$)/Chlorine ($Cl^-$)). A schematic diagram of the ion replacement process is demonstrated in figure **2a**. Since interaction between a soft Lewis acid and a soft Lewis base is more than that between soft Lewis acid and borderline/hard Lewis base. Thus, extra $I^-$ coming from DIO can replace $Br^-$/$Cl^-$ and chelate with $Pb^{+2}$ during crystal growth.[20] Images of MAPbBr$_3$ based perovskite precursor solutions and thin films made from those solutions show a distinct color difference (Figure **2b**), the color of solution changes from transparent (pure MAPbBr$_3$) to light yellow (MAPbBr$_3$ with extra $I^-$) with the addition of DIO. The color of perovskite films also changes from yellow to orange with increased wt.% of DIO in the precursor solution.

To further test whether the observed perovskite crystal domain growth is because of Ostwald ripening phenomenon, we examined the effect of DIO on different perovskite system. Figure **3** represents the microstructures of different wt.% of DIO additive MAPbI$_{3-x}$Cl$_x$ perovskite-based systems. We observed an improved (compact and smooth) morphology with bigger domains in 1 wt. % DIO additive thin film of MAPbI$_{3-x}$Cl$_x$. The trend of increasing domain size with respect to increasing DIO wt.% is similar to MAPbBr$_3$ based films. Figure **4** shows the microstructure of MAPbI$_3$ thin films with (1 wt.%) and w/o DIO additive. It is observed that domain size is increased with addition of DIO but domains are not connected to each other. Figure **5** represents the cross-section FESEM images of the different wt.% of DIO additive MAPbBr$_3$ film, which found to be pretty flat and smooth. We note that thickness of the perovskite films decreases with increase in DIO wt.% in the perovskite precursor solution. Figure **6** and **7** represent the atomic force microscopy (AFM) images of MAPbBr$_3$ and MAPbI$_{3-x}$Cl$_x$ based perovskite films with different wt.% of DIO into the perovskite precursor solution, respectively. It is observed that 1 wt. % DIO additive films are smoother than all other films in both cases, which are consistent with FESEM images (figure **1 and 3**). The composition of perovskite thin film with different wt.% of DIO is determined by energy dispersive X-ray spectroscopy (EDS) and listed in **Table 1**. In MAPbI$_{3-x}$Cl$_x$ films, I:Cl ratio is found to be higher in DIO perovskite film (180:1) than in film w/o DIO additive (47:1). The same trend is observed in MAPbBr$_3$ based DIO additive films. I:Br ratio increases with increase in the DIO wt.% in the perovskite precursor solution. These results will be further analyzed in discussion section in terms of nucleation site *vs* growth kinetics due to addition of different wt.% of DIO.

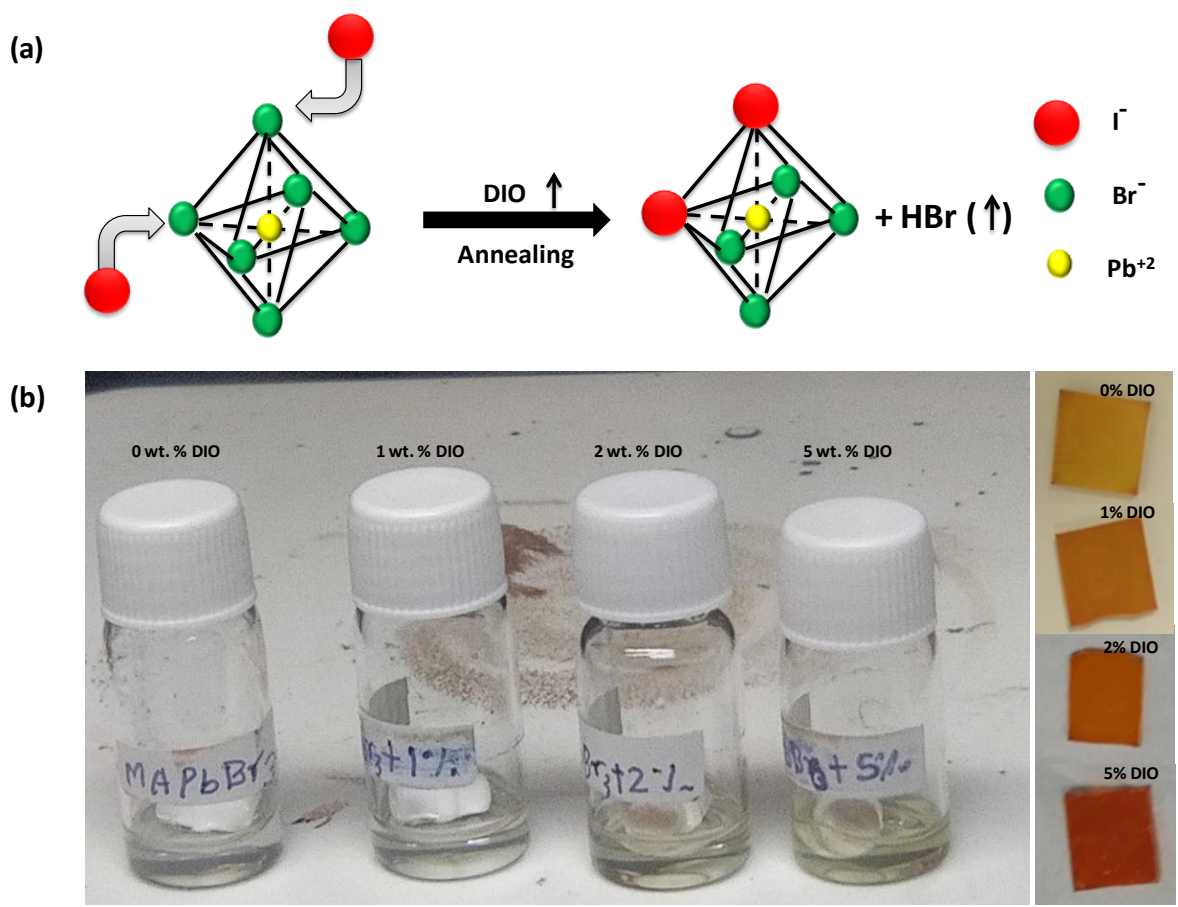

**Figure 2:** *(a) Schematic diagram of the ion exchange process by addition of DIO into the perovskite precursor solution. (b) Images of MAPbBr₃ based perovskite precursor solutions and thin films with different DIO wt. %.*

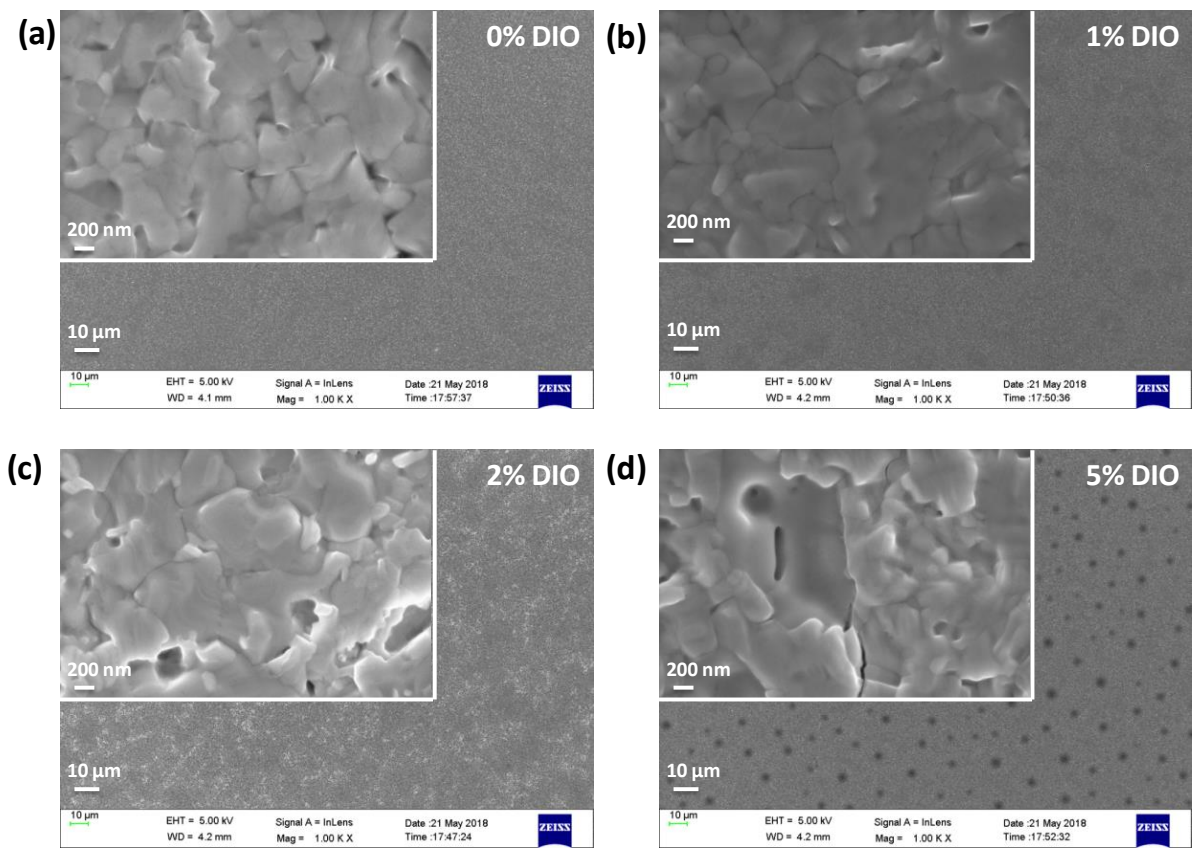

**Figure 3:** *Top-view FESEM images of (a) 0 wt.%, (b) 1 wt.% , (c) 2 wt.% and (d) 5 wt.% DIO additive MAPbI$_{3-x}$Cl$_x$ based perovskite films. Scale bars, 10µm and 200 nm (inset).*

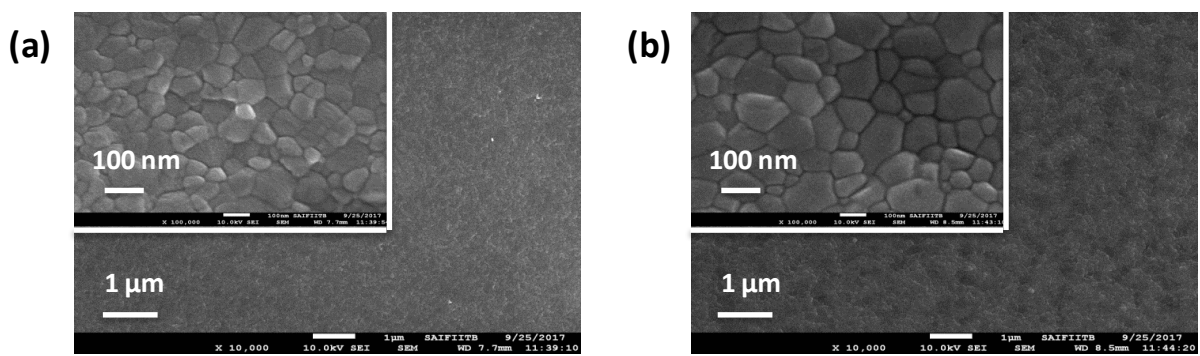

**Figure 4:** *Top-view SEM images of (a) 0 wt.%, and (b) 1 wt.% DIO additive MAPbI$_3$ based perovskite films. Scale bars, 1µm and 100 nm (inset).*

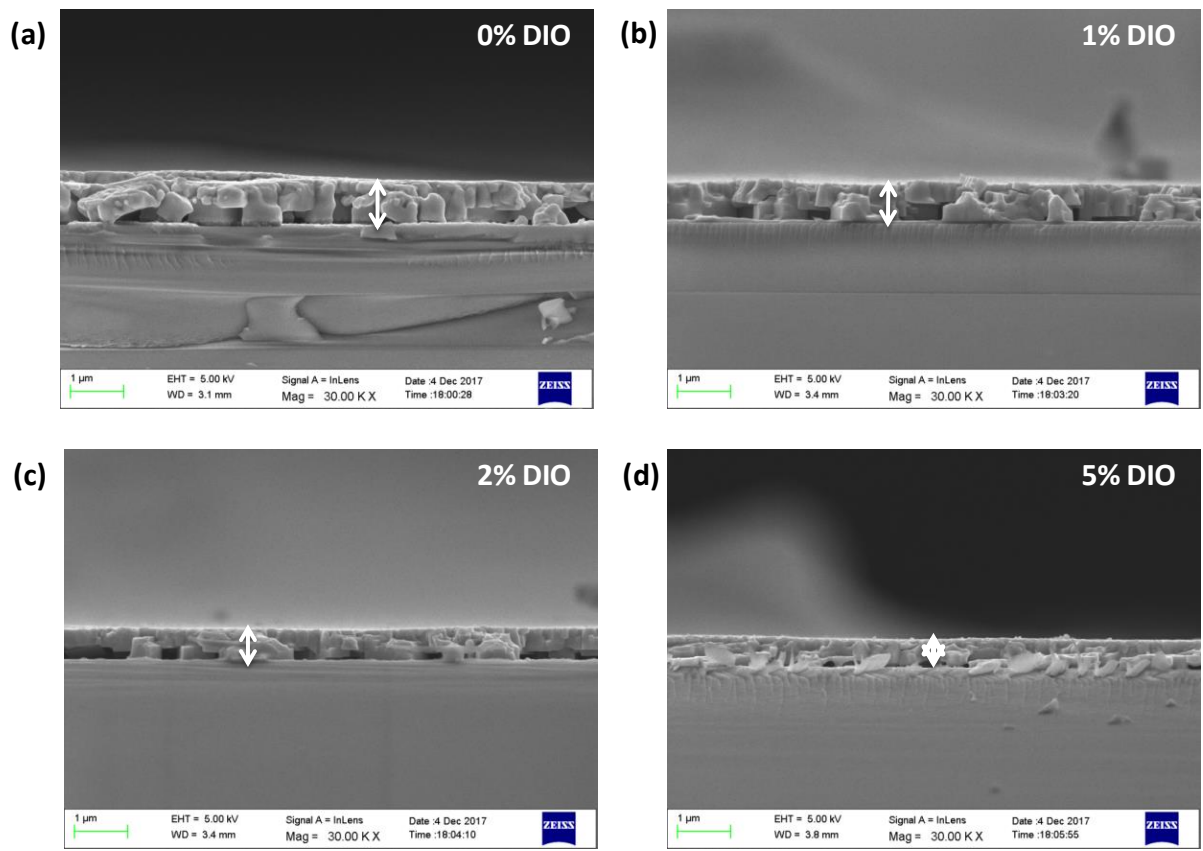

**Figure 5:** *FESEM cross-section images of (a) 0 wt.%, (b) 1 wt.%, (c) 2 wt.% and (d) 5 wt.% DIO additive MAPbBr₃ based thin films prepared on glass substrates.*

**Table 1:** *Elements present in with and w/o DIO additive perovskite thin films obtained through EDS measurement. DIO is mentioned in wt.%.*

| Perovskite Films | C (Atomic%) | N (Atomic%) | O (Atomic%) | Br (Atomic%) | Cl (Atomic%) | I (Atomic%) | Pb (Atomic%) |
|---|---|---|---|---|---|---|---|
| MAPbBr3 | 46.48 | 10.64 | 23.31 | 15.77 | - | - | 3.79 |
| 1% DIO | 49.59 | 17.27 | 7.81 | 19.55 | - | 0.46 | 5.31 |
| 2% DIO | 53.05 | 12.78 | 9.66 | 18.74 | - | 0.77 | 4.99 |
| 5% DIO | 38.67 | 10.79 | 34.84 | 11.90 | - | 0.85 | 2.95 |
| MAPbI$_{3-x}$Cl$_x$ | - | - | - | - | 1.62 | 76.43 | 21.95 |
| 1% DIO | - | - | - | - | 0.43 | 77.55 | 22.05 |

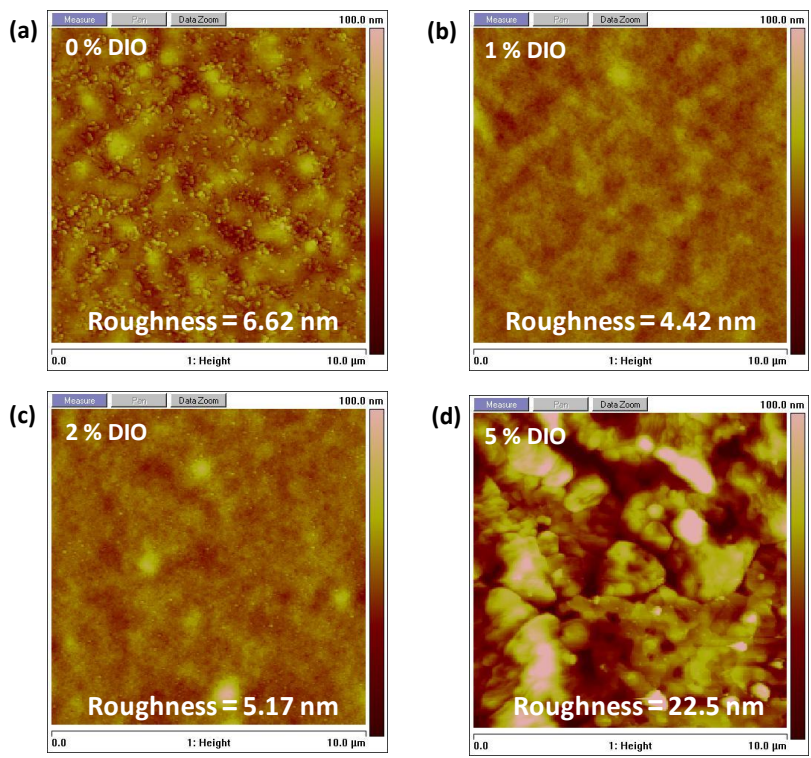

*Figure 6: Atomic force microscopy (AFM) images of MAPbBr₃ with (a) 0 wt.%, (b) 1 wt.%, (c) 2 wt.% and (d) 5 wt.% DIO additive based perovskite thin films.*

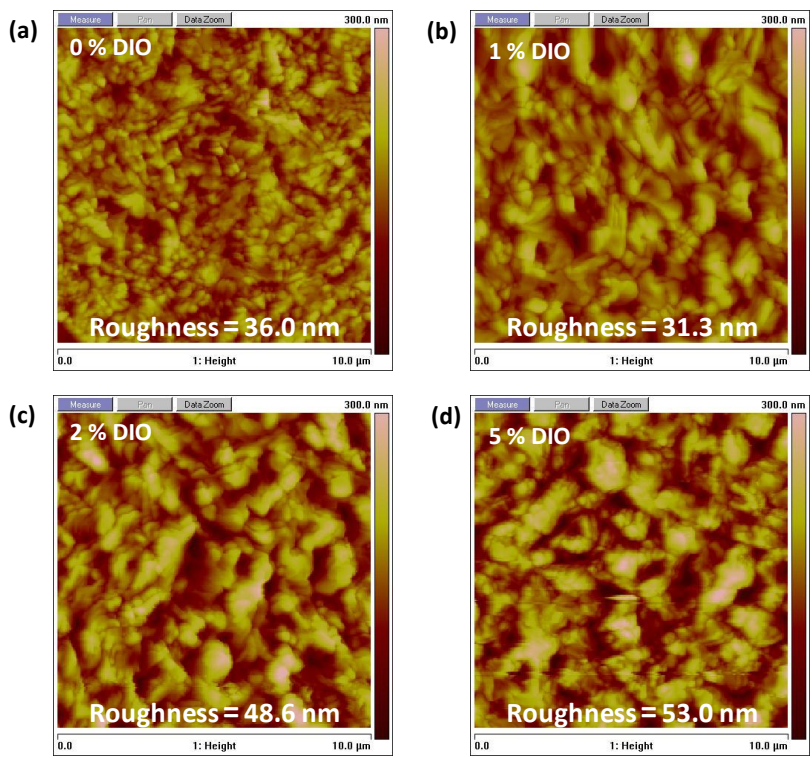

*Figure 7: Atomic force microscopy (AFM) images of MAPbI₃₋ₓClₓ with (a) 0 wt.%, (b) 1 wt.%, (c) 2 wt.% and (d) 5 wt.% DIO additive based perovskite thin films.*

### 5.4.2   Chemical Analysis:

Apart from EDS measurement (**table 1**), the atomic percentage of different elements present in different perovskite films is investigated by X-ray photoelectron spectroscopy (XPS) and listed in **table 2**. The XPS survey scan of with (1 wt.%) and w/o DIO additive MAPbBr$_3$ films coated over ITO/PEDOT:PSS substrate are shown in figure **8a**. We note that I 3d is present in XPS survey scan of DIO additive MAPbBr$_3$ perovskite film, which confirms the incorporation of I⁻ by addition of small amount of DIO in the perovskite precursor solution. Surface morphology of pure perovskite film is changing with addition of DIO; which can be due to change in chemical environment of the DIO additive perovskite films. In order to investigate the chemical states of perovskite film, we performed the high resolution (HR) XPS on with and w/o DIO additive based perovskite films. To eliminate the surface charging effect, we use core level C 1s (C-H bond = 284.6 eV) as a reference to calibrate the energy position of other spectra (figure **8b**). Presence of I 3d in with DIO additive MAPbBr$_3$ perovskite films are clearly seen in HRXPS spectra (figure **8c**). We observed from **table 1** that atomic weight percent of I 3d is increasing and Br 3d is decreasing with increase in DIO concentration. This indicates that with increase in the DIO concentration, more number of Iodine ions is incorporated in the perovskite structure by replacing Bromine. This is also evident by chemical shift of I 3d core level spectra towards lower binding energy for DIO additive based perovskite film.[21,22] Figure **9a** and **9b** show the HRXPS spectra of Br 3d and Pb 4f of with (1 wt.%) and w/o DIO additive perovskite films, respectively. The core level spectra of Br 3d and Pb 4f consist of doublet components due to spin-orbit splitting. Br 3d splits into Br 3d$_{5/2}$ (68.02 eV) and Br 3d$_{3/2}$ (69.05 eV), and the components are separated by 1.03 eV (figure **9a**).

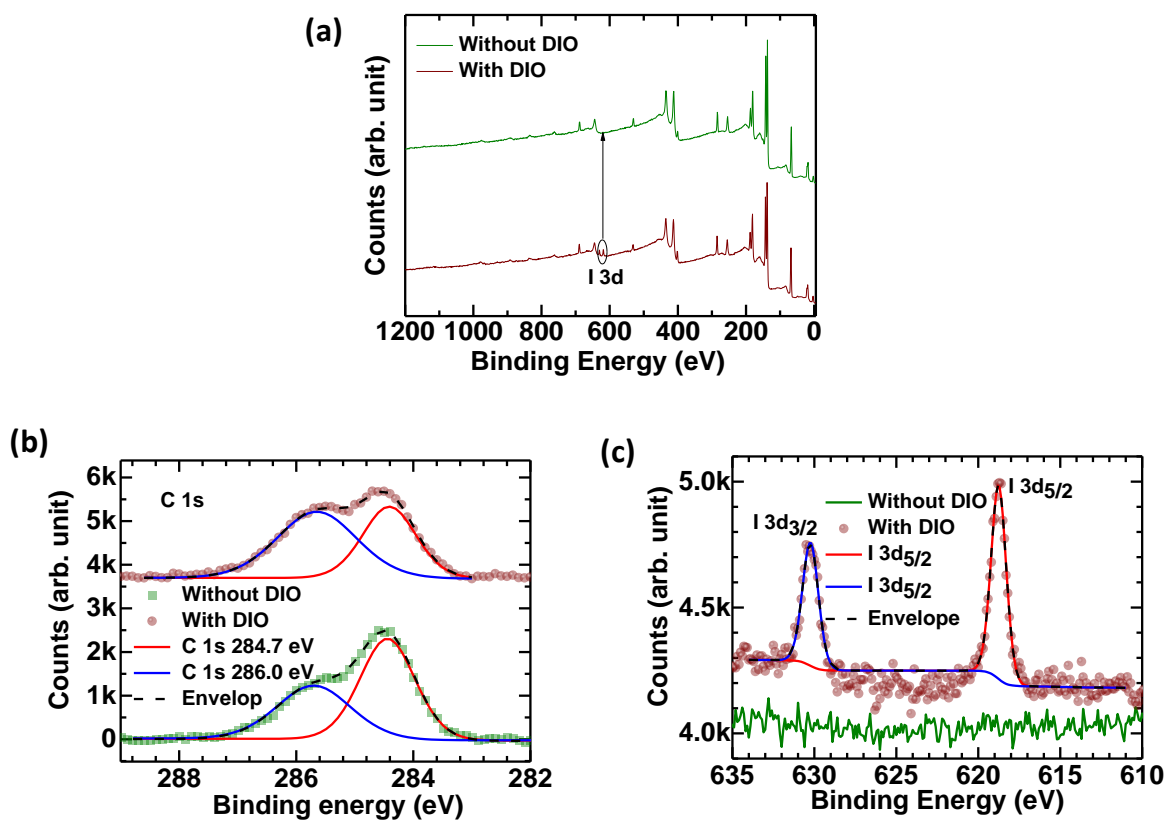

***Figure 8:*** *(a) XPS full scan spectra of with (1 wt.%) and w/o DIO additive MAPbBr₃ based perovskite thin films. Core level HRXPS spectra of (b) C 1s and (c) I 3d for MAPbBr₃ based perovskite thin films fabricated with and w/o DIO.*

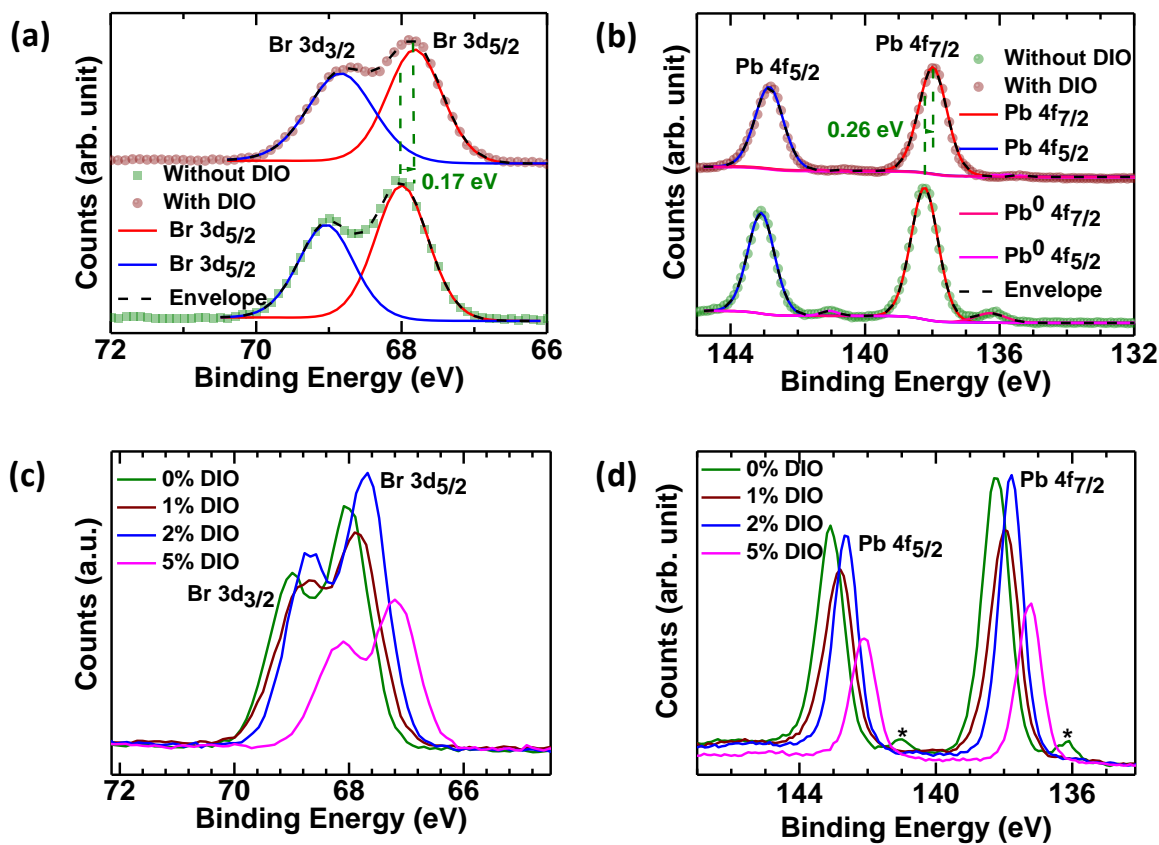

**Figure 9:** *Core level HRXPS spectra of (a) Br 3d and (b) Pb 4f for MAPbBr₃ based perovskite thin films fabricated with and w/o DIO. HRXPS core level photoemission lines for (c) Br 3d and (d) Pb 4f of MAPbBr₃ based perovskite thin films with different wt.% of DIO.*

Similarly, the core level spectra of Pb 4f splits into Pb $4f_{7/2}$ (138.23 eV) and Pb $4f_{5/2}$ (143.11 eV). The separation between the doublets of Pb 4f is 4.88 eV (figure **9b**), which is in good agreement with literature.[23] Along with this, we observed that there is a chemical shift of 0.17 eV for Br 3d and 0.26 eV for Pb 4f towards lower binding energy in DIO additive perovskite film. In the core level spectra of Pb 4f, pristine solution based thin film has an additional Pb feature at binding energy 1.8 eV lower than the main doublet.[24,25] Interestingly, we do not observe any significant additional metallic Pb feature in the DIO additive (1 wt.%, 2 wt.% and 5 wt.%) perovskite films (figure **9d**). Figure **9c** and **9d** represent the HRXPS of Br 3d and Pb 4f core level spectra for different wt.% of DIO in the perovskite precursor solution. We observe that the core level photoemission lines of Br 3d and Pb 4f shifts towards lower binding energy by increasing the wt.% of DIO.

***Table 2:*** *Elements present in with and w/o DIO additive perovskite thin films obtained through XPS measurement.*

| Perovskite Films | C (Atomic%) | N (Atomic%) | O (Atomic%) | Br (Atomic%) | Cl (Atomic%) | I (Atomic%) | Pb (Atomic%) |
|---|---|---|---|---|---|---|---|
| $MAPbBr_3$ | 44.59 | 7.96 | 6.09 | 28.11 | - | - | 13.25 |
| 1% DIO | 44.58 | 10.15 | 4.56 | 27.87 | - | 0.49 | 12.35 |
| 2% DIO | 28.04 | 15.69 | 2.00 | 38.42 | - | 1.55 | 14.30 |
| 5% DIO | 48.75 | 10.53 | 8.11 | 21.45 | - | 2.95 | 8.22 |
| $MAPbI_{3-x}Cl_x$ | 41.27 | 8.67 | 5.38 | - | - | 34.08 | 10.59 |
| 1% DIO | 38.09 | 9.17 | 4.84 | - | - | 36.58 | 11.32 |

In order to study the chemical shift in more depth, we perform HRXPS on $MAPbI_{3-x}Cl_x$, and $MAPbI_3$ thin films. We observe that there are no chemical shifts in I 3d and Pb 4f core level spectra of DIO additive (1 wt.%) $MAPbI_3$ based thin film with respect to w/o DIO additive film and it is shown in figure **10**. Figure **11a** shows the XPS full scan spectra of both with (1 wt. %) and w/o DIO additive $MAPbI_{3-x}Cl_x$ films coated over ITO/PEDOT: PSS substrate. The HRXPS core level photoemission line of C 1s has been normalized for better visibility and it is shown in figure **11b**. The ratio of Iodine to lead is nearly same (3.22:1) in both the cases which are consistent with the EDS result (**table 1**) and earlier reports.[26] We observed that there is a chemical shift of 0.5 eV for I 3d (figure **11c**) and 0.6 eV for Pb 4f (figure **11d**) towards lower binding energy in DIO additive perovskite films. Since, there is a small amount of Cl is present in case of $MAPbI_{3-x}Cl_x$ thin film (figure **11e** and **11f**), we can expect a chemical shift in the core level spectra of with DIO additive film. In the core level spectra of Pb 4f, pristine solution based thin film has an additional Pb feature at binding energy 1.8 eV lower than the main doublet. However, we do not observe any additional Pb feature in the DIO additive perovskite film. We shall discuss the chemical shift and the additional Pb feature in the discussion part.

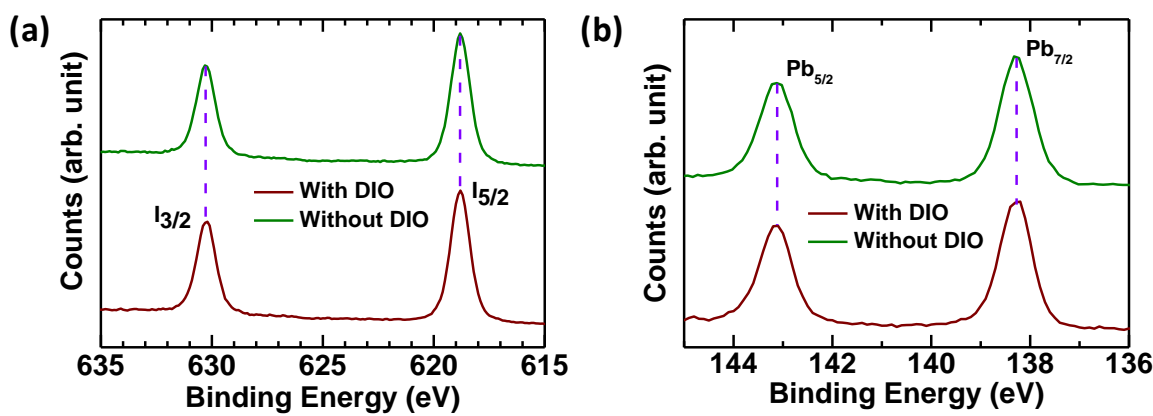

***Figure 10:*** *HRXPS core level photoemission lines for (a) I 3d, and (b) Pb 4f of MAPbI₃ based with (wine solid line) and without (olive solid line) DIO additive perovskite thin films.*

### 5.4.3   Electronic State Study:

We investigated the electronic states of with and w/o DIO additive MAPbBr₃ based perovskite films deposited on PEDOT: PSS/Glass substrate using ultra-violet photoelectron spectroscopy (UPS). Figure **12** represents the evolution of UPS spectra as a function of different concentration of DIO additive in the perovskite precursor solution. As shown in fig. **12a**, the fermi level of pure MAPbBr₃ film is 4.96 eV, which is in good agreement with literature.[27] Sargent *et al.* shows an investigation on MAPbBr₃₋ₓClₓ based perovskite system, where incorporation of Cl⁻ into MAPbBr₃ film can tune the electronic states.[28] By adding 1 wt.% DIO in the perovskite precursor solution, we found that the fermi level shifts above the fermi level of the w/o DIO additive MAPbBr₃ by 0.15 eV. On increasing the wt.% of DIO from 1 wt.% to 5 wt.%, we observed a continuous upward shift in the fermi level of DIO additive perovskite films with respect to fermi level of w/o DIO additive perovskite film. This results in a transition from p-type pure MAPbBr₃ based perovskite film to n-type DIO additive MAPbBr₃ perovskite film. Upward shift in the valence band maximum (VBM) position of DIO additive MAPbBr₃ based films is also observed with respect to pure perovskite film. In case of 1 wt.% DIO additive film, the shift in VBM is very small (0.03 eV) and there is no change in the conduction band minimum (CBM) position. However, on further increasing the wt.% of DIO in the perovskite precursor solution, we observed a significant shift in the VBM and CBM positions. Thus, we can tune the electronic states of pure perovskite film by addition of different wt.% of DIO. These results will be useful to understand structural, UV-Vis and PL spectral studies on these films.

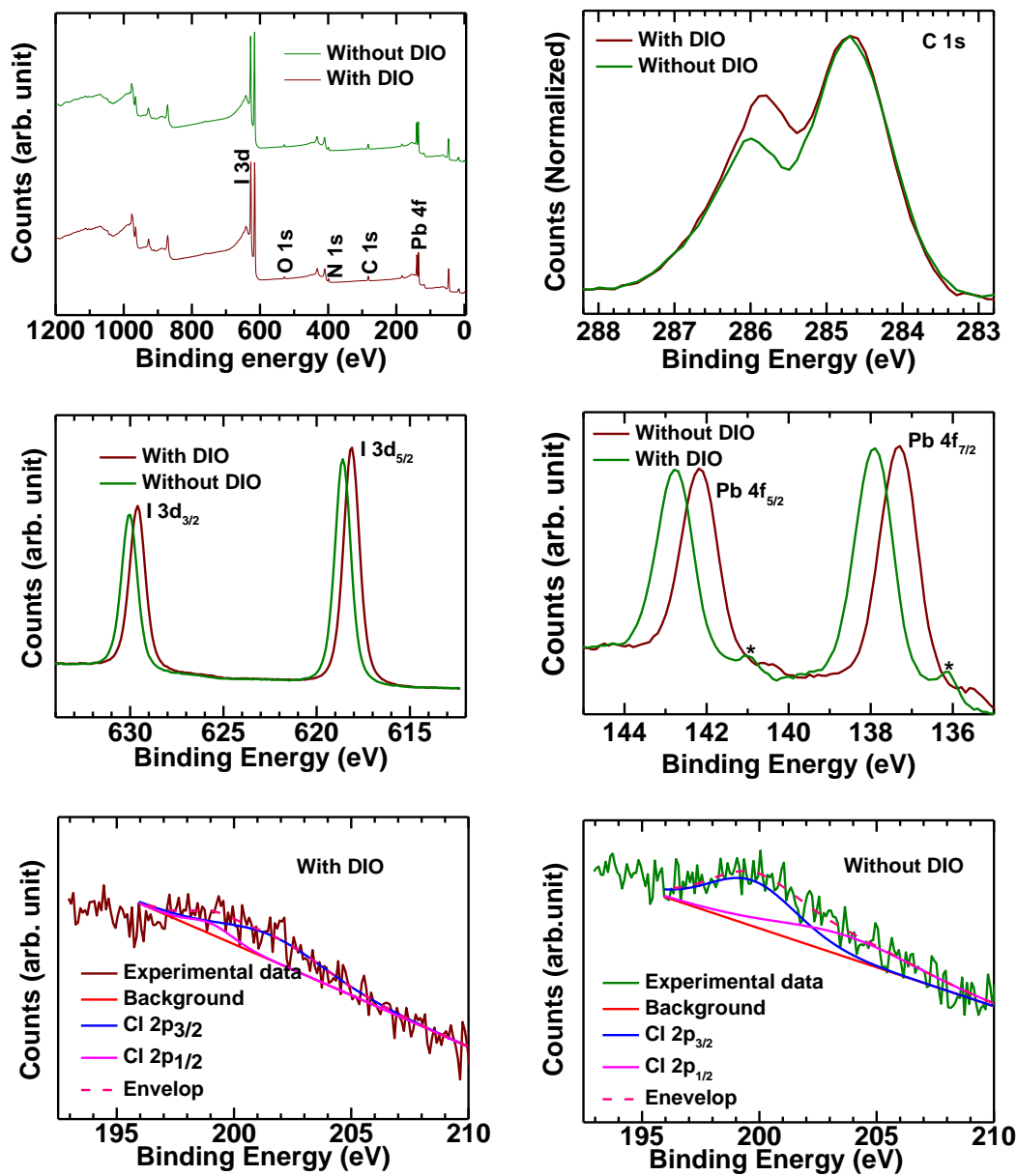

***Figure 11:*** *(a) Wide XPS spectra of MAPbI$_{3-x}$Cl$_x$ based with and without DIO additive thin films and corresponding HRXPS photoemission lines of (b) C 1s, (c) I 3d, and (d) Pb 4f. HRXPS core level photoemission line for Cl 2p for (e) with (1 wt.%) and (f) without DIO additive MAPbI$_{3-x}$Cl$_x$ based perovskite thin films.*

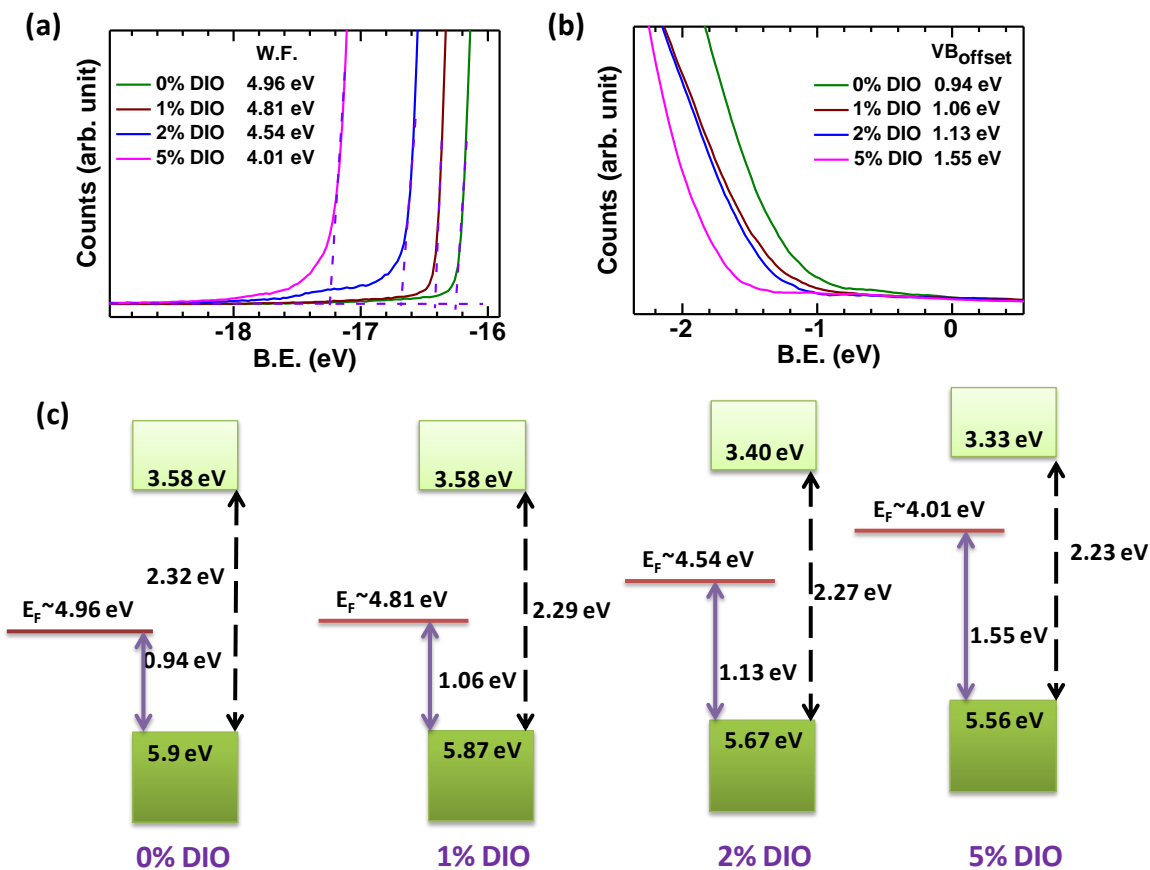

***Figure 12:*** *(a) Work function, (b) valence band maximum position of MAPbBr₃ based perovskite thin films with different wt.% of DIO and (c) corresponding electronic states.*

### 5.4.4   Structural and Optical Studies:

The evidence of enhanced crystallization in MAPbBr$_3$ is observed in the X-ray diffraction (XRD) study of with and w/o DIO additive based perovskite films (figure **13a**). Figure **13b** represents the XRD (100) peak of MAPbBr$_3$ based with and w/o DIO additive films with different wt.% of DIO in the perovskite precursor solution on PEDOT: PSS/Glass substrate in the range of $14.8^0$ to $15.3^0$. We observe from figure **13b**, that the first order diffraction peak shifts toward the lower angle of diffraction (2θ) with an increase in the wt.% of DIO. It is relatively lower for 1 wt.% DIO additive film, however, pretty evident for 2 wt.% and 5 wt.% additive DIO based films, which are consistent with UPS studies. The second and third order XRD peak position of MAPbBr$_3$ with different concentration of DIO is listed in **Table 3**. The shift in XRD peak position toward lower angle suggests the replacement of smaller atom (Br) by bigger atom (I). The (110) peak location in $MAPbI_{3-x}Cl_x$ based film decreases slightly but noticeably from ~ $14.12^0$ to $14.10^0$ with addition of 1 wt.% DIO (figure **13c**). However, we do not observed any significant shift in XRD peak position of the film made with DIO additive in comparison to w/o DIO additive $MAPbI_3$ based perovskite system (figure **13d**).

Figure **14a** represents the optical absorption of $MAPbBr_3$ perovskite film with different wt.% of DIO. With increase in the wt.% of DIO, the UV-vis absorption shows a systematic red shift (decrease in the band gap ($E_g$)) in the absorption edge for the DIO additive $MAPbBr_3$ films. The red shift in absorption edge suggests the incorporation of $I^-$ in the pure $MAPbBr_3$ perovskite film to form $MAPbBr_{3-x}I_x$. The band gap of the perovskite film with different wt.% of DIO is estimated using the relation

$$(\alpha E)^2 = A(E - E_g)$$

Where, $\alpha$ and E are the absorption coefficient and photon energy, respectively (Figure **14b**). Apart from that, we also observe a decrement in the exciton feature with an increase in DIO wt.%.[29] We also observe that optical density of the films decreases with increase in the DIO wt.%, which suggest that increase in DIO wt.% make films thinner than the w/o DIO additive perovskite film, which is consistent with the FESEM cross-section data (figure **5**). So, the change in the color of the perovskite films (figure **2**) is mainly because of chelation of $I^-$ with $Pb^{+2}$ in consistent with UPS, XRD and red-shifted absorption edge based studies. Effect of morphology on shape of UV-Vis spectrum is also evident in Figure **14a**, where a compact film shows enhanced absorption at higher energies as compare to films with voids, as discussed in literature.[30]

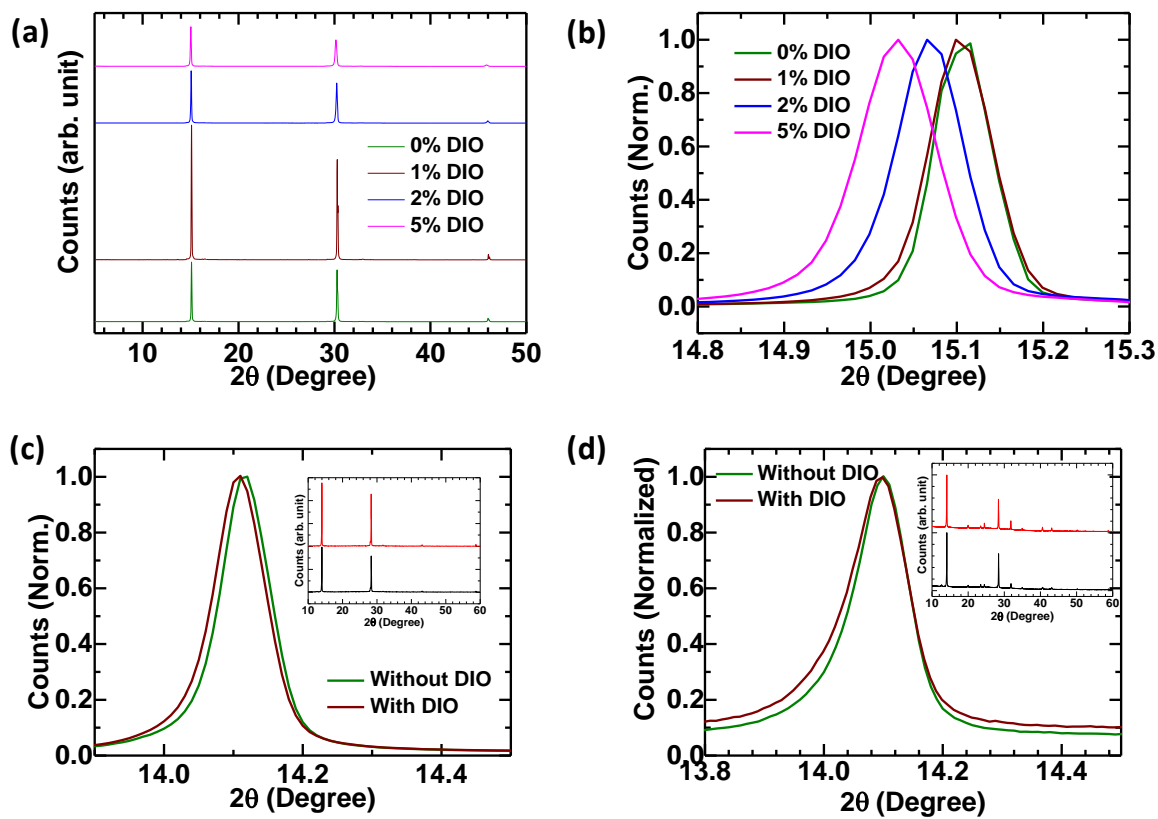

***Figure 13:*** *(a) XRD pattern of MAPbBr₃ based perovskite thin films with different wt.% of DIO into the perovskite precursor solution. First order diffraction peak of (b) MAPbBr₃, (c) MAPbI₃₋ₓClₓ, and (d) MAPbI₃ with and without DIO additive. The inset of (c) and (d) represents the XRD pattern in the range of 5⁰ to 60⁰. 1 wt.% DIO is used as an additive in (c) and (d).*

The same trend of red shift is also evident in steady-state photoluminescence (PL) study of MAPbBr₃ based films. PL peak position is red shifted with an increase in DIO concentration and it is shown in figure **14c**. Red shift in PL is significant for 2 wt.% and 5 wt.% DIO additive case, as compare to 1 wt.% DIO additive film, again in agreement with UPS, XRD and UV-Vis studies. Figure **14d** shows the time-resolved PL spectra of with (1 wt.%) and w/o DIO additive MAPbBr₃ perovskite films. The PL lifetime of both films was fitted using a mono-exponential decay function. The lifetime of the DIO additive based perovskite film is longer by a factor of two than w/o DIO additive perovskite film.

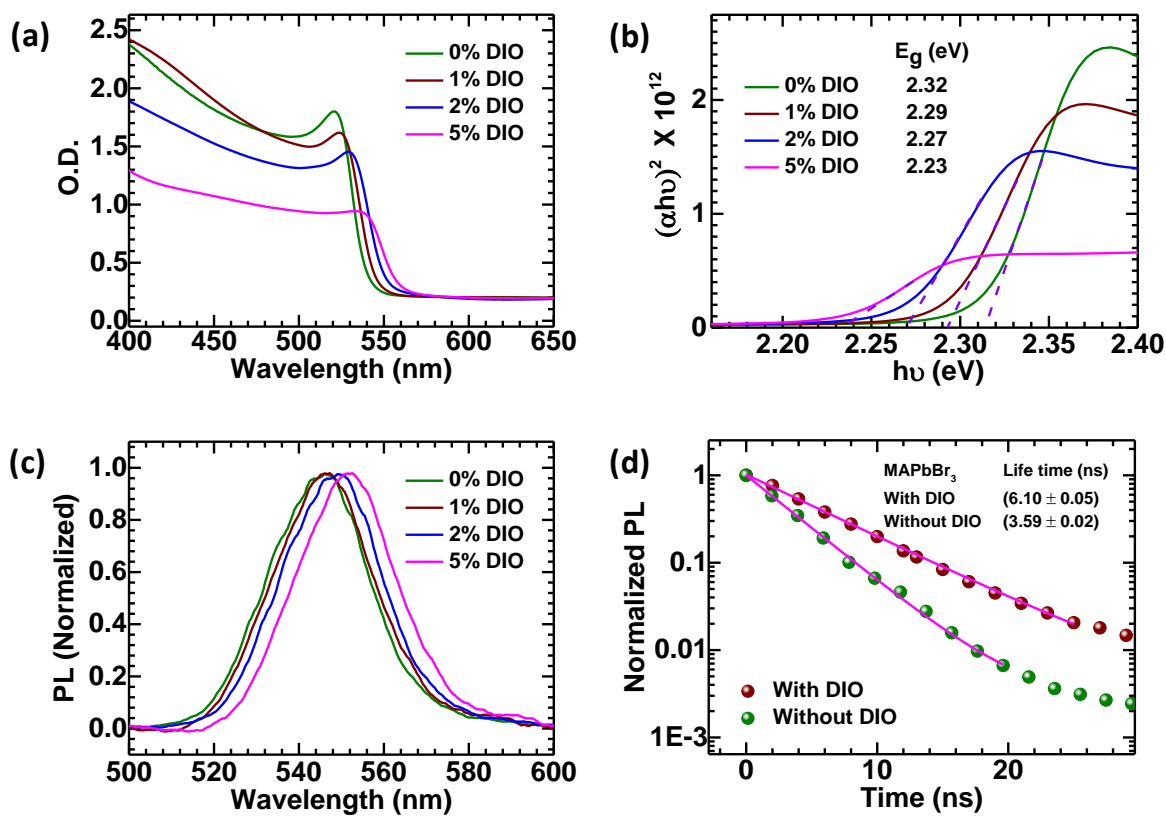

**Figure 14:** (a) Absorption and (b) Tauc plot in order to derive the optical band gap of MAPbBr₃ based perovskite thin films with different wt.% of DIO into the perovskite precursor solution. (c) Steady-state PL and (d) time-resolved PL spectra of MAPbBr₃ based perovskite thin films with different wt.% of DIO.

**Table 3:** First, second and third order diffraction peak position of DIO additive (with different wt.%) MAPbBr₃ based perovskite thin films.

| DIO (wt.%) | 1st order 2θ (Degree) | 2nd order 2θ (Degree) | 3rd order 2θ (Degree) |
|------------|------------------------|------------------------|------------------------|
| 0 | 15.11 | 30.29 | 46.05 |
| 1 | 15.09 | 30.27 | 46.02 |
| 2 | 15.06 | 30.24 | 45.97 |
| 5 | 15.03 | 30.15 | 45.83 |

### 5.5.5   Photovoltaic Performance:

On the basis of above studies, we have fabricated $MAPbI_{3-x}Cl_x$ based perovskite solar cells with *p-i-n* configuration. Figure **15a** represents the crystal structure of $MAPbBr_3$. The device structure (ITO/ poly(3,4-ethylene-dioxythiophene):polystyrene sulfonate (PEDOT: PSS) / $MAPbBr_3$/ phenyl-C61-butyric acid methyl ester ($PC_{61}BM$) / bathocuproine (BCP) /Ag) with corresponding energy level diagram of each layer used in device is shown in figure **15b**. Figure **15c** shows the Dark J-V characteristics of with and w/o DIO additive $MAPbBr_3$ perovskite solar cells. Illuminated J-V characteristics for three devices are shown in figure **15d**. The devices made with 1 wt.% DIO additive have an improved PCE of 7.71 % with open circuit voltage ($V_{oc}$) of 1.44 V,  short-circuit current density ($J_{sc}$) of 8.25 mA/cm$^2$ and fill factor (FF) of 64.87 %. In contrast, the devices made from pristine precursor solution (w/o additive) showed a PCE of 2.39% only with $V_{oc}$ of 1.17 V, $J_{sc}$ of 4.15 mA/cm$^2$ and FF of 49.07 %. We observed a significant decrease in PCE by increasing the DIO concentration to 2 wt. %, which suggest that 1 wt. % DIO is the optimized concentration for fabrication of better perovskite solar cell. The average PCE values over 16 devices in a single batch process are listed in **Table 4**. The increase in $J_{sc}$ for DIO added device is estimated using external quantum efficiency (EQE) and is shown in figure **15e**. It is worth to notice the red shift in the EQE spectra of DIO additive cells with respect to w/o DIO additive perovskite solar cells. This also gives the evidence of replacement of Br$^-$ with I$^-$ by adding different concentration of DIO in the pure $MAPbBr_3$ based perovskite solution, which is consistent with absorption, steady-state PL, XRD, and chemical analysis.

We have also fabricated $MAPbI_{3-x}Cl_x$ based perovskite solar cells with *p-i-n* device structure (similar to $MAPbBr_3$). Figure **16a** and **16b** shows the illuminated and dark J-V characteristics of with and w/o DIO additive $MAPbI_{3-x}Cl_x$ perovskite solar cells, respectively. The similar trend in all photovoltaic parameters is observed as it is found in $MAPbBr_3$ based perovskite solar cells and listed in **Table 4**. The devices made with 1 wt.% DIO additive have an improved PCE of 14.70 % with open circuit voltage ($V_{oc}$) of 1.04 V,  short-circuit current density ($J_{sc}$) of 20.62 mA/cm$^2$ and fill factor (FF) of 68.53 %, which is higher than previous reports for similar processed condition of $MAPbI_{3-x}Cl_x$with DIO in *p-i-n* configuration.[31,32] The increase in $J_{sc}$ for DIO added device is estimated using EQE and internal quantum efficiency measurement (IQE) studies. The EQE of the devices with DIO additive is higher than that of the device w/o DIO additive (figure **16c**). This is consistent with the enhancement in $J_{sc}$ of DIO additive devices. We also calculated the $J_{SC}$ by integrating the EQE spectrum with the AM 1.5 spectrum and obtained a value of 16.22 mA/cm$^2$ , 19.66 mA/cm$^2$ and 18.96 mA/cm$^2$ for 0 wt. %, 1 wt. % and 2 wt. % DIO additive solar cells, respectively. The small difference (less than 10 %) in the current density calculated from J-V and EQE measurement is due to edge effects from the small cells having active area of 4.5 mm$^2$.[33] Figure **16c** also shows the IQE spectrum which is calculated from EQE and reflective absorption spectrum. The increased collection efficiency at longer wavelength for 1 wt. % DIO added solar cell suggests that the carrier mobility is larger for DIO added solar cell. This is a consequence of larger grain size in the DIO additive films. The $J_{sc}$ from IQE is 19.19 mA/cm$^2$, 21.92 mA/cm$^2$ and 21.33 mA/cm$^2$ for w/o, with 1 wt.%, and 2 wt.% DIO additive devices, respectively, validating the $J_{sc}$ values obtained from J-V curve.

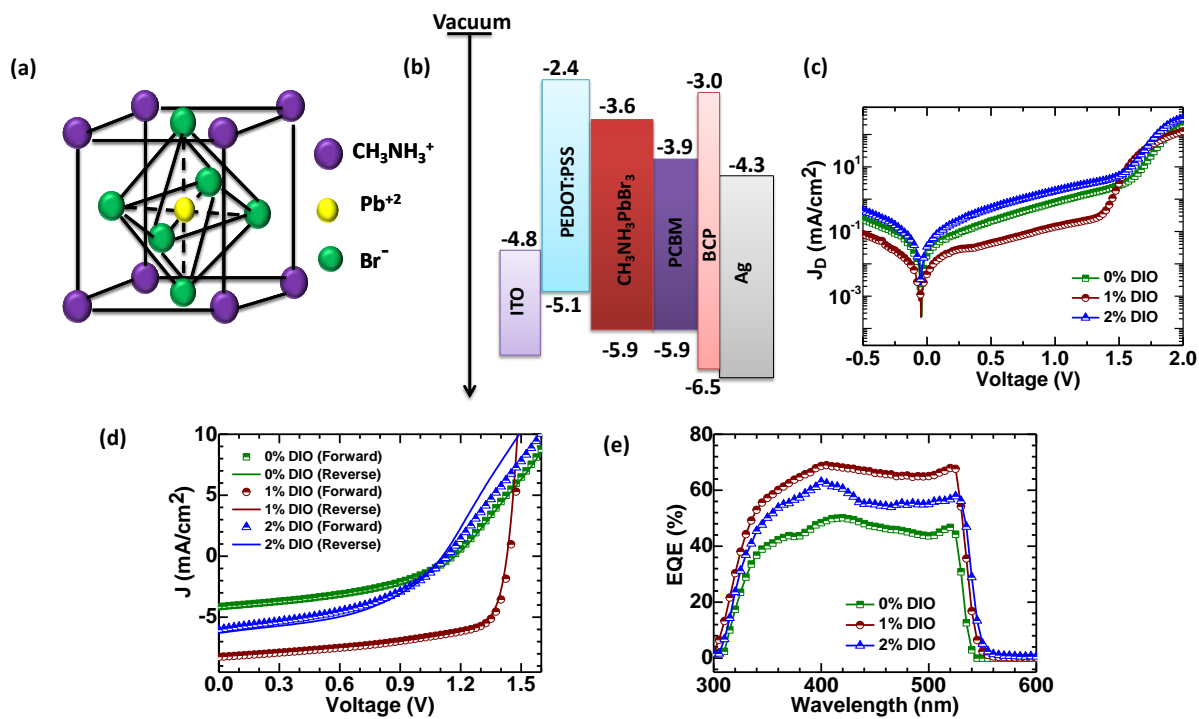

***Figure 15:*** *(a) Crystal structure of MAPbBr₃. (b) Energy level diagram of each layer used in the fabrication of MAPbBr₃ based perovskite solar cells. (a) Dark & (d) illuminated current-voltage characteristics under AM1.5 illuminations, and (e) external quantum efficiency (EQE) of with & w/o DIO solvent additive MAPbBr₃ based perovskite PV devices.*

In order to understand the charge recombination process in MAPbI$_{3-x}$Cl$_x$ and MAPbBr$_3$ perovskite based PV devices, light-intensity dependent J-V characteristics of solar cells was measured. Figure **17a** represents the variation of V$_{oc}$ with incident light intensity on with (1 wt. % DIO) and w/o DIO additive MAPbI$_{3-x}$Cl$_x$ based PV devices. The slope calculated from V$_{oc}$ versus logarithmic light-intensity plot provides the ideality factor "η" which gives information about the charge carrier recombination processes in open-circuit condition.[34] In the case of MAPbI$_{3-x}$Cl$_x$ w/o DIO additive PV device, the slope of V$_{oc}$ vs ln(intensity) gives a value of 1.69(kT/q) (k is Boltzmann constant, T is temperature and q is elementary charge). It indicates (η = 1.69) Schokley-Read-Hall (SRH) type of trap assisted recombination occurring in the control device. However, for DIO additive MAPbI$_{3-x}$Cl$_x$ perovskite based device, we observed a value of 1.05 (kT/q). It indicates that recombination takes place within the bulk (occurrence of bimolecular recombination). Figure **17b** demonstrates the power law dependence of J$_{sc}$ with respect to incident light intensity. The DIO additive device shows α = 1.01, which is close to unity, similar to control device with α = 0.96 with slight deviation.

**Table 4:** *The photovoltaic parameters achieved for with and w/o DIO additive perovskite thin film based solar cells over an area of 4.5 mm². Average PCE over 16 devices are also mentioned.*

| Devices | DIO wt.% | V$_{OC}$ (V) | J$_{SC}$ (mA/cm2) | J$_{SC}$ (EQE) (mA/cm2) | FF (%) | PCE (%) | Av. PCE (%) |
|---|---|---|---|---|---|---|---|
| MAPbI$_{3-x}$Cl$_x$ | 0% | 0.92 | 17.71 | 16.22 | 64.59 | 10.53 | (9.47±1.06) |
| MAPbI$_{3-x}$Cl$_x$ | 1% | 1.04 | 20.62 | 19.66 | 68.53 | 14.70 | (14.34±0.36) |
| MAPbI$_{3-x}$Cl$_x$ | 2% | 0.66 | 20.33 | 18.96 | 51.32 | 6.89 | (4.32±2.57) |
| MAPbBr$_3$ | 0% | 1.17 | 4.15 | 3.87 | 49.07 | 2.39 | (1.64±0.75) |
| MAPbBr$_3$ | 1% | 1.44 | 8.25 | 7.69 | 64.87 | 7.71 | (7.16±0.55) |
| MAPbBr$_3$ | 2% | 1.10 | 6.06 | 5.59 | 45.27 | 3.02 | (2.14±0.88) |

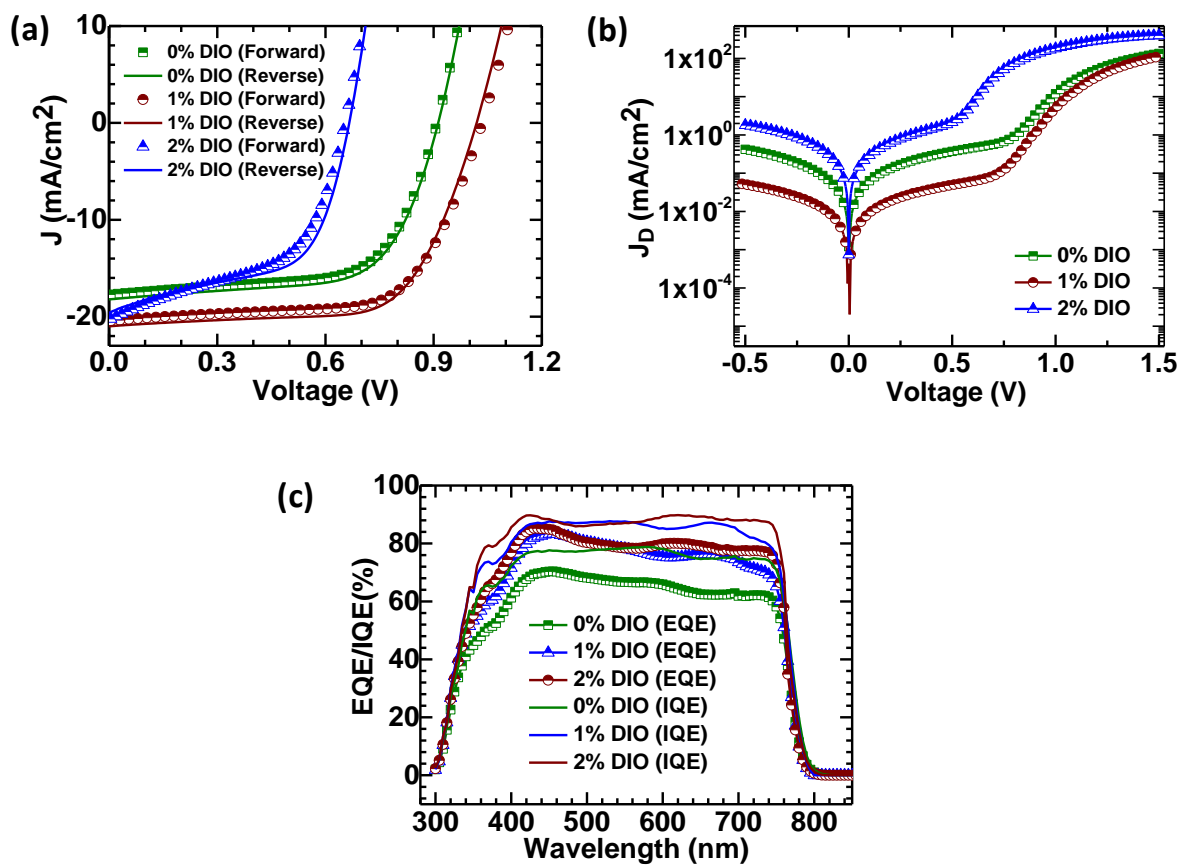

***Figure 16:*** *(a) Illuminated & (b) Dark current-voltage characteristics under AM1.5 illuminations and (b) external/internal quantum efficiency (EQE & IQE) of with & w/o DIO additive MAPbI$_{3-x}$Cl$_x$ based perovskite PV devices.*

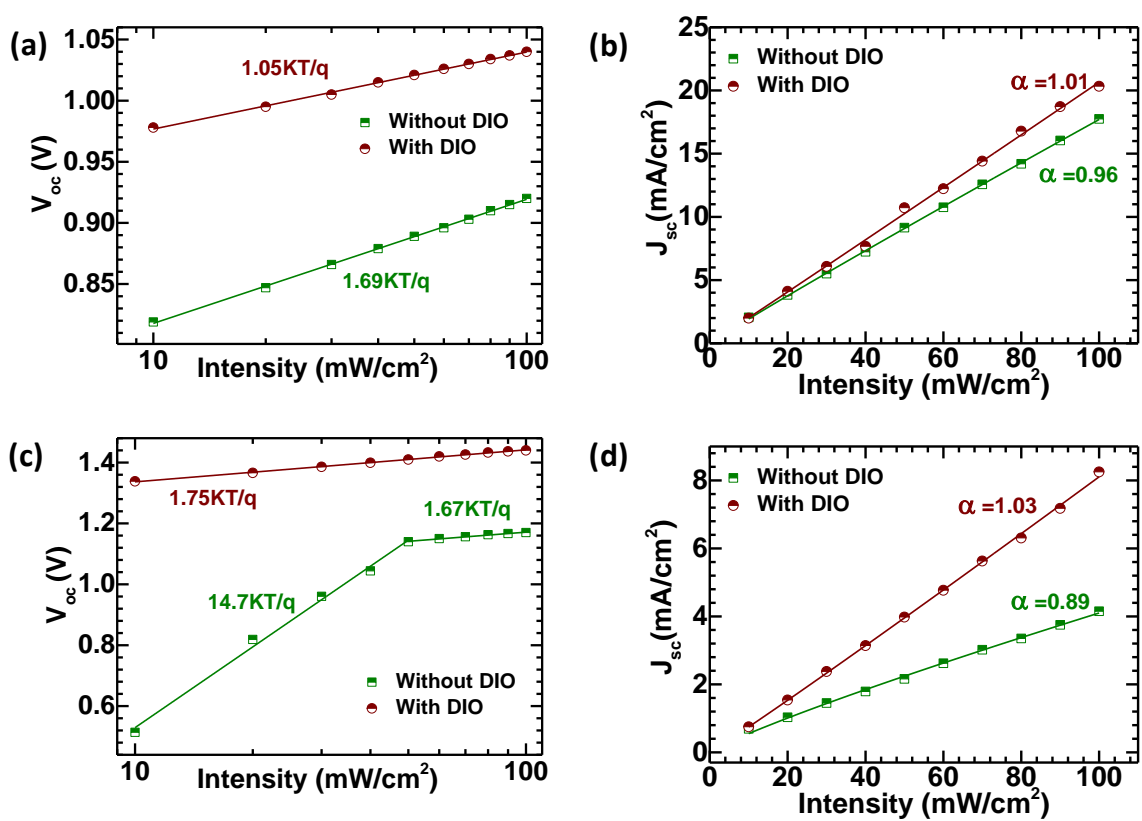

***Figure 17:*** *(a and c) Open-circuit voltage ($V_{OC}$) and (b and d) short-circuit current density ($J_{SC}$) vs light intensity of with & w/o DIO additive MAPbI$_{3-x}$Cl$_x$ and MAPbBr$_3$ based perovskite PV devices, respectively.*

Figure **17c** represents the variation of $V_{oc}$ with incident light intensity on with and w/o DIO additive MAPbBr$_3$ based devices. In w/o DIO additive MAPbBr$_3$ perovskite based device, we observed two different slopes, i.e., two ideality factors in different intensity range. For I > 50 mw/cm$^2$, the value of n~1.67 which indicates (η = 1.67) SRH type of trap-assisted recombination. At lower intensity, n is very large. Such large values of n can arise at low intensity when the current flows through internal shunts[35]. However, for DIO additive MAPbBr$_3$ perovskite-based device, we observed a value of 1.75 (kT/q) throughout all the intensities. It also indicates SRH type of trap-assisted recombination process in this device as well. Figure **17d** demonstrates the power law dependence of $J_{sc}$ with respect to incident light intensity. The DIO additive device shows α = 1.03, which is close to unity. The control device shows deviation with α = 0.89.

### 5.4.6 Electroluminescence Studies

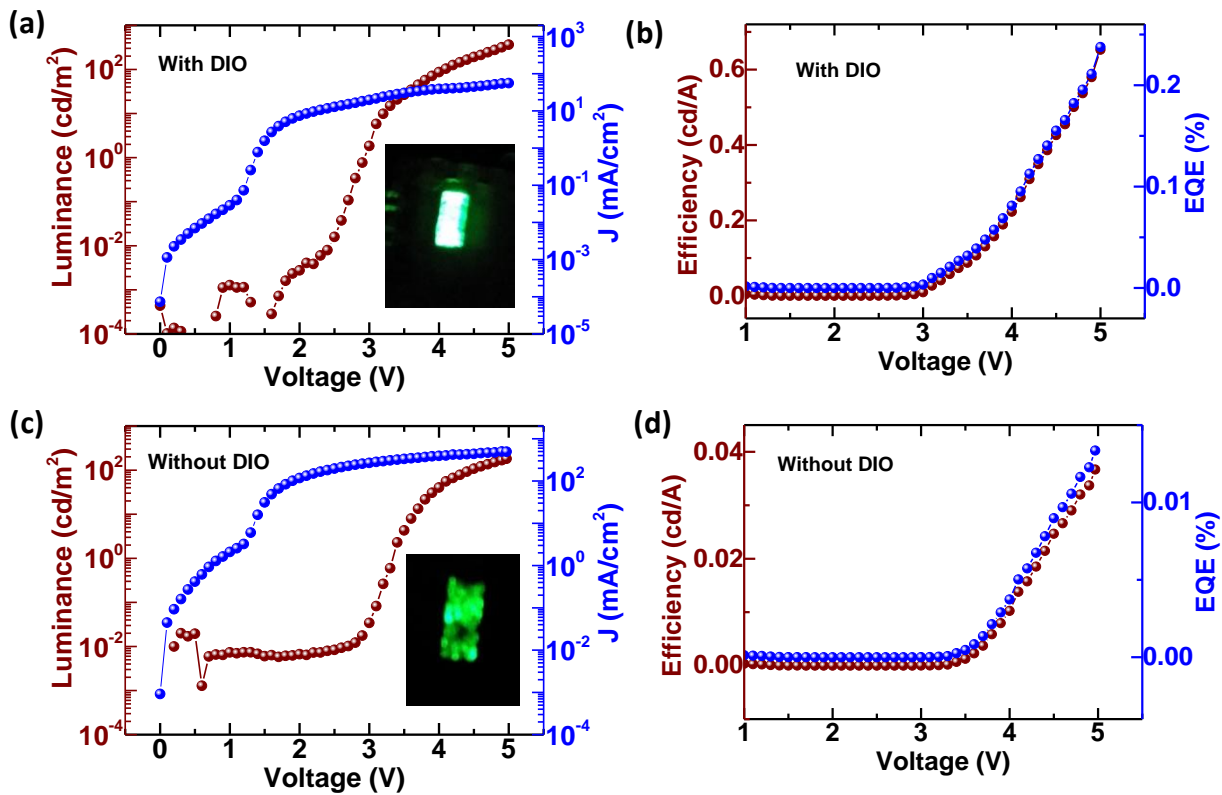

***Figure 18:*** *J-V-L characteristics of (a) with and (c) w/o DIO additive MAPbBr₃ PeLEDs. EQE and luminescence efficiency (cd/A) for (b) with and (d) w/o DIO additive MAPbBr₃ based PeLEDs, respectively. Inset of (a) and (c) represents the picture of glowing PeLED fabricated from with and without DIO additive perovskite films, respectively.*

Figure **18a** and **18c** represents the J-V-L characteristics of with and w/o DIO additive MAPbBr₃ based perovskite light emitting diodes (PeLEDs), respectively. Figure **18b** and **18d** represents the external quantum efficiency (EQE %) and luminescence efficiency (cd/A) for with and w/o DIO additive MAPbBr₃ based PeLEDs, respectively. The current density for w/o DIO additive MAPbBr₃ based PeLEds is almost one order higher than that of with DIO based PeLEDs. We expect that the lower current density is due to improved morphology by addition of 1 wt.% DIO. We observe that 1 wt.% DIO additive MAPbBr₃ based PeLEDs has the highest luminescence of 350 cd m⁻² and 0.67 cd A⁻¹ at 5V. However, w/o DIO additive PeLEDs has the highest luminescence of 142 cd m⁻² and 0.037 cd A⁻¹ at 5V.

### 5.4.7   Thermal Stability:

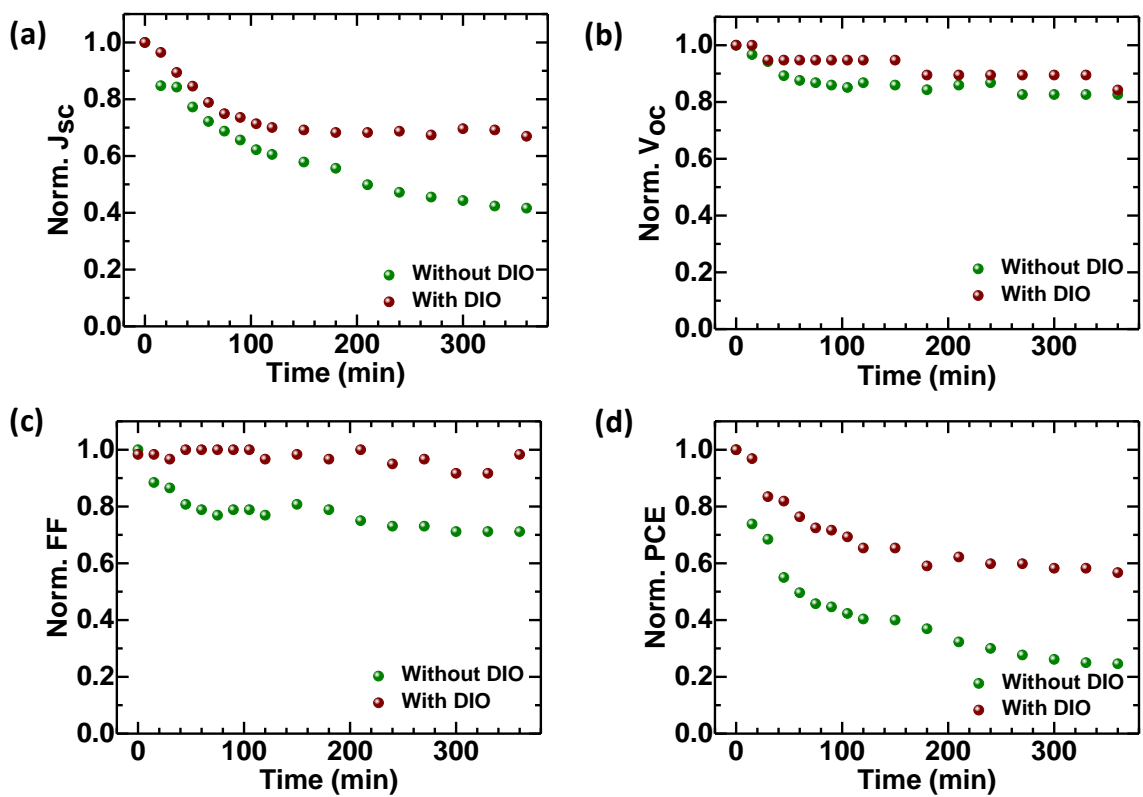

***Figure 19:*** *Thermal stability for photovoltaic parameters of with and w/o DIO additive MAPbBr₃ based perovskite solar cell without encapsulation under continuous exposure to 1Sun irradiation.*

We observed that DIO additive devices have better stability than w/o DIO additive devices. We observed that after 360 minutes of continuous illumination of 1 Sun irradiation, w/o DIO additive MAPbBr₃ device degraded by 80% of its original value. However, with DIO additive MAPbBr₃ device degraded by 40% only (figure **19**). The stability is further more in DIO additive MAPbI$_{3-x}$Cl$_x$ solar cells and the degradation is even less than 20% after 360 minutes (figure **20**). The probable reason behind the thermal degradation of perovskite solar cells will be discussed in the **chapter 6** in good details.

We have also investigated the stability of with (1 wt.%) and w/o additive devices for few months. The devices were encapsulated with glass and kept inside the glove box under dark condition and measured after every 10 days. The changes in different photovoltaic parameters with number of days are shown in figure **21**. The stability data shows that DIO additive devices are more stable than the w/o additive devices. There is only reduction of 30 % in PCE of DIO additive perovskite PVs after 90 days. However, w/o DIO additive devices are degraded very rapidly and show a reduction of nearly 60 % in PCE after 30 days only. The reduction in PCE of both devices is mainly due to reduction in $J_{sc}$. With DIO additive devices were still retain 65% PCE after six months. But, control devices were not showing any diode characteristics after one month.

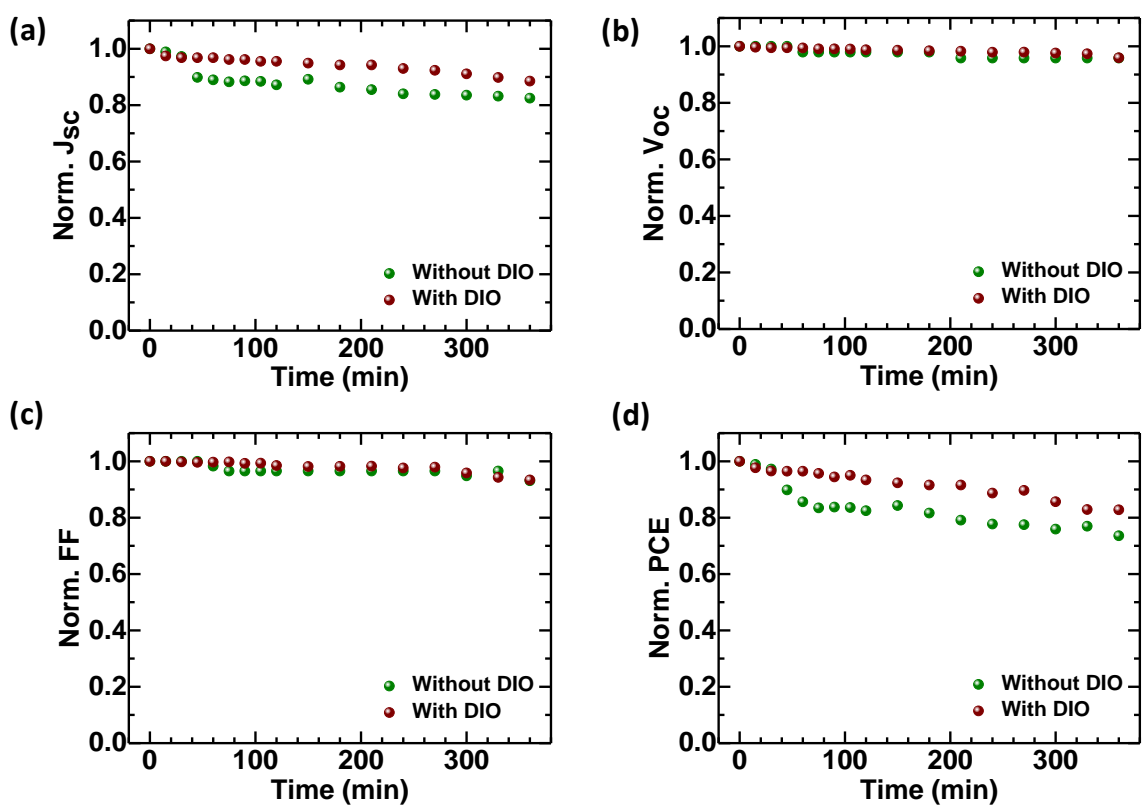

**Figure 20:** *Thermal stability for photovoltaic parameters of with and w/o DIO additive MAPbI$_{3-x}$Cl$_x$ based perovskite solar cell without encapsulation under continuous exposure to 1Sun irradiation.*

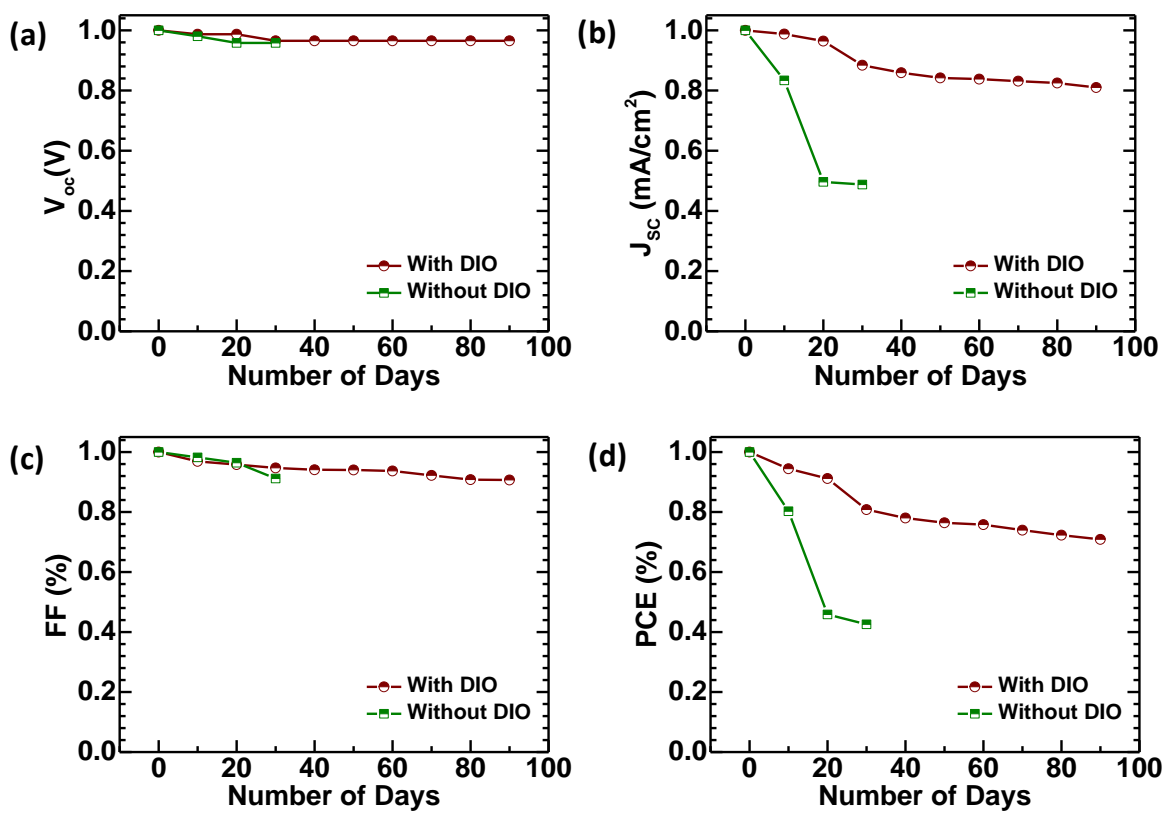

**Figure 21:** *Stability of MAPbI$_{3-x}$Cl$_x$ based PV with (1 wt.%) and w/o additive over few months.*

## 5.5    Discussion:

Based on the above findings, the improved morphology and enhanced crystallization is depend upon the higher boiling point of the solvent additive and its tendency to increase the solubility of perovskite in the host solution. And later on, during thermal annealing, smaller domains (due to the presence of DIO, whose boiling point is higher than the host solvent DMF) fuses to form a compact film with large domains (figure **1**).[36] Such obvious growth in perovskite domain with addition of DIO is consistent with a typical Ostwald ripening process. Ostwald ripening is a phenomenon in solid solutions that describes the inhomogeneous changes in the structure over the time i.e. small crystal dissolves and grows into larger crystals. The concentration of dissolved components near the small particle boundary is always higher than that near the bigger particle boundary. Hence, there is always a concentration gradient of the dissolved component between the surface/boundary of large crystal and small crystal. This is known as Fick's law, which describes the mass transfer of the dissolved components from the smaller particle to larger particle. [37] This can be understood by the relation between the particle radius and chemical potential on the surface of the particle.

$$\mu = \mu_0 + \frac{2\gamma V}{r} \qquad (1)$$

Where, μ is the chemical potential on the surface of the particle, $\mu_0$ is the chemical potential for a flat surface, r is the radius of particle, V is the molar volume and γ is the surface energy. The schematic diagram of nucleation time and growth rate *vs* temperature is shown in figure **22**. Thus, smaller particle have high chemical potential at the surface than the larger particles. The dissolution and growth of perovskite crystals can also be understood by equation (1). The small domains having a large chemical potential continuously undergoes the dissolution process. However, large domains having small chemical potential undergo growth process. The large domain continuously grows until the inverse flux of the dissolved components from the small domain stops. And later on during thermal annealing, smaller domains (due to presence of DIO, whose boiling point is higher than the host solvent DMF) fuses to form compact film with large domains (figure **1**) and less nucleation sites. Whereas, w/o DIO film have more nucleation sites with fast crystal growth, providing relatively smaller domains as seen in morphological studies (figure **1**). However, rate of crystallization is further retarded due to increased wt.% of DIO (2 wt.% and 5 wt.%) in the perovskite precursor solution. This results in even lesser nucleation sites with bigger domains.

The additional metallic $Pb^0$ 4f feature (Figure **9b**) at binding energy 1.8 eV lower than the main doublet (Pb $4f_{5/2}$ & Pb $4f_{7/2}$) is found to be significant in w/o DIO thin film only, which acts as a recombination center.[24,25] This indicates that with DIO thin film has less recombination center than pristine thin film, which facilitate better charge transport in with DIO perovskite based solar cells with higher FF. (figure **15** and **16**). In addition, there is a chemical shift in the core level XPS spectra of Br 3d and Pb 4f towards the lower binding energy in with DIO film (figure **9c** and **9d**) due to different electronegativity banckground/surrouding.[21,22] Since the exact binding energy of an electron is not only depends upon the level from which photoemission is occurring but it also depends upon the local chemical environment. When an atom is bonded to another atom of an element having different electronegativity, the electron binding energy may decrease or increase with the electronegativity of neighbouring atom.

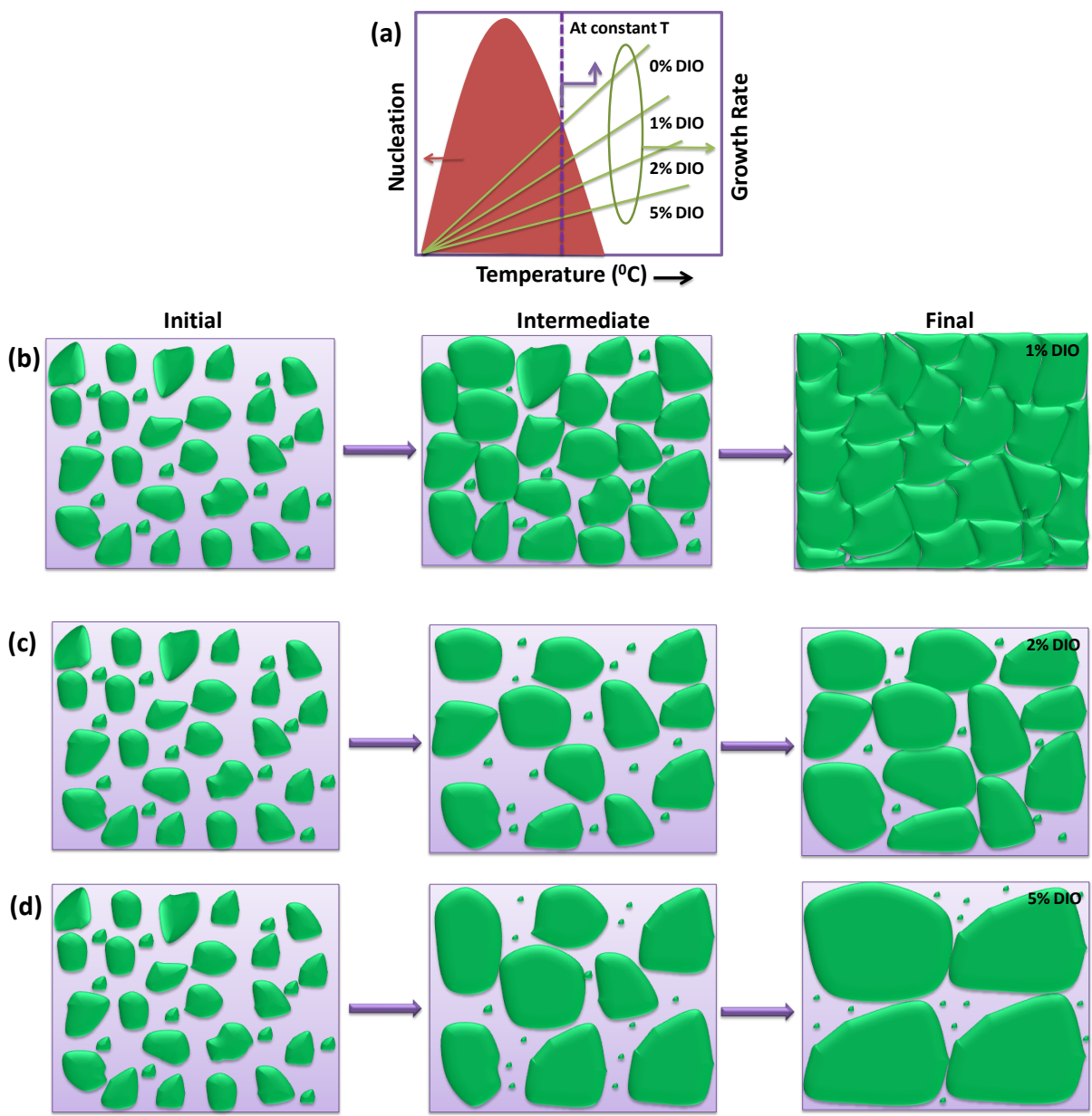

**Figure 22:** *(a) Schematic of thin film nucleation and growth rate of perovskite film with different concentration of DIO in the perovskite precursor solution. Schematic diagram of Ostwald ripening occurs with (b) 0 wt.%, (c) 1 wt.%, (d) 2 wt.% and (e) 5 wt.% DIO additive perovskite films.*

EDS result suggests that iodine to chlorine ratio in with DIO thin film is more than that in w/o DIO thin film (table **1**). Since, $I^-$ is less electronegative than $Br^-$. Thus, we can expect a chemical shift in the photoelectron lines of Pb 4f and Br 3d towards lower binding energy. We observed a reduced ionization potential via upward shift of the VBM of the MAPbBr$_3$ perovskite film with an increase in DIO concentration, which basically reduces the hole injection-barrier at PEDOT:PSS/perovskite interface. The VBM of perovskite is mainly composed of mixed Pb 6s-orbital and halide $n$p-orbital ($n$ is principle quantum number).[38] Thus, the VBM position of perovskite is dominated by halide p-orbital. However, the CBM is mainly composed of Pb 6p orbital. The VBM position is depending on the choice of halide: Cl to Br to I, because of the valence atomic orbitals changes from 3p to 4p to 5p, respectively. I is the largest atom among all the above three mentioned halides, having a higher n value and thus lower binding energy. This can be understood by UPS result, in which incorporation of $I^-$ via DIO in the MAPbBr$_3$ perovskite film causes upward shift in the VBM position from pure perovskite to 5 wt.% DIO additive perovskite film (figure **12**). Along with XPS and UPS results, XRD, absorption, and steady state PL (figure **13** and **14**) results also support the incorporation of $I^-$ in the MAPbBr$_3$ perovskite film. Thus, all above finding confirms that replacement of the hard Lewis base ($Br^-$) by soft Lewis base ($I^-$), which basically prolonged the film evolution time. Along with that, Ostwald ripening phenomenon helps in formation of bigger domain crystallite. Further, we fabricated perovskite solar cells from MAPbBr$_3$ (figure **15**) (also for MAPbI$_{3-x}$Cl$_x$ (figure **16**)) based perovskite films and PeLEDs using MAPbBr$_3$ perovskite film (figure **18**).

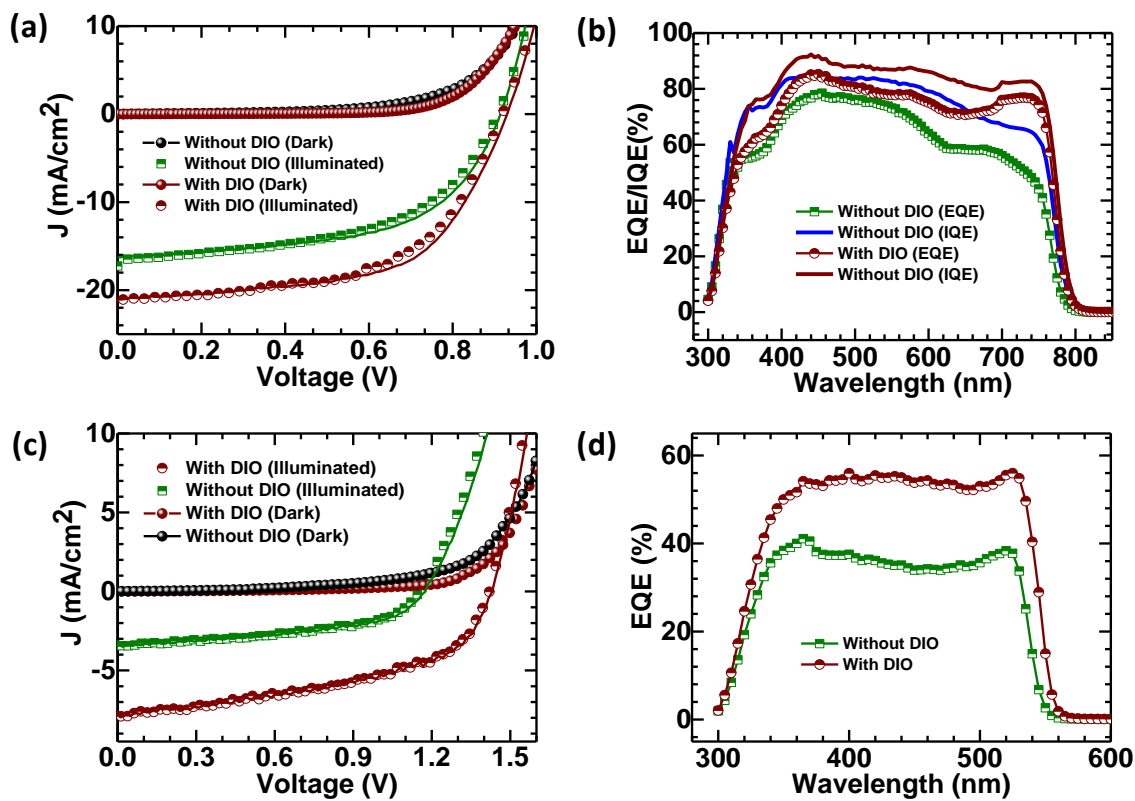

**Figure 23:** *Current-voltage characteristics of (a) MAPbI₃₋ₓClₓ and (c) MAPbBr₃ based with (1 wt.%) and w/o DIO additive solar cells under 1 sun illumination without using BCP as an interlayer between PC₆₁BM and Ag. External quantum efficiency of (b) MAPbI₃₋ₓClₓ and (d) MAPbBr₃ based with (1 wt.%) and w/o DIO additive solar cells.*

We obtained an improved performance of both perovskite solar cells and PeLEDs by addition of 1 wt.% DIO into the precursor solution. These enhanced performances of optoelectronics devices are attributed to improved quality of thin films by DIO additive. We note that thin-film morphology is an important aspect for device performance, however, a lot of optimization also needed on charge extraction layers to get the high efficiency devices. These two are well known separate issues in perovskite device community.[35] Figure **23** represents the device performance of MAPbI₃₋ₓClₓ and MAPbBr₃ based perovskite solar cells without BCP as an interlayer between PC₆₁BM and Ag. We observe that the trend in increase of photovoltaic parameters with addition of DIO is similar in both cases i.e. with or without using BCP as an interlayer (Figure **15** and **16**). However, interlayer helps in improving the charge extraction property. The role of interlayer will be discussed in **chapter 6** in good details. Hence, these device results with two different structures clearly suggest that 1% DIO solvent additive improves the thin-film's optoelectronic properties to achieve the higher efficiency devices.

**Table 5:** *The photovoltaic parameters achieved for with and w/o DIO additive perovskite thin film based solar cells without BCP as an interlayer between PC$_{61}$BM and Ag over an area of 4.5 mm$^2$.*

| Devices | V$_{oc}$ (V) | J$_{sc}$ (mA/cm$^2$) | J$_{sc}$ EQE (mA/cm$^2$) | FF (%) | PCE (%) | cd/A | EL$_{EQE}$ (%) |
|---|---|---|---|---|---|---|---|
| MAPbBr$_3$ | 1.16 | 3.50 | 3.20 | 49 | 2.00 | 0.037 | 0.013 |
| With DIO | 1.42 | 7.91 | 5.86 | 53 | 6.0 | 0.67 | 0.24 |
| MAPbI$_{3-x}$Clx | 0.908 | 16.84 | 16.17 | 51.05 | 8.00 | - | - |
| With DIO | 0.922 | 21.2 | 19.47 | 56.6 | 11.1 | - | - |

## 5.6    Conclusion

In conclusion, we have described a comparative study of the perovskite thin films fabricated from wide band gap MAPbBr$_3$ with and w/o DIO as an additive to the perovskite precursor solution. The DIO plays an important role in providing extra iodine, which participates in the perovskite formation via ion exchange process and thus, it alters the kinetics of crystalline film formation by making the film more iodine-rich with less metallic Pb. The core level XPS photoemission lines of Br 3d and Pb 4f shows a chemical shift towards lower binding energy due to surrounding chemical environment is less electronegative. The EDS results also validate the iodine-rich film by addition of DIO into the perovskite precursor solution. Along with that, steady-state PL, UV-Vis, XRD, XPS and UPS results also confirm the exchange of Br$^-$ by I$^-$. Herein, we show that the addition of DIO in pristine solution can significantly improve the film quality and reduces the density of metallic lead, i.e., non-radiative recombination center, leakage currents and improved charge transport, which leads to more efficient and reproducible photovoltaics and PeLEDs devices.

**Part II: Phosphinic acid as a solvent additive**

## 5.7    Introduction

Low defect and high purity materials are the key to achieve high performance semiconductor based opto-electronic devices. The solution processable method of perovskite solar cells creates impurities and defects in the bulk as well as at the grain boundaries, which acts as recombination centers and thus it affects the optoelectronic properties of perovskite film.[39] Control over the nucleation rate and crystal growth plays a key role in deciding the defect density in the perovskite thin films.6 Advancement in the semiconductor technology such as copper indium gallium selenide (CIGS) and CdTe, based on thin film technology along with polycrystalline silicon are mainly due to passivation of electronic defects at the grain boundaries.[40,41,42] The defects at the grain boundaries created energetic potential barrier which affect the charge transport and thus, finally the PCE of the solar cells. deQuilettes et al. have shown the quenching of PL at grain boundaries in perovskite thin film, which indicates that grain boundaries have large number of non-radiative centers.[43] There are several reports in the literature to improve the morphology of perovskite thin films by using small quantity of additives in the perovskite precursor solution.[44] Most of the studies deal mainly with altering the kinetics of crystal growth, which decides the elemental composition (stoichiometry) and morphology of the final perovskite films. Inorganic acids such as hydroiodic acid (HI), hydrochloric acid (HCl), hydrobromide acid (HBr), hypophosphorous acid (HPA), zwitter-ionic sulfamic acid ($NH_3SO_3$) etc. are widely used as an additive in the perovskite precursor solution to improve the efficiency of PSCs.[45] The reasons behind using such additives are: (a) increase the solubility of perovskite in host solvent and leads to a higher saturation point which promotes the nucleation and ended into bigger domains, (b) they prevent oxidation of I- into $I_2$ which improves the stability and (c) they can form intermediate states with $PbI_2$ which alter the rate of crystallization. There are also few reports on the effect of such additives on the defect density as well.[44,35,46] Shunt paths, charge imbalance, non-radiative recombination, bulk and interfacial defects are the main losses which affect the efficiency of PSCs.[33,47] Hence, the understanding of the parameters that can influence the density of defects in the bulk and grain boundaries of the perovskite thin films is necessary to achieve high performance solar cells.

## 5.8    Experimental details

*Materials:* Lead acetate trihydrate (Pb(Ac)$_2$.3H$_2$O), phosphinic acid and BCP were purchased from Sigma Aldrich. Methyl ammonium iodide (MAI) was purchased from Greatcell Solar. PC$_{61}$BM and PEDOT:PSS were purchased from Solenne BV and Clevios, respectively. All the materials are used as received.

***Solution Preparation:*** 40 wt % perovskite precursor solution was prepared using MAI and Pb(Ac)$_2$.3H$_2$O in the ratio of 3:1 in DMF and stirred at room temperature for overnight. 3 μl of phosphinic acid is added to per mL of the perovskite precursor solution for solvent additive perovskite solar cells. 15mg of PC$_{61}$BM is dissolved in 1 ml of 1,2, dichlorobenzene for overnight stirring at room temperature. 1mg/ml of BCP in isopropanol (IPA) solution was kept under stirring for 2 hours in the glove box at 70$^o$C.

***Device fabrication and Characterization:*** *p-i-n* configuration based perovskite solar cells were fabricated on indium tin oxide (ITO) coated glass substrates. Substrates were cleaned sequentially with soap solution, deionized water, acetone and isopropanol. After oxygen plasma treatment for 10min, PEDOT:PSS was spincoated at 5000rpm for 30 sec on the ITO coated substrates and annealed at 120$^o$C in N$_2$ atmosphere for 20 min. Spincoating of perovskite films and electron transporting layers (ETL) were performed inside the N$_2$ filled glove box with the maintained H$_2$ and O$_2$ levels. PEDOT:PSS coated ITO substrates were transferred into the glove box for perovskite spincoating. Perovskite precursor solution was spincoated on the prepared PEDOT:PSS/ITO substrate at 2000 rpm for 45 sec. The prepared perovskite films undergo drying and annealing. The spincoated light brownish perovskite films were dried at room temperature for 10 min on workbench and then annealed at 100$^o$C on hotplate for 5 min. The PC$_{61}$BM solution was spincoated at 1000 rpm for 60 sec and workbench dried for 5min. After bench drying of the PC$_{61}$BM, BCP was spincoated at 5000rpm for 20sec. The prepared films were transfers into the thermal evaporation chamber to deposited 100 nm of Ag.  Shadow mask of area 4.5mm$^2$ was used to decide the active area of the cell.

Photocurrent density-voltage (J-V) measurements were carried out using a Keithley 4200 semiconductor characterization system and LED solar simulator (ORIEL LSH-7320

ABA) after calibrating through a reference solar cell provided from ABET. All the J-V measurements were performed using scan speed of 40 mV/s. Quantum efficiency measurements were carried out to measure the photo-response as a function of wavelength using Bentham quantum efficiency system (Bentham/PVE300).

The XRD measurements were carried out in Rigaku smart lab diffractometer with Cu $K_\alpha$ radiation ($\lambda$=1.54Å). $\Theta$-2$\Theta$ scan has been carried out from $10^o$-$50^o$ with step size of $0.001^o$. Structural and morphological analysis was done using a Rigaku lab spectrometer X-ray diffraction machine using Cu K$\alpha$ = 1.54 Å and FESEM, respectively. Optical studies were carried out using a spectrometer (PerkinElmer LAMBDA 950) and photoluminescence (PL) spectrometer (Horiba FluoroMax 4). Steady state PL measurement was done on thin-film in vacuum at a pressure of $10^{-5}$ mbar in a custom made chamber. Excitation energy and wavelength were 30 nJ and 355 nm, respectively. For transient PL, a gated i-CCD (ICCD, Andor iStar) was used for detection with excitation laser (3rd harmonic Nd:YAG with emission wavelength of 355 nm) pulse width of 840 ps pulse with 1 kHz repetition rate. Steady state current–voltage–light characteristics were measured using a Keithley 2400 source meter, Keithley 2000 multimeter, and calibrated Si photodiode (RS components). Surface chemical states were obtained from high resolution X-ray photoelectron spectroscopy (XPS, KRATOS) with an Al K$\alpha$ (1486.6 eV) X-ray source. Electronic states are derived from ultra-violet photoelectron spectroscopy (UPS, KRATOS) with He (I) source (21.22 eV). Chamber pressure was below 5.0E-9 Torr for XPS measurement.

## 5.9    Results

We have fabricated a conventional planar heterojunction (*p-i-n*) structure of indium tin oxide (ITO) / poly(3,4-ethylene-dioxythiophene):polystyrenesulfonate (PEDOT:PSS) / $CH_3NH_3PbI_3$ (MAPI) / phenyl-C61-butyric acid methyl ester ($PC_{61}BM$) / Bathocuproine (BCP) / silver (Ag). Figure **24a** shows the crystal and chemical structure of MAPI and phosphinic acid, respectively. Figure **24b** represents the device configuration of planar perovskite solar cells and energy level diagram of each layer.

### 5.9.1 Morphological, Structural and Optical Studies

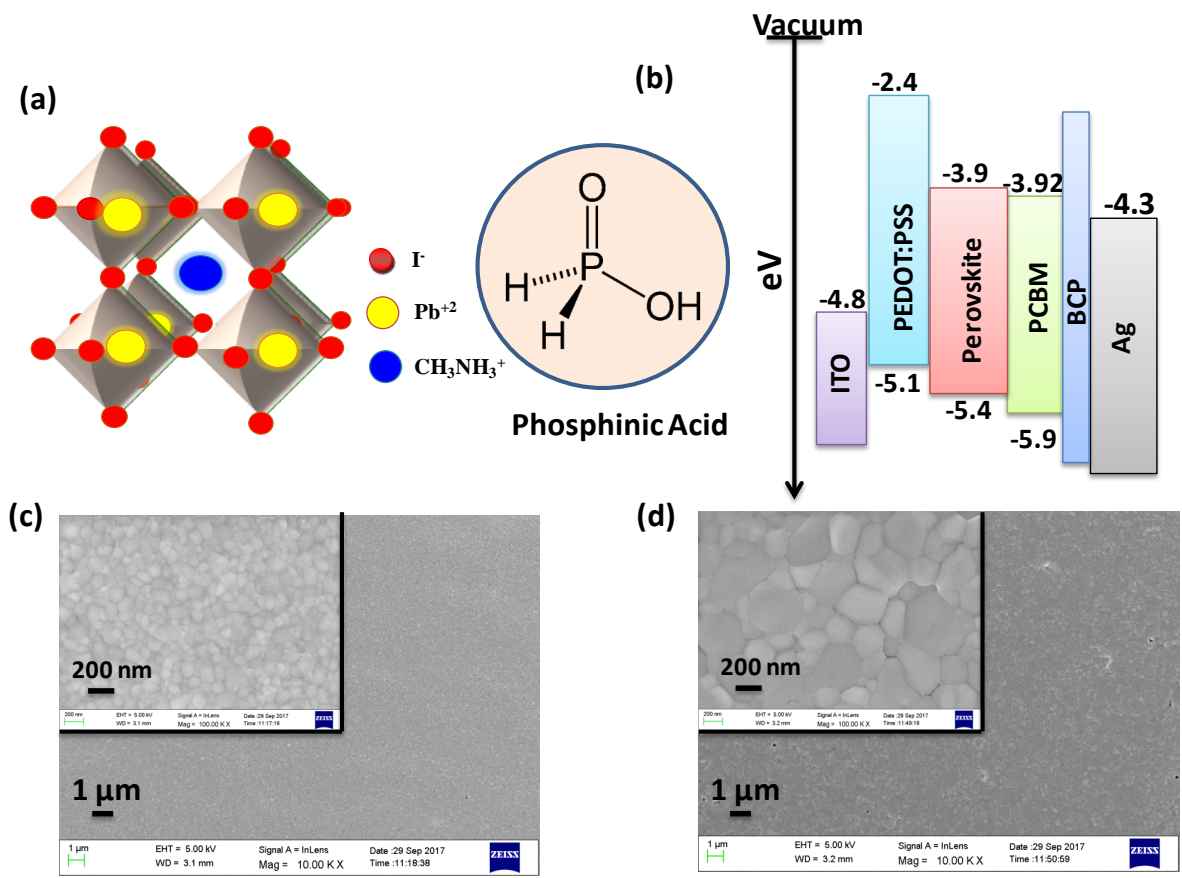

**Figure 24:** *(a) Crystal structure and chemical structure of CH₃NH₃PbI₃ (MAPI) and phosphinic acid (solvent additive), respectively. (b) Energy level diagram of perovskite solar cell. Morphology of (c) control and (d) solvent additive based perovskite thin films over Glass/PEDOT:PSS substrate.*

The perovskite (MAPI) thin films top-view scanning electron microscopy (SEM) images of without additive (termed as control) and with phosphinic acid (termed as solvent additive) over PEDOT:PSS/Glass substrate at different length scale are shown in Figure **24c and 24d,** respectively. The SEM images reveal that control film is compact with small domains (~ 100 nm). By incorporating solvent additive into the perovskite solution, the film morphology is still compact with increased grain size (~500 nm).[48,49,50] Figure **25a** represents the X-ray diffraction (XRD) pattern of control and solvent additive thin films. Inset of figure **25a** shows that full width half maxima (FWHM) of control perovskite film (2Θ = 0.117⁰) is higher than that of solvent additive perovskite film ($2\Theta = 0.105^0$), which supports the increased domain size of solvent additive perovskite film as shown in FESEM images (figure **24d**). Figure **25b** represents the optical density (O.D.) of control and solvent additive perovskite films. The improved absorption due to solvent additive is in good agreement with FESEM (figure **24d**) and XRD (figure **25a**).

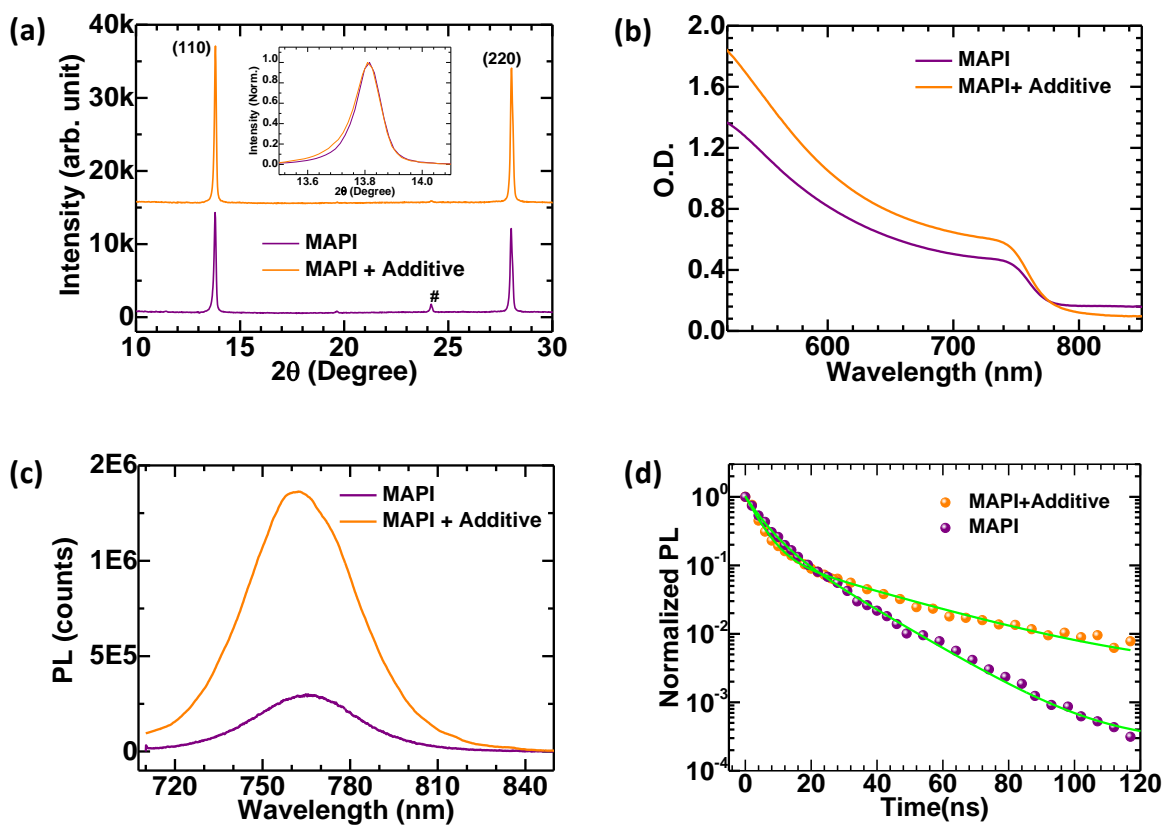

***Figure 25:*** *(a) XRD pattern and (b) absorption spectra of control and solvent additive perovskite thin films. Inset of (a) represents the FWHM of (110) peak for both kinds of films. # represents the orientation of crystals in (222) direction in control film. However, it is completely removed in solvent additive film. (c) Steady-state and (d) time-resolved PL spectra of control and Solvent additive perovskite thin films.*

Urbach energy ($E_u$) is extracted from absorption, which is a parameter of disorder in the bulk, by using the equation

$$A = A_0 \exp\left(\frac{E}{E_u}\right) \qquad (2)$$

Here, A is absorption, $A_0$ is constant and E is the excitation energy. We observe sharper band-edge rise in absorption of solvent additive film in comparison to control film. Solvent additive perovskite film shows a lower urbach energy ($E_u$ = 39 meV) than control film ($E_u$ = 56 meV).

Figure **25c** represents the steady-state photoluminescence (PL) spectra of control and solvent additive perovskite film. Enhanced PL is observed in solvent additive perovskite film in comparison to control film. This suggests that solvent additive suppressed the non-radiative recombination. Figure **25d** represents the time-resolved PL of control and solvent additive perovskite films. Time resolved PL spectra of solvent additive based perovskite films shows an improved average lifetime 18.23 ns in comparison to control film having average lifetime of 10.82 ns. The PL lifetime of both films are fitted through bi-exponential decay function having a fast decay and a slow decay lifetime. The parameters related to PL decay are listed in table **6**. For control film, the fast decay lifetime is 4.98 ns and its fraction is 70%. Slow decay lifetime for control perovskite film is 15.25 ns with fraction of 30%. But, solvent additive perovskite based film has fast decay lifetime of 4.28 ns with fraction of 86% and slow decay lifetime of 30.45 ns with fraction of 14%. The weight fraction of fast decay process is increased from 70% to 86% by addition of solvent additive in the perovskite precursor solution. The higher slow decay lifetime and increased weight faction of fast decay in solvent additive film promotes radiative recombination.

***Table 6:*** *Time-Resolved Photoluminescence Characterization for control and solvent additive perovskite Thin Films.*

| Perovskite Films | $A_1$ | $\bar{\tau}_1$ (ns) | $A_2$ | $\bar{\tau}_2$ (ns) | $\bar{\tau}_{av}$ (ns) |
|---|---|---|---|---|---|
| MAPI | 0.70±0.05 | 4.98±0.28 | 0.30±0.05 | 15.25±0.62 | 10.82 |
| MAPI+ Additive | 0.86±0.04 | 4.28±0.30 | 0.14±0.03 | 30.45±0.13 | 18.23 |

### 5.9.2    Chemical and Electronic States Analysis

To investigate the changes in chemical environment of solvent additive based perovskite film, we have performed X-ray photo-electron spectroscopy (XPS) on both films. Figure **26a** represents the full scan spectra (1200 eV-0 eV) of both control and solvent additive films coated over PEDOT:PSS/Glass substrate. Phosphorous is one of the main components of phosphinic acid (figure **24a**). It is observed that in survey scan spectra of solvent additive perovskite film there is no trace of phosphorous element (figure **26a**). We have calculated I/Pb ratio from Pb 4f and I 3d spectra of survey scan and it was found that the ratio for control film is much more away (2.37) from the required ratio (3.00). However, the ratio of I/Pb is 2.83 in solvent additive perovskite thin film. This indicates that there is a I deficiency in both kinds of perovskite films but it is higher for control one. Thus, phosphinic acid helps to maintain the stoichiometry of I/Pb in perovskite film. Figure **26b** represents the core level spectra of Pb 4f and it is observed that control additive based perovskite film has an additional Pb 4f feature at binding energy 1.8 eV lower than the main doublet, which is assigned as metallic lead ($Pb^0$). However, we do not observe any metallic lead feature in the solvent additive based perovskite film.

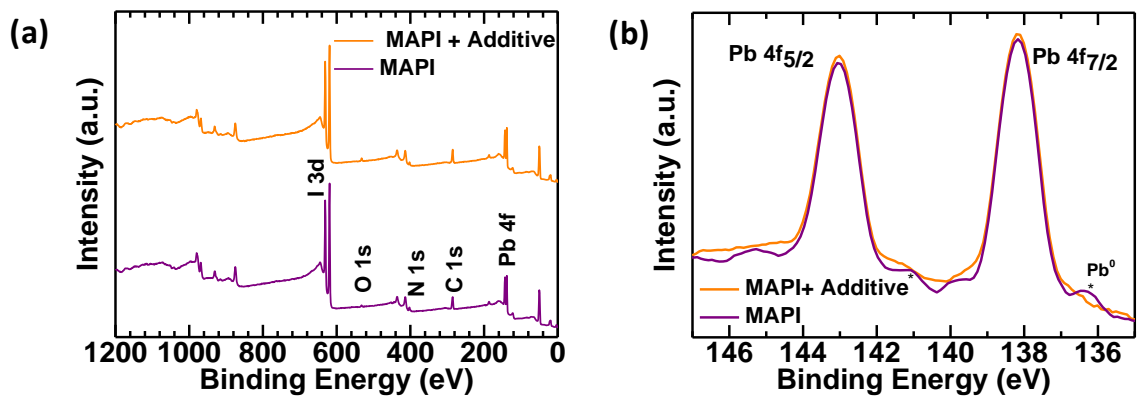

***Figure 26:*** *(a) XPS full scan spectra for control and solvent additive perovskite thin films. (b) Core level high resolution XPS spectra of Pb 4f for control and solvent additive perovskite thin films.*

We further investigated the electronic structure of both control and solvent additive films through ultra-violet photoelectron spectroscopy (UPS) and it is shown in figure **27**. An upward shift in work-function of solvent additive film (3.99 eV) is observed in comparison to control film (4.10 eV). Hence, the fermi-level ($E_F$) shifts by 110 meV towards the conduction band minimum. This represents a more n-type behavior for the solvent additive film.[51] However, we did not observe any significant shift in the energy levels towards the vacuum.

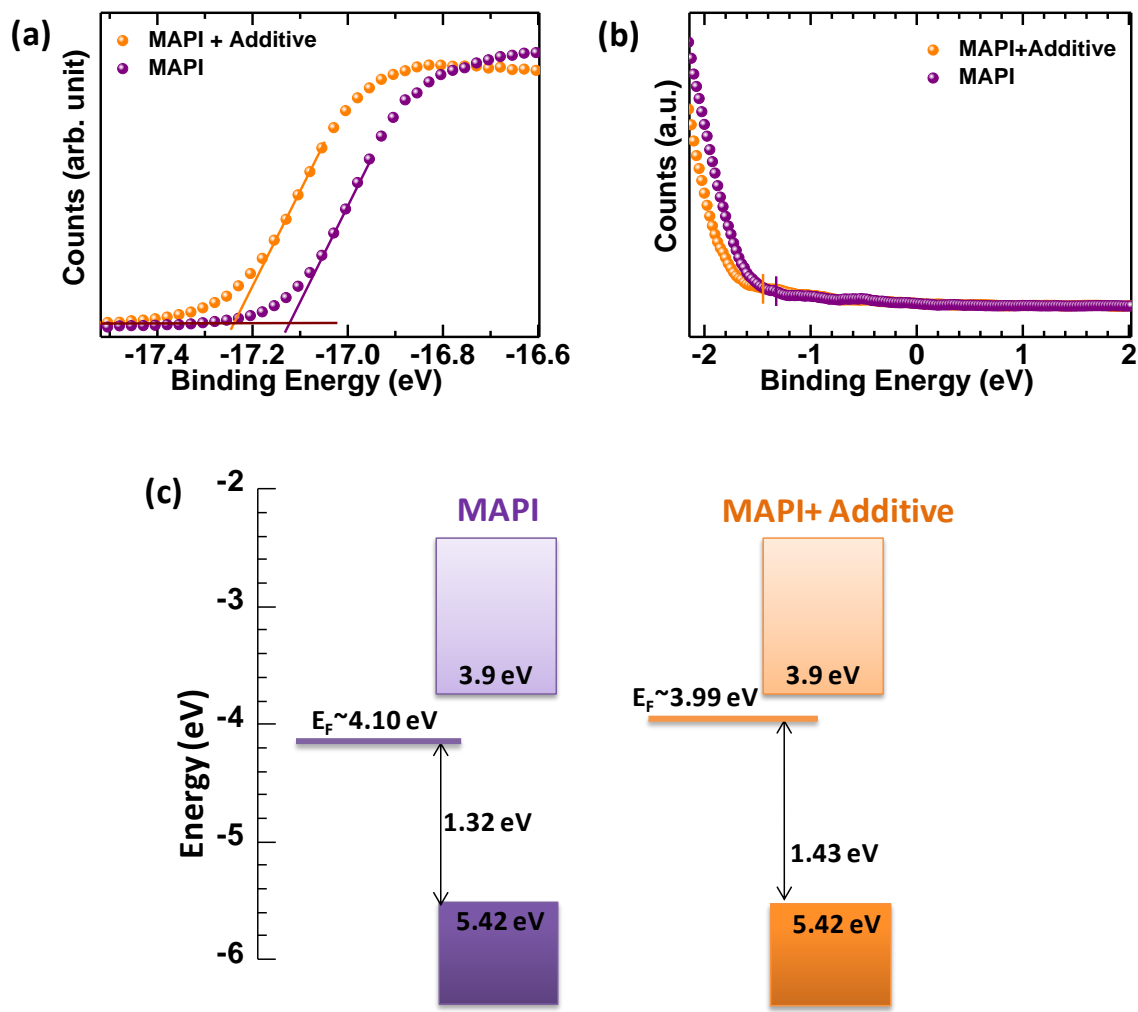

***Figure 27:*** *(a) and (b) UPS spectra of control and solvent additive perovskite thin films. (c) Electronic states derived from UPS spectra for both kinds of perovskite films.*

### 5.9.3 Photovoltaic and Light Emitting Diode Performance:

Figure **28a** shows the current density vs. voltage (J-V) characteristics of MAPI based perovskite solar cells under 1 Sun condition (100 mW/cm$^2$). The photovoltaic parameters of control and solvent additive perovskite solar cells are listed in table **7**. The devices made with solvent additive shows better power conversion efficiency (PCE) of 16.22% with open circuit voltage ($V_{oc}$) of 1.01 V, a short-circuit current density ($J_{sc}$) of 21.13 mA/cm$^2$ and fill factor (FF) of 76 %. In contrast, the control devices showed a PCE of 11.40% only with $V_{oc}$ of 0.89 V, a $J_{sc}$ of 18.57 mA/cm$^2$ and FF of 69 %. The increased $J_{sc}$ of solvent additive device can be understood by external quantum efficiency (EQE) and it is shown in figure **28b**.

Figure **29a** represents the J-V-L characteristics of control and solvent additive PSCs. We observe that solvent additive devices have lower dark current density than the control one. Figure **29b** represents the electroluminescence (EL) spectra of both control and solvent additive PSCs. We note that there is a slight blue shift in the solvent additive device in comparison to control PSC. We also notice that the EL quantum efficiency (QE) of solvent additive devices is even more than two orders of magnitude higher than that of the control device (figure **29c**). In control device, EL$_{QE}$ represents a bias dependent characteristic. However, a solvent additive device shows bias independent EL$_{QE}$ characteristics. It started saturating even at low applied bias of 3 V. We will discuss about it in detail in discussion section.

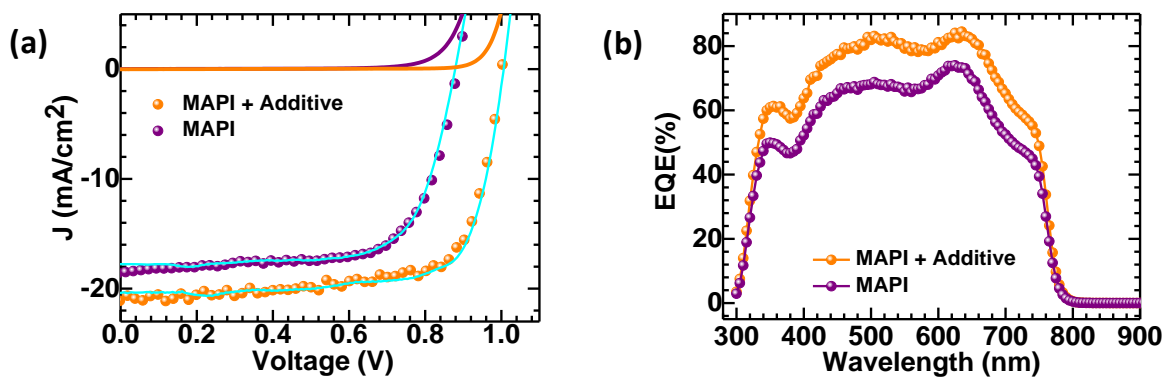

***Figure 28:*** *(a) J-V characteristics and (b) EQE of with and w/o solvent additive based MAPI perovskite solar cells under 1Sun illumination.*

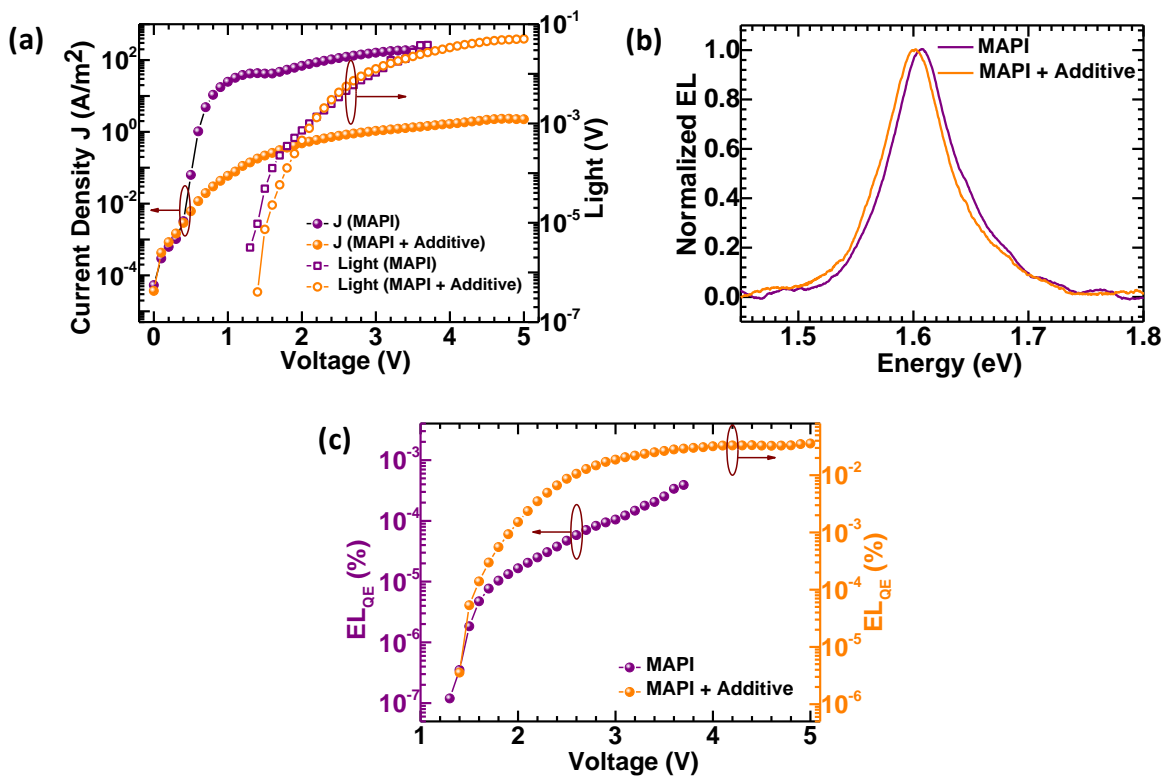

**Figure 29:** *(a) J-V-L, (b) electroluminescence (EL) and (c) Electroluminescence quantum efficiency of control and solvent additive perovskite solar cells.*

**Table 7:** *Photovoltaic parameters of the best performing control and solvent additive MAPI based perovskite solar cells. Average PCE over 16 devices are also listed.*

| Devices | $V_{OC}$ (V) | $J_{SC}$ (mA/cm²) | FF (%) | PCE (%) | Avg. PCE (%) |
|---------|------|--------|-----|-----|-----|
| MAPI | 0.89 | 18.57 | 69 | 11.4 | (9.2 ± 2.2) |
| MAPI + Additive | 1.01 | 21.63 | 74 | 16.2 | (14.9 ± 1.3) |

## 5.10    Discussion

The correct stoichiometry of MAI vs Pb is one of the potential candidates to decide the numbers of defects in a polycrystalline perovskite thin film. Since, MAI is hygroscopic material; there is also a probability of oxidation of $I^-$ to $I_2$ for a long term storage. Even, during thermal crystallization, there is also a chance for occurrence of oxidation of $I^-$. Thus, we need a strong reducing agent which can retard the oxidation of $I^-$ to $I_2$. We have demonstrated that addition of phosphinic acid into the perovskite precursor solution to retard the oxidation of $I^-$ to $I_2$, which can be understood by increased stoichiometry ratio of I/Pb for solvent additive perovskite thin film (2.83). However, the vapor pressure of phosphinic acid is lower than that of host solvent DMF. Hence, it will enhance the rate of crystallization and ends up with bigger domains (figure **24d**). The same observation was reported in case of water additive perovskite domains. Higher crystallinity and increased grain size also reflects from absorption and XRD results (figure **25**). The reduced energetic disorder is also understood by lower urbach energy for solvent additive perovskite film (figure **25b**). The $EL_{QE}$ measurement shows the amount of light emitted from the PSC by applying a forward bias to the number of charge carriers injected in the device. If non-radiative trap assisted states are available in the device, the injected charge carriers are recombine via non-radiative channels and shows a bias dependent $EL_{QE}$, which is the case in control device (figure **29c**). However, $EL_{QE}$ of solvent additive device starts saturating at nearly 3V. This indicates that addition of phosphinic acid into precursor solution helps in reduction of non-radiative channels and promotes radiative recombination. Reduction of defect sites can also be understood by core level photoemission lines of Pb 4f, where fraction of metallic lead is reduced by addition of phosphinic acid (figure **26b**). Improved $V_{oc}$ suggest that solvent additive device has lower tail state density enabling an increased separation of the quasi fermi level under illumination. Enhanced PL and longer average life-time of solvent additive based perovskite film also support the reduction of trap states. Thus, $EL_{QE}$ and PL measurements validate the improved $V_{OC}$ by addition of phosphinic acid in perovskite solution. Hence, the above study suggests that the role of phosphinic acid in enhancing the efficiency of $CH_3NH_3PbI_3$ by reduction of trap states and thus, improving the charge transport.

## 5.11    Conclusion

A comparative study was performed between best-performing control and solvent additive $CH_3NH_3PbI_3$ perovskite solar cells. Higher crystallinity and improved morphology of solvent additive film are evident from XRD and SEM measurement, respectively. As a result of it, higher knee voltage (dark J-V), $V_{oc}$ and better charge transport property ($J_{sc}$ and FF) in illuminated J-V curves are observed. Addition of small amount of solvent additive in the perovskite solution reduces the number of recombination centers. The enhanced steady state PL intensity and transient PL average lifetime shows that phosphinic acid makes the perovskite film more defect free than the control one, which is in good agreement with XPS results. Thus the increased crystallinity enhanced optical absorption, bigger domain size, improved charge transport property, and less number of recombination centers explaining the superior performance of solvent additive devices compared to control devices.

## 5.12    Post script

Overall, this chapter helps to understand the crystal growth kinetics of perovskite thin films using two different kind of solvent additives (DIO and phosphinic acid). We correlate the possible mechanism to validate the enhancement in the PCE of different halide based perovskite solar cells ($MAPbI_{3-x}Cl_x$, $MAPbBr_3$ and $MAPbI_3$) by reducing the number of defect density in the bulk and at the grain boundaries with the help of different opto-electronic techniques such as steady state PL, time resolved PL, photoelectron spectroscopy, electroluminescence quantum efficiency etc. Apart from efficiency number, moisture stability is another big issue in the field of perovskite solar cells. Phosphinic acid and DIO as an additive in the perovskite precursor solution enhances the PCE of the PSCs, but, stability is still an issue in these devices. Hence, there is need of such kind of additives which can improve the efficiency and stability both at the same time, will be presented in the next chapter.

# Chapter 6

# Investigation on Organic Molecule Additive for Moisture Stability and Defect Passivation via Physisorption in $CH_3NH_3PbI_3$ Based Perovskite.

# CHAPTER 6

# Investigation on Organic Molecule Additive for Moisture Stability and Defect Passivation via Physisorption in CH$_3$NH$_3$PbI$_3$ Based Perovskite.

**Abstract:** Perovskite semiconductors are known to have dynamic disorder which is the main cause to affect its recombination dynamics. These defects seem to be controlled via various processing conditions in literature. Here, we studied the role of small organic molecule, bathocuproine (BCP), as an additive in the CH$_3$NH$_3$PbI$_3$ (MAPbI$_3$) thin films based devices using optoelectronic measurements and first principle calculation. A perovskite solar cell with power conversion efficiency (PCE) of ~16% and fill-factor of 82% with no significant hysteresis is achieved, where BCP not only passivate the bulk and interfacial defects but also shows highly improved electroluminescence efficiency. Addition of BCP into perovskite precursor does not cause any structural change in the 3D structure of perovskite, which is confirmed using time-delayed emission spectroscopy and first principle electronic structure calculations. Furthermore, calculations suggest that a physisorption type of interaction has been found in between MAPbI$_3$ and BCP with an average distance of 2.9 Å. Contact angle measurement suggest that BCP provides a high moisture resistivity which is consistent with the stability results. Hence, hydrophobic organic molecule plays an important role in defect passivation and moisture stability to achieve efficient and stable perovskite solar cells. This study has also shown to work for light emitting diodes (LEDs) application, hence can be utilized for wider bandgap semiconductors LED application and/or triple cation based 3D perovskite solar cells. Apart from moisture stability, this study also deals with thermal stability of perovskite solar cells by using double hole and double electron extraction layers. PEDOT:PSS is mostly known for its acidic & hydrophilic nature and is one of the main reason for degradation at the interfaces. MoO$_3$ is used as a double hole extraction layer along with PEDOT:PSS to improve the charge extraction as well as thermal stability of the perovskite solar cell.

## 6.1 Introduction

Recently, organic-inorganic halide perovskites have received tremendous attention from the photovoltaic community due to their low cost and easy fabrication technique[1,2]. For the very first time, power conversion efficiency (PCE) of 3.9% for perovskite solar cell was reported by Miyasaka et. al. in 2009[3]. Within a short-time period, PCE of perovskite solar cells has exceeded 20%[4,5]. Methylammonium lead iodide (MAPbI$_3$) is the most commonly used perovskite in the community. Its small binding energy, higher diffusion length, high optical absorption and carrier mobility has created wide interest for the material in the perovskite photovoltaic community.[6,7] For the synthesis of MAPbI$_3$, lead halide has been the most common lead source. There are some reports in the community where conventional lead halide source (lead iodide or lead chloride) has been replaced with lead acetate[8]. Use of lead acetate results in smooth and pin-hole free perovskite films without the need of any further additives or post treatment. There is a report from Zhang *et. al.*, which shows the superiority of lead acetate over the conventional lead halide sources8. But, there are defects in the perovskite films even using lead acetate route, which need to be eliminated in order to improve the solar cell efficiency. Zhang *et. al.* have shown that addition of hypophosphorous acid in the precursor solution can significantly improve the PCE of perovskite solar cells[9].

Perovskite solar cells have been fabricated by using one step[10,11,12], two step[13,14], vapor assisted deposition[15], anti-solvents[16] and additive methods.[17,18] But, there are very few reports on the perovskite-organic blend in an inverted device structure.[19,20] This may be due to lower solubility of the organic component in the solvents used for perovskite precursor such as N,N-dimethylformamide (DMF) and dimethylsulphoxide (DMSO). Degradation of perovskite solar cells also suffers due to internal ion migration, which is triggered by thermal treatment. In case of MAPbI$_3$, the accumulation of I$^-$ at the PC$_{61}$BM/Ag interface, results in decomposition of MAPbI$_3$ by formation of AgI at the interface.[21] This internal cross layer ion migration caused an irreversible damage on the perovskite layer and metal electrode. In perovskite solar cells with *p-i-n* configuration, suitable buffer layers are incorporated between PC$_{61}$BM and electrode[18] Bathocuproine (BCP) has been widely used as a buffer layer in *p-i-n* structure based perovskite solar cell to enhance the device performance by establishing an ohmic contact between $PC_{61}BM$ and metal electrode, which basically prevent the perovskite film to form metal induced charge state.[21,22,23]

Solution-processed thin film based perovskite solar cells in general suffers from pin-holes and defects at the grain boundaries and perovskite surface. There are various reports in the community to overcome from these challenges by introducing other organic/inorganic molecules in the perovskite thin films. But, it is found that addition of such organic/inorganic molecules reduces the dimensionality (3D to 2D) of the perovskite in order to passivate the defect sites and to improve the moisture resistive property of the perovskite.[24,25] The major drawback of these quasi 2D perovskite solar cells is their relatively poor photovoltaic performance in comparison to 3D perovskite based solar cells.[26,27,28] Hence, there is a necessity of a novel method, which improves the moisture stability and defect passivation in the perovskite solar cells without compromising with the photovoltaic performance. In this chapter, we investigate the role of BCP as an additive in the perovskite precursor solution. We explain the mechanism by which BCP passivate the defects in the perovskite thin film in addition to enhanced the perovskite's resistivity to the moisture without altering the 3D structure of $MAPbI_3$. However, the solubility of BCP is limited in DMF and thus there is a need to optimize the amount of BCP addition into the perovskite precursor solution. Moreover, we demonstrate that this is equally important for light emitting diodes and hence can be further extended to wider band gap perovskite semiconductors too.[29]

## 6.2 Experimental details

***Materials:*** Lead acetate trihydrate and methyl ammonium iodide (MAI) were purchased from Sigma Aldrich and used as received. $PC_{61}BM$, BCP and PEDOT:PSS were purchased from Solenne BV, Sigma Aldrich, and Clevios, respectively.

***Solution preparation:*** 40 wt. % perovskite precursor solution was prepared by mixing MAI and lead acetate trihydrate in 3:1 ratio in anhydrous DMF and stirred for overnight at room temperature. 0.02 wt.% of BCP is added to the perovskite precursor solution for perovskite-BCP solar cells. 15 mg $PC_{61}BM$ was dissolved in 1 mL of 1,2-dichlorobenzene and kept for overnight stirring at room temperature. 1 mg BCP was dissolved in 1 mL of isopropanol and kept at 70$^o$C for 15 min before use. BCP solution was used as a buffer layer in between PC$_{61}$BM and Ag.

***Perovskite solar cell fabrication and characterization:*** ITO coated glass (10 Ω sq$^{-1}$) was cleaned by sequential sonication with soap solution, deionized water, acetone and isopropanol for 10 min each, respectively. After drying with nitrogen flow, heat the substrate at 110$^o$C for 10 min on hotplate to remove the organic solvents. After that, cleaned ITO substrate was treated with oxygen plasma for 10 min. Oxygen plasma treated ITO coated glass were spincoated with PEDOT:PSS at 5000 rpm for 30 sec and annealed at 120$^o$C for 30 min in nitrogen atmosphere. Then, PEDOT:PSS spincoated substrates were transferred into nitrogen filled glove box and perovskite precursor solution was spincoated over PEDOT:PSS/ITO at 2000 rpm for 45 sec. The thickness of both films (with and without BCP) is measured by Dektak profilometer and thickness of both films is ~ 400 nm. After that, perovskite coated substrate kept at working bench for 10 min for room temperature crystallization. Then, transfer the perovskite spincoated ITO substrate to the hotplate and keep it at 100$^o$C for 5 min. After that, PC$_{61}$BM was spincoated at 1000 rpm for 30 sec and then BCP was spincoated at 5000 rpm for 20 sec. Finally, 100 nm Ag was deposited under a vacuum of 1X10$^{-6}$ mbar. For all studied devices, the device area was defined as 4.5 mm$^2$ by metal shadow mask.

All the photovoltaic measurements were carried out in ambient atmosphere. Photocurrent density-voltage (J-V) measurements were carried out using a Keithley 2400 source meter and solar simulator after calibrating through a reference solar cell provided from Accreditation Board for Engineering and Technology (ABET). Light-intensity dependent measurements were carried out under the same solar simulator using neutral density (ND) filters. All the J-V measurements were performed using scan speed of 40 mV/s. EQE and IQE measurements were carried out to measure the photoresponse as a function of wavelength using Bentham quantum efficiency system. The XRD measurements were carried out in Rigaku smart lab diffractometer with Cu K$_\alpha$ radiation (λ=1.54Å). 2θ scan has been carried out from 10$^o$-60$^o$ with step size 0.001$^o$. Structural and morphological analysis was done using a Rigaku lab spectrometer X-ray diffraction machine using Cu Kα = 1.54 Å and FESEM, respectively. Optical studies were carried out using a spectrometer (PerkinElmer LAMBDA 950) and photoluminescence (PL) spectrometer (Horiba FluoroMax 4). Steady state PL measurement was done on thin-film in vacuum at a pressure of $10^{-3}$ mbar in a custom made chamber. Excitation energy and wavelength were 30 nJ and 355 nm, respectively. For transient PL, a gated i-CCD (ICCD, Andor iStar) was used for detection with excitation laser (3rd harmonic Nd:YAG with emission wavelength of 355 nm) pulse width of 840 ps pulse with 1 kHz repetition rate. Steady state current–voltage–light characteristics were measured using a Keithley 2400 source meter, Keithley 2000 multimeter, and calibrated Si photodiode (RS components). Surface chemical states were obtained from high resolution X-ray photoelectron spectroscopy (XPS, KRATOS) with an Al Kα (1486.6 eV) X-ray source. Chamber pressure was below 5.0E-9 Torr for XPS measurement.

## 6.3 Results

The typical crystal structure of the perovskite ($CH_3NH_3PbI_3$) and chemical structure of BCP are shown in figure **1a** and energy level diagram of the cell is given in figure **1b.**[30,31,32] We have fabricated an inverted perovskite solar cell using perovskite with BCP addition as an active layer instead of just as buffer layer.[21,22] The top views scanning electron microscopy (SEM) images of films prepared from with and without BCP additive perovskite are shown in figure **1c** and figure **1d,** respectively. Perovskite film made from pristine solution has small domains (~ 50 nm) and is compact. We used 0.02 wt.% BCP as an additive in the perovskite precursor solution. The corresponding SEM image (figure **1c**) shows more compact and smooth film than the pristine film. In addition to improved film morphology, the domain size (~ 150nm) in film prepared from perovskite containing BCP is also increased. The thickness of both films (with and without BCP) is measured by Dektak profilometer and shown in figure **2**. Thickness of with and without BCP additive films are almost identical and are ~ 400 nm.

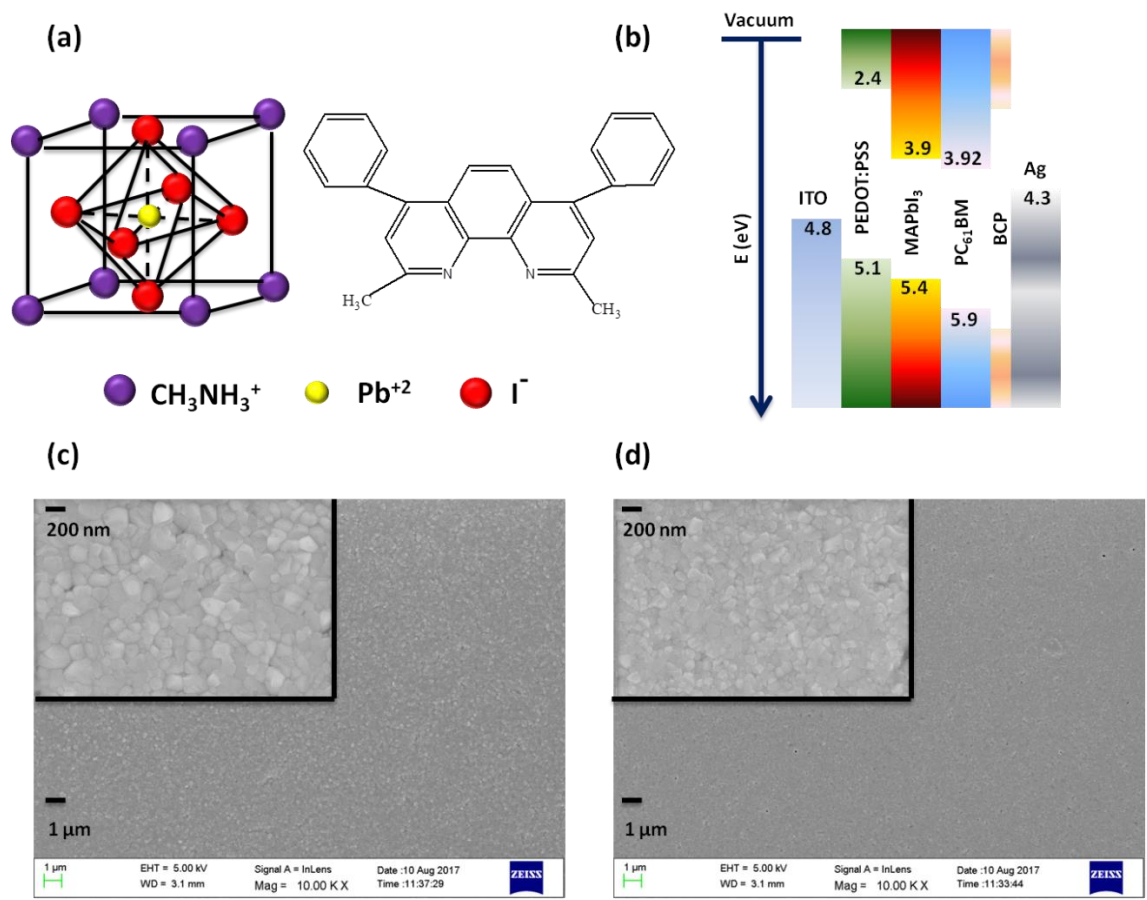

**Figure 1**: (a) Crystal and chemical structure of perovskite (MAPbI₃) and BCP, respectively. (b) Schematic of energy level diagram of perovskite solar cells, energy level values are taken from literature[32]. Microstructures of (c) with and (d) without BCP additive MAPbI₃ based perovskite film.

We have characterized the perovskite-BCP films with different BCP ratio via SEM and atomic field microscopy (AFM). **Figure 3** represents the SEM and AFM images of perovskite-BCP additive films with different BCP concentrations. These results reveal that when we increased the BCP concentration beyond 0.02 wt. %, BCP aggregates in the perovskite film. The size of aggregates increases with increase in BCP concentration (AFM image in figure **3**). However, both SEM and AFM data supports that no BCP aggregates are seen for 0.02 wt. % concentration of BCP. Hence for further characterization and device fabrication, we used 0.02 wt. % concentration of BCP as an additive in perovskite precursor solution. AFM results also show that roughness of perovskite film decreases with increase in the BCP concentration (figure **3**) and it is listed in Table **1**. Having a small quantity of organic molecule BCP in precursor solution facilitates the filling of any possible voids during the growth of perovskite film, which is very useful for device operations in order to prevent leakage current through pinholes.[33]

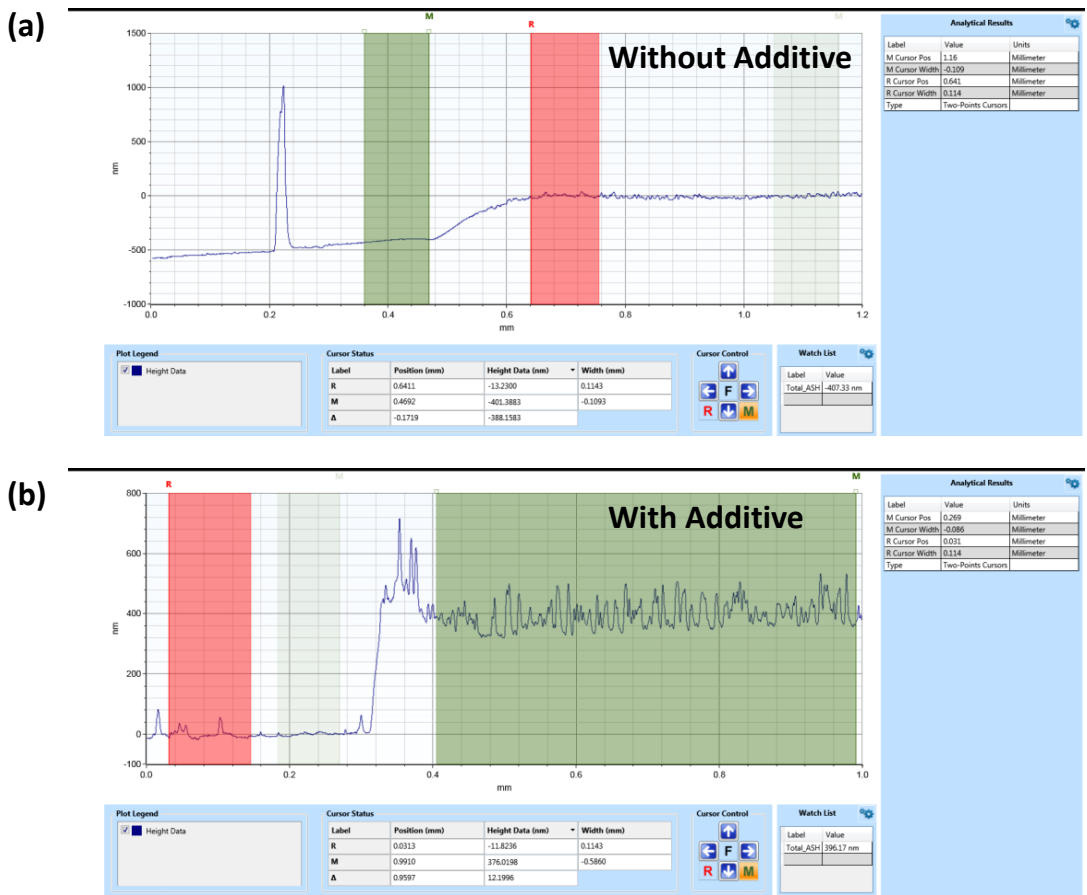

***Figure 2:*** *Thickness of with and without BCP additive Perovskite films measured by Dektak profilometer.*

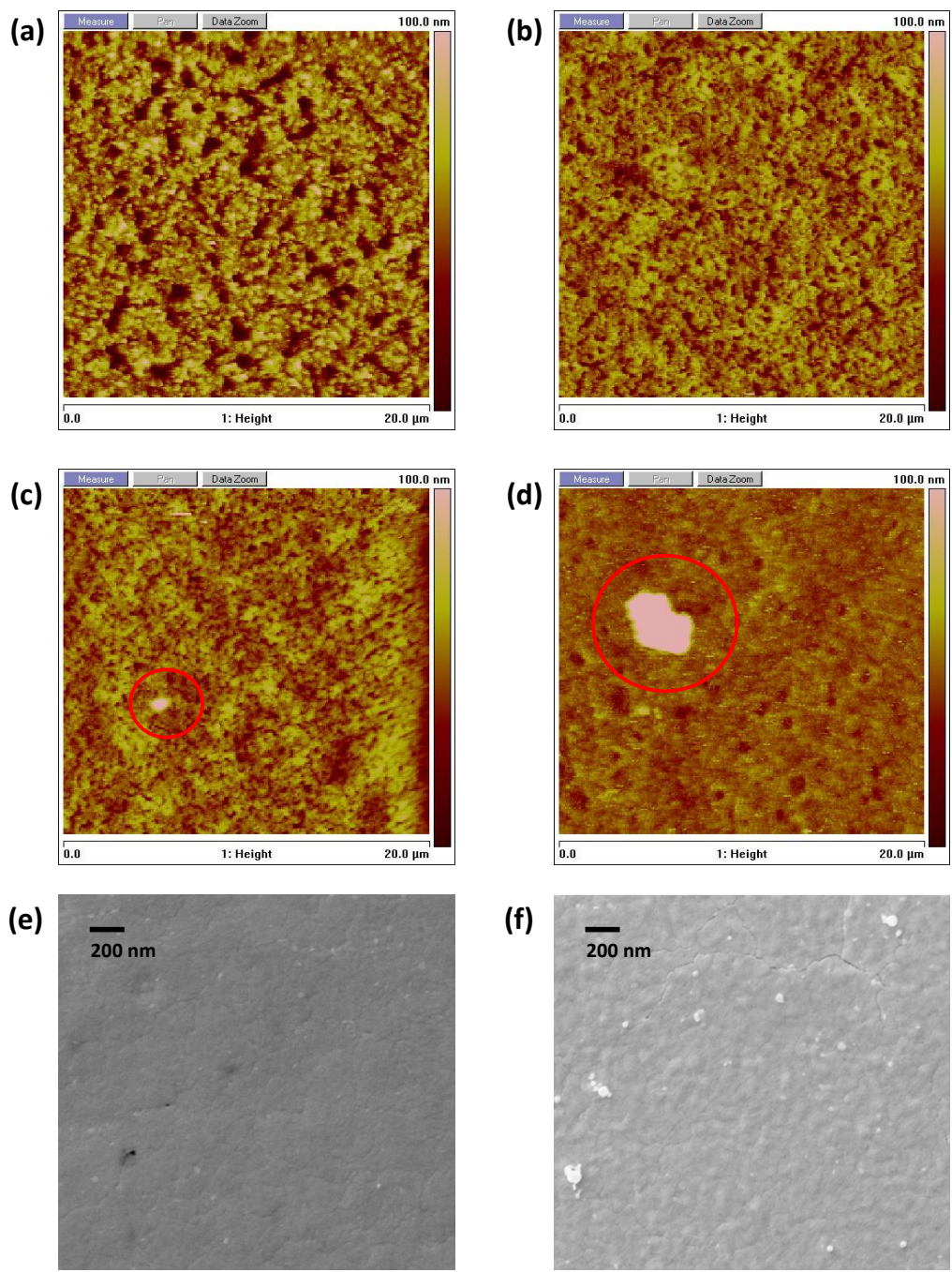

***Figure 3:*** *AFM images of (a) pristine and BCP additive perovskite thin films with (b) 0.02 wt.%, (c) 0.06 wt.% and (d) 0.1 wt.%. FESEM images of BCP additive perovskite thin films with (e) 0.06 wt.% and (f) 0.1 wt.%.*

**Table 1:** *Root mean square (RMS) roughness of pristine and BCP additive perovskite thin films with different wt. %.*

| MAPbI$_3$ + X wt. % of BCP | RMS Roughness (nm) |
|:---:|:---:|
| X = 0 | (18.2 ± 1.2) |
| X = 0.02 | (10.3 ± 0.9) |
| X = 0.06 | (7.6 ± 1.7) |
| X = 0.10 | (5.1 ± 2.1) |

Figure **4a** represents XRD pattern of with and without BCP additive freshly prepared films. We observe that addition of such a small quantity (0.02 wt.%) of organic molecule in the perovskite solution does not affect the crystallinity of perovskite film. In order to study the moisture resistive behavior of BCP, we performed contact angle measurement on with and without BCP additive perovskite films with water droplets. The contact angle θ (angle between water droplet and perovskite thin film) is higher (72$^0$/85$^o$) in BCP additive/interlayer based perovskite film (figure **4b**). In literature, high moisture stability is achieved by forming quasi 2D perovskites with larger contact angle (70$^0$ to 80$^0$).[34,35] The contact angle achieved in with BCP additive/interlayer based perovskite film suggest that there is a possibility of formation of 2D perovskites at interface of grain boundaries, which can result in observed moisture barrier. However, pristine perovskite film is soluble and hydrophilic with water and hence contact angle is relatively low (52$^o$) (figure **4b**). The wetting behavior of perovskite thin film with BCP as an additive/interlayer is significantly low. Since, BCP includes large pyridine and alkyl groups, which reduces the surface free energy of the substrate and hence results in low wettability of perovskite film. To understand the chemical changes in with and without BCP additive perovskite films with moisture, we carried out XRD of the aged films (15 days) and it is shown in figure **4c.** The films were kept in ambient condition under atmospheric humidity of (45 ± 5) %. We observe an additional peak near to second order diffraction peak (220) of without BCP additive film and it is attributed to PbO formation in the perovskite film[36]. Along with PbO formation, the peaks at 8.4$^0$ and 10.3$^0$ indicates the formation of monohydrate species CH$_3$NH$_3$PbI$_3$(H$_2$O)[36] (Inset of figure **4c**). While BCP additive perovskite film did not show any additional peaks after 15 days in ambient. In addition, high resolution XPS of O 1s peak in BCP additive perovskite film is reduced significantly with respect to without BCP additive perovskite film (figure **4d**). Hence, contact angle and O 1s high resolution XPS studies supports that BCP added film may be forming a capping layer to boost moisture barrier. These results indicate that BCP plays an important role as water resistive layer in order to prevent the degradation of perovskite film with ambient (figure **4c**). These observations suggest that either BCP forming a quasi 2D perovskite according to literature[34,35,36] or it is forming a capping layer over perovskite without altering its structure.

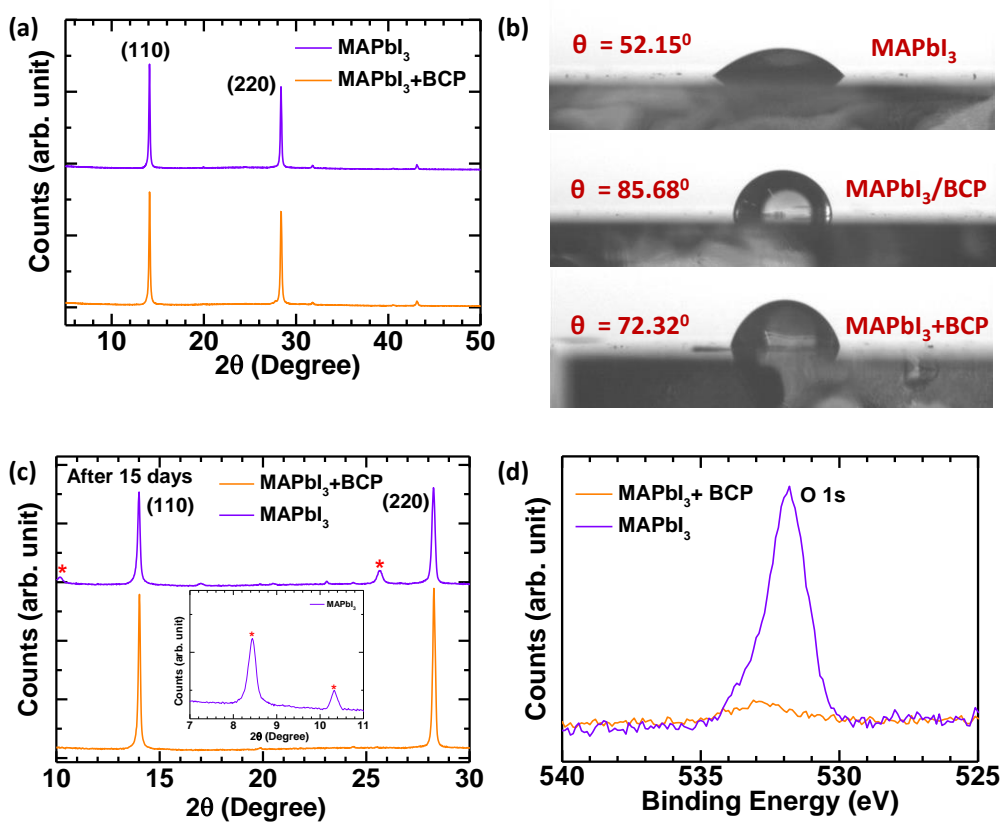

**Figure 4:** *(a) XRD pattern of with and without BCP additive as prepared perovskite films. (b) Contact angle between water droplet and pristine perovskite, perovskite with additional BCP layer & perovskite with BCP additive thin films on ITO coated glass. (c) XRD pattern of with and without BCP additive perovskite films after 15 days. The inset represents the XRD peak of monohydrate species associated with the pure perovskite film due to moisture. (d) Core level spectra of O1s for with and without BCP additive perovskite thin films.*

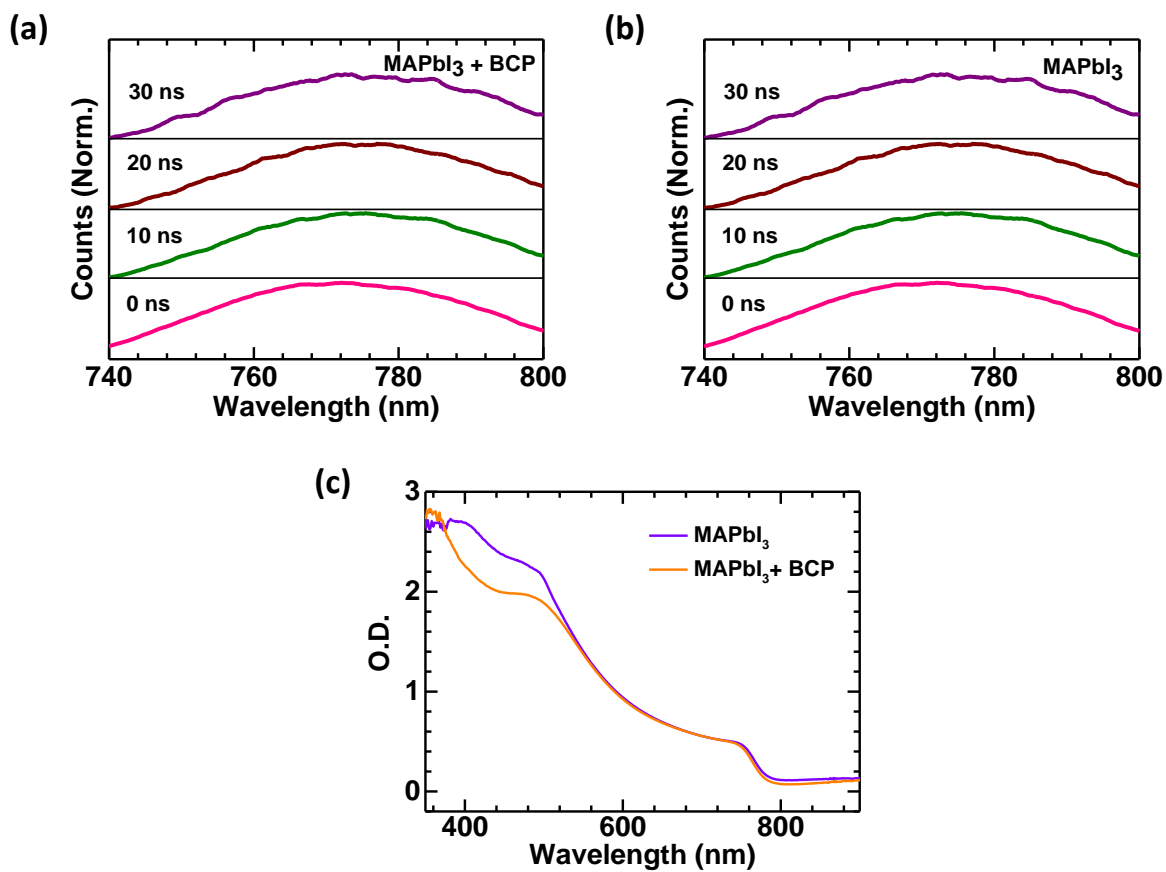

**Figure 5:** *PL spectra at different decay time for (a) with and (b) without BCP additive perovskite films. (c) UV-Vis spectra of with and without BCP additive perovskite thin films.*

In order to clarify whether there is quasi-2d layer is being formed then we might be able to see a blue shifted emission peak at early time scales in time-resolved spectroscopy, we performed delayed emission spectroscopy on with and without BCP additive perovskite films. Figure **5a** and figure **5b** represents the PL spectra at different delayed time for with and without BCP additive perovskite films, respectively. Figure **5a** does not show any blue shift peak or double peak emission in the PL spectra of with BCP additive perovskite film suggests no structural changes in 3D perovskite.[27] Slight shift in time dependent PL spectra towards longer wavelength occur due to spectral diffusion. Spectral diffusion is controlled by degree of energetic disorder in the system. The conduction band edge has a Gaussian density of states distribution. During the spectral relaxation, carrier follows the random walk through Gaussian density of states distribution in conduction band edge.[37] This reflects in the slight time dependent shift in the emission spectra. Along with that, UV-Vis (figure **5c**) and XRD studies (figure **4a**) also does not show any change/shift with respect to the without BCP additive perovskite film. These results suggest that 3D structure of perovskite is not disturbed; else at least early part of PL spectrum should have got blue shifted PL hump/peak in time resolved PL spectrum study.

In order to support our experimental findings, we carried out first principles electronic structure calculations based on DFT in MAPbI$_3$ after insertion of BCP. The detailed information regarding the computational methodology is given in supporting information. Figure **6** shows the Bird's eye and side perspectives of minimum energy configuration for MAPbI$_3$ + BCP system. Among different configurations of the combined system consisting of BCP and MAPbI$_3$, we have found that BCP cannot go completely inside MAPbI$_3$, therefore getting a 2D hybrid perovskite structure is not plausible in this scenario. We have also tried to adsorb BCP both in parallel and perpendicular direction to MAPbI$_3$ surface, which leads to the most stable configuration where BCP has been physisorbed on MAPbI$_3$ surface in parallel orientation. The interaction of BCP with MAPbI$_3$ has been found as physisorption type with an average distance between surface (MAPbI$_3$) and adsorbate (BCP) of 2.9 Angstrom. This has been confirmed after systematic calculations by varying the surface and adsorbate distance to find the minimum energy configurations of the complete system. This corresponds to the experimental outcome of delayed emission spectroscopy of no 2D structure formation in MAPbI$_3$ and outcome of contact angle study of making a capping layer over the MAPbI$_3$ surface at a distance of 2.9 A$^0$ after the inclusion of BCP in the system.

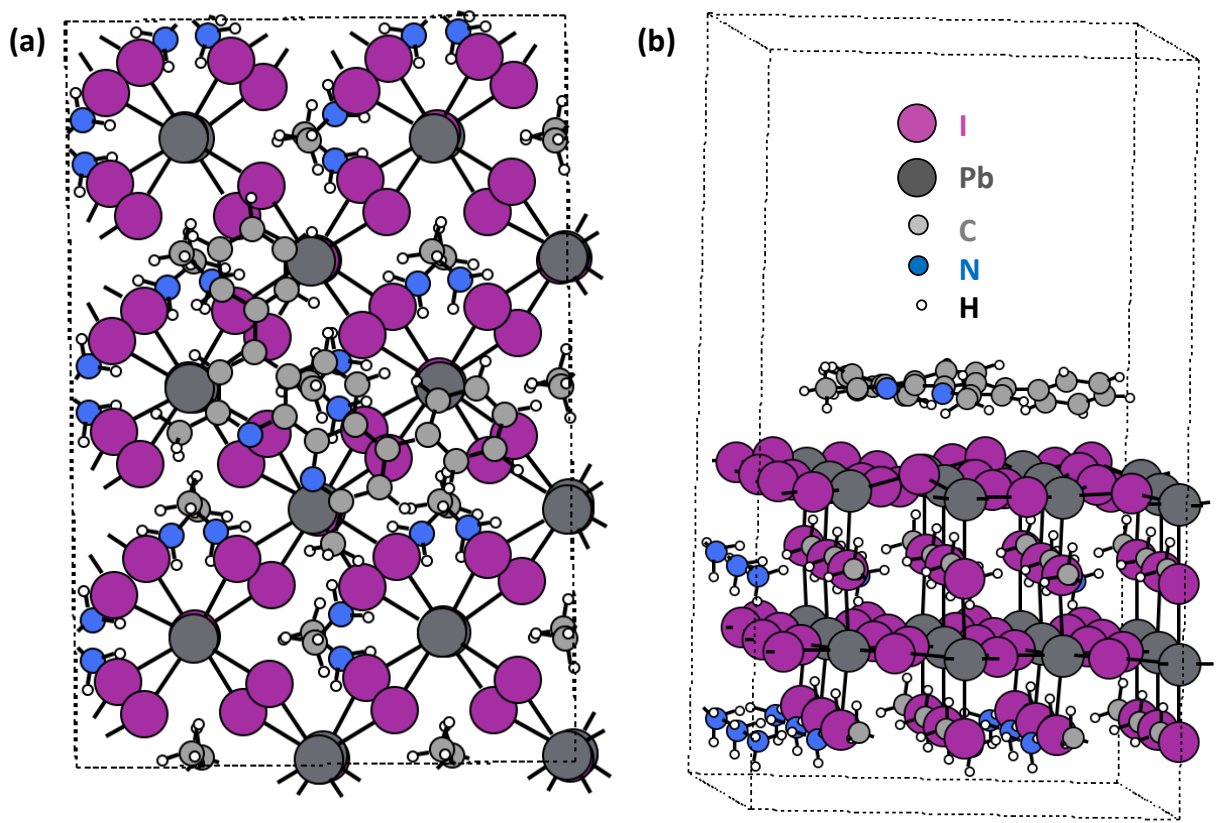

***Figure 6:*** *(a) Bird's eye and (b) side perspective of minimum energy configuration for* MAPbI$_3$ *+ BCP system. Pb, C, I, N,and H atoms are represented with as Dark Grey, Light Grey, Magenta, Blue and White respectively.*

We have also investigated the recombination process in with and without BCP additive based perovskite films by steady state photoluminescence (PL) and time resolved PL measurements. Figure **7a** represents the steady state PL of pure perovskite film with different layers. Addition of small amount of BCP (0.02 wt.%) in the perovskite film results in an enhancement of 51% in PL intensity with respect to the pure perovskite film. This suggests that BCP passivate the perovskite surface as well as grain boundaries and suppresses the non-radiative recombination. However, an extra layer of PC$_{61}$BM over the perovskite film leads to PL quenching with PL quenching efficiency of 72%. To understand the behavior of BCP, we spin-coated an extra layer of BCP on the top of the perovskite/PC$_{61}$BM film and observed slight enhancement in PL intensity, which indicates that BCP might be entering into the perovskite film through PC$_{61}$BM and suppresses the non-radiative recombination. Figure **7b** shows the time resolved PL spectra of the pure perovskite film and the film with BCP additive. The PL lifetime of both films was fitted using a bi-exponential decay function having two lifetime components: a fast decay and a slow decay lifetime. The parameters related to PL decay are listed in table **2**. Both the lifetime of the fast component and the average lifetime are longer for the film with the BCP additive by a factor of two. This is consistent with our hypothesis that BCP serves to passivate defects and grain boundaries in the perovskite film.

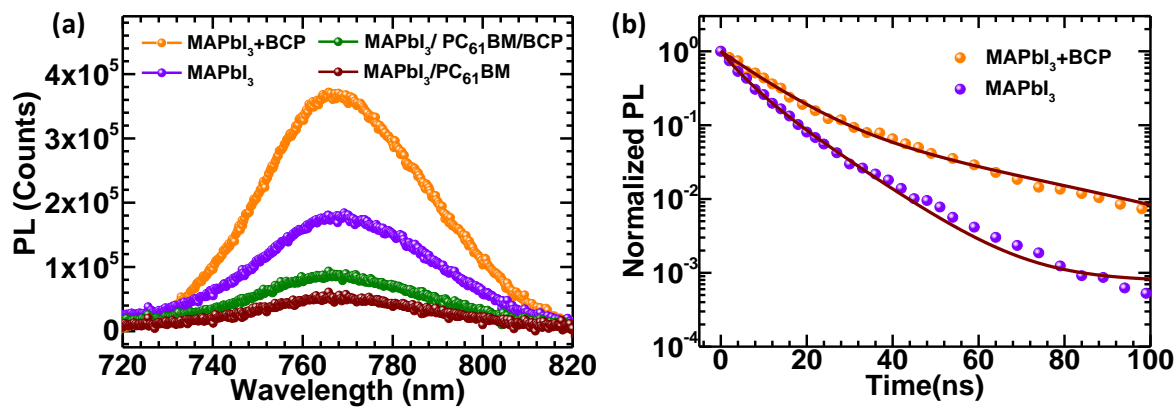

***Figure 7:*** *PL spectra of (a) MAPbI₃ (black line), MAPbI₃ + BCP (red), MAPbI₃/PC₆₁BM (wine line), and MAPbI₃/PC₆₁BM/BCP (green line) on PEDOT:PSS coated glass substrate. (b) TRPL decay curves of with and without BCP additive perovskite thin film. Solid line is bi-exponential decay fit to the experimental data.*

***Table 2:*** *Time resolved photoluminescence characterization for with and without BCP additive perovskite thin films.*

| Perovskite Films | $A_1$ | $\tau_1$ (ns) | $A_2$ | $\tau_2$ (ns) | $\tau_{av}$ (ns) |
|---|---|---|---|---|---|
| MAPbI₃ | (0.51 ± 0.08) | (4.44 ± 0.42) | (0.49 ± 0.08) | (11.00 ± 0.86) | 9.06 |
| MAPbI₃ + BCP | (0.90 ± 0.06) | (10.13±0.67) | (0.10±0.06) | (44.06 ± 2.5) | 21.18 |

Figure **8a** shows J-V characteristics of three devices under AM1.5 illumination: pure perovskite without BCP as buffer layer (termed as D1); pure perovskite with BCP as buffer layer (termed as D2) and perovskite-BCP additive with BCP as buffer layer (termed as D3) based solar cells. Dark J-V characteristics for three devices are shown in figure **8b**. Device D1 exhibits a leakage current density of the order of 1 mA/cm$^2$ even under a small reverse bias of 0.5 V. The dark current is substantially reduced by an order of magnitude in device D2 and it is in well agreement with the earlier reports[30]. The device D3 shows a further reduced leakage current. Hence, we conclude that BCP helps in reducing the shunts in perovskite film, though film morphology looks pretty similar in SEM images (figure **1c** and figure **1d**). Table **3** summarizes the parameters of the lighted J-V characteristics for the three devices. The addition of BCP improves all the optoelectronic factors like $V_{OC}$, EQE, FF , $J_{SC}$ and efficiency. The devices with BCP as a bulk additive have a high FF of 0.82 , $V_{OC}$ =0.95 V and an efficiency of about 16%. This is in sharp contrast with devices with no additive which have a FF=0.62, $V_{OC}$ =0.78V and an efficiency of 7.5%. In addition, the degree of hysteresis (DOH) is highly reduced from 17% (D1) to 2.6% (D2) and further reduced to 0.8% only (D3), which is negligible.

The average values (over 16 devices in single batch process) of all photovoltaic parameters with their standard deviation for three devices are shown in figure **9** and also listed in table **3**. Figure **8c** shows the external quantum efficiency (EQE) for the three devices. The increase in $J_{SC}$ for BCP added device is consistent with the enhancement in $J_{SC}$ from light J-V measurement (figure **8a**). We have also calculated the $J_{SC}$ by integrating the EQE spectrum with the AM 1.5 spectrum and obtained a value of 14.44 mA/cm$^2$, 17.69 mA/cm$^2$ and 18.72 mA/cm$^2$ for D1, D2 and D3 devices, respectively. The small difference (less than 10 %) in the current density calculated from J-V and EQE measurement is due to edge effects from the small cells having active area of 4.5 mm$^2$, which will be discuss later in this chapter only.[38] From our experiments, we conclude that BCP has two roles: (a) Reduce shunt paths by deposition on areas of Poly(3,4-ethylenedioxythiophene)-poly(styrenesulfonate) (PEDOT:PSS) which are not covered by the perovskite**.** (b) Passivate defect states in the bulk of the perovskite resulting in a large $V_{oc}$ for device D3 and improved transport and charge collection for devices D2 and D3.

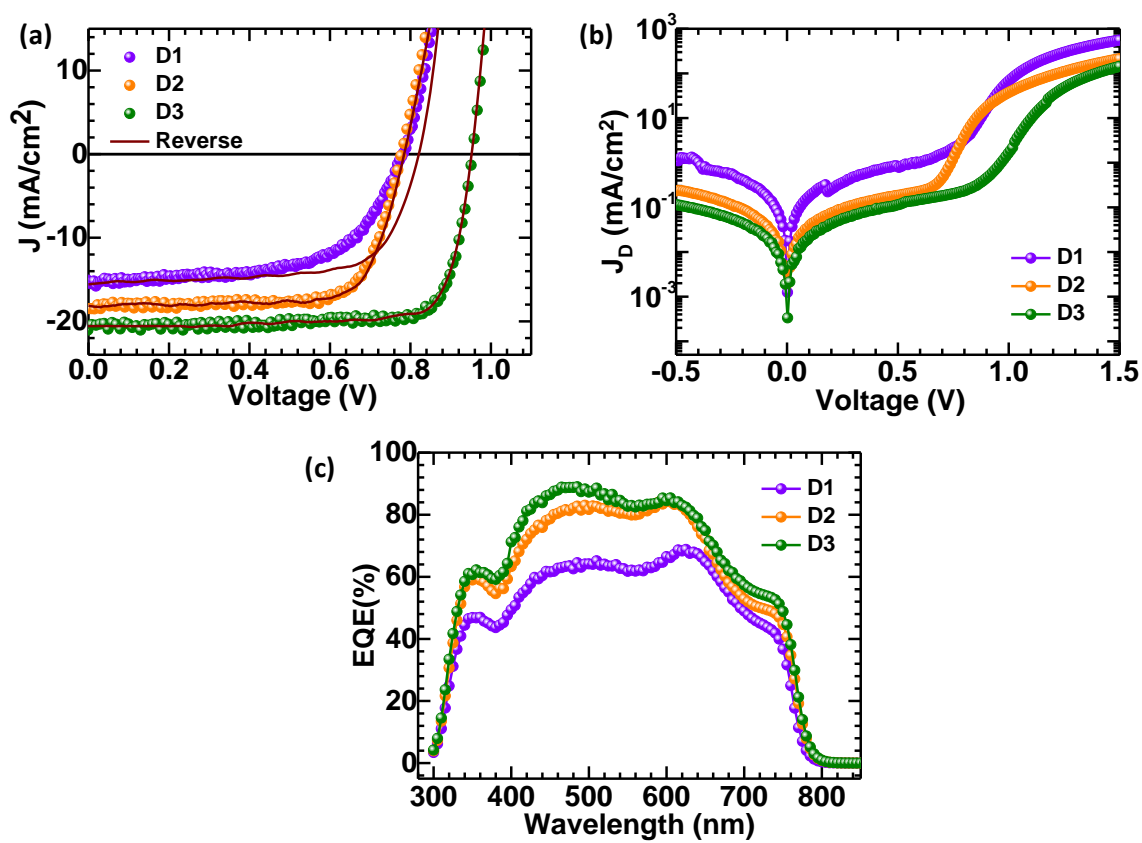

**Figure 8:** *J-V curves of the best performing devices D1, D2 and D3 under (a) simulated AM 1.5 sunlight of 100 mW/cm² irradiance and (b) dark condition. Solid scatters represent the scan from forward bias to short circuit (Forward), solid lines represent the scan from short-circuit to forward bias (Reverse). (c) External quantum efficiency of D1, D2 and D3 perovskite solar cells.*

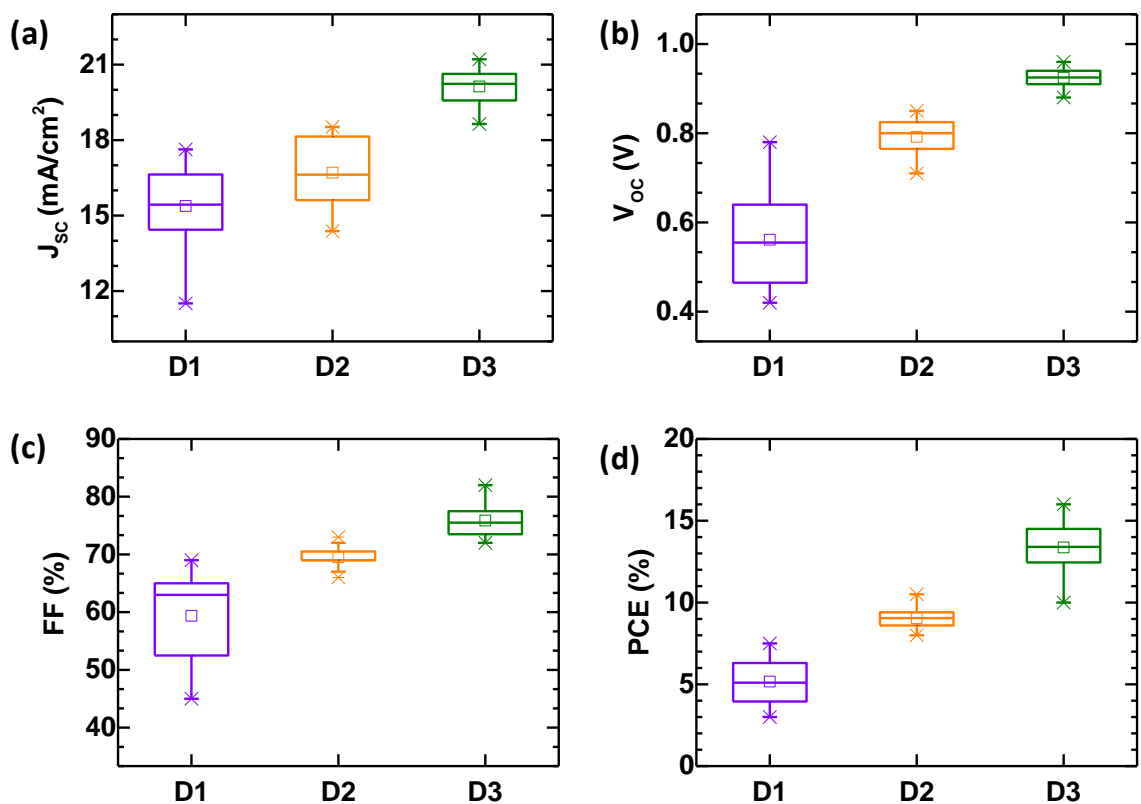

***Figure 9:*** *Histogram for different photovoltaic parameters such as (a) Jsc, (b) Voc, (c) FF and (d) PCE of three different perovskite solar cells (D1, D2 & D3).*

***Table 3:*** *Photovoltaic Parameters Achieved for Best Performing with and without BCP Additive Perovskite Thin Film Based Solar Cells over an Area of 4.5 mm². Average and standard deviation of photovoltaic parameters are listed in bracket for device D1, D2 and D3.*

| Perovskite Solar Cells | $V_{oc}$ (V) | $J_{sc}$ (mA/cm²) | $J_{sc}$ (EQE) (mA/cm²) | FF (%) | PCE (%) | $R_S$ (Ω/cm²) | $R_{Sh}$ (KΩ/cm²) |
|---|---|---|---|---|---|---|---|
| D1 | 0.78 (0.55 ± 0.09) | 15.5 (14.31 ± 1.76) | 14.44 | 62 (59.68 ± 8.56) | 7.5 (5.10 ± 1.44) | 13.69 | 0.43 |
| D2 | 0.78 (0.78 ± 0.05) | 18.5 (16.50 ± 1.21) | 17.69 | 73 (69.75 ± 2.48) | 10.5 (8.93 ± 0.46) | 6.29 | 2.52 |
| D3 | 0.95 (0.92 ± 0.02) | 20.54 (19.93 ± 1.58) | 18.72 | 82 (76.00 ± 2.21) | 16.0 (12.87 ± 1.88) | 4.40 | 5.32 |

Figure **10a** shows the steady state EL spectra of device D1, D2 and D3. We observed that EL peak position is similar for all devices. Figure **10b** and **10c** represents the J-V-L characteristics of device D1, D2 and D3. The current density number for device D1 is higher than that for device D2 and D3. We expect that the lower current density in device D2 and D3 is due to radiative recombination losses in BCP additive/interlayer perovskite based device. Figure **10d** shows the electroluminescence (EL) EQE of devices D1, D2 and D3. Due to lower leakage current in device D2 and D3 in comparison to device D1, we observed that EL quantum efficiency in BCP additive/interlayer (D2 and D3) devices is higher than that in pure perovskite based device (D1). These results again confirm that BCP passivate the defects in the perovskite film and promote radiative recombination.

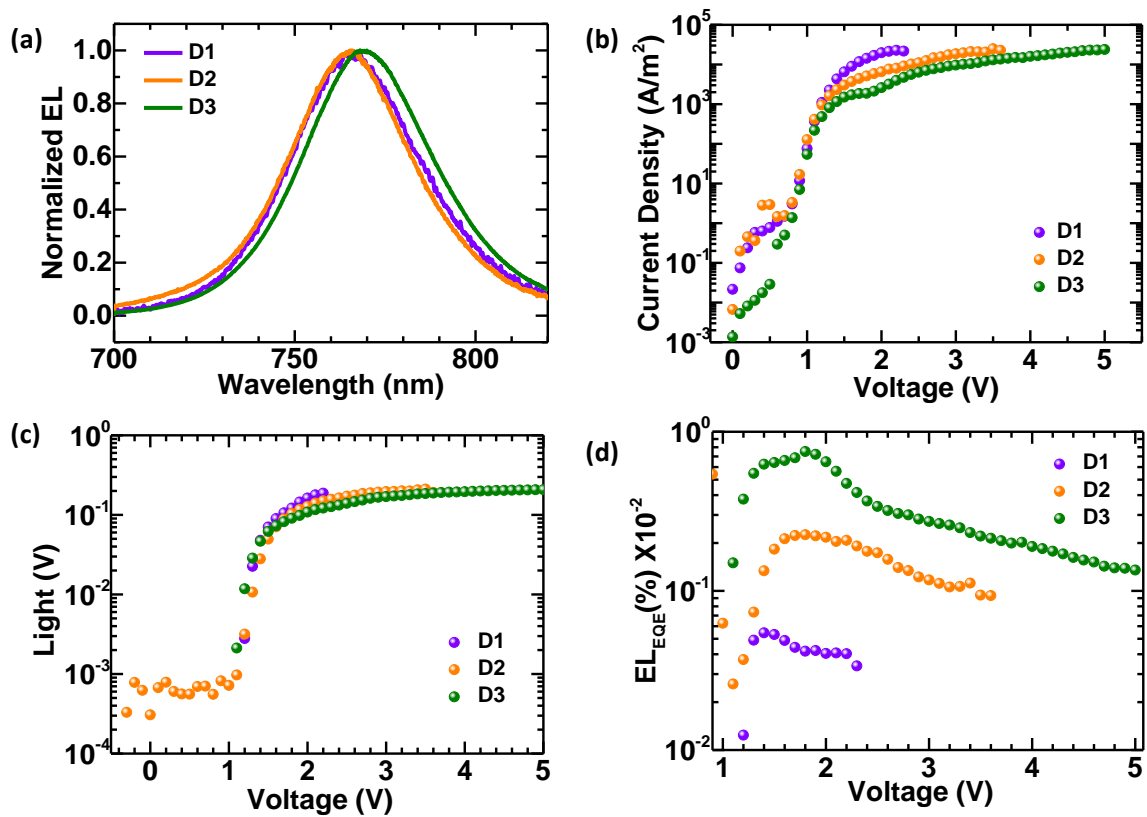

**Figure 10:** *(a) Normalized electroluminescence (EL) spectra (b), (c) J-V-L characteristics and (d) EL quantum efficiency for all three devices; D1, D2 and D3.*

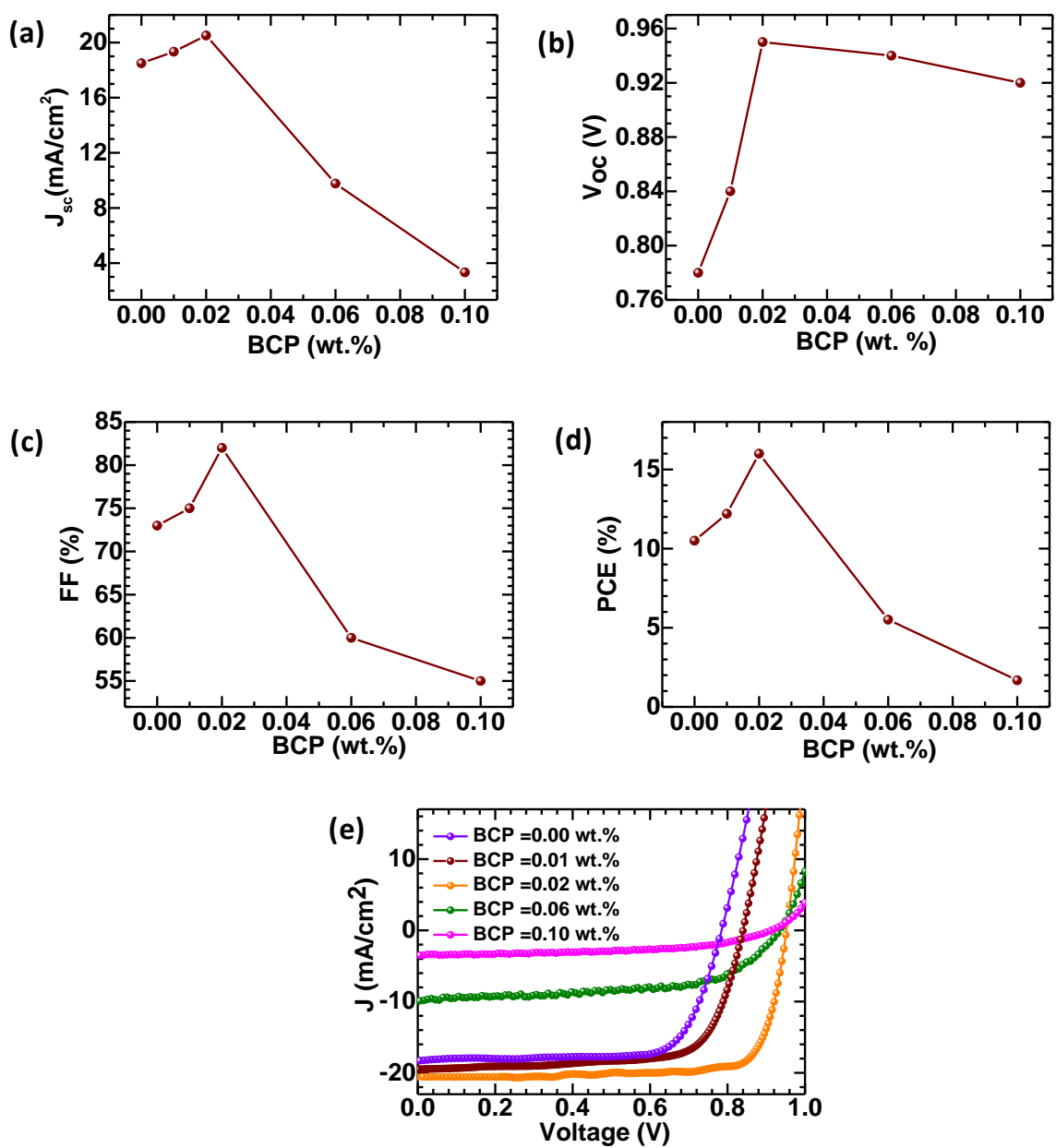

**Figure 11:** *Variation of photovoltaic parameters such as (a) Jsc, (b) Voc, (c) FF, and (d) PCE with BCP concentration. (e) The J-V characteristics of the highest efficiency devices with different BCP concentration under 1 Sun illumination.*

We found that the concentration of BCP in the perovskite precursor solution is a crucial parameter to obtain smooth perovskite films with comparatively large domains and no BCP aggregates. When we increased the BCP concentration from 0.02 wt % to 0.06 wt.%, $J_{SC}$ reduced from 20.52 mA/cm$^2$ to 9.77 mA/cm$^2$. It further reduced to 3.33 mA/cm$^2$ on increasing the BCP concentration to 0.1 wt.%. The FF also reduced in the similar manner (82% (0.02 wt.%) to 60% (0.06 wt.%) to 55% (0.10 wt.%). This is due to lower solubility of BCP in perovskite solution leading to BCP aggregate formation with increased BCP concentration in the perovskite films. J-V characteristics and different photovoltaic parameters with respect to BCP concentration are shown in figure **11**.

The intensity dependence (from 100 mW/cm$^2$ to 10 mW/cm$^2$) of the J-V characteristics for devices D1-D3 is shown in figure **12**. Assuming superposition, the relation between $V_{oc}$ and the short circuit current $J_{SC}$ can be written as

$$V_{OC} = \frac{nkT}{q} \ln\left(\frac{J_{SC}}{J_0} + 1\right) \qquad (1)$$

**Figure 13a** shows that $J_{SC}$ is linearly proportional to the light intensity.

**Figure 13b** shows a plot of $V_{OC}$ vs light intensity. Devices D2 and D3 are characterised by a single value of n in equation 1, i.e. 0.84 and 1.78 for D3 and D2 respectively. n= 1.78 suggests that recombination is dominated by Shockley Read Hall (SRH). n~1 suggests that recombination is in the bulk of the device. This is consistent with the fact that in device D3, both surface and bulk defect states are passivated by BCP in addition to reduction of shunt paths in the device. In device D1, for I > 50 mw/cm$^2$, the value of n~1.55. At lower intensity n is very large. Such large values of n can arise at low intensity when the current flows through internal shunts.[39] Again assuming superposition, under illumination

$$J(V) = -J_{SC} + J_0 \exp\left(\frac{qV}{nkT}\right) + \frac{V}{R_{sh}} \qquad (2)$$

where $R_{sh}$ is the shunt resistance. At $V_{oc}$, J ($V_{oc}$) =0, equation (2) becomes,

$$\left(\frac{J_{SC}R_{sh} - V_{OC}}{R_{sh}}\right) = J_0 \exp\left(\frac{qV_{OC}}{nkT}\right) \qquad (3)$$

At low intensity most of the current goes through the shunt, rather than through the diode (figure **12d**). Hence equation (1) is not valid. At high light intensity, the current through the diode dominates and equation (1) is recovered. This is consistent with the fact that the morphology of D1 is poor and has large leakage represented by a low shunt resistance. A low shunt resistance will also lead to a low FF which will be a function of light intensity. This is seen in figure **13c**, where the FF varies slightly for devices D2 and D3 (higher shunt resistance) in contrast to device D1 where FF is a strong function of light intensity (low shunt resistance).

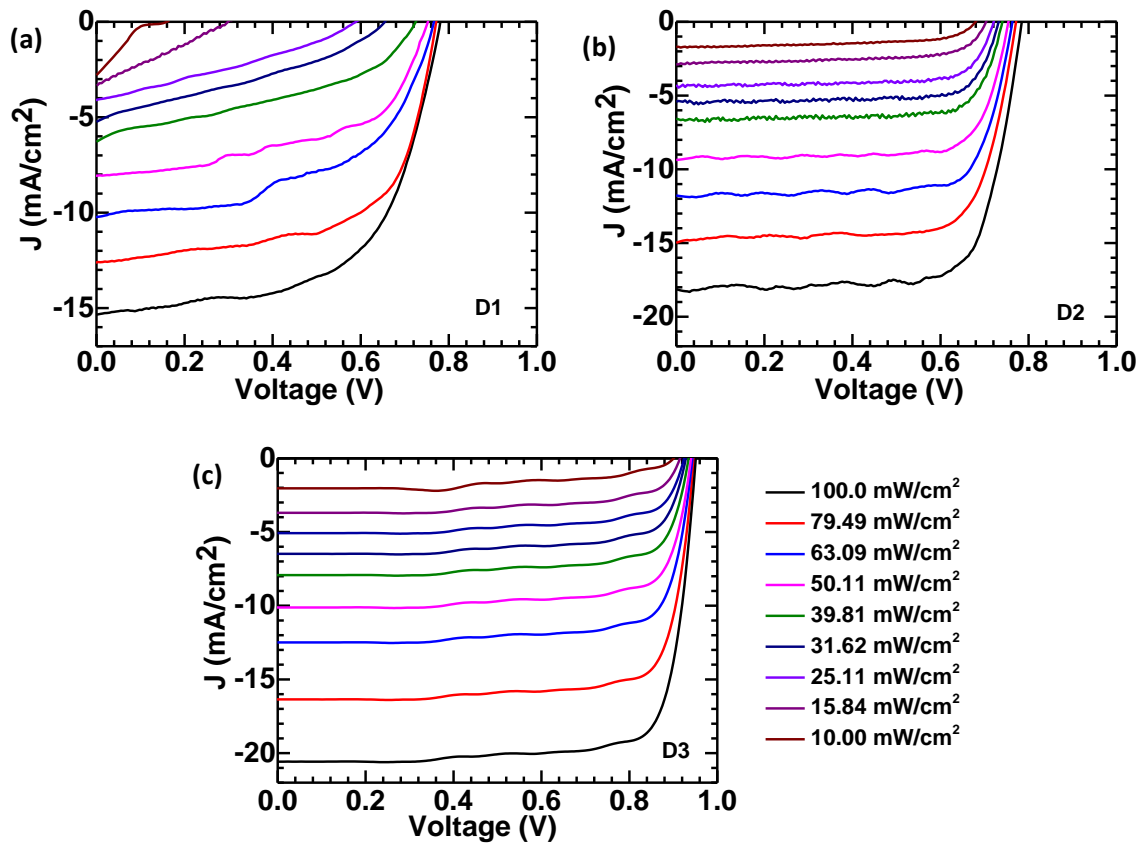

***Figure 12:*** *Intensity dependent J-V measurement of perovskite solar cells: (a) D1, (b) D2 and (c) D3.*

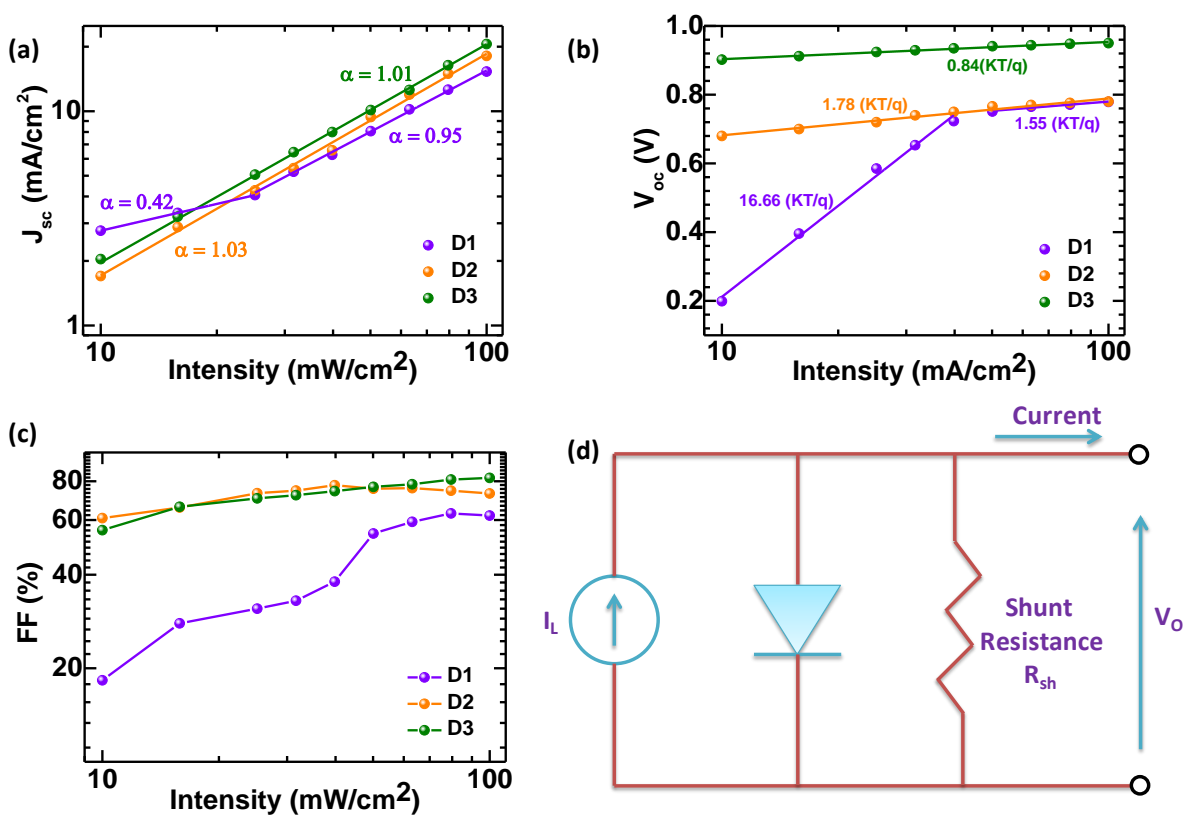

**Figure 13:** (a) Power law dependence of $J_{SC}$ with light intensity for three different perovskite solar cells. (b) Open-circuit voltage and (c) Fill factor of three different perovskite PV devices plotted against light intensity. (d) Circuit diagram of a solar cell including the shunt resistance.

Surface morphology of pure perovskite film is changing with BCP additive; one can expect that there might be change in chemical states of the perovskite surface elements in the two films. The device performance is highly influence by the impurities at the surface and on the grain boundaries. In order to investigate the surface impurities and chemical states of perovskite film, we performed the high resolution X-ray photo-electron spectroscopy (HRXPS) on the two films. In HRXPS of C 1s (shown in figure **14a**), two peaks at 286.5 eV and 285.0 eV are assigned to C 1s peak. In literature, the energy peak at 286.5 eV is reported to come from $CH_3NH_3^+$ and energy peak at 285.0 is assigned to some non-perovskite species formed at the grain boundaries of perovskite domain. The amount of carbon related impurities is decided by the intensity ratio of the two peaks of C 1s ($\alpha$ = area of low energy peak/ area of high energy peak)[5]. The value of α is 0.96 in the case of pure perovskite film, but is reduces to almost half of its value in the BCP additive perovskite film (α = 0.56). It demonstrates that carbon related impurities are significantly removed from the grain boundaries of perovskite film by incorporation of small amount of BCP in the perovskite precursor solution.

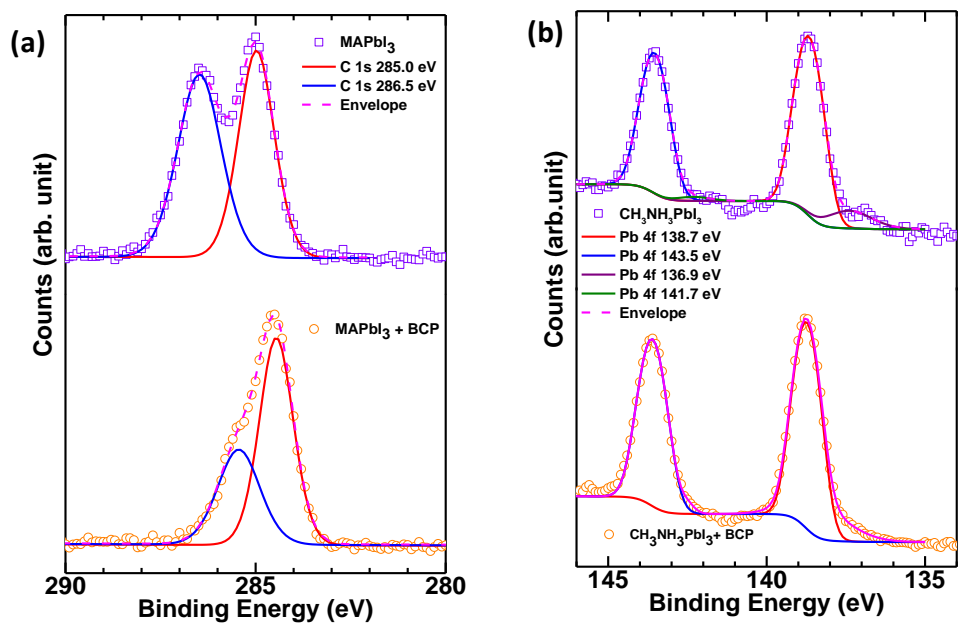

***Figure 14:*** *HR core level spectra and corresponding Gauss-Lorentz fit to the experimental data of (a) C 1s and (b) Pb 4f.*

Both pure and BCP additive perovskite film shows two symmetric peaks of $Pb^{2+}$ with energy peak at 143.5 eV and 138.6 eV and they are assigned to Pb $4f_{5/2}$ and Pb $4f_{7/2}$, respectively (figure **14b**). We observed another pair of $Pb^0$ (metallic lead) peaks at energy of 1.8 eV lower than the main peaks (141.7 eV and 136.9 eV) in pure perovskite film[6]. However, the metallic lead peaks in BCP additive perovskite film do not exist on the perovskite surfaces. In addition, O 1s peak in BCP additive perovskite film is reduced significantly and it is shown in **Figure 14b**. Since, oxygen is not an intrinsic element of perovskite or not of BCP. Hence, appearance of O 1s peak in both films is due to surface contamination. Thus, BCP reduces the significant amount of lead, carbon and oxygen related surface impurities in the perovskite film. It is a well-known phenomenon that surface impurities are act as non-radiative recombination centers. Therefore, from HRXPS of C 1s, Pb 4f and O 1s, we conclude that non-radiative recombination losses are suppressed by addition of small amount of BCP in the perovskite precursor solution.

**Table 4:** *Quantification of different elements present in with and without BCP additive MAPbI₃ based perovskite thin films using XPS wide spectra.*

| Elements | $MAPbI_3$ (Atomic %) | $MAPbI_3$ + BCP (Atomic %) |
|:---:|:---:|:---:|
| C 1s | 61.65 | 66.45 |
| N 1s | 7.42 | 5.21 |
| O 1s | 3.65 | 1.90 |
| Pb 4f | 8.26 | 6.69 |
| I 3d | 19.01 | 19.75 |

Apart from controlling the crystallization and nucleation growth rate to obtain a high quality perovskite thin film, interlayer induced interfacial properties and morphology are inevitable too. Interfacial losses such as interface barriers, recombination by the defects, Schottky contact and shunting paths needs to be reduced to improve device performances.[40,41,42] Materials including organic and inorganic material,[43,44,45] Fullerene derivatives,[46] self-assembled monolayers[47] and functional polymers[48] has been used as interlayers. These interfacial layers increase the film coverage and passivate the charge trap states in the active perovskite layer.[49] In addition, hysteresis phenomenon between the forward and reverse bias of the perovskite solar cell, are attributed to the charge accumulation at the interfaces.[50,51,52] MAPbI₃ based perovskite films are well understood in the past few years, but interfacial engineering in the perovskite solar cell needs more insight and improvement.[53] Under illumination, charge carriers are generated in the perovskite layer and transported to the anode and cathode by respective transport layers (EEL or HEL).[54] Such well-coordinated layers prevent the carriers from being quenched or recombined due to their balanced electronic states alignment. On the other side, defects in the interlayer contribute to charge loss leading to reduction in fill factor, $V_{OC}$, $J_{SC}$ and thus results in poor PCE.[55] Thus, modification of the interfaces in perovskite solar cell is an effective approach to maximize electron-hole separation, collection and minimize charge recombination. In addition, such kind of extraction interlayers greatly reduces the device hysteresis.[56,57,58]

We have explored bathocuproine (BCP), and $MoO_3$ as an additional electron and hole extraction layers making a hybrid double extraction layer in a perovskite solar cell with $PC_{61}BM$) and PEDOT:PSS, respectively. The double HEL film is fabricated by thermal evaporation of $MoO_3$ and spin-coating of PEDOT:PSS. The double EEL is formed by spin-coating of $PC_{61}BM$ and BCP. The role of BCP and $MoO_3$ in perovskite solar cells is to (a) increase the charge transfer rate from the perovskite to the electrode, (b) reduces shunts in perovskite thin-films and (c) provides the stability over few hours under constant illumination at 1 sun intensity. We further extended our study to area dependent devices in order to study the underlying mechanism for higher PCE of small area perovskite solar cells than the larger ones. This study demonstrates the working mechanism behind improved charge extraction property and high thermal stability due to additional second EEL and HEL in the perovskite solar cells.

We and others have shown the superiority of BCP additive $MAPbI_3$ film over pristine $MAPbI_3$ film in terms of morphology and device performance.[59.60] Thus, in this study, we are using BCP additive $MAPbI_3$ film as an active layer and call it $MAPbI_3$ as default. Figure **15a** and figure **15b** represents the scanning electron microscopy (SEM) images of $MAPbI_3$ based films prepared on PEDOT:PSS and $MoO_3$/PEDOT:PSS coated glass substrate, respectively. We observe that perovskite film coated over $MoO_3$/PEDOT:PSS is compact and having slightly bigger domains than the perovskite film coated over PEDOT:PSS. Since, lead acetate trihydrate [$Pb(Ac)_2.3H_2O$] is being used as a lead source instead of $PbI_2$ for making $MAPbI_3$ films, we observed relatively smaller crystallites size. The reason behind using $Pb(Ac)_2.3H_2O$ is its uniform coverage without any further treatment like anti-solvent, solvent additive etc with high reproducibility.[61] Whereas, by using anti-solvent treatment, big domains are obtained but reproducibility is a challenge known in perovskite community.[62] Moreover, the aim of this study is to demonstrate the role of extracting layers in terms of morphology, optical losses, shunting effect etc, which will be applicable to films prepared via other processes with an added advantage of being tolerant on any possible minute pin-holes issues commonly observed in thin film devices. Figure **15c** represents the x-ray diffraction pattern of MAPbI$_3$ film coated over PEDOT:PSS and MoO$_3$/PEDOT:PSS substrate and it is observed that full width half maxima (FWHM) of MAPbI$_3$ film over MoO$_3$/PEDOT:PSS substrate is slightly lower than that of MAPbI$_3$ film fabricated over PEDOT:PSS coated substrate only (inset of figure **15c**), which is in good agreement with morphology results (figure **15b** & **15c**). Figure **15d** represents the photoluminescence (PL) spectra of MAPbI$_3$ perovskite film with different extraction layers. It is observed that MoO$_3$/PEDOT:PSS/MAPbI$_3$ interface shows slightly more PL quenching than only PEDOT:PSS/MAPbI$_3$ interface, which seems to be detrimental for solar cell performance, however, it is not the case as will be discussed later. A layer of PC$_{61}$BM over the MoO$_3$/PEDOT:PSS/MAPbI$_3$ shows further quenching in PL intensity, which suggest a better electron extraction at MAPbI$_3$-PC$_{61}$BM interface. An additional layer of BCP over MoO$_3$/PEDOT:PSS/MAPbI$_3$/PC$_{61}$BM results in slight enhancement of PL intensity. It suggest that BCP as a buffer layer passivate the EEL interface with perovskite as well as it also helps in suppressing the non-radiative channels at the surface by entering through PC$_{61}$BM in the MAPbI$_3$ perovskite layer.[59]

Figure **16** represents the atomic force microscopy (AFM) images of MAPbI$_3$ based films spin-coated over PEDOT:PSS and MoO$_3$/PEDOT:PSS coated glass substrate. It is observed that perovskite film fabricated over MoO$_3$/PEDOT:PSS is relatively smoother than the perovskite film spin-coated over PEDOT:PSS (table **5**). This can be understood by enhanced wettability of PEDOT:PSS over MoO$_3$ coated glass substrate during spincoating (figure **17**). The contact angle between water droplet and ITO coated glass is higher ($36^0$) than that of ITO/MoO$_3$ ($09^0$) substrate. It suggest that having an additional layer of MoO$_3$ over ITO coated glass substrate helps PEDOT:PSS to spread completely over the substrate and provide a smooth PEDOT:PSS film (figure **18**). Higher roughness of PEDOT:PSS films results in relatively poorer morphology of MAPbI$_3$ film. Using uniform and smooth MAPbI$_3$ perovskite thin films (figure **15**), we studied the performance of *p-i-n* architecture based perovskite solar cells with different extraction layers.

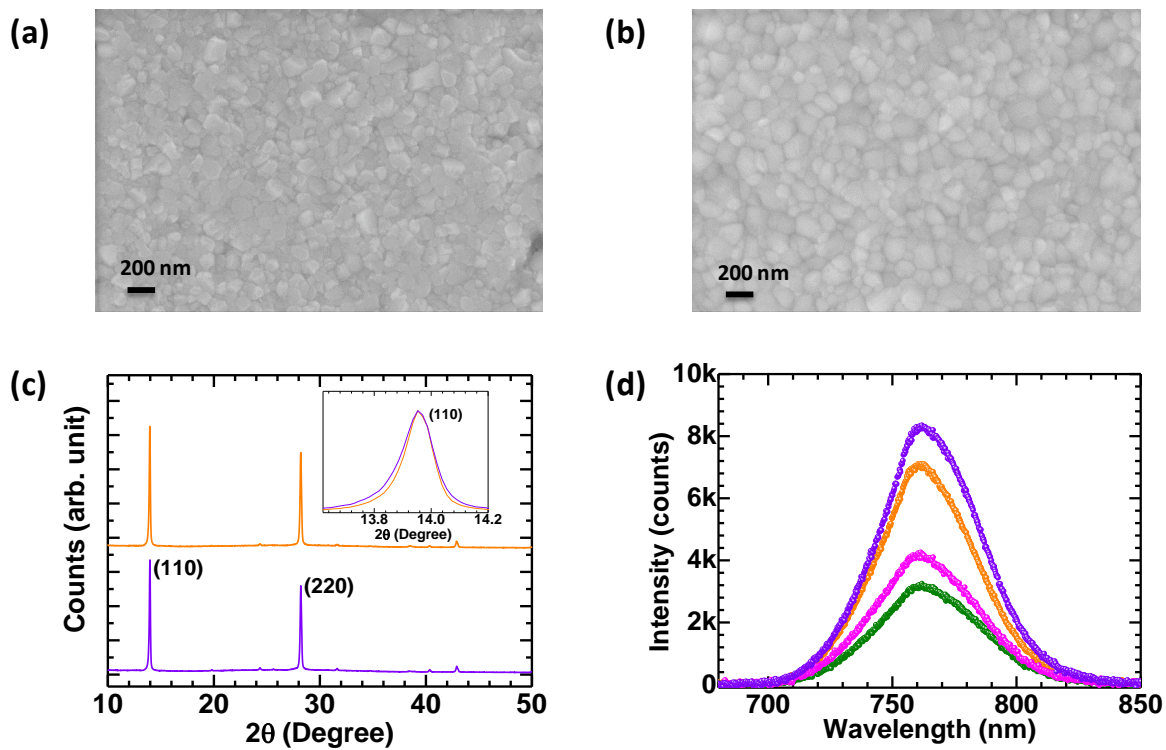

***Figure 15:*** *Morphology of BCP additive perovskite film over (a) Glass/PEDOT:PSS and (b) Glass/MoO₃/PEDOT:PSS. (c) XRD of perovskite films spincoated over Glass/PEDOT:PSS (violet) and Glass/MoO₃/ PEDOT:PSS (orange). Inset represents the (110) peak of perovskite film spincoated over Glass/PEDOT:PSS (violet) and Glass/MoO₃/PEDOT:PSS (orange). (d) Photoluminescence (PL) of PEDOT:PSS/MAPbI₃ (violet), MoO₃/PEDOT:PSS/MAPbI₃ (orange), MoO₃/PEDOT:PSS/MAPbI₃/PC₆₁BM (olive) and MoO₃/PEDOT:PSS/MAPbI₃/PC₆₁BM/BCP (magenta).*

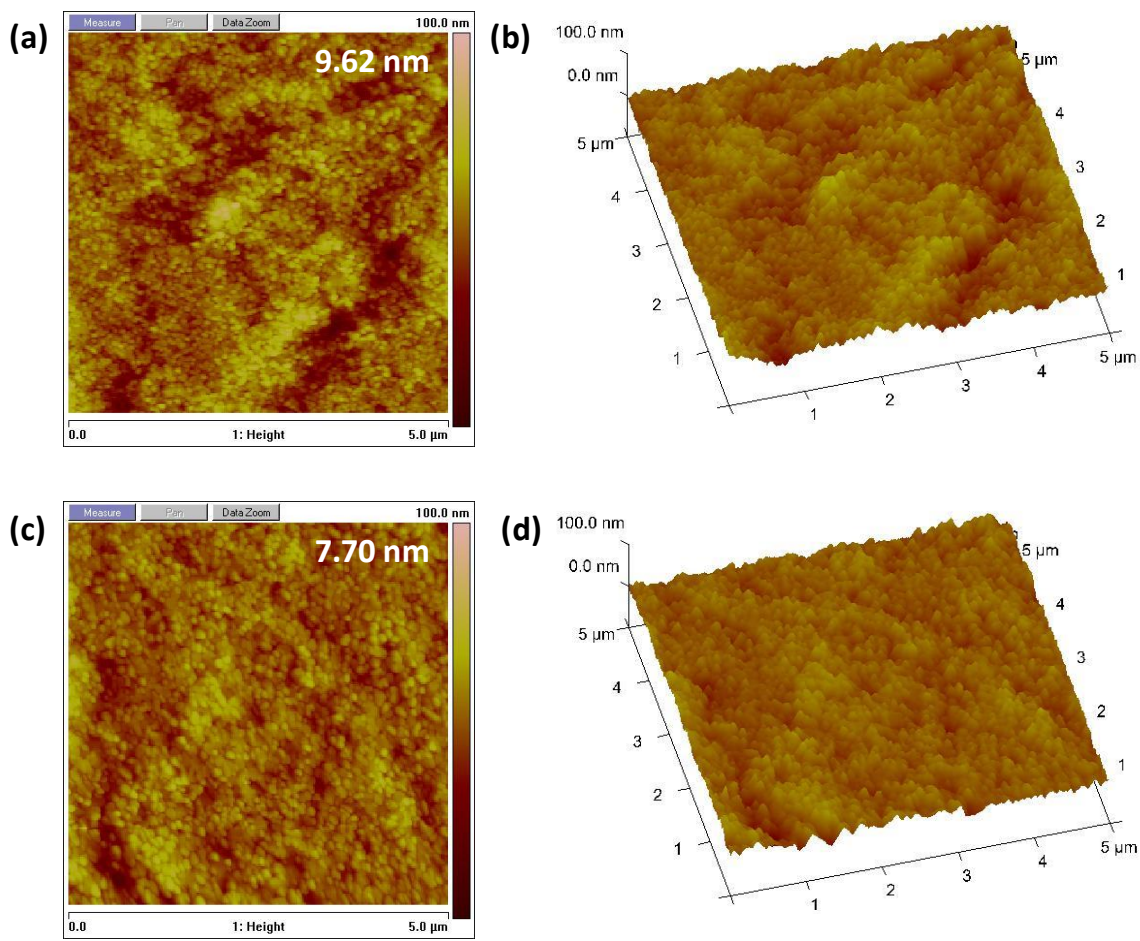

**Figure 16:** *2D AFM images of (a) ITO/ PEDOT:PSS/ MAPbI₃+BCP and (c) ITO/ MoO₃/ PEDOT:PSS/ MAPbI₃+BCP based perovskite films and corresponding 3D AFM images are shown in figure (b) and (d), respectively.*

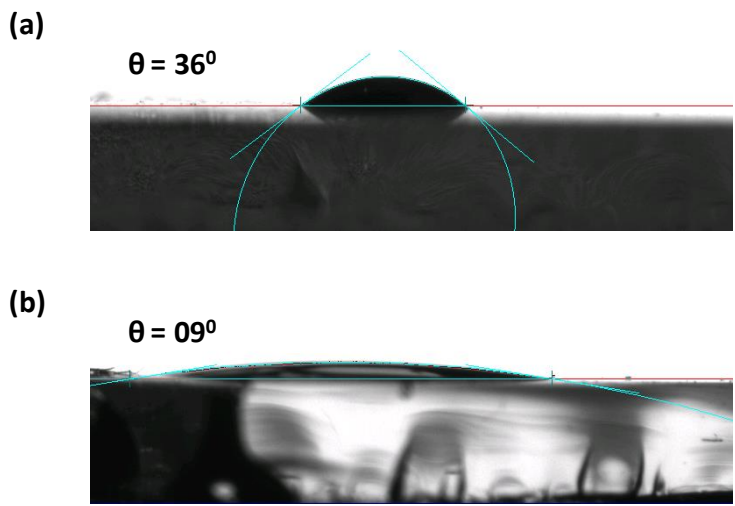

*Figure 17:* Contact angle between the water droplet and (a) ITO coated glass and (b) MoO₃/ITO coated glass to understand the wettability of PEDOT:PSS over ITO and MoO₃/ITO coated glass substrate.

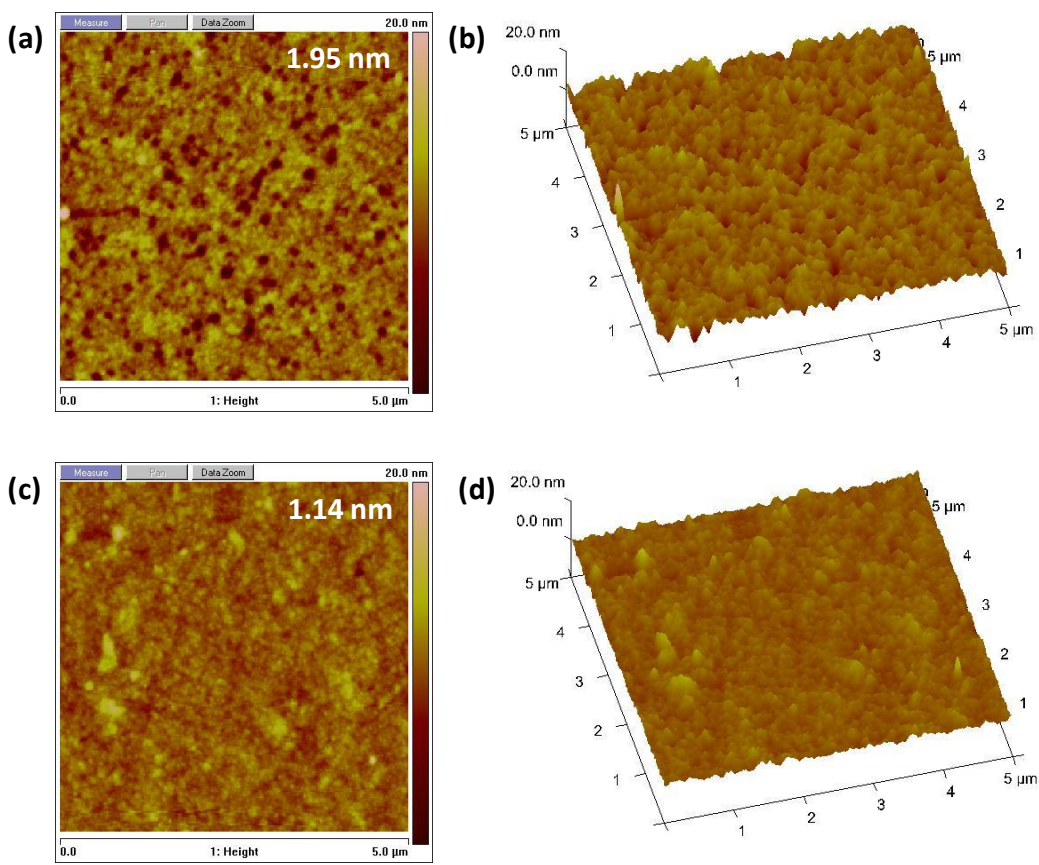

*Figure 18:* 2D AFM images of (a) ITO/ PEDOT:PSS and (c) ITO/ MoO₃/ PEDOT:PSS films and corresponding 3D AFM images are shown in figure (b) and (d), respectively.

Figure **19a** represents the energy level diagram of inverted (*p-i-n*) architecture based perovskite solar cell with double extraction layers.[63,59] Figure **19b** shows J-V characteristic of *p-i-n* configuration based perovskite solar cells under 1 sun illumination: ITO / PEDOT:PSS / MAPbI$_3$ / PC$_{61}$BM / Ag (termed as S1), ITO / PEDOT:PSS / MAPbI$_3$ / PC$_{61}$BM / BCP / Ag (termed as S2), and ITO / MoO$_3$ / PEDOT:PSS / MAPbI$_3$ / PC$_{61}$BM / BCP / Ag (termed as S3). As shown in figure **19b**, device S1 having BCP as an additive in the perovskite precursor solution with single HEL and EEL represents PCE of 10.63% with Voc of 0.85 V, Jsc of 18.13 mA/cm$^2$, and FF of 69%. Further improvement in the device occurred while using a buffer layer of BCP as an interlayer between PC$_{61}$BM and Ag (S2). Having a buffer layer of BCP promotes the radiative recombination as well as decrease the interfacial resistance (figure **15d**), which yields into a Voc of 0.90 V, Jsc of 20.40 mA/cm$^2$,FF of 72 % and PCE of 13.22 %. By using BCP as an interlayer, we passivate the perovskite-EEL interface, which can be understood by improved fill factor of device S2. Figure **20** represents the AFM images of Glass/ MoO$_3$/ PEDOT:PSS/ MAPbI$_3$+BCP/ PC$_{61}$BM and Glass/ MoO$_3$/ PEDOT:PSS/ MAPbI$_3$+BCP/ PC$_{61}$BM/ BCP. It is observed that an additional layer of BCP over PC$_{61}$BM layer makes the film smoother than with only PC$_{61}$BM layer (table **5**). Thus, it can make a good contact with Ag and it clearly seen in figure **21**. Ag (8 nm) is thermally evaporated over the two films under same condition and it is observed that roughness of silver over the two films are different. Perovskite film having additional BCP layer gives smooth layer of Ag than the perovskite film having only PC$_{61}$BM (table **5**).  Due to hydrophilic nature of PEDOT:PSS, a good coverage over ITO coated glass is crucial. This can be understood by AFM images of PEDOT:PSS and MoO$_3$/PEDOT:PSS over glass substrate (figure **18** and table **5**). Thus, we fabricated device S3, in which there is an additional layer of MoO$_3$ (10 nm) in between ITO and PEDOT:PSS. This device configuration (S3) forms a double hole and electron extraction layers, lead to reach a champion device with Voc of 0.98 V, Jsc of 19.92 mA/cm$^2$, FF of 77 % and PCE of 15.03 %. Figure **19c** represents the dark J-V characteristics of perovskite solar cells with different extraction layers based devices. The device S1 shows a leakage current of nearly 0.2 mA/cm$^2$ at a reverse bias voltage of 0.5 V. Introduction of double EEL and HEL in the *p-i-n* based devices reduces the leakage current up to one order of magnitude. This clearly indicates that the improvement in the performance is due to the interfacial passivation by BCP (figure **15d**)

and remarkable reduction in shunt path by MoO$_3$ (figure **19c**), leading to efficient charge transport and collection. The external quantum efficiency (EQE) spectra and corresponding integrated J$_{SC}$ for all the three devices are shown in figure **19d**. However, the J$_{SC}$ calculated from EQE is ~ 10% lower than that is calculated from J-V characteristic under 1 Sun illumination. We will discuss this issue later during the comparison of area dependent PV performance. But, the improved J$_{SC}$ of the devices with additional extraction layers are consistent with higher EQE values in devices S2 and S3. The higher EQE values of the devices suggest that the interlayers more efficiently collect the charges and successfully transport it to the respective electrodes by reducing the energy barrier between the interfaces.

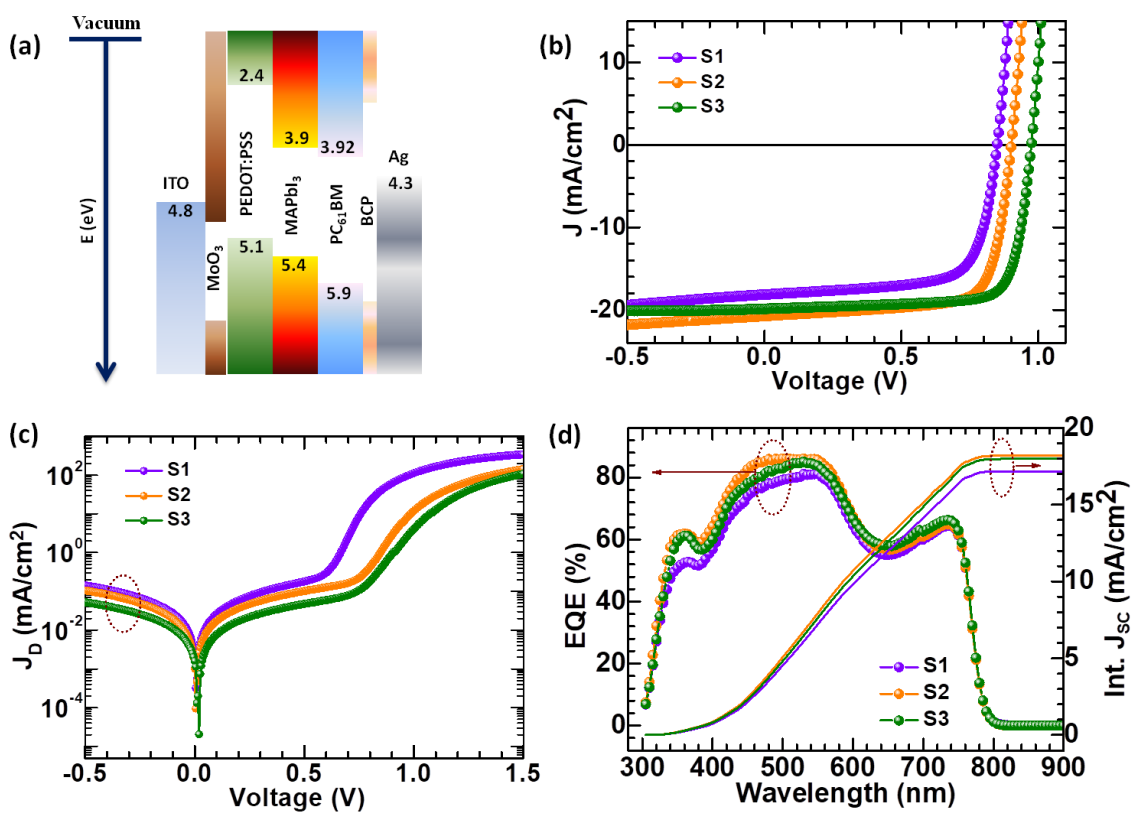

***Figure 19:*** *(**a**) Energy level diagram of p-i-n based perovskite solar cell with double extraction layers. (**b**) J-V curve under 1 Sun illumination & (**c**) dark and (**d**) corresponding EQE & calculated integrated J$_{SC}$ for p-i-n configuration based perovskite solar cells with different extraction layers.*

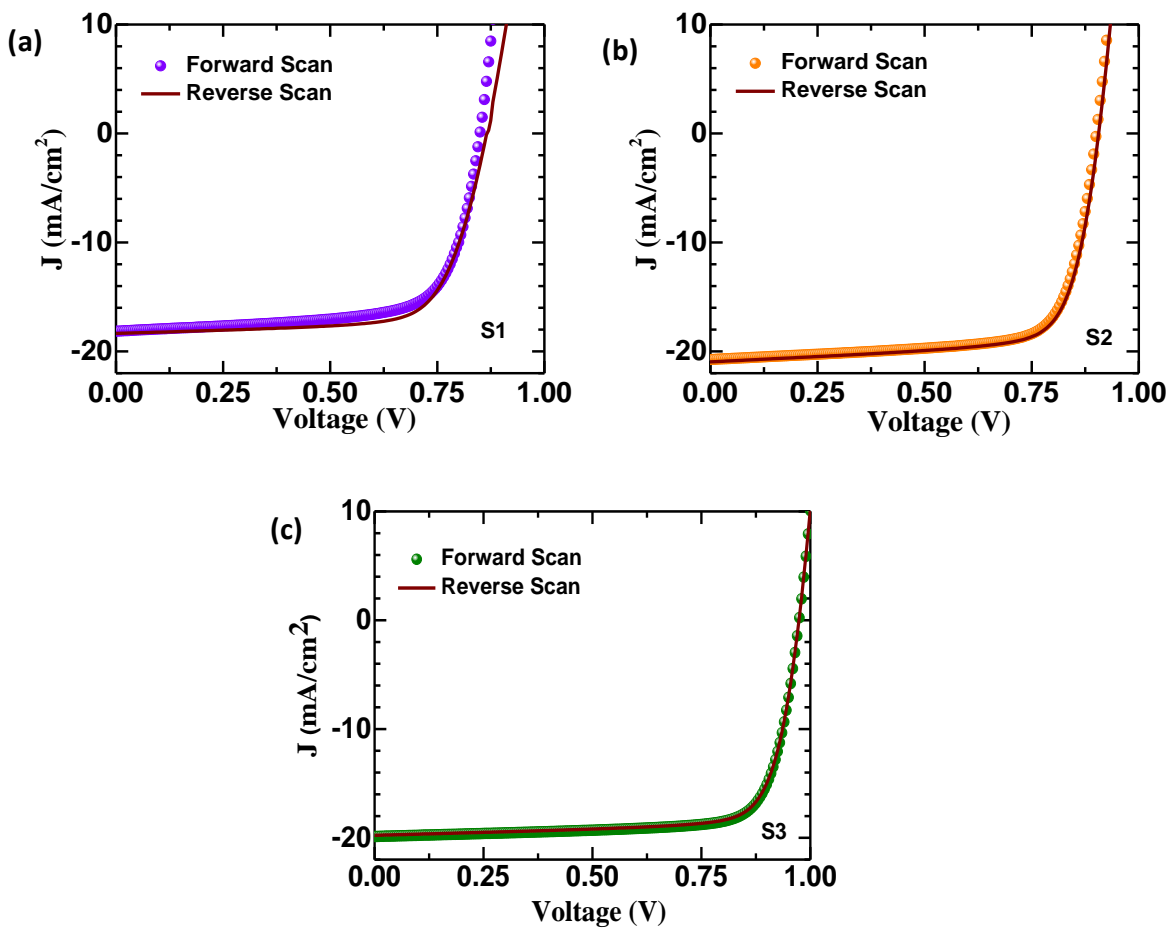

***Figure 20:*** *Illuminated I-V characteristics of devices (a) S1, (b) S2 and (c) S3 with active area of 4.5 mm² in forward and reverse scan direction.*

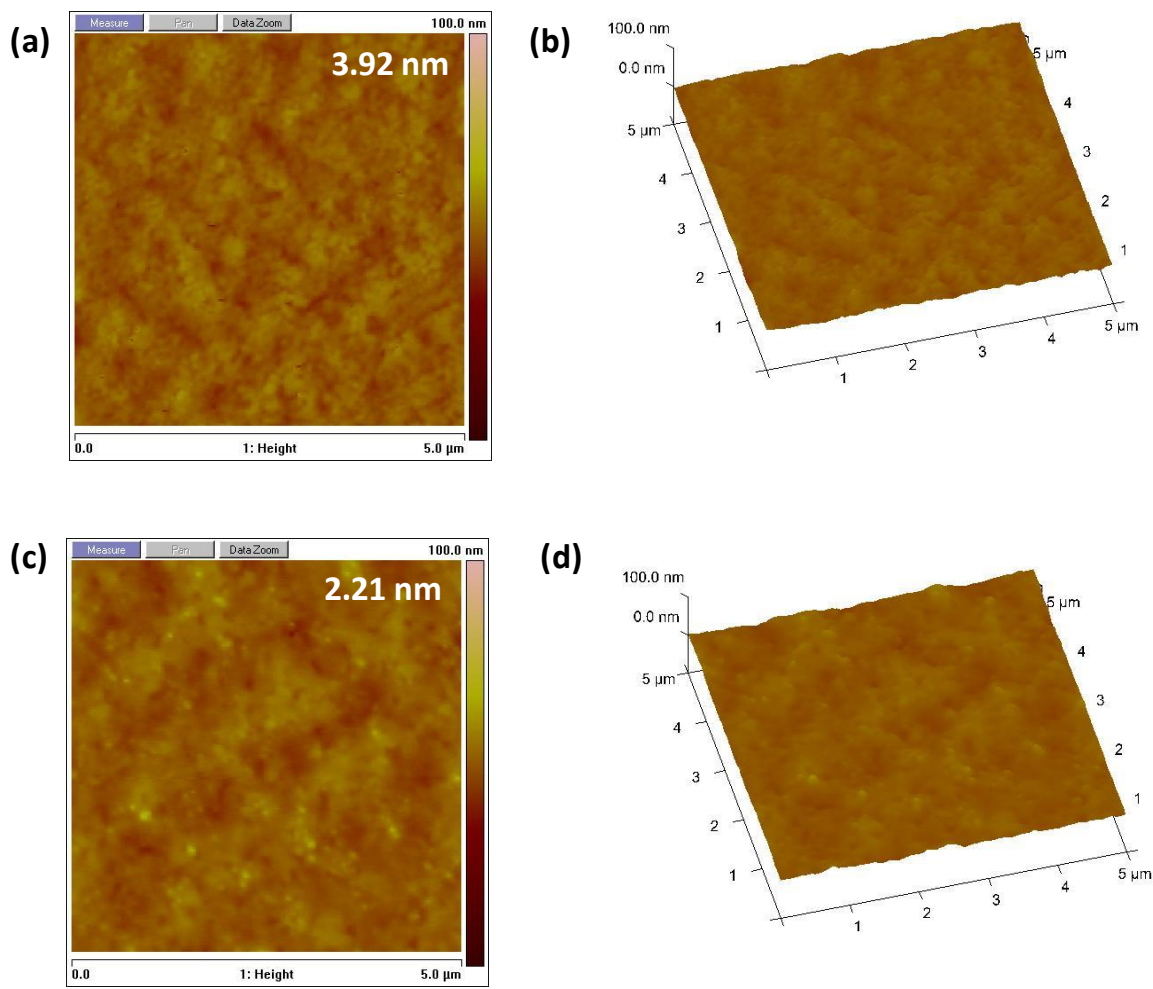

**Figure 21:** *2D AFM images of (a) ITO/ MoO₃/ PEDOT:PSS/ MAPbI₃+BCP/ PCBM and (c) ITO/ MoO₃/ PEDOT:PSS/ MAPbI₃+BCP/ PCBM/ BCP films and corresponding 3D AFM images are shown in figure (b) and (d), respectively.*

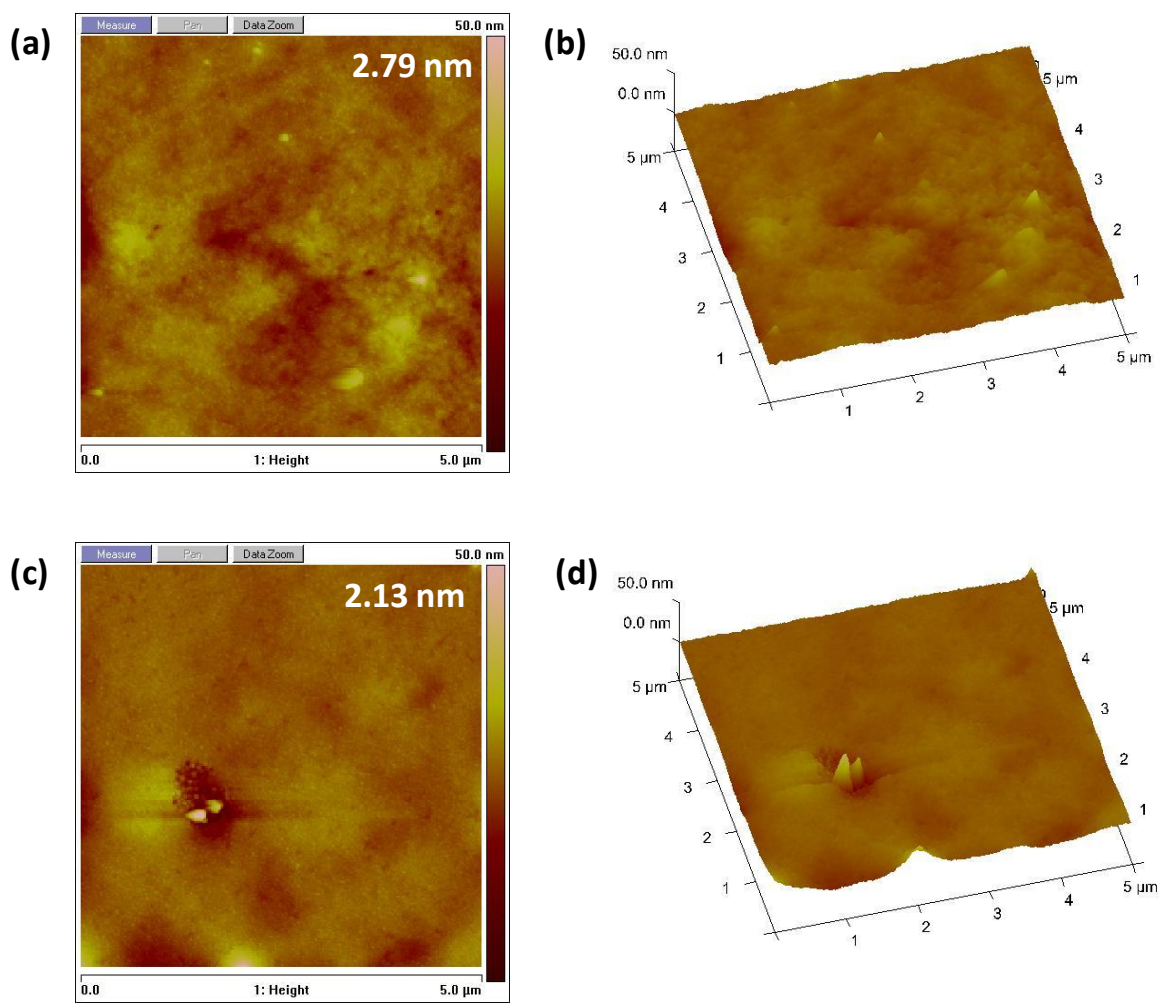

*Figure 22: 2D AFM images of (a) ITO/ MoO₃/ PEDOT:PSS/ MAPbI₃+BCP/ PCBM/ Ag (8 nm) and (c) ITO/ MoO₃/ PEDOT:PSS/ MAPbI₃+BCP/ PCBM/ BCP/ Ag (8 nm) films and corresponding 3D AFM images are shown in figure (b) and (d), respectively.*

As mentioned earlier, $MoO_3$ introduces slight PL quenching at PEDOT:PSS/ $MAPbI_3$ interface, however, remarkable reduction in shunts overall increases Voc with significant amount. Thickness of $MoO_3$ layer has been optimized in order to observe its effect over device performance and it is shown in figure **22**. An improvement in the overall efficiency using double EEL and/or HEL is observed due to suppression of interfacial defect states, shunts (higher $V_{oc}$) and improved charge transport (higher FF). The PV parameters of *p-i-n* configuration based perovskite solar cells are listed in table **6**. It is reported by our group[59] and others[64] that using BCP as an interlayer in between $PC_{61}BM$ & Ag reduces the hysteresis significantly by interfacial defects passivation (figure **23**). Figure **24a** represents the transient photovoltage (TPV) measurement carried out on three devices.[65] The enhanced average lifetime of perturbed charge carrier for devices S2 and S3 supports the suppression of interfacial defects and it is in good agreement with improved $V_{OC}$ and steady-state PL data.[66] The parameter related to TPV measurement is listed in table **7**. Figure **24b** represents the transient photocurrent (TPC) measurement of the three devices held at short circuit condition under 1 sun illumination, which shows the effect of second EEL and/or HEL on the charge transport of the MAPbI$_3$ based perovskite solar cells. It is found that charge transport time for device S3 is faster than that of device S1 and S2, which supports the enhanced FF of device S3.

***Table 5:*** *Root mean square roughness of perovskite based films with different extraction layers.*

| Films | Roughness (nm) |
|---|---|
| ITO/PEDOT:PSS | 1.95 |
| ITO/MoO$_3$/PEDOT:PSS | 1.14 |
| ITO/PEDOT:PSS/MAPbI$_3$+BCP | 9.62 |
| ITO/MoO$_3$/PEDPT:PSS/MAPbI$_3$+BCP | 7.70 |
| ITO/MoO$_3$/PEDPT:PSS/MAPbI$_3$+BCP/PCBM | 3.92 |
| ITO/MoO$_3$/PEDPT:PSS/MAPbI$_3$+BCP/PCBM/BCP | 2.21 |
| ITO/MoO$_3$/PEDPT:PSS/MAPbI$_3$+BCP/PCBM/Ag | 2.79 |
| ITO/MoO$_3$/PEDPT:PSS/MAPbI$_3$+BCP/PCBM/BCP/Ag | 2.13 |

***Table 6:*** *Device performance of the p-i-n configuration based perovskite solar cell with different extraction layer.*

| Device Configuration | $V_{OC}$ (V) | $J_{SC}$ (mA/cm$^2$) | | FF (%) | PCE (%) |
|:---:|:---:|:---:|:---:|:---:|:---:|
| | | From JV | From EQE | | |
| S1 | 0.85 | 18.13 | 16.87 | 69 | 10.63 |
| S2 | 0.90 | 20.40 | 17.87 | 72 | 13.22 |
| S3 | 0.98 | 19.92 | 17.66 | 77 | 15.03 |

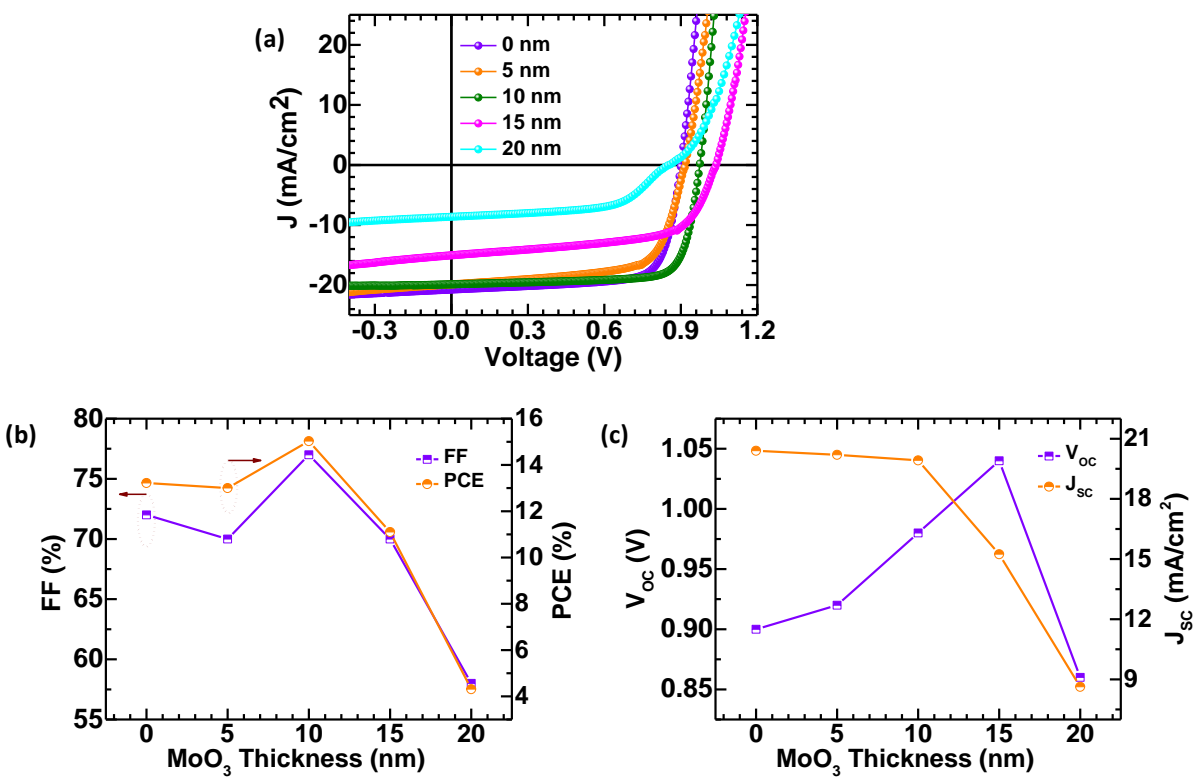

***Figure 23:*** *(a) Illuminated J-V characteristics of device S3 with different thickness of MoO₃. Variation of (b) FF & PCE and (c) $V_{OC}$ & $J_{SC}$ with different thickness of MoO₃ for device S3.*

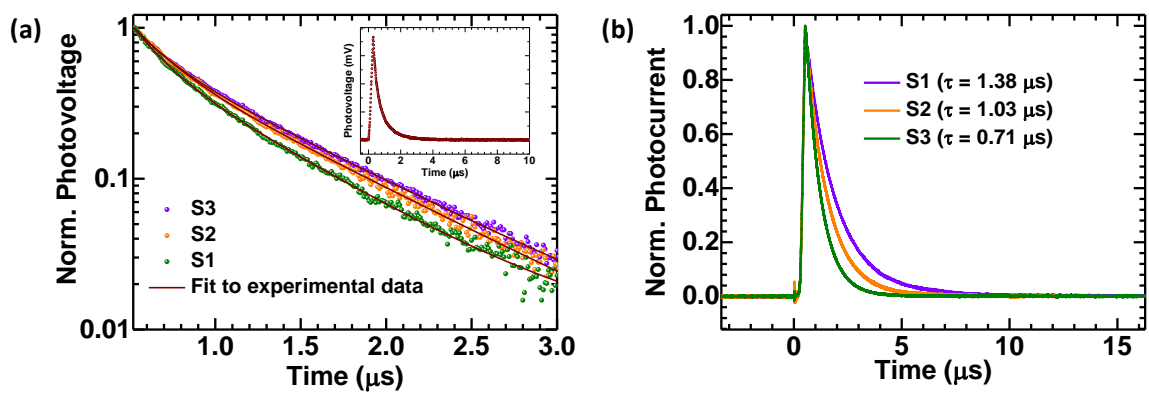

***Figure 24:*** *Normalized transient (a) photo-voltage and (b) photo-current profile of device S1, S2 and S3. Inset of figure 23(a) represents the transient photo-voltage decay profile in linear scale. TPV data are fitted with bi-exponential equation and related parameters are listed in table **7**.*

***Table 7:*** *Perturbed charge carrier lifetime of device S1, S2 and S3 calculated through transient photo-voltage profile.*

| Devices | $A_1$ | $\tau_1$ (µs) | $A_2$ | $\tau_2$ (µs) | $\tau_{av.}$ (µs) |
|---------|-------|---------------|-------|---------------|-------------------|
| S1 | 0.59 | 0.64 | 0.41 | 0.17 | 0.57 |
| S2 | 0.58 | 0.77 | 0.42 | 0.21 | 0.67 |
| S3 | 0.64 | 0.77 | 0.36 | 0.18 | 0.70 |

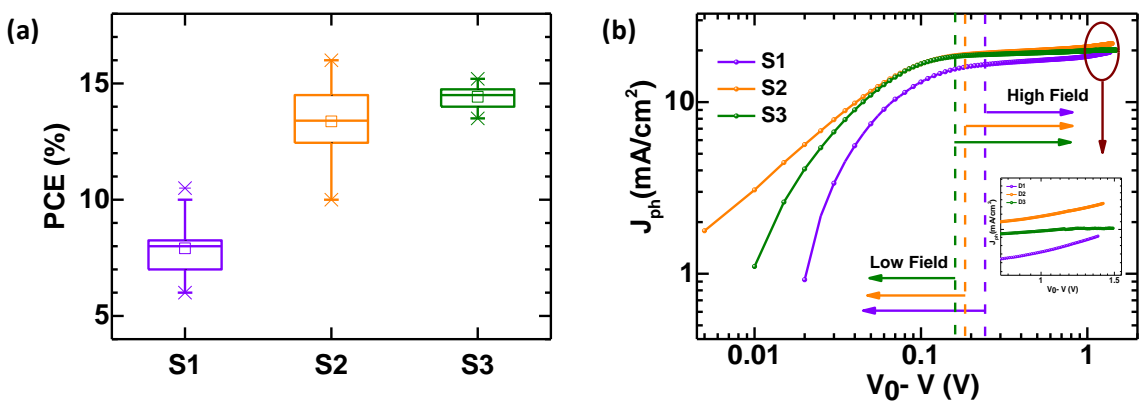

***Figure 25:*** *(a) Histogram for PCE of device S1, S2 and S3 over 16 devices. (b) The plot of photocurrent ($J_{ph} = J_L - J_D$) verses effective applied bias voltage ($V_0 - V$) for three devices.*

Figure **25a** represents the histogram for PCE of device S1, S2 & S3 for 16 devices, which indicates the higher reproducibility of device S3 than device S2.[59] Figure **25b** shows the plot between the photocurrent ($J_{ph}$), which is calculated by the difference of illuminated ($J_L$) and dark ($J_D$) current, verses effective bias voltage ($V_0$-V). Here, $V_0$ is defined as compensation voltage at which $J_{ph}$ is zero. $V_0$ for devices S1, S2 and S3 are 0.87V, 0.91V and 0.99V, respectively. Thus, device S1 saturates at higher effective bias voltage as compared to devices S2 and S3. In the low effective field (figure **25b**), Jph gradually increases with the voltage and shows a weak dependence in the high effective field. It is also observed that at very high field ($V_0$-V > 1), device S1 and S2 starts showing field dependence (inset of figure **25b**). However, device S3 is showing field independent behaviour even effective bias voltage is higher than 1. Further, we have calculated second derivative of Jph ($d^2Jph/dV^2$) in the region $V_{sat}<V<V_{OC}$ for all three devices. Where, $V_{sat}$ is the voltage at which $J_{ph}$ starts saturating. $d^2Jph/dV^2$ for all three devices are greater than zero but higher for device S3, which shows that extent of convexity in J-V curves is higher in device S3 than S1 and S2 (i.e. enhanced FF).[67] The higher value of $d^2Jph/dV^2$ for device S3 also suggests that homogeneity of the interfaces is higher with double ETL and HTL layers, which is consistent with AFM results (**figure 18, 21 and 22**). Hence, improved charge transport (higher FF) in the device S3 is due to better morphology over MoO$_3$/PEDOT:PSS substrate and interfacial defect passivation by BCP at MAPbI$_3$-PC$_{61}$BM interface.

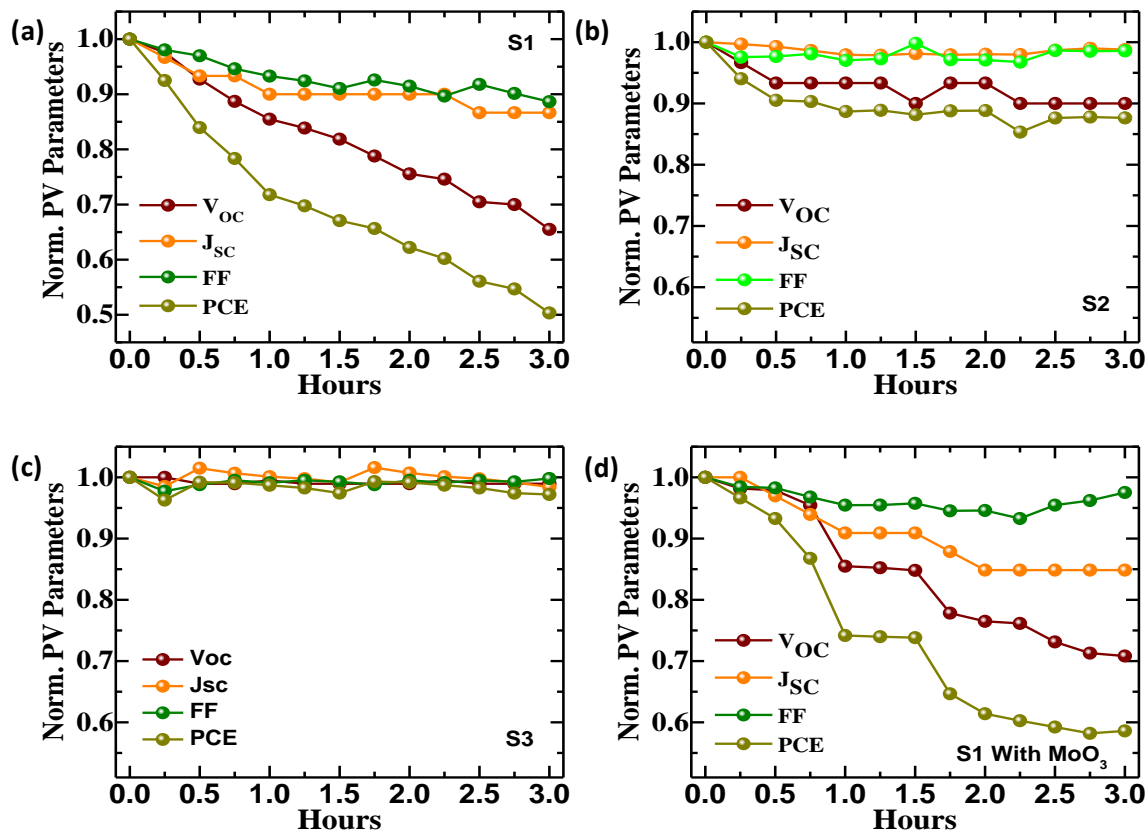

**Figure 26:** *Thermal stability of devices (a) S1, (b) S2 (c) S3 and (d) S1 with additional MoO₃ interlayer without encapsulation under continuous exposure to 1 Sun irradiation.*

In addition to efficient charge transport and PCE, stability is another crucial and heavily studied parameter for the perovskite solar cells in recent times. Stability of the devices was evaluated under continuous illumination of 1 Sun intensity and the J-V curves were measured after every 15 minutes in an ambient condition. However, the devices were encapsulated with epoxy and glass inside the glove box. The measured PV parameters for all the three devices with different extraction layers over a span of 3 hours are shown in figure **26**. The performance of the device S3 at t=0 maintained even after t=3 hours aging test. We did not observe any significant decay in the PV parameters with respect to time and lies well within the error limits (figure **26c**). Such improved stability of the devices is also been attributed to double extraction layers BCP and MoO₃. Thermal stability of device S1 and S2 are shown in figure **26a** and **26b,** respectively. We observe that device S1 shows decrement of 50% in PCE after 3 hours. $V_{OC}$ of device S1 decreases almost linearly with respect to time, which is the main factor for decrement in the PCE. However, having an extra buffer layer of BCP over $PC_{61}BM$ enhances the thermal stability of device S2. It shows decrement of ~ 12% in PCE after 3 hours. It is also worth to note that this 12% decrement is only in the first 30 min, after that device S2 is quite stable. In literature, it is reported that accumulation of iodine ions takes place at the $PC_{61}BM$ /Ag interface due to thermal decomposition of perovskite layer, which leads to formation of metal induced charge states (AgI) at the interface.[59] BCP is used as a buffer layer in between $PC_{61}BM$ and Ag in *p-i-n* structure in order to overcome this issue.[68,69] It is pretty evident in the figure **26c**. Since, there is no BCP layer in the device S1, thermal degradation occurred due to formation of AgI *via* thermal decomposition of perovskite layer. However, device S2 and S3 have BCP buffer layer before Ag. Hence, it resists the accumulation of iodine ion and thus the PCE remains same after 3 hours (figure **26b** & **26c**). We also carried out thermal stability test of device S1 with additional $MoO_3$ layer in between ITO and PEDOT:PSS (figure **26d**). We found a slight improvement of ~10% in the thermal stability of the device D1 by having an additional layer of $MoO_3$. However, origin behind this improvement is under study. We speculate that $MoO_3$ helps in blocking the migration if In ion from ITO to perovskite, which helps in providing better thermal stability to device S3. Thus, double EEL and HEL improve the charge extraction by passivation of the charge trap states, and protect the fast degrading $MAPbI_3$ from the thermal heat under continuous illumination. The temperature of devices after 3 hour continuous illumination was around ~$55^0$C. To confirm the reliability of PCE for perovskite solar cells, we have measured the steady state photocurrent of the device S3 with an applied voltage of 0.8 V at the maximum power point under AM 1.5 illumination in air (figure **27**). We observed that steady state photocurrent and PCE are constant up to 300 seconds.[70,71]

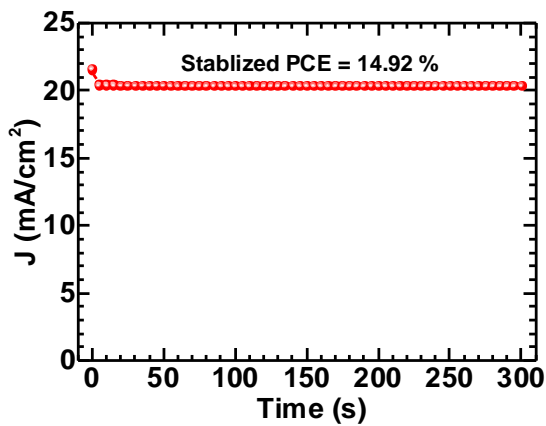

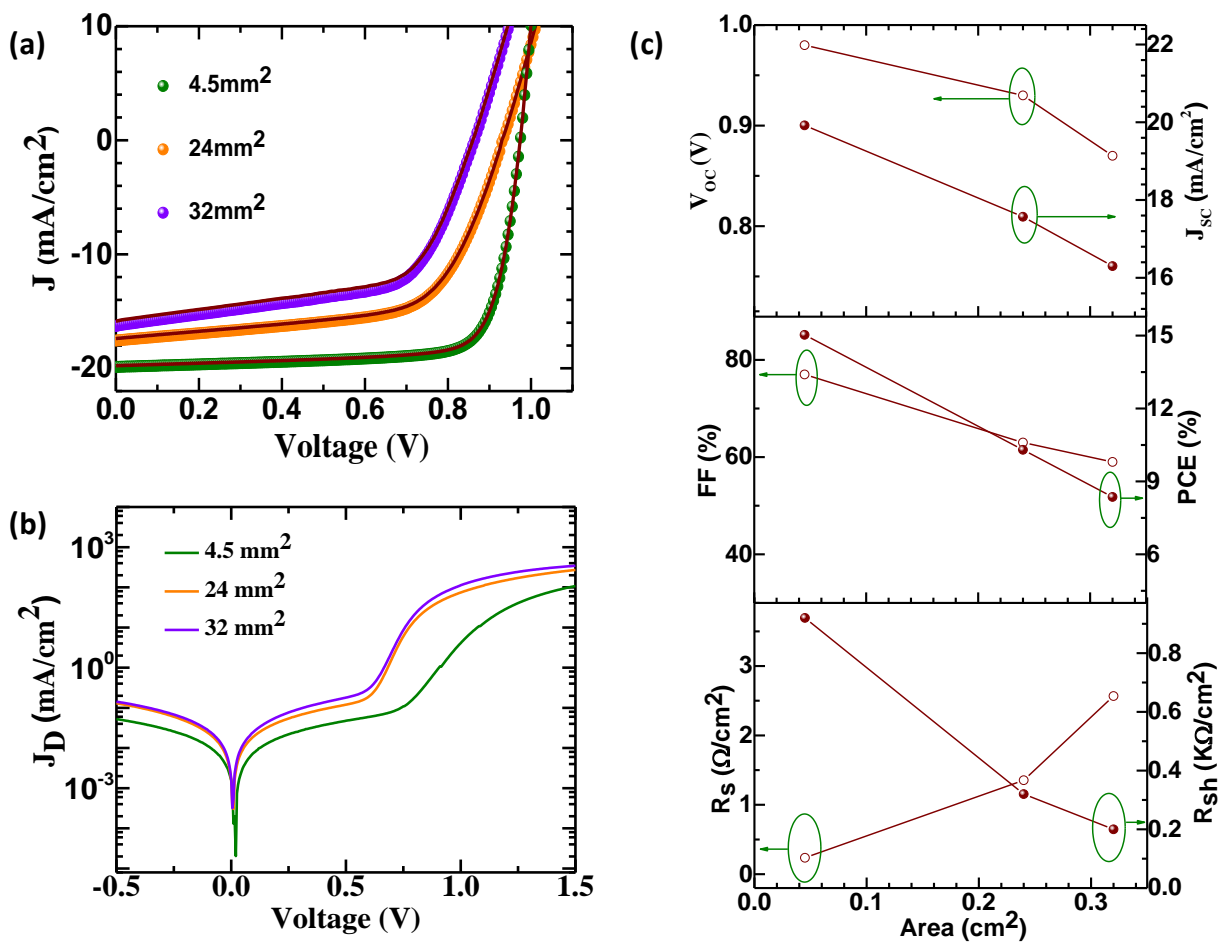

*Figure 27:* Steady state photocurrent measurement of the device S3 with an applied voltage of 0.8 V at the maximum power point under AM 1.5 illumination in air.

*Figure 28:* Area dependent J-V characteristics of S3 device configuration (a) under 1 Sun illumination (scatter – forward scan, solid line- reverse scan) and (b) dark condition. **(c)** Statistics of area dependent PV parameters and series/shunt resistance of S3 device configuration with areas of 4.5mm², 24mm², 32mm².

Further, we extend our study to different area solar cells in order to elucidate the relationship between different PV parameters and active cell area. Devices with additional extraction layers of BCP and $MoO_3$ (S3) were fabricated and studied with area of 4.5 mm$^2$, 24 mm$^2$, and 32 mm$^2$. Figure **28a** and figure **28b** shows the illuminated and dark J-V characteristics of *p-i-n* structure based MAPbI$_3$ perovskite solar cells with different active cell area, respectively. Figure **28c** shows the statistics of area dependent PV parameters which decreases with increase in device area.

To determine the underlying mechanism for the lower PCE, FF, J$_{SC}$ and V$_{OC}$ with increasing active area, we analyze the series resistance R$_S$ of the perovskite solar cell with double extraction layers from their J-V curves using the approach of single diode model [72,73]. By using the single diode model in J-V characteristics equations

$$J = J_{SC} - J_0 \left[ \exp q \left( \frac{V + JR_S}{\eta K_B T} \right) - 1 \right] - \frac{V + JR_S}{R_{Sh}} \qquad (4)$$

Assuming that the shunt resistance R$_{Sh}$ (figure **28c**) is much larger than the series resistance (R$_S$), the value of R$_S$ can be determined by

$$-\frac{dV}{dJ} = \frac{K_B T}{q} (J_{SC} - J)^{-1} + R_S \qquad (5)$$

Where V is the applied voltage, J is the current density under the applied bias, K$_B$ is the Boltzmann constant, T is the absolute temperature and q is the elementary charge. The value of R$_S$ can be obtained by linear fitting –dV/dJ versus (J$_{SC}$-J)$^{-1}$, near the Voc region. R$_S$ decreases with decrease in active cell area with Rs of 2.6 $\Omega$cm$^2$ for 32mm$^2$, 1.4 $\Omega$cm$^2$ for 24mm$^2$ and 0.24 $\Omega$cm$^2$ for 4.5mm$^2$ (figure **28c**). Series resistance is a parameter which impacts directly on the PV characteristics (mainly on fill factor) with increased cell area under illumination. Devices with larger area exhibit a loss of approximately 40% in overall performance because of higher series resistance and lower shunt resistance that may be responsible for lower FF and J$_{SC}$. Resistance of the transparent conductive oxide (TCO) like FTO and ITO might also be the reason for high R$_S$ values. Employing high conducting TCO's might further contribute in improving the R$_S$ values and the FF in larger area devices.[74] It suggests that the origin of losses in FF and J$_{SC}$ can be attributed to a poor film quality over larger area and higher sheet resistance of ITO. Such analysis gives insight into the photo-generated power loss in understanding the area dependent device performance.[75]

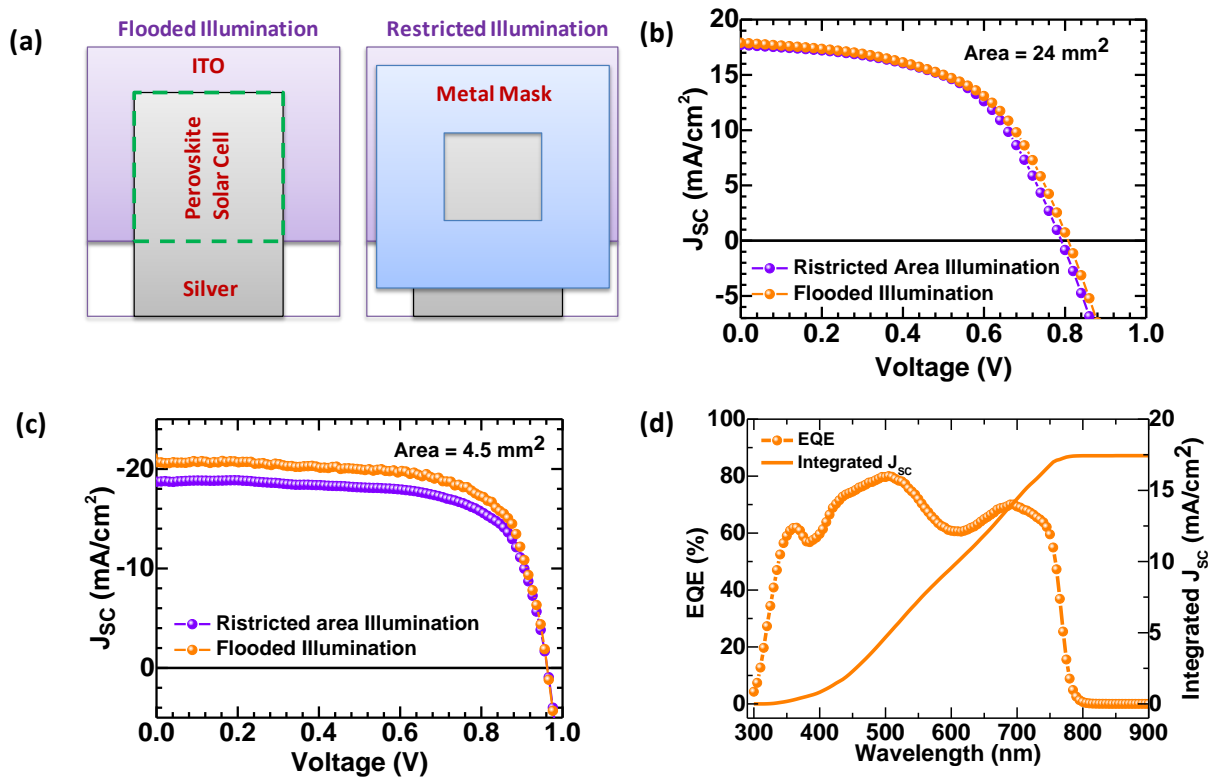

***Figure 29:*** *(a) Device layout of perovskite solar cells with flooded and restricted area illumination (via metal mask). The J-V characteristics of device S3 with area of (b) 24 mm² and (c) 4.5 mm² under flooded and restricted area illumination condition using 1 Sun irradiation. (d) External quantum efficiency (EQE) and calculated integrated $J_{SC}$ for 0.24 mm² area based perovskite solar cells with double extraction layers (Device S3).*

Additional experiments were carried out to identify the contribution from peripheral region on $J_{SC}$ of perovskite solar cells with different active area. The J-V characteristics were carried out using restricted area illumination with the help of metal mask (i.e. stopping the generation of charge carriers outside device area; mask size = 2 mm² for active area of 4.5 mm² & mask size = 6 mm² for active area of 24 mm²) which covers the periphery of the active region and have negligible shadow effect. Figure **29a** represents the schematic of perovskite solar cells with flooded and restricted area illumination (*via* metal mask). Figure

**29b** represents the J-V characteristics of device S3 with active area of 24 mm$^2$ under flooded and restricted illumination condition.

We do not observe any significant reduction in the J$_{SC}$ (~2%) from restricted area illumination condition as compared to the flooded illumination. However, for smaller active area perovskite solar cells (4.5 mm$^2$), we observe a substantial drop in J$_{SC}$ (~11%) under restricted area illumination condition as compared to the flooded illumination (figure **29c**). This can be also understood by J$_{SC}$ calculated from EQE for small and large area devices (figure **19d** and **29d**). The difference in J$_{SC}$ calculated from EQE and lighted J-V characteristic (flooded illumination condition) is ~10% for smaller area solar cells which is in good agreement with J$_{SC}$ obtained from restricted area illumination (table **6**). However, J$_{SC}$ matches well for both measurements in case of large area device (figure **29d**) and there is almost no change in J$_{SC} \approx 17.83$ mA/cm$^2$ for 24 mm$^2$ area PSC under flooded and restricted area illumination. Hence, large area devices or to measure illuminated I-V characteristics through restricted area will give right estimation of perovskite solar cells.[76]

## 6.4 Discussion

Adding the BCP interlayer not-only improved the film morphology to reduce the leakage current (figure **8b**) but also results in a huge improvement in the optical properties of perovskite film by passivating interfacial (figure **7a**) and bulk (figure **7b**) defect sites. A similar approach was adopted by mixing the fullerene into perovskite precursor to improve the perovskite solar cell performance[19]. However, our results and previous studies suggest that PC$_{61}$BM introduces non-radiative channels and quenches PL of perovskite/PC$_{61}$BM interface.[77,78] Our study suggests that BCP has a dual role, where perovskite solution processed with BCP not only reduces leakage current but also passivate defects. These defect bands are also seen in chemical analysis of these films using X-ray photo-electron spectroscopy (figure **13b**), which confirmed that addition of BCP passivate the bulk defects. These results were further confirmed in intensity dependent photo-voltage measurements (figure **12**). We note that the amount of BCP addition (0.02 wt.%) is very crucial in getting the optimal performance, lesser amount does not passivate all the defect sites and higher amount creates aggregates BCP in the films, as seen in the SEM images (figure **3**). An optimal concentration of BCP in bulk of perovskite film most likely goes to grain-boundaries of polycrystalline perovskite film considering the size of BCP molecule (a rough estimation indicates that its width will be ~ 15 Angstrom, if we use hydrogen for terminating the rings then its height will be ~8 Angstrom).[79] This study also provides a route to use double HEL and EEL in perovskite solar cells, which perform their work individually to gain better control over the device performance in terms of reproducibility and stability. We attribute this enhanced PV performance to BCP which passivate the perovskite-EEL interface and $MoO_3$, which provides a better coverage of perovskite resulting in a reduced leakage current.

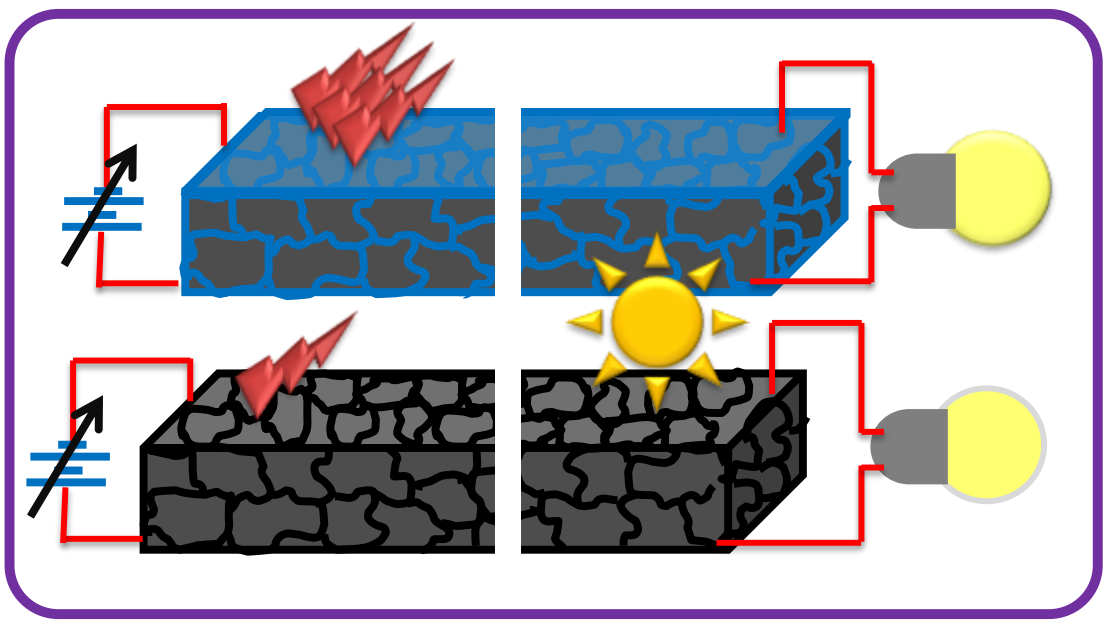

**Figure 30:** *Enhanced optoelectronic properties of perovskite thin-film by passivating the grain-boundary and interfacial defects using organic semiconductor based small molecule.*

## 6.5 Conclusion

We demonstrate high fill-factor (0.82) and high efficiency (~ 16%) perovskite solar cells *via* addition of BCP in $MAPbI_3$ perovskite solution without disturbing the 3D structure of perovskite. Moreover, BCP also provides additional moisture barrier for perovskite films, which can improve the stability of these solar cells. In addition to improved PV performance, there is significant improvement in LED efficiency too due to BCP additive. The BCP added $MAPbI_3$ perovskite has reduced leakage current and defects sites at interface and in bulk.

Steady-state and transient optical studies in addition to PV and EL results confirmed the role of BCP additive in $MAPbI_3$ bulk and at interface. We provide a new design strategy to improve the performance and stability of perovskites without disturbing its 3D structure using organic molecules, which can be used in higher efficiency multi-cation, based 3D perovskites ($MA^+$, $FA^+$ and $Cs^+$) too for photovoltaics and LED applications. We also demonstrate that double EEL and HEL based device configurations have much better reproducibility and thermal stability even in under continuous illumination conditions, which is attributed to the reduced defect states or the recombination centers at the interfaces. This study also suggests working with large area solar cells in order to understand the real performance of perovskite solar cells as the contribution from the peripheral region to the short-circuit current is higher in the case of small area based PVs than the larger one. Thus, our work provides insight into the importance of molecular additive engineering and double extraction layers, which enhance both the PV performance and the stability (moisture and thermal) of perovskite solar cells.

## 6.6    Post Script

Overall, this chapter provides a novel approach of small organic molecule addition in the perovskite precursor solution in order to improve moisture stability without altering the 3D structure of perovskite. Chapter **5** and **6** demonstrate an improvement in the PCE of perovskite solar cell *via* solvent and molecular additive engineering, respectively. It is observed that additive based perovskite solar cells have better charge transport than the without additive solar cells, which is attributed to the reduced number of defect states in the former one. Hence, there is need of some experimental techniques which can provide a direct proof of improved charge transport in terms of ambipolarity of charge carriers in perovskite solar cells. Chapter **7** deal with the transport length scale of charge carriers (hole and electron) in with and without additive (phosphinic acid) based perovskite ($MAPbI_3$) solar cells. The improved fill factor and current density is correlated with the degree of ambipolarity in solvent additive perovskite solar cells.

# Chapter 7

# Correlation between Charge Transport Length Scales and Dielectric Relaxation Time Constant in Hybrid Halide Perovskite Semiconductor

# CHAPTER 7

# Correlation between Charge Transport Length Scales and Dielectric Relaxation Time Constant in Hybrid Halide Perovskite Semiconductor

**Abstract:** The charge carrier diffusion length and dielectric relaxations are important parameters which decide the performance of various optoelectronic devices, in particular for photovoltaic devices. A comparative study is carried out on charge transport length scale ($L$) for passivated and pristine $CH_3NH_3PbI_3$ (MAPI) thin-film based perovskite solar cells (PSCs) through scanning photocurrent microscopy (SPM). The SPM study suggested an improved $L$ and degree of ambipolarity of photo-generated charge carriers (electron and hole) in a passivated as compared to pristine MAPI based PSCs. These results found to be in correlation with frequency dependent photocurrent measurement, which shows that relaxation time of charge carrier is relatively lower in passivated MAPI based PSCs. This mechanism could be explained by trap-assisted recombination, where trap states are induced by ion migration in halide perovskite films. Furthermore, passivation of traps showed an increased degree of ambipolarity in perovskite semiconductor thin-film.

## 7.1 Introduction

The fundamental understanding of charge transport and photo-physics is still under study due to exotic defects dynamics of these dynamically disordered semiconductors.[1] Recently, $CH_3NH_3PbI_3$ (MAPI) is the most explored material used in the field of photovoltaic (PV) community because of its ease of processing via solution route at low temperature with lower exciton binding energy, higher optical absorption, long diffusion length, higher carrier mobility.[2] However, the solution processable method creates impurities and defects in the bulk as well as at the grain boundaries, which acts as recombination centres and thus it affects the transport length scales of charge carriers and optoelectronic properties of the perovskite film.[3] Control over the nucleation rate and crystal growth plays a key role in deciding the defect density in the perovskite thin films.[4,5] Selection of hole and electron extraction layers decides the charge transport in the PV devices.[6,7] Since, the charge carrier diffusion length ($L$) is one of the important parameters which decide the performance of perovskite solar cells (PSCs). There are various techniques reported in the literature to determine $L$ in semiconductor materials such as Si, GaAs, CdTe/CdS, P3HT:PCBM etc. using surface photovoltage,[8] electron beam induced currents[9], optical beam induced currents,[10] etc. Terahertz spectroscopy was used to measure $L$ in MAPI films and it is reported ~ 1-3µm.[11] Surface photo-voltage measurements provide $L$ in the order of ~ 10 µm.[12] The value of $L$ calculated from transient fluorescent spectroscopy in MAPI film is even low (~ 100 nm).[13] Steady-state photo-carrier grating (SSPG)[14] measurement provides $L$ ~ 500 nm for MAPI. Femto-second transient optical spectroscopy over MAPI film also provides the $L$ in the range of 100 nm to 1 µm.[15,16] Other techniques such as microwave photo-conductance transients,[17] ultrafast microscopy,[18] etc. also provides a similar value of $L$ for MAPI based thin films. Interestingly, $L$ for MAPI based single crystals is quite higher than that of thin films due to few orders of difference in defect density. [19,20] However, the interpretation made on the basis of such experiments is in the presence of both kinds of charge carriers (electrons and holes) in the same region, which gives a transport length of carriers in the background of recombination processes such as trap assisted and/or free electron-hole. Hence in order to determine the ambipolarity character of semiconductor one would be interested in individual carrier $\mu\tau$ product, i.e., diffusion length or transport length.

The origin of enhanced charge transport of PSCs has been explained by different models such as Rashba splitting,[21] defect tolerance,[22] ferroelectric polarons, photon recycling, etc.[23] MAPI exhibit features of ferroelectric material and have a strong frequency dependence of the dielectric constant. The frequency dependent dielectric response of perovskite is mainly due to displacement of ions at lower frequency.[24,25] However, the polarization of ferroelectric domains is heavily dependent on surface defects and thus it affects the charge transport in the PSCs. It has been observed that larger perovskite crystals shows larger spontaneous polarization under an external electric field and retention behavior of larger crystals are longer than the smaller one.[26] Some computational studies were carried out over the MAPI and its interfaces shows the iodine vacancy and interstitials can easily travel through the perovskite crystals.[27] Snaith *et al.* have reported that excess iodine or methylammonium ions may be present at the interstitial sites and they can migrate on either side of the film under influence of an external electric field.[28] Hence, polarization by ion migration and ferroelectricity both can affect the charge transportation in PSCs.

In this chapter, we estimate the charge transport length of photoinduced carriers in MAPI based solar cells using scanning photocurrent microscopy (SPM)[29] and relate transport length scales with dielectric relaxation time.[30] For better understanding, this chapter presents a comparative study of *L* between passivated and pristine MAPI based PSCs, where phosphinic acid as an additive can passivate the defects states in perovskite thin film to facilitate *L* and promotes bimolecular recombination.[31,32] SPM provides direct insight on such as degree of ambipolarity and lateral transport length of hole and electron, separately from a single bipolar device. Relaxation time constant ($\tau$) of charge carriers is determined experimentally in a separate study of frequency dependent photocurrent and found to be in good agreement with SPM studies of *L*.[33] We observed higher *L*, a higher degree of ambipolarity, lower value of relaxation time in passivated PSCs.

## 7.2 Experimental section

*Materials:* Lead acetate trihydrate ($Pb(Ac)_2.3H_2O$), phosphinic acid and BCP were purchased from Sigma Aldrich. Methyl ammonium iodide (MAI) was purchased from Greatcell Solar. $PC_{61}BM$ and PEDOT:PSS were purchased from Solenne BV and Clevios, respectively. All the materials are used as received.

***Solution preparation:*** 40 wt % perovskite precursor solution was prepared using MAI and Pb(Ac)$_2$.3H$_2$O in the molar ratio of 3:1 in dimethylformamide (DMF) and stirred at room temperature for overnight. 3 µl of phosphinic acid was added to per mL of the perovskite precursor solution for solvent additive perovskite solar cells. 15 mg of PC$_{61}$BM was dissolved in 1 ml of 1,2,dichlorobenzene and kept for overnight stirring at room temperature. 1mg/ml of BCP in isopropanol (IPA) solution was kept under stirring for 15 min in the glove box at 70$^o$C before use.

***Device fabrication and characterization:*** *p-i-n* configuration based perovskite solar cells were fabricated on indium tin oxide (ITO) coated glass substrates. Substrates were cleaned sequentially with soap solution, deionized water, acetone and isopropanol. After oxygen plasma treatment for 10min, PEDOT:PSS was spincoated at 5000rpm for 30 sec on the ITO coated substrates and annealed at 150 $^0$C in N$_2$ atmosphere for 30min. Spincoating of MAPI films and electron transporting layers (ETL) were performed inside the N$_2$ filled glove box with the maintained H$_2$ and O$_2$ levels. PEDOT:PSS coated ITO substrates were transferred into the glove box for MAPI spincoating. Perovskite precursor solution was spincoated on the prepared PEDOT:PSS/ITO substrate at 2000 rpm for 45 sec. The prepared MAPI films undergo drying and annealing. The spincoated light brownish perovskite films were dried at room temperature for 10 min on workbench and then annealed at 100$^o$C on hotplate for 5 min. The PC$_{61}$BM solution was spincoated at 1000 rpm for 60 sec and workbench dried for 5min. After bench drying of the PC$_{61}$BM/MAPI/PEDOT:PSS/ITO substrate, BCP was spincoated at 5000rpm for 20sec. The prepared films were transferred into the thermal evaporation chamber for deposition of 80 nm of Ag. Shadow mask of area 4.5mm$^2$ was used to decide the active area of the cell.

Photocurrent density-voltage (J-V) measurements were carried out using a Keithley 4200 semiconductor characterization system and LED solar simulator (ORIEL LSH-7320 ABA) after calibrating through a reference solar cell provided from ABET. All the J-V measurements were performed using scan speed of 40 mV/s. Quantum efficiency measurements were carried out to measure the photo-response as a function of wavelength using Bentham quantum efficiency system (Bentham/PVE300). Morphological study was done using field emission scanning electron microscopy (FESEM).

***Scanning Probe Microscopy (SPM):*** Figure **1a** represents the picture of experimental setup established in our lab. A 490 nm TOPTICA class 3B laser diode was mechanically chopped (1000 Hz) and focused on the sample through 60X (0.7 NA) objective in an inverted Olympus IX73 microscope. PCB sample holder was designed to hold the PSCs and it will get fixed in the slot provided in the scanning stage as shown in figure **1b**. The spot size of 490 nm diode laser is also shown in the figure **1b** and it is ~450 nm. The intensity of the laser was decided in such a way that the laser spot will not form any circular fringes around the central spot. The power of laser used in this experiment is ~100 nW and it was measured through Si based photo-detector and lock-in amplifier. When the laser source is incident on the device (black portion; figure **1b**), the holes and electrons are generated and simultaneously collected to their respective electrodes and results in short-circuit photocurrent. When the laser spot is scanned in extended ITO direction (orange portion; figure **1b**), holes are easily collected by the ITO. But electrons have to travel in the lateral direction in order to reach the electron collecting electrode (Ag). Hence, as the laser spot moves away from the Ag electrode, the efficiency of electrons to reach the electrode reduces. As a result of this, we observed an exponential decay in the photocurrent profile as a function of distance from the Ag electrode towards the film, which is used to calculate the lateral diffusion length of electrons in PSCs. In a similar way, one can calculate the diffusion length for holes by scanning the laser spot in extended Ag direction. SR830 lock-in amplifier coupled with the same mechanical chopper mentioned previously is used to measure the electron and hole photocurrent. SR830 lock-in amplifier measures R and θ simultaneously for an incoming signal. Here, R represents the photocurrent and θ represents phase delay between the chopped reference signal and device signal. The lock-in amplifiers and scanning stage are interfaced via LabView VI. LabView software was customized to collect the photocurrent and phase delay simultaneously with respect to (x or y) position of the device

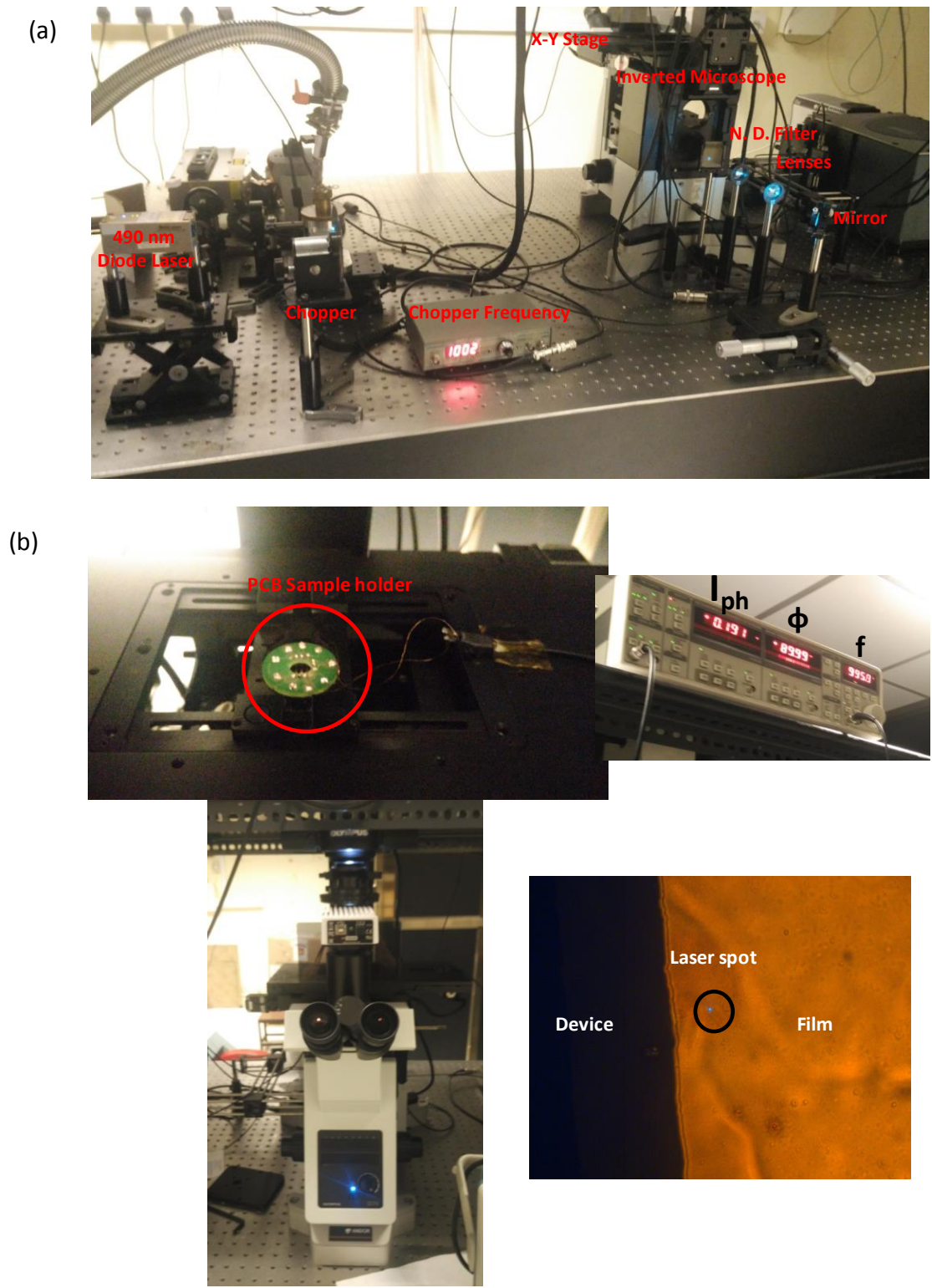

**Figure 1:** *Experimental setup of scanning probe microscopy in our laboratory to determine the lateral transport length of charge carriers in perovskite solar cells.*

***Dielectric measurement:*** A voltage $V_0$ with a fixed frequency $\omega/2\pi$ is applied to the sample. $V_0$ causes a current $I_0$ at the same frequency in the sample. In addition, there will generally be a phase shift between current and voltage described by the phase angle $\phi$. The ratio between $V_0$ and $I_0$ and the phase angle $\phi$ are determined by the sample material electric properties (permittivity and conductivity) and the sample geometry.

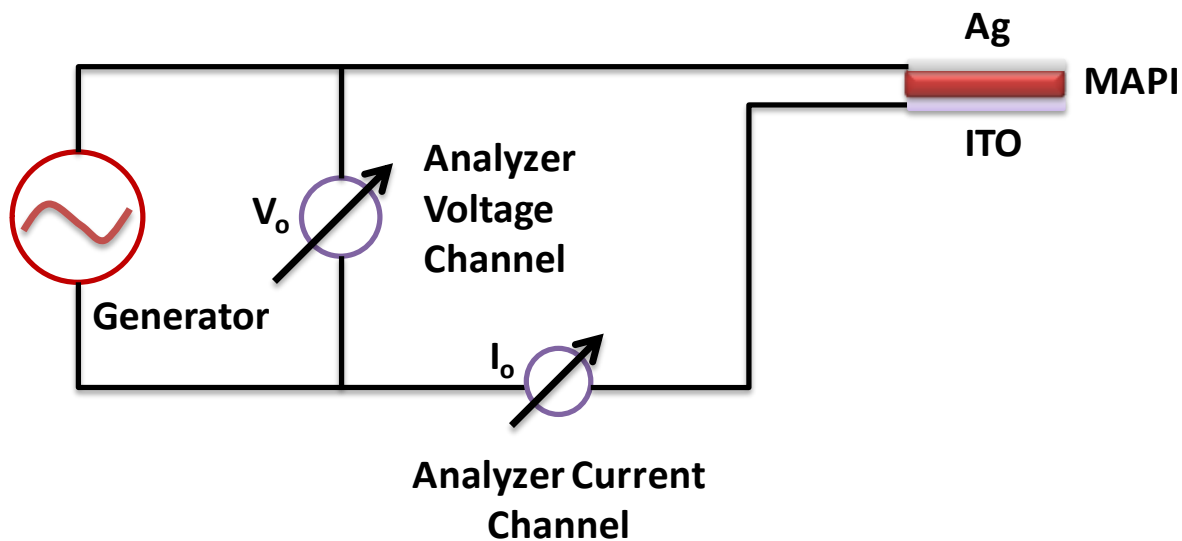

***Figure 2:*** *Schematic of dielectric measurement.*

### 7.3 Results and Discussion

A conventional planar heterojunction (*p-i-n*) structure of indium tin oxide (ITO) / poly(3,4-ethylene-dioxythiophene):polystyrenesulfonate (PEDOT:PSS) / MAPI / phenyl-C61-butyric acid methyl ester ($PC_{61}BM$) / Bathocuproine (BCP) / silver (Ag) was fabricated. Figure **3a** shows the crystal and chemical structure of MAPI and phosphinic acid (solvent additive), respectively. Figure **3b** represents the device configuration of planar perovskite solar cells and energy level diagram of each layer. Figure **3c** and **3d** represents the top view microstructures of pristine and passivated perovskite thin films fabricated over PEDOT:PSS coated ITO substrate. Both the films are quite compact; however, passivated perovskite films have lesser number of grain boundaries than the pristine film which is required for better charge transport in the PSCs.[34]

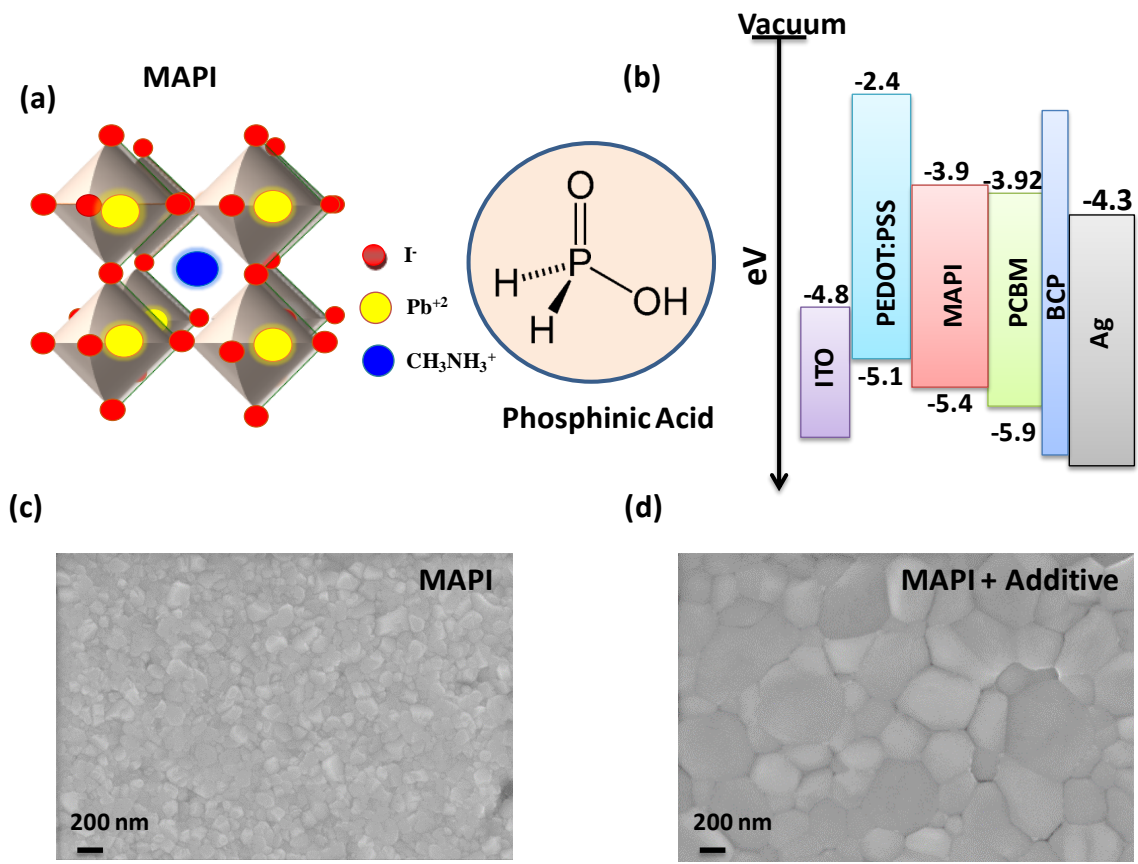

*Figure 3:* (a) Crystal structure and chemical structure of CH₃NH₃PbI₃ (A = CH₃NH₃⁺, B = Pb⁺², X=I⁻) and phosphinic acid (solvent additive), respectively. (b) Energy level diagram of perovskite solar cell. Top view morphology of (c) pristine and (d) passivated MAPI based perovskite thin films.

Table **1** listed the photovoltaic parameters of passivated (MAPI + additive) and pristine MAPI based PSCs with average PCE over 16 devices. Corresponding J-V and external quantum efficiency (EQE) curves for pristine and passivated MAPI based PSCs are shown in figure **4a** and **4b**, respectively. The devices made with solvent additive (passivated) shows better power conversion efficiency (PCE) of 16.22% with open circuit voltage (V$_{oc}$) of 1.01 V, a short-circuit current density (J$_{sc}$) of 21.13 mA/cm² and fill factor (FF) of 76 %. In contrast, the control devices showed a PCE of 11.40% only with V$_{oc}$ of 0.89 V, a J$_{sc}$ of 18.57 mA/cm² and FF of 69 %. The increased J$_{sc}$ of solvent additive device can be understood by external quantum efficiency (EQE) and it is shown in figure **4b.**

***Table 1:*** *Photovoltaic parameters of the best performing pristine and passivated MAPI based perovskite solar cells. Average PCE is also listed over 16 devices.*

| Devices MAPI | $V_{OC}$ (V) | $J_{SC}$ (mA/cm$^2$) | FF (%) | PCE (%) | Average PCE (%) |
|---|---|---|---|---|---|
| Pristine | 0.89 | 18.57 | 69 | 11.4 | (9.2 ± 2.2) |
| Passivated | 1.01 | 21.43 | 75 | 16.2 | (15.3 ± 0.9) |

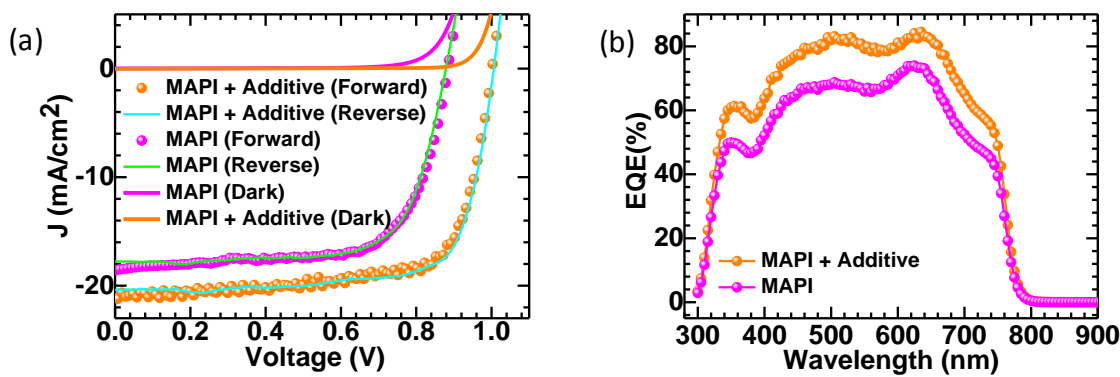

***Figure 4:*** *(a) Dark and illuminated J-V characteristics under 1Sun illumination and (b) EQE of control and solvent additive based MAPI perovskite solar cells.*

. Figure **5a** shows the schematic of spatial photocurrent ($I_{ph}$) measurement in x-y direction. A chopped 490 nm laser diode having spot size ~450 nm is scanned over the 2D active area of both pristine and passivated MAPI based PSCs and corresponding $I_{ph}$ is measured through lock-in amplifier. Figure **5b** and **5c** represent the contour plot for a 2D scan over the area of 20 μm X 20 μm (figure **5a** with blue dotted rectangle area) with a step size of 2 μm for photocurrent in the active region of pristine and passivated PSCs, respectively. The passivated MAPI PSCs shows a uniform distribution of photocurrent in the active area than pristine, which is consistent with homogeneous morphology of passivated perovskite thin film due to lesser grain boundaries than the pristine one (figure **3c and 3d**). Although, the field emission scanning electron microscopy (FESEM) shows both the film quite compact, but, 2D scan of photocurrent reveals the in-homogeneity in the pristine film based PSCs. This suggests the superiority of photocurrent scanning over the FESEM.

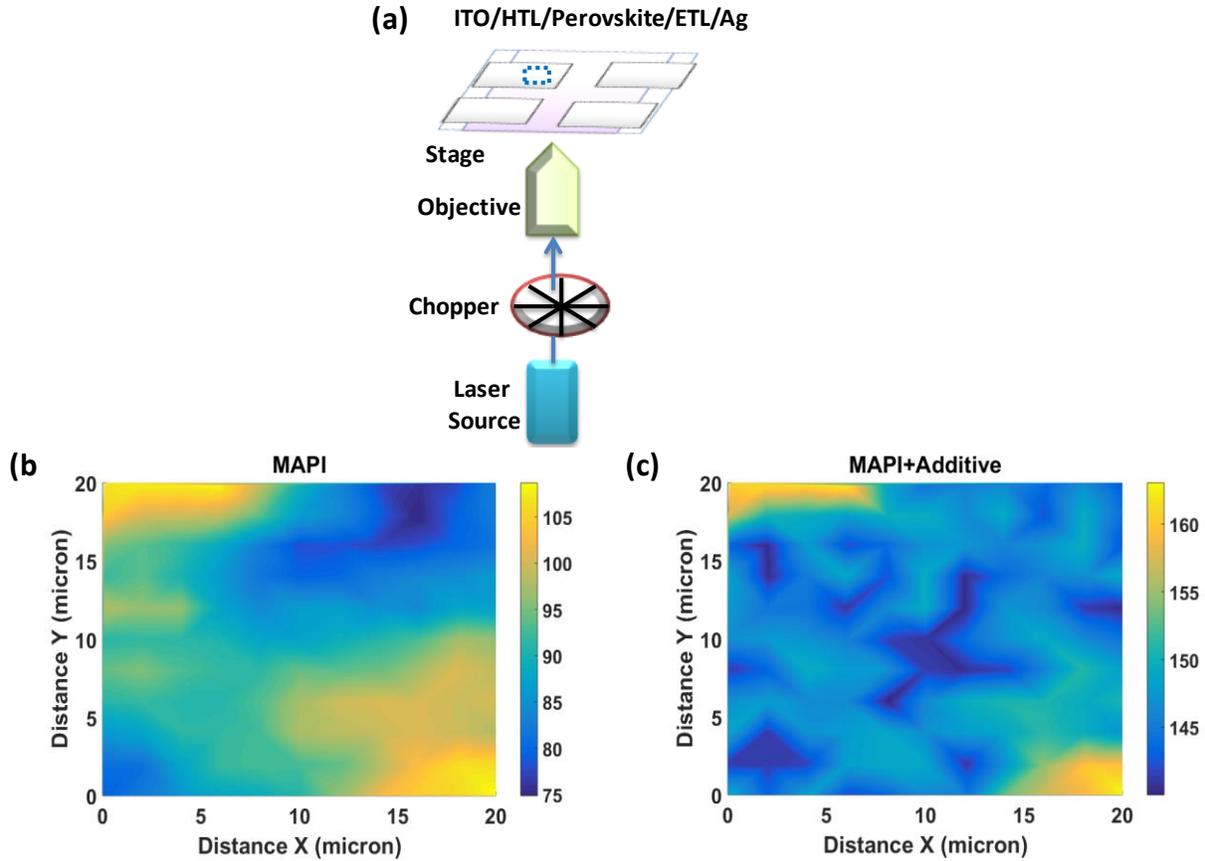

***Figure 5:*** *(a) Schematic of the 2D scan of photocurrent to determine the homogeneity in the perovskite thin films. Contour plot for the 2D scan (20 μm X 20 μm) of photocurrent in the active region (blue dotted square region in figure 2a) of (b) pristine and (c) passivated MAPI based PSCs.*

The schematic diagram of SPM is shown in figure **6a**. In this technique, we probe either extended Ag or extended ITO region at a time outside the active area using a focused spot of 490 nm diode laser to estimate *L* of electron or hole, respectively, using spatially resolved photocurrent measurement. The measurement is carried out using a low intensity (~100 nW) laser source, high chopper frequency (~ 1KHz) and an over-filled 60X microscope objective (NA = 0.7) with a spot size of ~ 450 nm. The stage translation is carried out with a step size of 2 μm in order to maintain the resolution limit ($\lambda/2*NA = 0.35$ µm). Figure **6b** represents a typical electron photocurrent decay profile in different regions of PSCs. The grey region (region 1) represents the active area of PSC (ITO/PEDOT:PSS/MAPI/PC$_{61}$BM/BCP/Ag), and it is observed that the photocurrent remains constant in the active area (figure **6b**). The Magenta region in figure **6b** represents the interface between the active area and perovskite film without Ag (region 2). Photo-generated electrons are collected by Ag and holes are collected by ITO due to band alignment of perovskite film with respective electron and hole transporting layer. A sudden drop in the photocurrent is observed in the region 2 because of the absence of silver electrode to collect the electron; however, holes are collected by ITO. Hence, electrons have to travel laterally in order to get collected by the Ag electrode and thus, it is termed as electron photocurrent decay profile. The wine region (region 3) represents the perovskite films without Ag (ITO/PEDOT:PSS/MAPI/PC$_{61}$BM/BCP) and this region is used to determine the electron diffusion length, which will be discussed below. Similarly, the extended Ag region will provide hole photocurrent decay profile. The lock-in amplifier provides an absolute magnitude of the photocurrent signal and also a phase delay ($\phi$). Figure **6c - 6f** represents the absolute electron and hole photocurrent signals ($I_e$ or $I_h$) and corresponding phase delay ($\phi e$ or $\phi_h$) for pristine and passivated PSCs. Fluctuation in the $\phi$ of pristine device (region 1) corroborated the in-homogeneity in the perovskite film itself. A sharp drop in the $\phi$ of both the devices is observed at the interface between the active area and the perovskite film without Ag, which is consistent with drop in the $I_{ph}$. Once the effect of interface vanishes, $\phi$ becomes relatively flatten in the region 3. The reflection maps follow the same trend of photocurrent decay and are shown in figure **7**. However, the constant phase region is used for the selection of region 3 to determine the charge transport length scale.

The decay profile of $I_e$ and $I_h$, respectively, for pristine and passivated PSCs in region 3 are shown in figure **8**. The region 3 is used to estimate the charge transport length scale of hole and electron in pristine and passivated PSCs. The short-circuit photocurrent ($I_{ph}$) arises from extended ITO region that is outside the Ag electrode boundary is denoted by $I_e$ (x) (electron photocurrent) and $I_h$ (x) represents the hole photocurrent which arises from extended Ag region that is outside the ITO electrode (figure **6a**). We observe that $I_{ph}$ decreases exponentially as the laser spot is scanned from the overlap region to either extended ITO region or extended Ag region (figure **6a**).

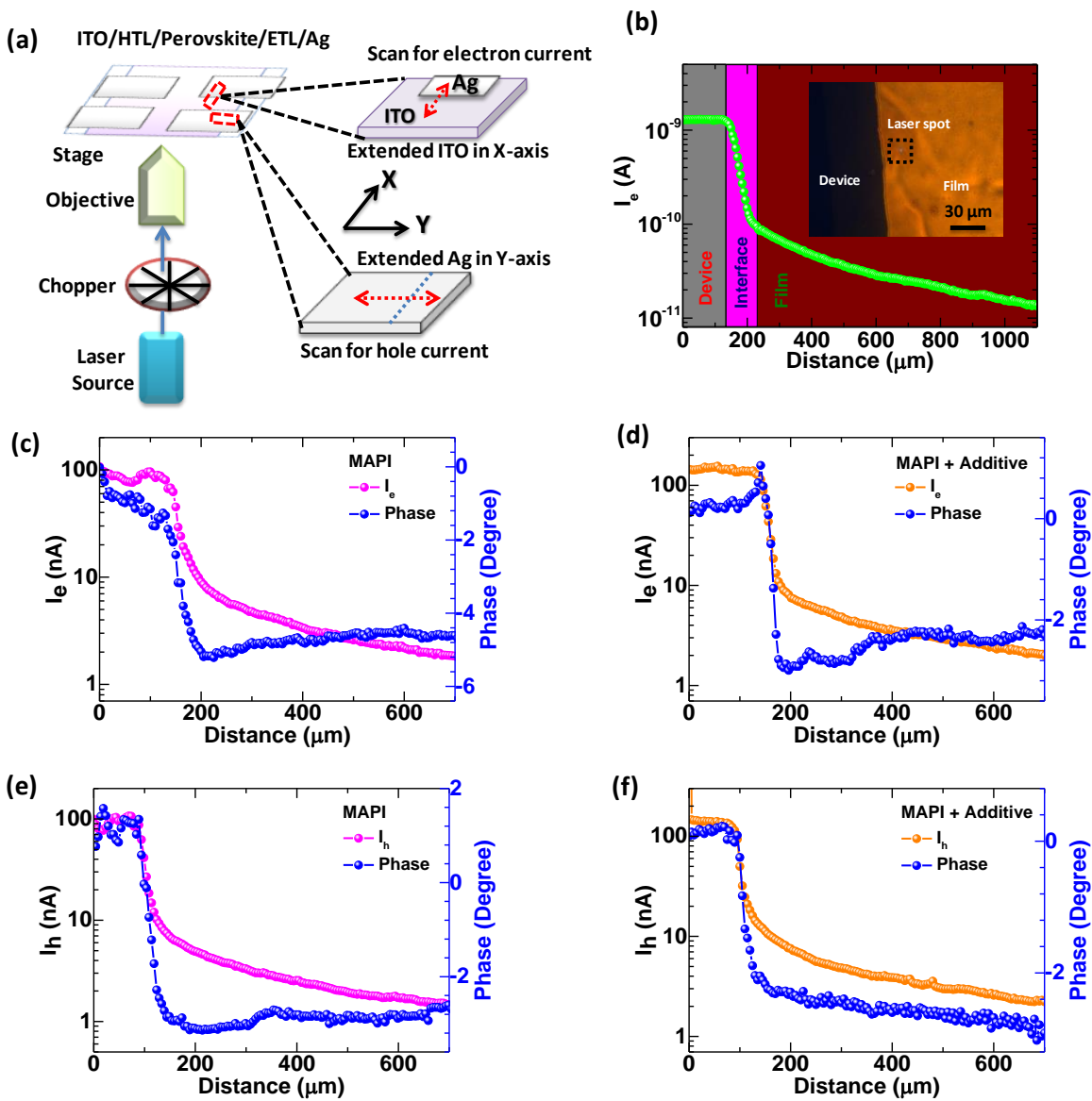

**Figure 6:** *(a) Schematic of the scanning probe microscopy to estimate the charge transport length of photo-generated electron and hole, separately. (b) Spatially resolved short-circuit photocurrent profile of the perovskite solar cells. Inset represents the image of device (black region) – perovskite film (orange region) interface through 60X (NA =0.7) objective microscope. The blue dot inside the black dotted square shows the 490 nm diode laser spot. (c, d, e & f) Photocurrent decay profile and phase delay of electron and hole for pristine and passivated MAPI based perovskite PSCs.*

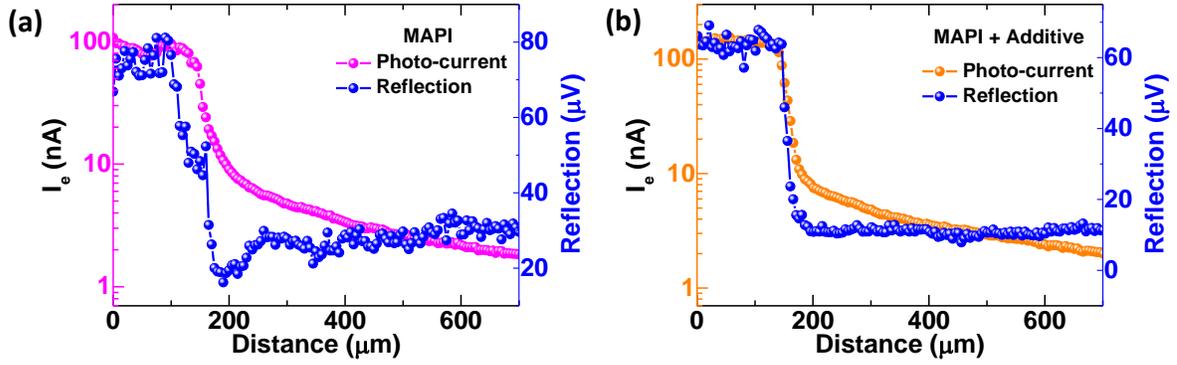

***Figure 7:*** *Photocurrent decay profile and corresponding reflection map for (a) MAPI and (b) MAPI + Additive based perovskite solar cells.*

The absorption of photons leads to the generation of electron-hole pair in the bulk of the perovskite semiconductor, which causes an excess minority charge carriers. The final current in the device depends upon the charge recombination and collection mechanism. The charge collection is determined by the contribution of both drift and diffusion based components. The drift based transport depends upon the electric field in the semiconductor; however, the charge carrier concentration in the semiconductor will define the diffusion based transport. The transport equation can be written as

$$\frac{\partial n}{\partial t} = D\frac{\partial^2 n}{\partial x^2} + \mu E\frac{\partial n}{\partial x} + \mu n\frac{\partial E}{\partial x} - \frac{\Delta n}{\tau_n} + G \qquad (1)$$

Where, D is the diffusion coefficient of charge carriers, n is the carrier concentration, G is the generation rate, μ is the mobility, E is the electric field in x-direction, Δn = n-n$_0$ is the excess carrier concentration, t is the time and τ is the carrier recombination lifetime. The first term in equation **1** represents the diffusion component of charge transport. Consider a system in which the electric field in the material is zero and there is no charge generation, then equation **1** becomes

$$\frac{\partial n}{\partial t} = D\frac{\partial^2 n}{\partial x^2} - \frac{\Delta n}{\tau_n} = 0 \qquad (2)$$

The solution of equation **2** is

$$n = n_0 \exp\left(-\frac{x}{L}\right) \qquad (3)$$

Where, $L$ is the diffusion length of minority charge carriers.

However, $Ls$ measured here have an additional influence of LPV effect, which is drift-assisted diffusion process. Hence, this is an effective transport length scale which has a lateral-drift component too and is different from the commonly used term "diffusion length" in the semiconductor community.

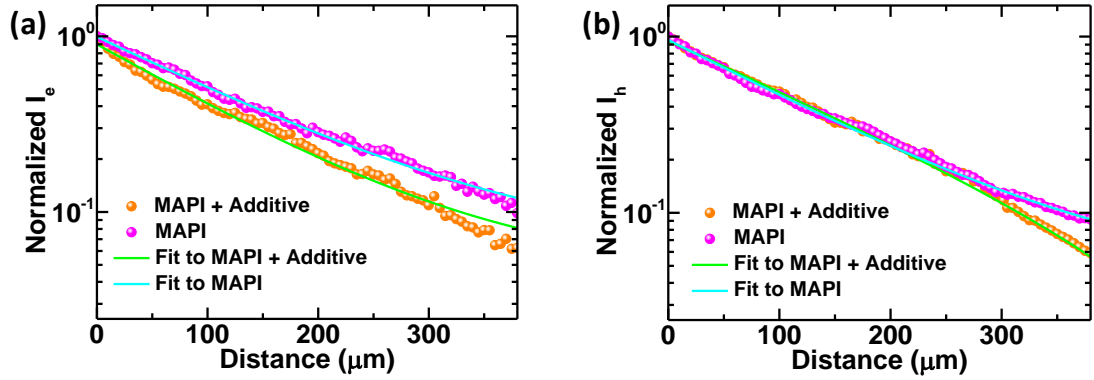

**Figure 8:** *Spatially resolved short-circuit (a) electron and (b) hole photocurrent profile of pristine and passivated MAPI based PSCs.*

The $I_{ph}$ decay profiles outside the overlapping region are fitted through the following equation:

$$I_{ph}(x) = I_{ph_0} \exp\left(-\frac{x}{L}\right) \qquad (4)$$

Where,

$$L = \sqrt{D\tau} \qquad (5)$$

$x$ is the distance from the overlap region either on extended ITO or extended silver region and $D$ is the diffusion coefficient. The decay profile of $I_e$ reveals $L_e$ of (135±4) μm and (166±3) μm for pristine and passivated PSCs, respectively (figure **8a**). However, the decay profile of $I_h$ gives $L_h$ of (152±2) μm and (163±2) μm for pristine and passivated PSCs, respectively (figure **8b**). It is observed that $L_e$ for passivated PSC is 1.23 times higher than that of pristine device. A comparison of grain boundary length estimation using ImageJ software suggests that the passivated perovskite films have lesser length of grain boundaries than the pristine MAPI film (figure **3c** & **3d**).

$$\frac{Average\ perimeter\ of\ pristine\ MAPI\ grain}{Average\ perimeter\ of\ passivated\ MAPI\ grain} = \frac{0.49\ \mu m}{1.88\ \mu m} = \frac{1}{3.83}$$

We did not find any linear relation between grain-boundary (GB) length scale *vs* measured charge transport length scales. However, we note that ratio of average transport length scale (i.e., $(L_e + L_h)/2$) and ratio of measured $J_{SC}$ are consistently correlate quantitatively. Hence, we believe that grain-boundary lengths indeed influence the charge-transport properties; however there is no linear relation between GB length scales to $J_{SC}$ and $L$.

$$\frac{\frac{L_e + L_h}{2}\ of\ pristine\ MAPI\ PSC}{\frac{L_e + L_h}{2}\ of\ passivated\ MAPI\ PSC} = \frac{143.5\ \mu m}{164.5\ \mu m} = \frac{1}{1.15}$$

$$\frac{J_{SC}\ of\ pristine\ MAPI\ PSC}{J_{SC}\ of\ passivated\ MAPI\ PSC} = \frac{18.57\ mA/cm2}{21.43\ mA/cm2} = \frac{1}{1.16}$$

However, the value of $L$ for perovskite reported here is higher than the values reported in the literature.[14,35] SPM technique is used to estimate lateral charge transport length scale instead of longitudinal diffusion length. During scanning of the light source in a particular direction (extended ITO region), one charge (hole) is collected by their respective electrode (ITO) and the other charge (electron) have to travel laterally in order to reach the other electrode (Ag). Since, all the holes are collected by the ITO and there is almost negligible recombination for the electron left over the surface. Hence, the separated electrons on the surface of the perovskite film distort the local built-in potential. This will set up a nonlocal electric field in the lateral direction, which allows the flow of separated electrons towards the Ag electrode due to induced potential drop between the locus of illumination and the Ag electrode. In pristine PSC, the ratio of electron to hole $L$ is 0.89. But, in passivated PSC, the ratio is near to unity (1.02). This represents a higher degree of ambipolarity in passivated PSC. The photocurrent scanning measurement is carried out on 4 different devices of both kind of PSCs (pristine and passivated) and it is observed that $L_e < L_h$ in case of pristine perovskite based PSC. This suggest that there is imbalance in the hole and electron transportation in pristine perovskite based PSC due to different mobility of hole and electron.

The lower value of $L_e$ for pristine MAPI based PSC than the passivated one suggests insuffcient 2D lateral transport of electrons due to deeper trap levels.

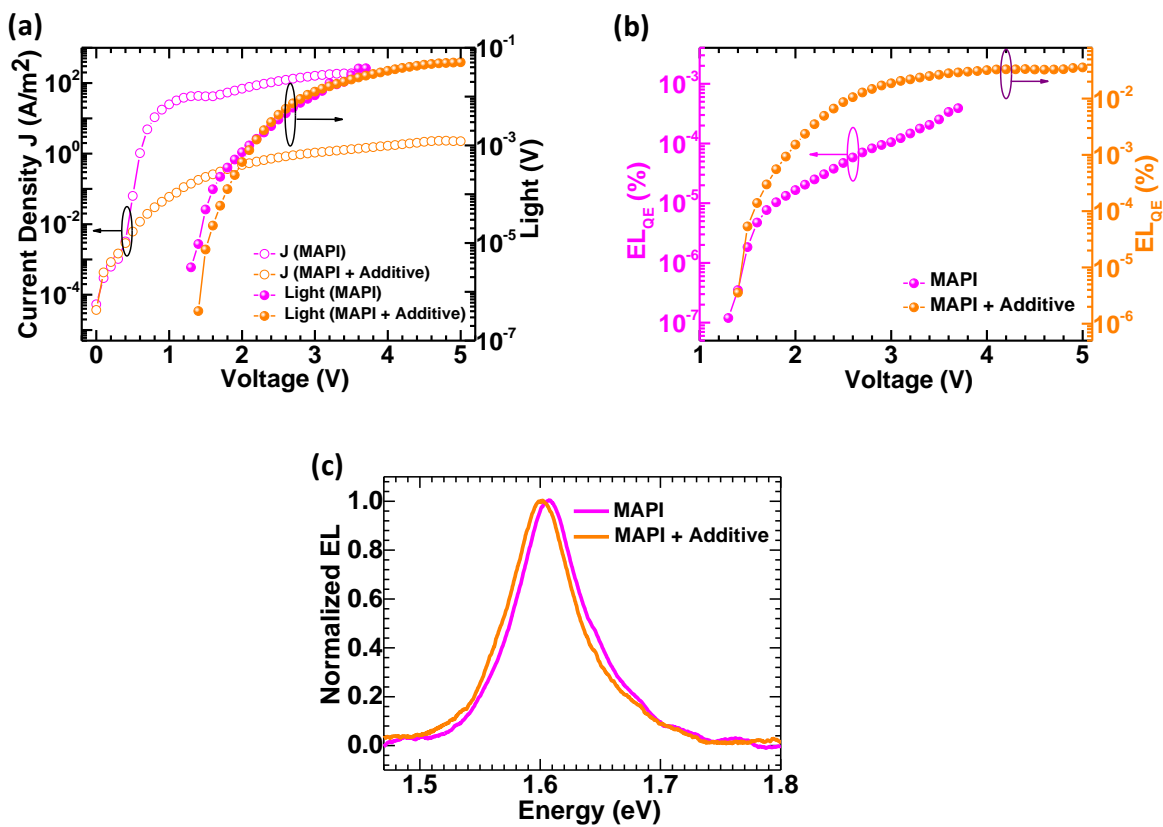

***Figure 9:*** *(a) J-V-L, (b) electroluminescence quantum efficiency ($EL_{QE}$) and (c) electroluminescence (EL) spectra of pristine and passivated MAPI based perovskite solar cells.*

In order to understand the difference in the diffusion length of passivated PSC from pristine one and connect it to defect states, electroluminescence quantum efficiency ($EL_{QE}$) measurement is carried out. Figure **9a** represents the J-V-L characteristics of pristine and passivated MAPI based PSCs. We observe that passivated MAPI devices have lower dark current density than the pristine one. We also notice that the $EL_{QE}$ of passivated devices is even more than two orders of magnitude higher than that of the pristine device (figure **9b**). The $EL_{QE}$ measurement shows the number of photons emitted from the PSC (by applying a forward bias) to the number of charge carriers injected in the device. If non-radiative trap assisted states are available in the device, the injected charge carriers are recombined *via* non-radiative channels and show a bias dependent $EL_{QE}$, which is the case in the pristine device (figure **9b**). However, $EL_{QE}$ of the passivated device starts saturating at nearly 3V. The reduction in trap density of passivated MAPI based PSC is also supported by enhanced lifetime of perturbed charge carriers in transient photovoltage measurement (figure **10**). This indicates that the addition of phosphinic acid into a precursor solution helps in reduction of non-radiative channels and promotes bimolecular recombination.[36]

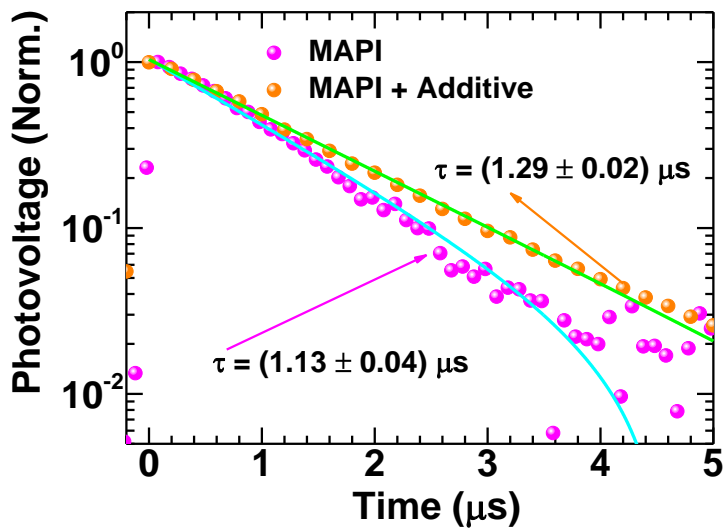

***Figure 10:*** *Transient photovoltage response of pristine (MAPI) and passivated (MAPI+ Additive) based perovskite solar cells (PSCs). Lifetime of perturbed charge carrier is higher for passivated MAPI (1.29 μs) based PSCs than the pristine one (1.13 μs) shows higher number of trap density in the later.*

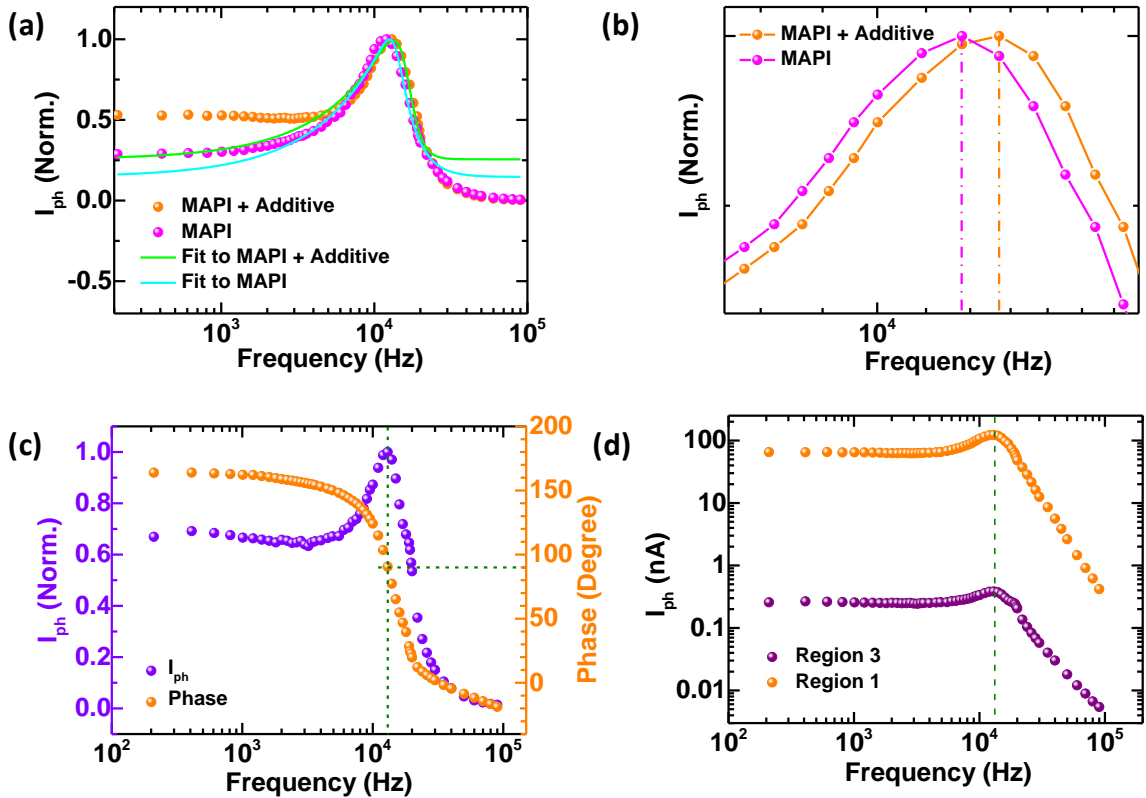

***Figure 11:*** *(a) Frequency dependent photocurrent spectrum from 210 Hz to 100 KHz for pristine and passivated MAPI based PSCs. (b) Magnified view of figure 5a to show distinct $I_{ph}$ maxima for pristine and passivated MAPI based PSCs. Frequency dependent photocurrent (violet solid sphere) and corresponding phase (orange solid sphere) plot of MAPI + additive based perovskite solar cell. Frequency dependent $I_{ph}$ measurement at region 1 (ITO/ PEDOT:PSS/ MAPI/ $PC_{61}BM$/ BCP/ Ag) and region 3 (ITO/ PEDOT:PSS/ MAPI/ $PC_{61}BM$/ BCP).*

In order to understand the relaxation process in the pristine and passivated MAPI based PSCs, we performed frequency dependent photocurrent $I_{ph}$ $(f)$ measurement on both devices from 210 Hz to 100 KHz.[37,38] Figure **11a** represents the typical frequency dependent photocurrent response of both kinds of PSCs. It is observed that for $f < 1$ $KHz$, the $I_{ph}$ is relatively independent of $f$ for both kind of devices and it is marginally changed at low frequency. For $f > 1$ $KHz$, the $I_{ph}$ started increasing and attains a distinct maximum value for pristine and passivated MAPI based PSCs. It is observed that the pristine PSC attains a maximum $I_{ph}$ value at relatively lower frequency than that of passivated PSCs (figure **11b**).

After attaining a maximum $I_{ph}$ value, it started decreasing with a power law having exponent of ~2. The corresponding phase value of $I_{ph}$ is shown in figure **11c** and is in good agreement with $I_{ph}$. We have also measured the $I_{ph}$ (f) outside the active area (region 3, figure **6b**) and observed that $I_{ph}$ (f) maximum is a material property and does not depend upon the spatial position. $I_{ph}$ (f) response at two different locations (region 1 and region 3) is shown in figure **11d**.

Since, the difference in the $I_{ph}$ (f) maximum between the two devices is relatively small; we carried out this experiment on several devices in order to confirm that the two devices have distinct maximum value of $I_{ph}$ (f). Khatavtar et. al. used frequency dependent modulated electroluminescence technique to determine the relaxation time ($\tau_d$) of Si and CIGS based solar cells.[39] Here, the frequency response of $I_{ph}$ is used to determine the $\tau$ of the photo-generated charge carrier in pristine and passivated MAPI based PSCs. The $\tau_d$ basically represents the response time of the system and it is normally presented as Debye relaxation expression, if the relaxation of photo-generated charge carrier can be characterized by a single $\tau_d$.[37,39] However, in our case, we used Havriliak–Negami (H-N) relaxation expression to fit the experimental data due to asymmetry and broadness in frequency dependent $I_{ph}$ response.[40] The H-N relaxation is a modification in the Debye relaxation expression and is given below

$$I_{ph}(f) = I_{ph\infty} + \frac{\Delta I_{ph}}{(1+(if\tau_d)^\alpha)^\beta} \qquad (6)$$

Where, $\tau_d = 1/f$ is the dielectric relaxation time, $I_{ph\infty}$ is the photocurrent at higher frequency, $\Delta I_{ph}$ is the difference between the photocurrent at low and high frequency. The parameters $\alpha$ (>0) and $\beta$ (<1) represents the asymmetry and broadening of the corresponding spectra, respectively. By fitting the experimental data of $I_{ph}$ (f) with equation (3), the calculated value of $\tau$ for pristine and passivated MAPI based PSCs are (72.7 ± 2.5) µs and (61.1 ± 5.5) µs, respectively. We have made a quantitative estimation of the dielectric relaxation time using experimental parameters and compared it with experimentally determined values. In semiconductors, the relaxation lifetime ($\tau_d$) is directly proportional to the resistivity ($\rho$) of the material and can be expressed as

$$\tau_d = \rho \varepsilon_0 \varepsilon_S = \frac{\varepsilon_0 \varepsilon_S}{\sigma} \qquad (7)$$

Where, $\rho$, $\sigma$, and $\varepsilon_S$ are the resistivity, conductivity and permittivity of the sample, respectively. $\varepsilon_0$ is the vacuum permittivity. The conductivity is given by:

$$\sigma = ne\mu \qquad (8) \text{ and}$$

$$\frac{D}{\mu} = \frac{K_B T}{e} \qquad (9)$$

Rearranging equation (5), (7), (8) and (9), the equation (7) becomes

$$\tau_d = \frac{K_B T}{e} \frac{\varepsilon_0 \varepsilon_S}{ne L^2} \tau \qquad (10)$$

This suggests that $\tau_d$ is directly proportional to $\tau$ (carrier lifetime) and inversely proportional to $L^2$ and $n$. The left-hand side (LHS) and right-hand side (RHS) of equation **10** are in good agreement for $L_h$ and slightly off for $L_e$. The disagreement in the LHS and RHS values of equation **10** for electrons can be attributed to the electron defect states and can be understood by significant difference in the transport length scale of pristine and passivated MAPI devices. However, the $L_h$ values of both kinds of devices are almost the same; thus, it shows a good agreement with the $\tau_d$ ratio of the two devices (Table **2**). A quantitative formulation of the $L_e$ versus $\tau_d$ ratio will require advanced semiconductor formulation with more rigorous knowledge about the defect density and energy landscape. In Table **2**, $\tau_d$ is measured through frequency dependent $I_{ph}$ measurement, $\tau$ is measured through TPV, $L$ is measured through SPM, $\varepsilon$ is measured through dielectric measurement and $n$ is extracted by integrating the EQE spectra for the two devices. $\tau_{d1}$ and $\tau_{d2}$ corresponds to dielectric relaxation time constant of pristine and passivated MAPI based perovskite films, respectively.

***Table 2:*** *Correlation between dielectric relaxation time constant ($\tau_d$) and charge carrier transport length scale (L) using Equation 10.*

| MAPI | $\tau_d$ (µs) | $\tau$ (µs) | $L_h$ (µm) | $L_e$ (µm) | $n$ (mA/cm$^2$) | $\varepsilon_s$ | Experimental $\tau_{d1}/\tau_{d2}$ | Estimation using $L_h$ $\tau_{d1}/\tau_{d2}$ | Estimation using $L_e$ $\tau_{d1}/\tau_{d2}$ |
|---|---|---|---|---|---|---|---|---|---|
| **Pristine** | 72.7 | 1.13 | 152 | 135 | 17.82 | 5.5 | 1.19 | 1.17 | 1.51 |
| **Passivated** | 61.1 | 1.29 | 163 | 166 | 20.31 | 5.5 | | | |

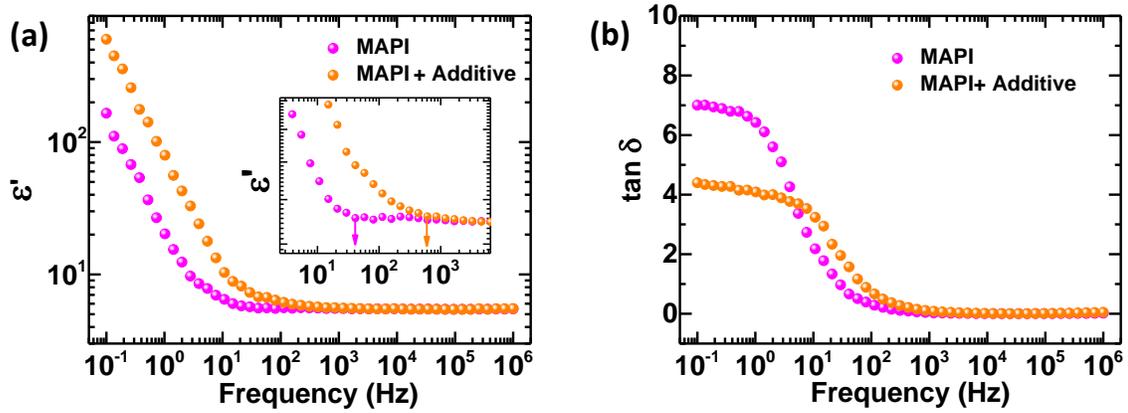

***Figure 12:*** *Frequency dependence of (b) dielectric constant and (c) loss tangent of pristine and passivated MAPI based PSCs.*

In order to validate the frequency dependent $I_{ph}$ measurement, we carried out frequency dependent dielectric measurement for the pristine and passivated MAPI based perovskite films sandwiched between ITO coated glass and Ag electrode.[25,41] Figure **12a** and **12b** represents the dielectric constant ($\varepsilon'$) and loss tangent (*tan δ*) of pristine and passivated MAPI based PSCs at room temperature, respectively. We observed with increase in frequency, the $\varepsilon'$ decreases rapidly in the range of 0.1 Hz to 50 Hz for pristine MAPI based films.[42] Similarly, $\varepsilon'$ decreases up to 600 Hz for passivated MAPI based films and becomes constant after that (figure **12a**). This behavior of MAPI based films refers to typical ferroelectric materials and can be explained by dipoles oscillating in an alternating electric field.[25] When $1/f >> \tau_d$, then the ferroelectric dipoles have enough time to respond the alternating electric field and thus responsible for such a large value of $\varepsilon'$ at lower frequency ($f$ <10 Hz). While further increase in the frequency (10 Hz $< f <$ 1000 Hz), dipoles started lagging behind the alternating field and thus, $\mathcal{E}'$ started approaching zero (relaxation process). However, the strong dependence of $\mathcal{E}'$ in the low frequency range is attributed to the ion migration. Here, the long distance displacement of $\Gamma$ and/or MA$^+$ refers to ion migration. Hence, the polarization due to ion migration exhibit slow response to the alternating field under higher frequency.[25] The dipoles of passivated MAPI based perovskite film follow the alternating field at comparatively higher frequency than the pristine one. The dependency of dipole oscillation over the frequency is in good agreement with the frequency dependent $I_{ph}$ measurement where it has been shown that $\tau_d$ is higher for pristine MAPI based PSCs. This indicates that any perturbed charge carrier in the pristine MAPI based PSC will take more time to come back into its equilibrium position than the passivated one. Along with that, bigger domains of passivated MAPI based thin film (figure **3d**) shows strong polarization in comparison to pristine MAPI with smaller domains (figure **3c**).[26] The tangential loss ($tan\ \delta = \mathcal{E}''/\mathcal{E}'$) for passivated MAPI is lower than that in pristine one (figure **12b**), which is in good agreement with homogeneity in $I_{ph}$ of passivated MAPI based PSCs (figure **4c**).

### 7.4 Conclusion

In summary, this chapter presents a comparative study of $L$ and $\tau_d$ between pristine and passivated MAPI based PSCs. We found that the passivated device has a longer carrier lifetime, higher degree of ambipolarity, lower relaxation time and lower tangential losses, which improves the charge transport. SPM and frequency dependent $I_{ph}$ measurements are the useful techniques which help to understand the quality of the perovskite film, relaxation process and estimate the charge length scale in perovskite solar cells.

### 7.5    Post Script

In recent years, Pb based perovskites are emerging rapidly in the field of photovoltaic's due to its unique opto-electronic properties. But the major challenge in the Pb based perovskite photovoltaics is to produce eco-friendly devices. The alternatives of Pb are addressed in the community to overcome the toxicity of the perovskite materials and tin (Sn) emerges as one of the best alternative. However, oxidation, efficiency and stability are the main concern with Sn. It is advisable to transfer the engineering and photo-physical knowledge of Pb based perovskite to Sn based perovskite to make it efficient and stable. In chapter 8, we compare the rate law for charge carrier decay dynamics of pure Pb, pure Sn and Pb-Sn mixed based perovskite solar cells. In addition, we tried to make an efficient and stable Sn based PSC by addition of small amount of Cs.

# Chapter 8

# Charge Carrier Recombination Dynamics in Pb *vs* Sn based Halide Perovskite Semiconductors

# CHAPTER 8

# Charge Carrier Recombination Dynamics in Pb *vs* Sn based Halide Perovskite Semiconductors

**Abstract:** In commercialization of perovskite solar cells (PSCs), the toxicity of lead (Pb) is a big challenge. Tin (Sn) based PSCs emerges as a promising candidate, however, complete replacement of Pb by Sn results into instability in the devices. Hence, a fractional substitution of Sn into Pb based PSCs gains much attention now a days due to application in all perovskite tandem solar cells as a top cell on the transparent substrate. However, charge carrier recombination dynamics in Sn based PSCs is still a debate in the community. Herein, we fabricated conventional *p-i-n* configuration based $FA_{0.95}Cs_{0.05}PbI_3$ (pure Pb), $MA_{0.20}FA_{0.75}Cs_{0.05}SnI_3$ (pure Sn) and $(MAPbI_3)_{0.4}$ $(FASnI_3)_{0.6}$ (Pb–Sn mixed) based PSCs and compare the charge carrier recombination dynamics of all the three PSCs through transient photovoltage (TPV) measurement. TPV is used to establish the relation between charge carrier density ($n$) and recombination rate constant ($k$) under different DC background intensity for all the three devices. This study establish the rate law of charge carrier decay in all the three devices and also reveals the nonlinear dependence of $k$ on $n$, which will be correlated to slow relaxation lifetime of the charge carrier and recombination through the defect states in the perovskite thin films.

## 8.1 Introduction

The power conversion efficiency (PCE) of lead (Pb) based perovskite solar cells (PSCs) approaching the PCE of Si, Cu(In,Ga), CdTe based convention solar cells which makes it to stand out in photovoltaic (PV) community.[1,2] However, the Pb based PSCs are currently suffering from two major issues: toxicity and stability.[3,4] Although, there are several efforts in the community to overcome the issue of stability for Pb based PSCs via encapsulation, improved device engineering and use of two dimensional (2D) perovskite.[5,6,7] However, efforts to reduce or completely remove the toxicity from PSCs without compromising on the PCE are in progress.[8] There are some less toxic options for Pb, which can make perovskite structures, are double perovskite, tin (Sn), Sn-Ga based halides, Bi-Sb based halide perovskite etc.[9,8] Among the all above mentioned low toxic candidates, Sn based halide perovskite have earned the most attention due to their interesting optical and electronic properties which is in good agreement with Pb based halide perovskite.[10] The band gap of Sn based perovskites are even narrower than that of Pb based perovskites, which are more useful in energy harvesting from the sun.[11,12] However, Sn is suffering from oxidation issue under ambient air condition; it degrades to $Sn^{+4}$ from $Sn^{+2}$ which ultimately forming $SnO_2$. Sn based perovskites have low formation energy which leads to create Sn vacancies. The electronic properties of Sn based perovskites are heavily dependent on its exposure to the atmosphere, hence the mismatch in the energy level of hole and electron transporting layers results into poor charge extraction in these devices.[10]

Hao et al. reported 6% PCE of the first Pb free or Sn based perovskite solar cell in 2014.[13] After that there are so many reports which focus on enhancing the PCE and stability of Sn based PSCs. Ethylenediammonium (*en*) cation is used in Sn based PSC to enhance the PCE and stability by forming a new kind of 3D hollow perovskite.[14,15] Other cations such as butylammonium (BA) and phenylethylammonium (PEA) are used in 3D Sn based perovskite to reduce the dimensionality in order to fabricate more stable PSCs.[16] The highest PCE for Sn based PSCs has been reported to achieve 9.6% by Jokar *et al.* in which 20% guanidinium (GA) cation is used along with 1% *en* cation in Sn based perovskite.[17] However, after so many efforts, achieving the stability and high performance simultaneously for Sn based PSCs is still a big challenge for the community. Hence, Pb-Sn mixed based perovskite open a new door to reduce the toxicity of Pb based PSCs and enhance the stability of Sn based PSCs. Recently; Prasanna *et al.* demonstrate a new design for Pb-Sn mixed based PSC by improving the morphology of the perovskite film and capping with sputtered indium zinc oxide (IZO) electrode to achieve high stability.[18] Lin et al. reported a novel method of introducing metallic Sn into the perovskite precursor solution to reduce the Sn vacancy in Pb-Sn mixed based PSC.[19] Photo-physics of lead based perovskites are well understood in last few years. There are also reports on the optical properties of Sn based perovskite which reveals some interesting phenomenon about the material such as red shift in photoluminescence (PL) with time, stoke shift in absorption *vs* PL peak position etc.[20] Handa et al. presents a quantitative comparison between the charge carrier life of the perovskite thin films and solar cells for Sn and Pb based perovskites. The $V_{OC}$ losses in the Sn based solar cells are quite higher than that in Pb-Sn mixed and Pb based PSCs due to non-radiative recombination through Sn vacancy.[21] Hence, the overall charge transfer rate in the device depends upon both non-radiative recombination and bimolecular recombination. Thus, it is important to study these recombination mechanisms and understand how the recombination dynamics changes with material.

In this study, we present a comparative study on recombination dynamics and rate law of charge carrier decay in $FA_{0.95}Cs_{0.05}PbI_3$, $MA_{0.20}FA_{0.75}Cs_{0.05}SnI_3$ and $(MAPbI_3)_{0.4}$ $(FASnI_3)_{0.6}$ based PSCs. we correlate transient photovoltage (TPV) and photocurrent (TPC) measurements to understand the rate law of charge carrier decay in all the three PSCs. This study shows a non-linear dependence of bimolecular rate constant ($k$) on charge carrier density ($n$) and the non-linearity is higher in case of pure Sn based PSC. Thus, this study provides an insight into the device physics of the $FA_{0.95}Cs_{0.05}PbI_3$, $MA_{0.20}FA_{0.75}Cs_{0.05}SnI_3$ and $(MAPbI_3)_{0.4}$ $(FASnI_3)_{0.6}$ based PSCs in terms of defect states and bimolecular recombination rate.

## 8.2 Experimental section:

***Materials:*** Lead iodide ($PbI_2$), tin iodide ($SnI_2$), tin fluoride ($SnF_2$), cesium iodide (CsI), lead thiocynate ($Pb(SCN)_2$) and bathocuproine (BCP) were purchased from Sigma Aldrich. Methyl ammonium iodide (MAI) and formamidinium iodide (FAI) were purchased from Greatcell Solar. $PC_{61}BM$ and PEDOT:PSS were purchased from Solenne BV and Clevios, respectively. All the materials are used as received.

*Solution preparation:*

*$MASnI_3$ solution:* 159 mg of MAI, 372 mg $SnI_2$ and 16 mg $SnF_2$ were dissolved in 800 µL dimethylformamide (DMF) and 200 µL dimethylsulfoxide (DMSO). The solution was stirred overnight at room temperature.

*$FASnI_3$ solution:* 172 mg FAI, 372 mg $SnI_2$ and 16 mg $SnF_2$ were dissolved in 800 µL DMF and 200 µL DMSO. The solution was stirred overnight at room temperature.

*$MAPbI_3$ solution:* 461 mg $PbI_2$, 159 mg MAI and 11.3 mg $Pb(SCN)_2$ were dissolved in 630 µL DMF and 70 µL DMSO. The solution was stirred overnight at room temperature.

CsI solution: 1M CsI was dissolved in DMSO and anneal at $150^0$C for 30 min.

*$MA_{0.20}FA_{0.75}Cs_{0.05}SnI_3$ {pure Sn} solution:* $MASnI_3$, $FASnI_3$ and CsI solutions were mixed in volumetric ratio of 20%, 75% and 5%, respectively and kept on stirring at room temperature for 4 hours.

*$(MAPbI_3)_{0.4}$ $(FASnI_3)_{0.6}$ {Pb-Sn mixed} solution:* $MAPbI_3$ and $FASnI_3$ solutions were mixed in volumetric ration of 40% and 60%, respectively and kept on stirring for 4 hours.

*$FA_{0.95}Cs_{0.05}PbI_3$ {pure Pb} solution:* 240.8 mg FAI, 646.8 mg PbI2 and 18.2 mg CsI were dissolved in 800 µL DMF and 200 µL DMSO. The solution was kept on stirring at $60^0$c for 4 hours.

*Electron transporting layers:* 15 mg of $PC_{61}BM$ is dissolved in 1 ml of 1,2,dichlorobenzene and kept for overnight stirring at room temperature. 1mg/ml of BCP in isopropanol (IPA) solution was kept under stirring for 15 min in the glove box at $70^o$C before use.

**Device fabrication and characterization:** *p-i-n* configuration based perovskite solar cells were fabricated on indium tin oxide (ITO) coated glass substrates. Substrates were cleaned sequentially with soap solution, deionized water, acetone and isopropanol. After oxygen plasma treatment for 10min, PEDOT:PSS was spincoated at 5000rpm for 30 sec on the ITO coated substrates and annealed at 150 $^0$C in $N_2$ atmosphere for 30min. Spincoating of perovskite films and electron transporting layers (ETL) were performed inside the $N_2$ filled glove box with the maintained $H_2$ and $O_2$ levels. PEDOT:PSS coated ITO substrates were transferred into the glove box for perovskite spincoating. Pure Sn and Pb-Sn mixed perovskite precursor solution were spincoated on the prepared PEDOT:PSS/ITO substrate at 4000 rpm for 60 sec and 100 µL of chlorobenzene was dropped at the end of last 10 seconds. The prepared perovskite films annealed at 100$^o$C on hotplate for 7 min. Pure Pb based perovskite precursor solution was spincoated at 1000 rpm for 15 sec and 5000 rpm for 45 sec and 100 µL of chlorobenzene was dropped at the end of last 10 seconds of step 2. The final perovskite film was annealed at 100$^0$C for 30 min. The $PC_{61}BM$ solution was spincoated at 1000 rpm for 60 sec and workbench dried for 5 min. After bench drying of the $PC_{61}BM$/perovskite/PEDOT:PSS/ITO substrate, BCP was spincoated at 5000 rpm for 20 sec. The prepared films were transfers into the thermal evaporation chamber to deposited 100 nm of Ag. Shadow mask was used to decide the active area of the cell to be 4.5mm$^2$.

Photocurrent density-voltage (J-V) measurements were carried out using a Keithley 4200 semiconductor characterization system and LED solar simulator (ORIEL LSH-7320 ABA) after calibrating through a reference solar cell provided from ABET. All the J-V measurements were performed using scan speed of 40 mV/s. Quantum efficiency measurements were carried out to measure the photo-response as a function of wavelength using Bentham quantum efficiency system (Bentham/PVE300).

The XRD measurements were carried out in Rigaku smart lab diffractometer with Cu $K_\alpha$ radiation (λ=1.54Å). 2θ scan has been carried out from 10$^o$-30$^o$ with step size of 0.001$^o$. Structural and morphological analysis was done using a Rigaku lab spectrometer X-ray diffraction machine using Cu Kα = 1.54 Å and FESEM, respectively. Optical studies were carried out using a spectrometer (PerkinElmer LAMBDA 950) and photoluminescence (PL) spectrometer (Horiba FluoroMax 4). Steady state PL measurement was done on thin-film in vacuum at a pressure of 10$^{-5}$ mbar in a custom made chamber. Excitation energy and wavelength were 70 nJ and 355 nm, respectively. Electronic states are derived from ultra-violet photoelectron spectroscopy (UPS, KRATOS) with He (I) source (21.22 eV). Chamber pressure was below 5.0E-9 Torr for XPS and UPS measurement.

## 8.3 Results

### 8.3.1 Morphological, structural and optical studies

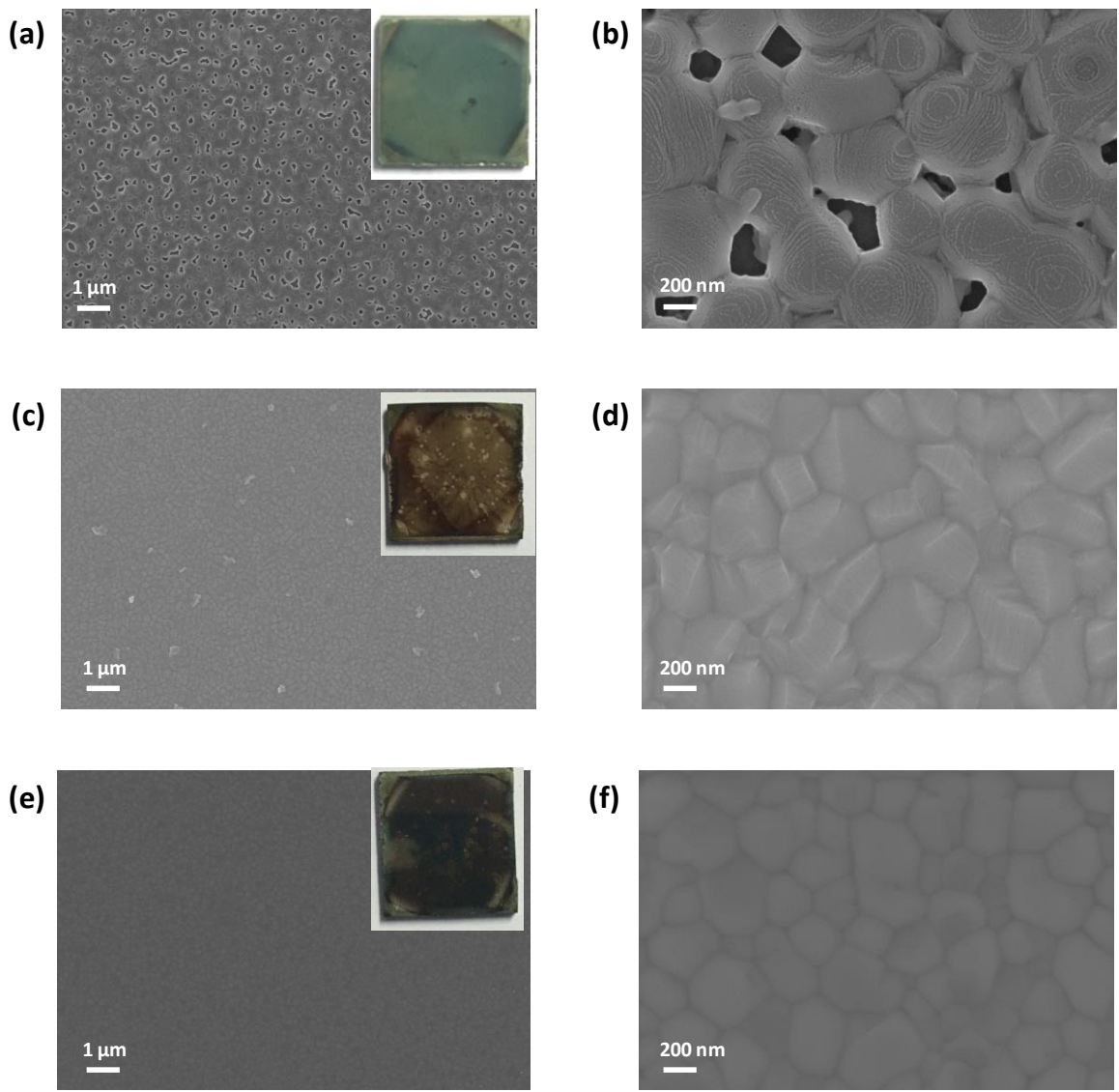

**Figure 1:** *Field emission scanning electron microscopy (FESEM) images of (a) & (b) MASnI$_3$, (c) & (d) FASnI$_3$, and (e) & (f) MA$_{0.25}$FA$_{0.75}$SnI$_3$. Scale bar is 1 µm for (a), (c) & (e) and 200 nm for (b), (d) & (f). Inset of figure (a), (c) & (e) represents the pictures of MASnI$_3$, FASnI$_3$, and MA$_{0.25}$FA$_{0.75}$SnI$_3$ based perovskite films stored in dark for one day under humidity of (40 ± 5) %.*

FASnI$_3$ suffers from poor stability (inset of figure **1a**) but film morphology is quite compact as shown in the top-view microstructure of FASnI$_3$ (figure **1a** and **1b** on a scale bar of 1µm and 200 nm). However, MASnI$_3$ suffers from poor morphology but relatively better moisture stability (inset of figure **1c**) and is shown in figure **1c** and **1d**. Inset of figure **1a** and **1c** represents the pictures of FASnI$_3$ and MASnI$_3$ based perovskite film coated over glass after one day, respectively. Both the films were stored in dark for one day under humidity ~ (40 ± 5) %.We observed that moisture stability of MASnI$_3$ is better than the FASnI$_3$. However, the surface coverage of the FASnI$_3$ is better than the MASnI$_3$. Hence, we fabricate a new composition of Sn based perovskite film with mixture of 1M MASnI$_3$ and 1M FASnI$_3$ based perovskite precursor in the volumetric ratio of 1:3, respectively. MA-FA mixed double cation based Sn halide perovskite (MA$_{0.25}$FA$_{0.75}$SnI$_3$) shows improved morphology than MASnI$_3$ (figure **1e** and **1f**) and enhanced moisture stability than FASnI$_3$ (inset of figure **1e**). It is well known in the perovskite community that having a little amount of Cs in the perovskite precursor solution can enhance the thermal stability of perovskite solar cells.[22] Thus we have added 5 volume % of CsI (from a stock solution of 1M CsI in dimethylsulfoxide (DMSO) solution) in MA-FA double cation based halide precursor solution with the final volumetric concentration of MA$_{0.20}$FA$_{0.75}$Cs$_{0.05}$SnI$_3$ and we call it pure Sn based PSC from here. The top view microstructure of pure Sn based perovskite thin film is shown in figure **2a** and **2b** on a scale of 1µm and 200 nm, respectively. The domain size of pure Sn based perovskite thin film is ~600 nm, which is bigger than that of MA$_{0.25}$FA$_{0.75}$SnI$_3$ based perovskite film. However, increasing the volumetric concentration of CsI from 5% to 10% results into even more bigger domains but the morphology gets poorer in this case and it is shown in figure **2c** and **2d**. Thus, we choose MA$_{0.20}$FA$_{0.75}$Cs$_{0.05}$SnI$_3$ precursor solution to fabricate pure Sn based perovskite thin film. It is worth to notice that pure Sn (MA$_{0.20}$FA$_{0.75}$Cs$_{0.05}$SnI$_3$) based perovskite film (inset of figure 2a) is more stable than MASnI$_3$, FASnI$_3$ and MA$_{0.25}$FA$_{0.75}$SnI$_3$ films. We used a recipe from Zhao et al. to fabricate (MAPbI$_3$)$_{0.4}$ (FASnI$_3$)$_{0.6}$ based perovskite films and name it Pb-Sn mixed perovskite from now onwards.[23] The microstructures of Pb-Sn mixed perovskite films are shown in figure **2e** and **2f**.

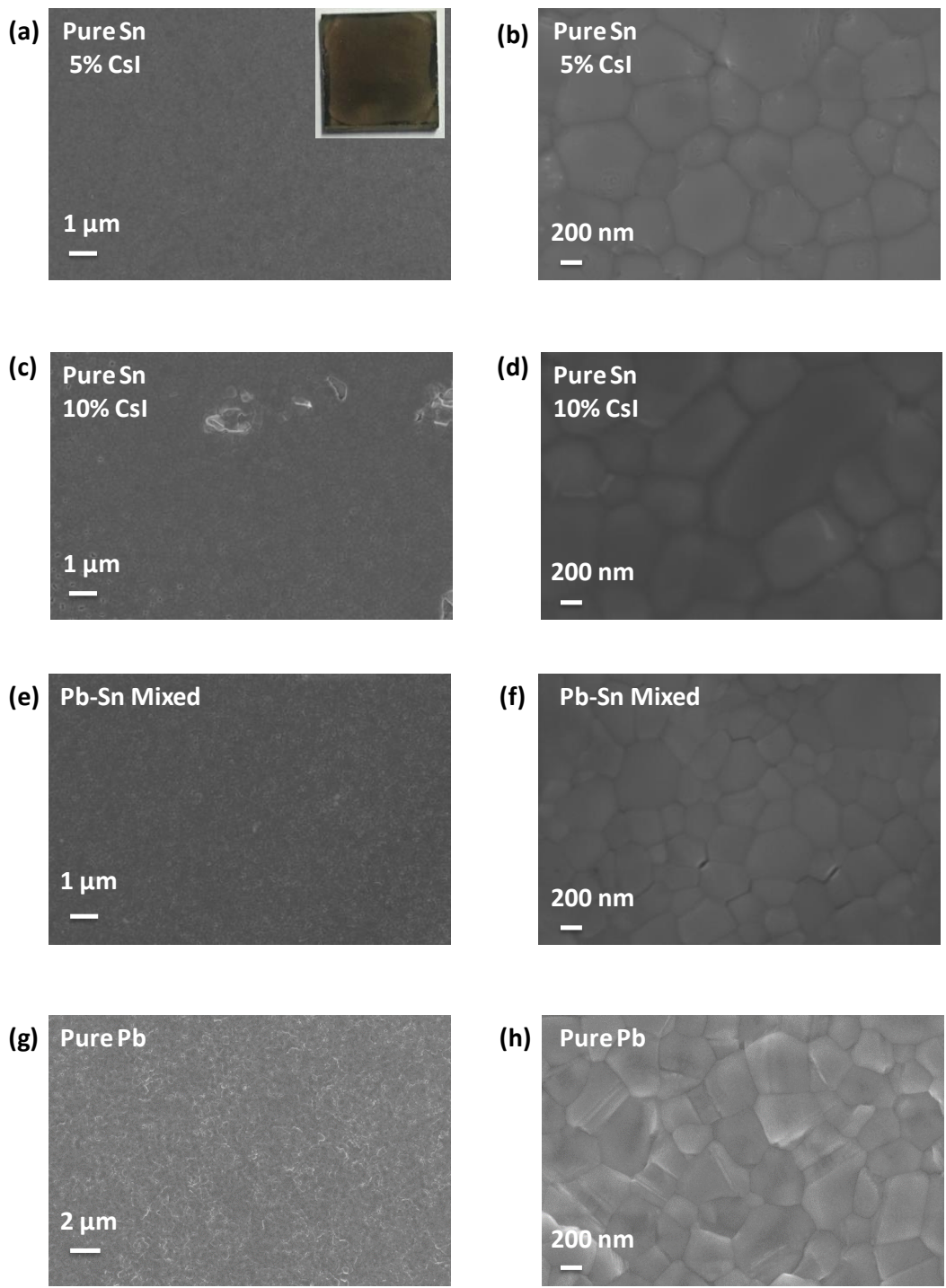

**Figure 2:** *FESEM images of (a) & (b) MA$_{0.20}$FA$_{0.75}$Cs$_{0.05}$SnI$_3$ {pure Sn with CsI 5%}, (c) & (d) MA$_{0.20}$FA$_{0.75}$Cs$_{0.05}$SnI$_3$ {pure Sn with CsI 10%}, e) & (f) (MAPbI$_3$)$_{0.4}$ (FASnI$_3$)$_{0.6}$ {Pb-Sn mixed} and (g) & (h) FA$_{0.95}$Cs$_{0.05}$PbI$_3$ {pure Pb}.  Scale bar is 2 µm for (a), (c), (e) & (g) and 200 nm for (b), (d), (f) & (h).*

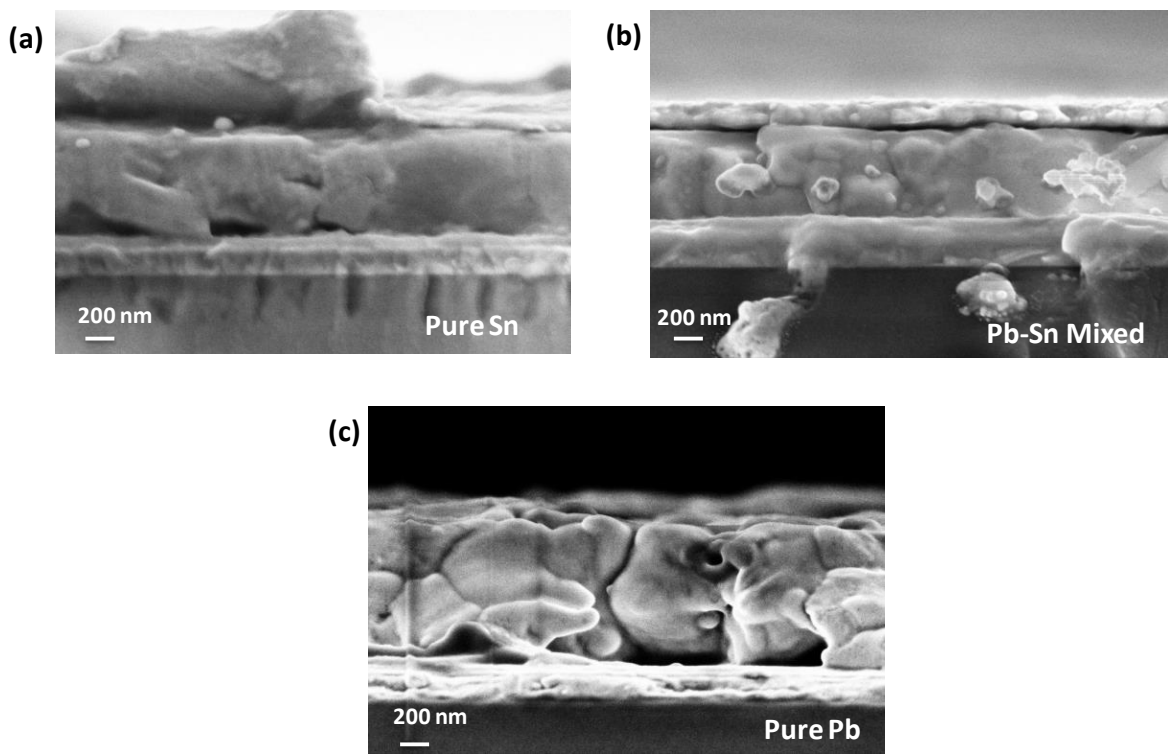

***Figure 3:*** *FESEM Cross-sectional view of $MA_{0.20}FA_{0.75}Cs_{0.05}SnI_3$ {pure Sn}, $(MAPbI_3)_{0.4}$ $(FASnI_3)_{0.6}$ {Pb-Sn mixed} and $FA_{0.95}Cs_{0.05}PbI_3$ {pure Pb} based perovskite films.*

The reason behind using this composition is its low band gap of ~1.25 eV and is well established in all perovskite tandem solar cells.[23] However, the tuning of band gap with different volumetric ratio of $MAPbI_3$ and $FASnI_3$ precursor solution mixture is still a topic of debate in the community. Considering the band gap of $MAPbI_3$ (~1.6 eV) and $FASnI_3$ (~1.36 eV), it is expected that band gap of the mixture of these two perovskite should lie in between 1.36 eV and 1.6 eV. However, for a particular composition, the resultant band gap is smaller than that of the two end perovskite material due to two competing mechanism: tilting of metal halide octahedral and lattice contraction.[24] We use $FA_{0.95}Cs_{0.05}PbI_3$ as a pure Pb based perovskite film due to its higher moisture stability than the conventional Pb based perovskite, $MAPbI_3$.[25] Figure **2g** and **2h** represents the microstructure of pure Pb based perovskite thin film and is compact with domain size of ~500 nm. Figure **3** represents the cross-section field emission scanning microscopy (FESEM) images of pure Sn, pure Pb and Pb-Sn mixed perovskite thin films and corresponding thicknesses are ~520 nm, ~760 nm and ~490 nm, respectively. The thickness of the perovskite layers are optimized in order to get high performance PSCs.

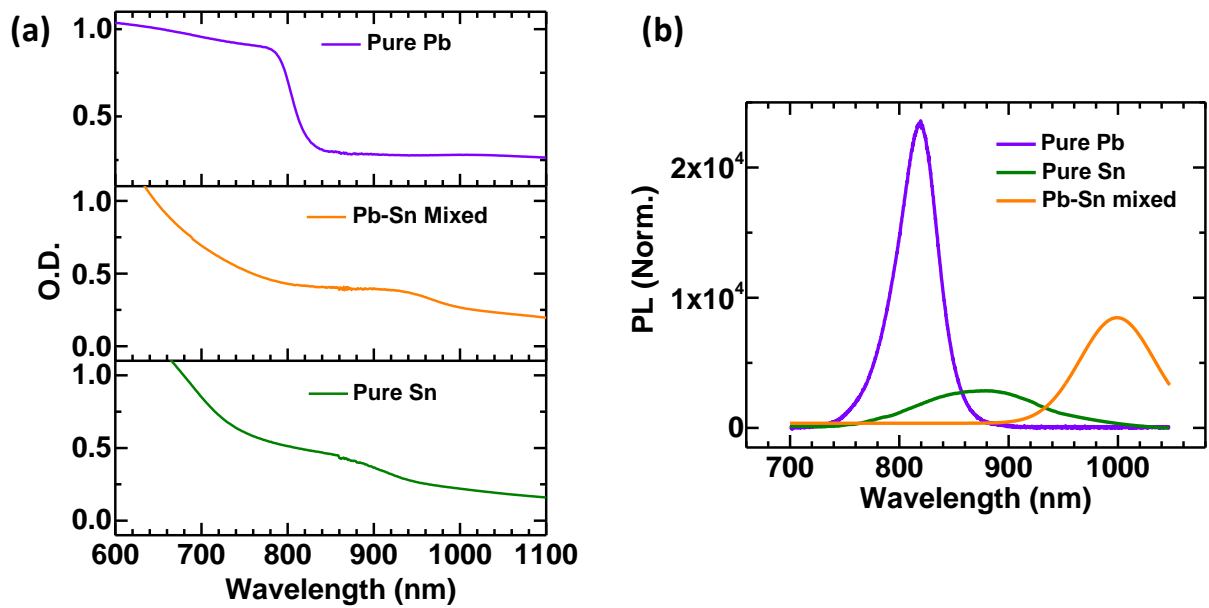

**Figure 4:** *(a) Absorption and (b) relative PL spectra of $MA_{0.20}FA_{0.75}Cs_{0.05}SnI_3$ {pure Sn}, $(MAPbI_3)_{0.4}$ $(FASnI_3)_{0.6}$ {Pb-Sn mixed} and $FA_{0.95}Cs_{0.05}PbI_3$ {pure Pb} based perovskite films. Without Cs based film shows double peak emission (oxidation of Sn from $Sn^{+2}$ to $Sn^{+4}$).*

The optical densities (OD) of all the three perovskite thin films are shown in figure **4a**. The slope of OD at the absorption edge is lowest for pure Sn and highest for pure Pb based perovskite film and Pb-Sn mixed film have intermediate slope. The inverse of the slope determines the Urbach energy ($E_u$) which is a measure of disorder (chapter **2**). Eu for pure Pb, pure Sn and Pb-Sn mixed based perovskite films are 43 meV, 176 meV and 108 meV, respectively. The higher value of $E_u$ represents higher number of defect states in pure Sn based perovskite film. [23] Figure **4b** represents the photoluminescence (PL) spectra of direct band gap pure Pb (PL$_{peak}$ emission ~ 819 nm), pure Sn (PL$_{peak}$ emission ~875 nm) and Pb-Sn mixed (PL$_{peak}$ emission ~1000 nm) based perovskite thin films. All the three films were encapsulated with 10 mg/mL Poly(methyl methacrylate) (PMMA) in chlorobenzene. PL emission spectra of all the three films were taken under same condition mentioned in experimental section. The lower PL intensity of pure Sn based perovskite films suggests higher defect states in the film even after using SnF₂ and Cs. It is well known that SnF₂ reduces the oxidation of Sn but it cannot completely stop the oxidation process. However, we do not observe any oxidation (double emission peak) in the Cs added pure Sn based perovskite film.

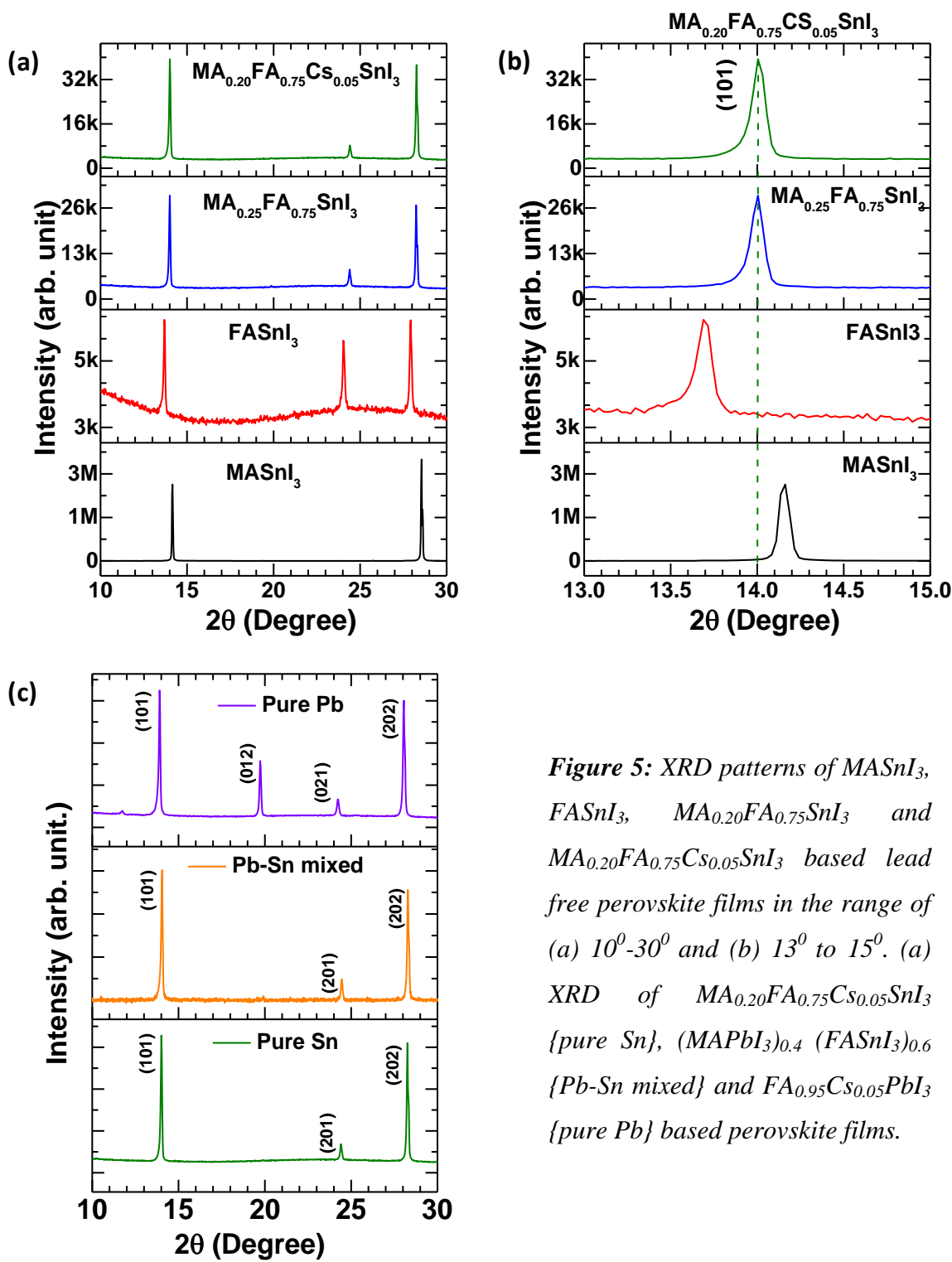

**Figure 5:** *XRD patterns of MASnI₃, FASnI₃, MA₀.₂₀FA₀.₇₅SnI₃ and MA₀.₂₀FA₀.₇₅Cs₀.₀₅SnI₃ based lead free perovskite films in the range of (a) 10⁰-30⁰ and (b) 13⁰ to 15⁰. (a) XRD of MA₀.₂₀FA₀.₇₅Cs₀.₀₅SnI₃ {pure Sn}, (MAPbI₃)₀.₄ (FASnI₃)₀.₆ {Pb-Sn mixed} and FA₀.₉₅Cs₀.₀₅PbI₃ {pure Pb} based perovskite films.*

To investigate the effect of small amount of Cs on the crystal structure of perovskite, we performed x-ray diffraction (XRD) on the Sn based perovskite films (figure **5a**). The MASnI$_3$ based perovskite crystallized in pseudo-cubic, tetragonal *P4mm* space group at room temperature. The FASnI$_3$ based perovskite belongs to pseudo-cubic, orthorhombic structure in *Amm2* space group.[26] As expected, FA being a bigger cation, the XRD (101) peak position for FASnI$_3$ (13.69$^0$) is at lower angle than that of MASnI$_3$ (14.16$^0$) and is shown in figure **5b**.[27] The first order XRD peak position for MA$_{0.25}$FA$_{0.75}$SnI$_3$ is at 14.00$^0$ which lies in between (101) peak of MASnI$_3$ and FASnI$_3$ and crystallized into a pseudo-cubic structure.[26] A small amount of Cs in MA$_{0.25}$FA$_{0.75}$SnI$_3$ does not alter the crystal structure of the final perovskite film and it is validated by no significant shift in the first order XRD peak position (figure **5b**). Figure **5c** represents the XRD pattern of pure Pb (tetragonal), pure Sn (pseudo-cubic) and Pb-Sn mixed (pseudo-cubic) based perovskite thin films and the XRD peak positions are matches well with the literature validating the formation of perovskite structure.[26,28]

### 8.3.2 Photovoltaic performance

A conventional planar heterojunction (*p-i-n*) structure of indium tin oxide (ITO) / poly(3,4-ethylene-dioxythiophene):polystyrenesulfonate (PEDOT:PSS) / perovskite / phenyl-C61-butyric acid methyl ester (PC$_{61}$BM) / Bathocuproine (BCP) / silver (Ag) was fabricated. Figure **6a** represents the device configuration of planar perovskite solar cells and energy level diagram of each layer. Energy level values are determined by UPS. Figure **6b** and **6c** shows the dark and illuminated under 1 Sun condition (100 mW/cm$^2$) current density vs. voltage (J-V) characteristics of pure Pb, pure Sn and Pb-Sn mixed based PSCs, respectively. Table **1** listed the photovoltaic parameters for the three champion PSCs with average PCE over 16 devices. The pure Pb based device shows higher PCE of 17.25 % with open circuit voltage (V$_{oc}$) of 0.93 V, a short-circuit current density (J$_{sc}$) of 24.74 mA/cm$^2$ and fill factor (FF) of 75 %. However, due to lower band gap of Pb-Sn mixed perovskite, enhancement in J$_{SC}$ (29.52 mA/cm$^2$) is observed and V$_{OC}$ decreases to 0.72 V. In case of pure Sn based PSC, V$_{OC}$ losses becomes more prominent although having higher band gap than mixed Pb-Sn perovskite, which is in agreement with relatively lower PL intensity of pure Sn based perovskite film (figure **4b**). Pure Pb have lower leakage current $J_0$ ($8.83 \times 10^{-9}$ mA/cm$^2$) than the Pb-Sn Mixed ($1.22 \times 10^{-5}$ mA/cm$^2$) and pure Sn ($3.24 \times 10^{-3}$ mA/cm$^2$), which represents high quality of pure Pb based perovskite film. Figure **6d** represents the external quantum efficiency (EQE) of the pure Pb, pure Sn and Pb-Sn mixed based PSCs which validates the $J_{SC}$ values and are in good agreement with their absorption edge.

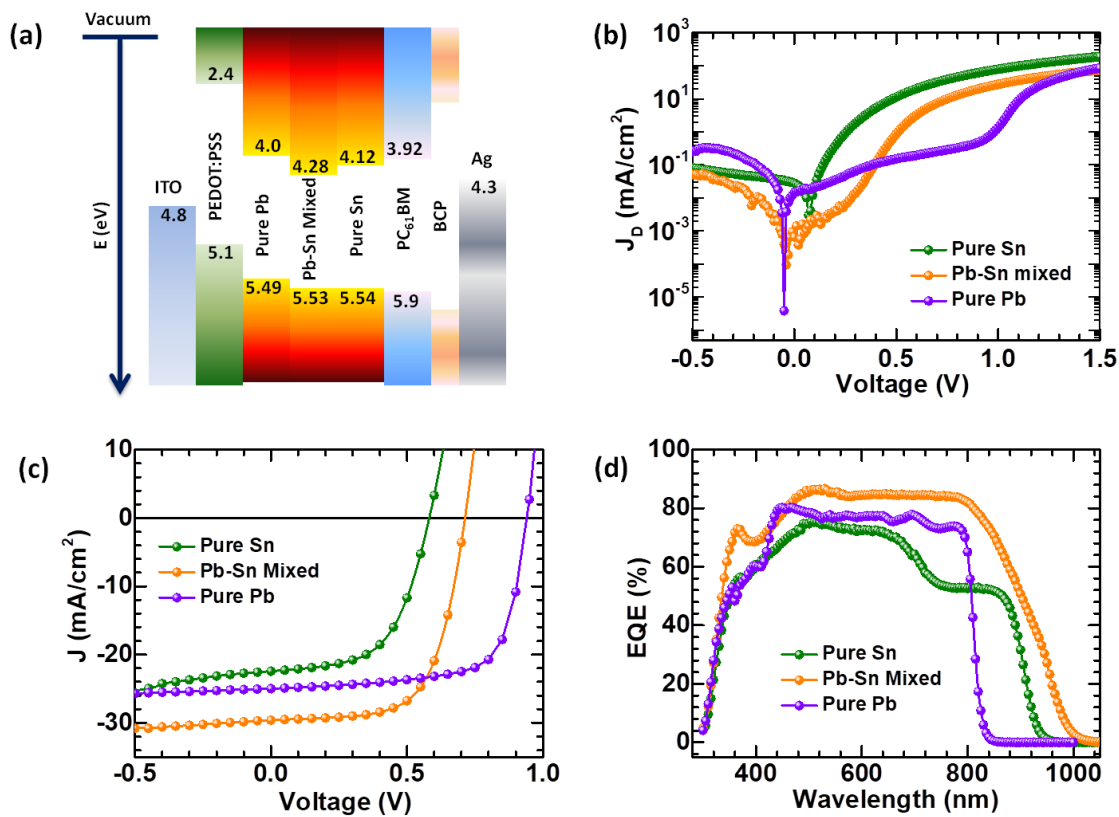

**Figure 6:** *(a) p-i-n device configuration and energy level diagram of each layer used in the perovskite solar cells. Valance band maxima and conduction band minima of perovskite films are extracted from UPS measurement. (b) Dark & (c) illuminated J-V and corresponding (d) EQE of $MA_{0.20}FA_{0.75}Cs_{0.05}SnI_3$ {pure Sn}, $(MAPbI_3)_{0.4}$ $(FASnI_3)_{0.6}$ {Pb-Sn mixed} and $FA_{0.95}Cs_{0.05}PbI_3$ {pure Pb} based perovskite solar cells.*

The $J_{SC}$ value calculated from EQE for pure Pb, Pb-Sn mixed and pure Sn are 22.79 mA/cm$^2$, 28.12 mA/cm$^2$ and 21.56 mA/cm$^2$, respectively. The $J_{SC}$ value calculated from the EQE measurement is within 10% of error bar than it is calculated from illuminated J-V curve due to small active area (4.5 mm$^2$) of the devices.[7] It can be clearly seen in figure **6d** that pure Pb based PSC have a sharp fall in EQE (lower urbach energy) after 780 nm which reflects high quality of perovskite film. However, EQE of Pb-Sn mixed based PSCs also starts falling down from 780 nm with a higher slope value and shows a small hump near the wavelength 930 nm. Similarly, the collection efficiency of pure Sn based PSC start decreasing in between 660 nm and 860 nm. This represents relatively poor charge extraction in Pb-Sn mixed and pure Sn based PSCs than the pure Pb based devices. The intensity dependent photovoltaic parameters such as V$_{OC}$, J$_{SC}$ and FF are shown in figure **7**. The ideality factor $\eta$ is around 2 for all the three devices which suggest that recombination is dominated by Shockley Read Hall (SRH). However, it is slightly higher than 2 for pure Sn based PSC, which is in good agreement with optical studies (figure **5**). J$_{SC}$ is linearly proportional to the light intensity for all the three devices. In pure Sn and Pb-Sn based PSCs, the FF is constant in between 100 mW/cm$^2$ and 50 mW/cm$^2$ light intensity. However, FF start decreasing at lower intensity and the decrement is more prominent in case of pure Sn than Pb-Sn mixed based PSC. In case of pure Pb based PSC, the FF is almost constant and decreases slightly at lower intensity. This reflects the high quality of pure Pb based perovskite film with lower number of defect states in the bulk. In general, *p-i-n* configuration based devices shows less hysteresis than *n-i-p* based PSCs. The illuminated J-V characteristics under forward and reverse scan for all the three devices are shown in figure **8**. It is observed that the hysteresis is not much significant in any of the three cases but relatively higher for Sn based devices. The hysteresis present in the case of Sn based PSC is attributed to the defect states in the perovskite bulk.

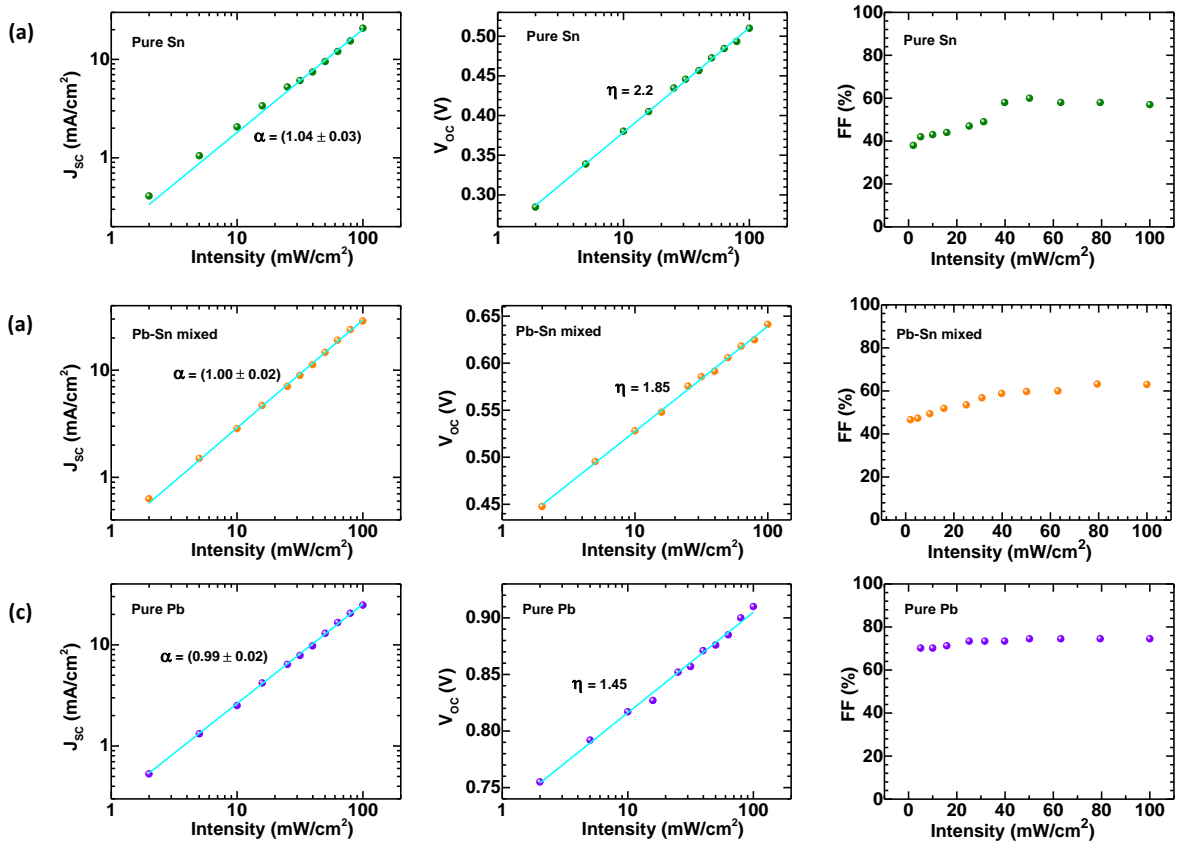

**Figure 7:** *Intensity dependent photovoltaic parameters $V_{OC}$, $J_{SC}$ and FF for (a) pure Sn, (b) Pb-Sn mixed and (c) pure Pb based PSCs.*

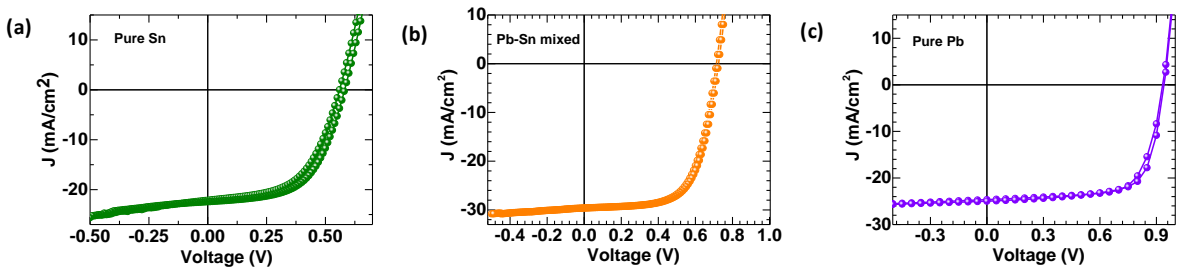

**Figure 8:** *Hysteresis in $MA_{0.20}FA_{0.75}Cs_{0.05}SnI_3$ {pure Sn}, $(MAPbI_3)_{0.4}$ $(FASnI_3)_{0.6}$ {Pb-Sn mixed} and $FA_{0.95}Cs_{0.05}PbI_3$ {pure Pb} based perovskite solar cells.*

**Table 1:** *Photovoltaic parameters of the best performing $MA_{0.20}FA_{0.75}Cs_{0.05}SnI_3$ {pure Sn}, $(MAPbI_3)_{0.4}(FASnI_3)_{0.6}$ {Pb-Sn mixed} and $FA_{0.95}Cs_{0.05}PbI_3$ {pure Pb} based perovskite solar cells. The average PCE is also mentioned over 16 devices.*

| Devices | $V_{OC}$ (V) | $J_{SC}$ (mA/cm$^2$) | FF (%) | PCE (%) | Average PCE (%) |
|---------|------|---------|------|------|------------|
| Pure Pb | 0.93 | 24.74 | 75 | 17.25 | (16.1 ± 1.2) |
| Pb-Sn Mixed | 0.72 | 29.52 | 64 | 13.60 | (11.4 ± 2.3) |
| Pure Sn | 0.59 | 22.41 | 59 | 7.81 | (5.1 ± 2.8) |

### 8.3.3 Charge carrier recombination dynamics

TPV measurement is a well established technique and widely used in organic & dye-sensitize solar cells to study the charge carrier recombination dynamics under open circuit condition.[29] TPV is used for quantitatively estimate of the charge carrier lifetime ($\tau$), charge carrier density ($n$) and bimolecular recombination rate constant ($k$). However, there are other techniques such as time of flight, transient absorption spectroscopy (TAS), charge extraction, photo-CELIV etc. are also used to study the charge carrier recombination dynamics.[30,31,32] But, TPV have advantage of simple instrumentation in comparison to other techniques and results are in good agreement with the TAS study.[33] It has been seen that $V_{OC}$ of PSCs depends upon the competition between radiative *vs* non-radiative recombination within the bulk and also upon the interfacial defects.6 TPV measurement gives an insight on the $V_{OC}$ losses *via* extracting the life time of perturbed charge carriers. The schematic representation of TPV experimental set up is shown in figure **9**. Different background light intensity condition for the devices was created by a constant DC LED light source and the devices were held at open circuit condition (1 MΩ termination). A small charge perturbation $\Delta n$ was added to the system with the help of an AC pulse laser (490 nm and pulse width of 500 ns with a repetition rate of 1 ms). The decay profile of the extra $\Delta n$ charges was fitted using a mono-exponential curve to get the lifetime $\tau_{\Delta n}$ of the transient charges. The normalized TPV curves at different DC background intensities are shown in the figure **10** for pure Pb, pure Sn and Pb-Sn mixed based PSCs. Figure **11a** represents the variation of $\tau_{\Delta n}$ with respect to different DC background intensities. The TPV profile is plotted against the laser profile with pulse width of 500 ns in figure **11b**. We observe that pure Sn based PSC have a higher lifetime (2.1 µs) than that of the Pb-Sn mixed (0.29 µs) and pure Pb (0.80 µs) based PCSs at 1 sun background intensity, which is in contrast to the $V_{OC}$ trend in these devices (table **1**). We will discuss about this discrepancy in discussion part with good details. The small perturbed charge carrier lifetime $\tau_{\Delta n}$ has an exponential relationship with the applied light bias (corresponding $V_{OC}$) as described in the below equation.

$$\tau_{\Delta n} = \tau_{\Delta n_o} \exp(-\beta V) \qquad (1)$$

where $\beta$ is the rate at which the charge carrier decays with respect to voltage (figure **12a**). The $\beta$ obtained for the pure Pb, pure Sn and Pb-Sn mixed based PSCs are 17.2 V⁻¹, 15.6 V⁻¹ and 17.4 V⁻¹, respectively. This shows that the pure Sn based PSCs has a slower decay rate when compared to the pure Pb and Pb-Sn mixed device. TPC profiles for all the three devices under approximately short circuit condition are shown in figure **10**.

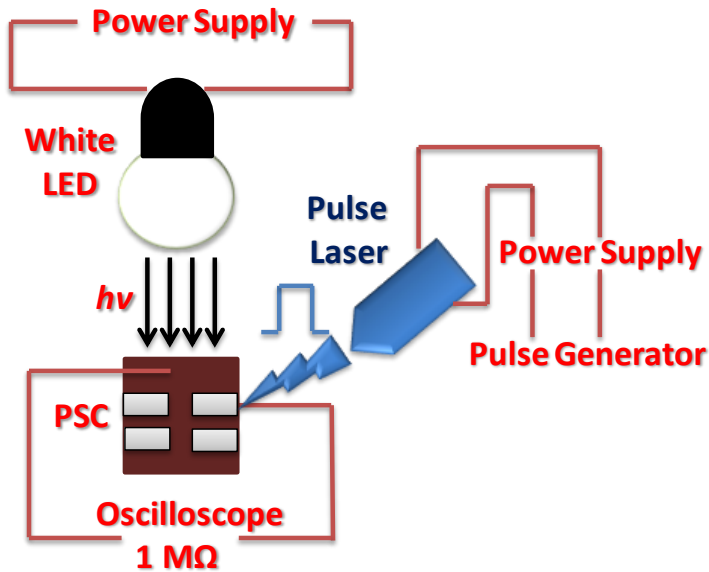

***Figure 9:*** *Schematic representation of transient photovoltage (TPV) experimental setup.*

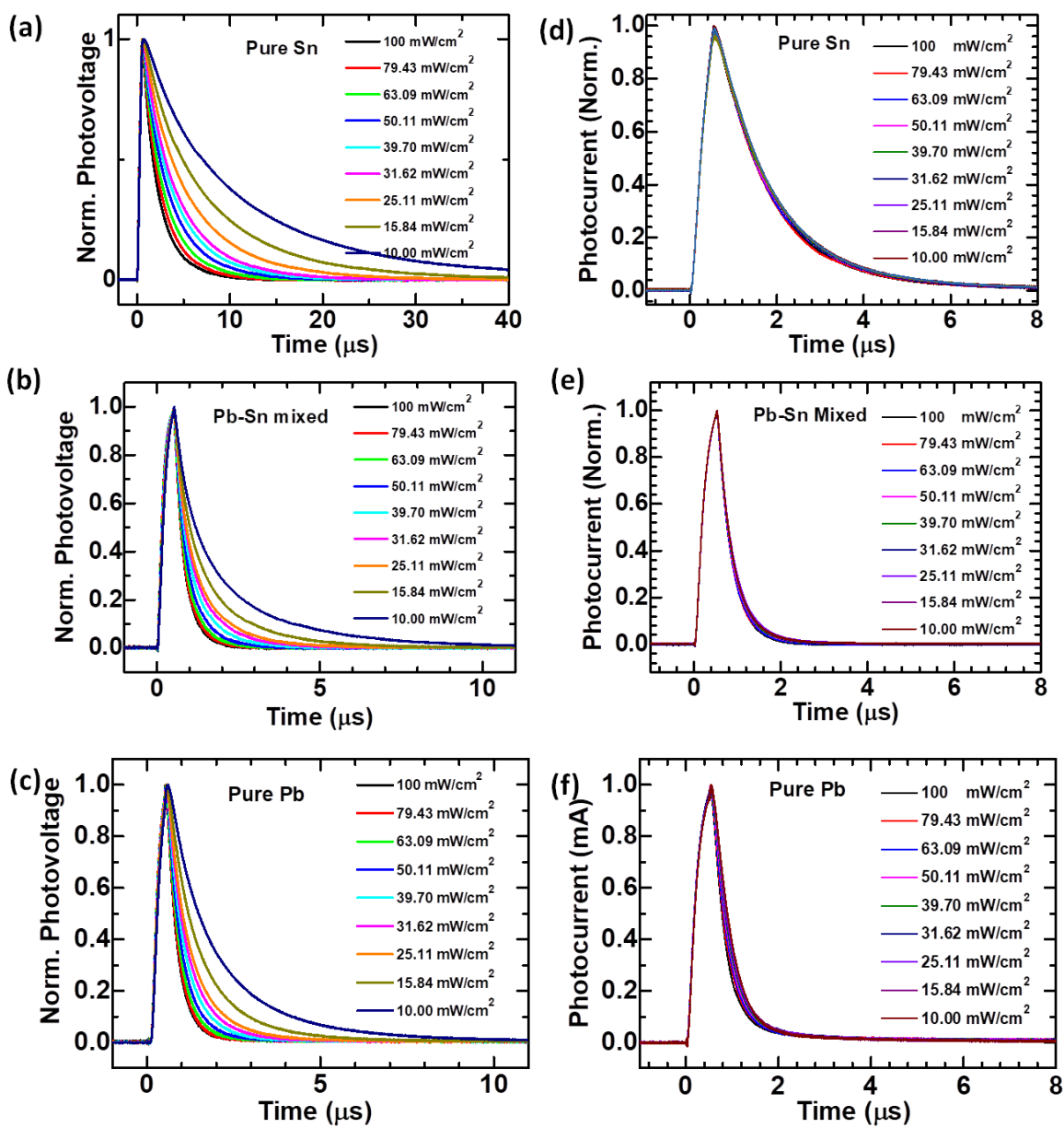

**Figure 10:** *Normalized TPV and TPC profile under different CW background intensity for (a), (d) MA$_{0.20}$FA$_{0.75}$Cs$_{0.05}$SnI$_3$ {pure Sn}, (b), (e) (MAPbI$_3$)$_{0.4}$ (FASnI$_3$)$_{0.6}$ {Pb-Sn mixed} and (c), (f) FA$_{0.95}$Cs$_{0.05}$PbI$_3$ {pure Pb} based perovskite solar cells, respectively.*

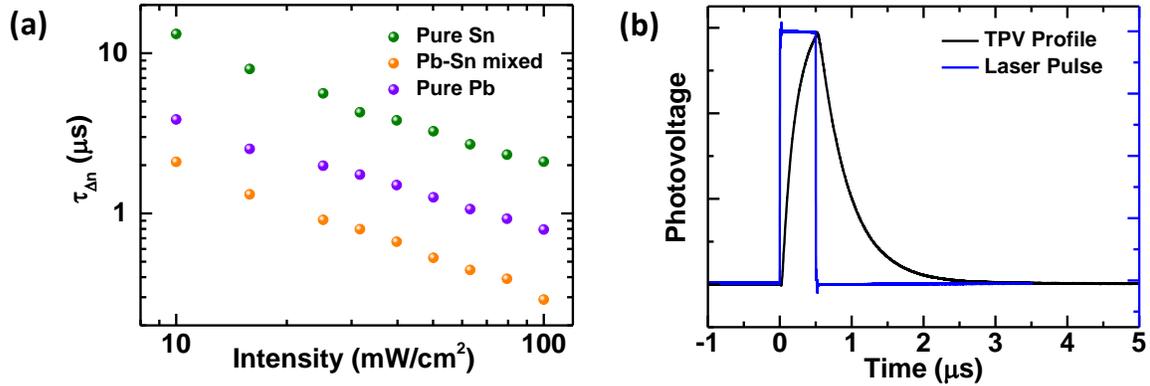

**Figure 11:** *(a) Variation of $\tau_{\Delta n}$ under different background intensity for $MA_{0.20}FA_{0.75}Cs_{0.05}SnI_3$ {pure Sn}, $(MAPbI_3)_{0.4}$ $(FASnI_3)_{0.6}$ {Pb-Sn mixed} and $FA_{0.95}Cs_{0.05}PbI_3$ {pure Pb} based perovskite solar cells. (b) A typical TPV profile is plotted with laser pulse width of 500 ns.*

TPV can be also used to calculate the carrier density *n* by measuring the photocurrent under approximately short circuit condition (50 $\Omega$) and the light intensity should be same as it was used during TPV measurement. By integrating the TPC curve, we can calculate total charge stored. Hence, the capacitance can be calculated as

$$c = \frac{\Delta Q}{\Delta V_0} \quad (2)$$

Where, $V_o$ is the voltage change at different background intensity. We found that *C* is exponentially increases with increase in the open circuit voltage ($V_{OC}$). The steady state carrier density *n* can be calculated by integrating *C* with respect to the voltage[33]

$$n = \frac{1}{Aed} \int\limits_0^{V_{OC}} C \, dV \quad (3)$$

Here, *A* is the area of the device (4.5 mm$^2$), *e* is the electronic charge and *d* is the thickness of the active layer.

The exponential increase of total $n$ in the cells as a function of light bias (corresponding $V_{OC}$) is shown in figure **12b**. The equation used for the fitting is

$$n = n_o \exp(\gamma V_{oc}) \quad (4)$$

A value of charge carrier density in the dark of $n_o$ = ~5.1 x $10^{12}$ cm$^{-3}$, ~8.1 x $10^{14}$ cm$^{-3}$ and ~1.2 x $10^{14}$ cm$^{-3}$ was used, with $\gamma$ = 9.5 V$^{-1}$, 6.4 V$^{-1}$ and 8.7 V$^{-1}$ for pure Pb, pure Sn and Pb-Sn mixed based PSCs, respectively. The parameter $\gamma$ is the rate of increase of $n$ with respect to bias. This value is typically 19 V$^{-1}$ ($\gamma \approx$ e/2K$_B$T) in an ideal semiconductor where traps are not present. The higher deviation in the value of $\gamma$ for pure Sn and Pb-Sn based PSCs from that of an ideal semiconductor suggests the presence of Sn vacancy or trap states. These trap states lead to increase the non-radiative recombination inside the perovskite layer and affect the charge transport. It is noted that the $\gamma$ value is lower in pure Sn based device suggesting a higher density of trap states.

The perturbed lifetime $\tau_{\Delta n}$ is related with the total charge carrier density $n$ in the cells with the power law which is depicted in the figure **12c**.

$$\tau_{\Delta n} \propto \tau_{\Delta n0} \left(\frac{n_0}{n}\right)^{\lambda} \quad (5)$$

The parameter $\lambda$ can also be obtained by dividing the decay time constant $\beta$ with the rate at which $n$ decays with respect to time i.e $\lambda = \beta/\gamma$ (rearrangement of equations 1, 2 and 3). The value of $\lambda$ is 1.81 (1.81), 2.58 (2.43) and 2.21 (2.00) for pure Pb, pure Sn and Pb-Sn mixed based PSCs ($\lambda = \beta/\gamma$, is given in bracket), respectively (figure **12c**). The power law parameter $\lambda$ provides information regarding the order of recombination in the bulk. The rate law for charge carrier decay dynamics is given by the following equation

$$\frac{dn}{dt} \approx -\frac{n^{\lambda+1}}{(1+\lambda)\tau_{\Delta n_0} n_0^{\lambda}} \quad (6)$$

For $\lambda \sim 2$, the charge carrier decay dynamics exhibits third order dependence on the charge carrier density (i.e. $dn/dt \; \alpha \; n^3$).

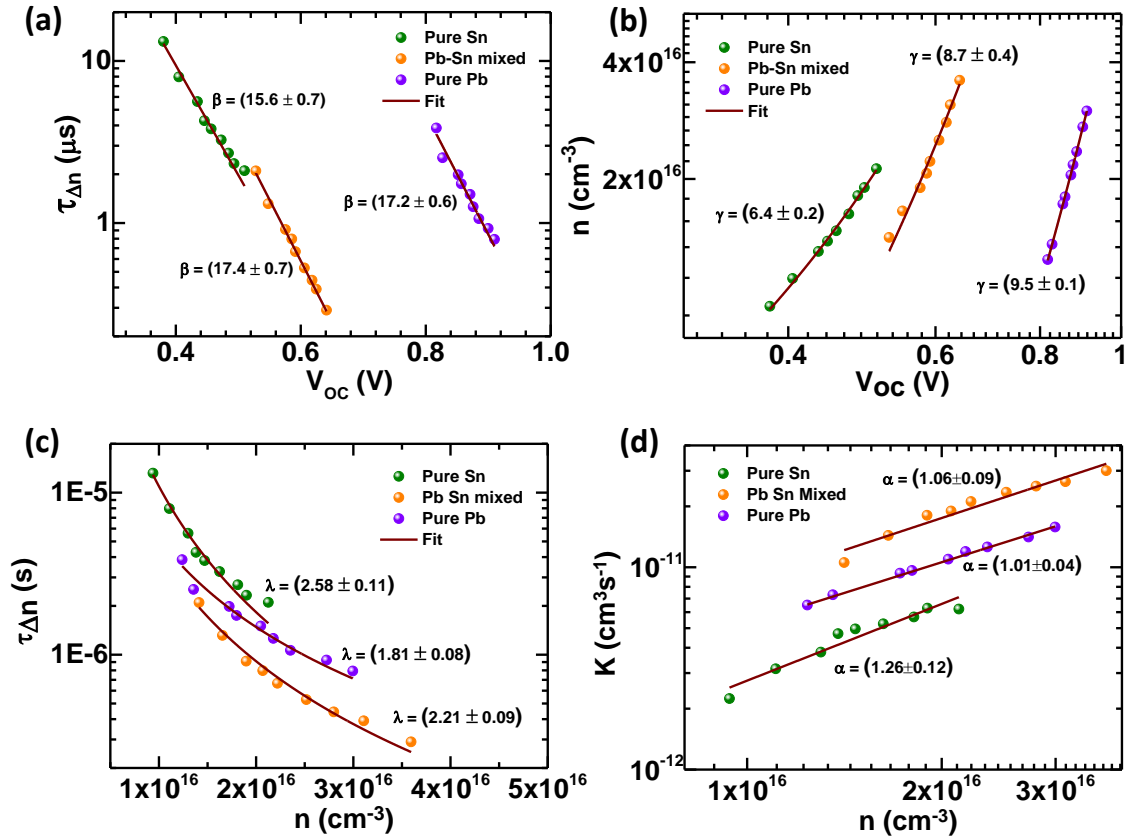

***Figure 12:*** *(c) Lifetime of perturbed charge carrier $\tau_{\Delta n}$ vs. open circuit voltage fitted with equation 1; (b) Charge carrier density n versus intensity dependent $V_{OC}$ in both cells. The solid lines represent fits of the data to equation 2. (c) Plot of $\tau_{\Delta n}$ as a function of charge carrier density in the cells fitted with power law; (d) bimolecular recombination rate k vs. charge carrier density n in $MA_{0.20}FA_{0.75}Cs_{0.05}SnI_3$ {pure Sn}, $(MAPbI_3)_{0.4}$ $(FASnI_3)_{0.6}$ {Pb-Sn mixed} and $FA_{0.95}Cs_{0.05}PbI_3$ {pure Pb} based PSCs.*

The charge carrier recombination dynamics are generally governed by rate law equation which is given below

$$\frac{dn}{dt} = G - k_1 n - k_2 n^2 - k_3 n^3 \qquad (7)$$

Where, G is the charge generation rate, $k_1$, $k_2$ & $k_3$ are monomolecular, bimolecular and auger recombination rate constant, respectively. It shows that bimolecular recombination have second order dependency on the $n$. However, pure Pb, pure Sn and Pb-Sn mixed based PSCs shows third order dependence on $n$; an auger recombination. But, this is not the case, we will discuss about it in discussion part. The total charge carrier lifetime $\tau$ in the device is obtained using the simple relation $\tau = \tau_{\Delta n}(\lambda + 1)$. The total charge carrier lifetime $\tau$ and charge carrier $n$ in the device is related by the equation

$$k(n) = \frac{n^{\lambda-1}}{(1+\lambda)\tau_{\Delta n_0}n_0^{\lambda}} \qquad (8)$$

Where $k(n)$ is the bimolecular recombination rate coefficient. $k$ has a power law dependence on charge carrier density $n$ ($k \propto n^{\lambda}$). Hence, $dn/dt$ exhibit third order dependence on charge carrier density (equation **6**) due to dependence of $k$ on $n$ in all the three PSCs.[33]

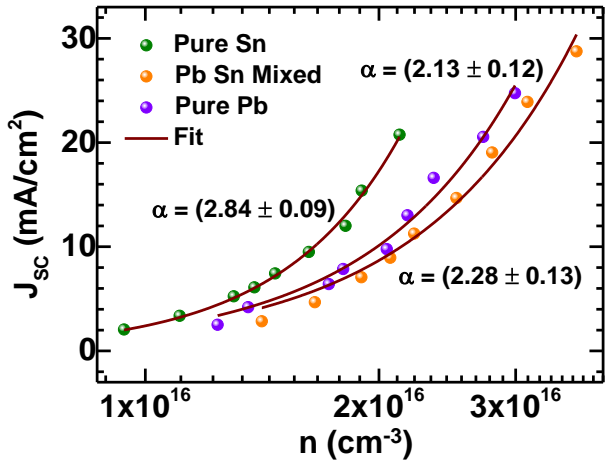

*Figure 13: Plot of current density ($J_{SC}$) at different intensity vs. corresponding charge carrier density ($n$) calculated through TPC under approximately short-circuit condition for pure Pb, pure Sn and Pb-Sn mixed PSCs.*

For bimolecular recombination dominated current,

$J \propto k <n> <p>$.

For typical condition, $<n> = <p>$     *(law of mass action)*

$J \propto k <n>^2$; which means, $J \propto n^2$.

However, $J$ has power law dependence of more than 2 with charge carrier density ($n$) in all kinds of devices (figure **13**). This clearly indicated that $k$ is also dependent on $n$.

Hence,                    $k \propto n^{0.13}$        (For pure Pb based PSC)

                         $k \propto n^{0.28}$        (For Pb-Sn mixed PSC)

                         $k \propto n^{0.84}$        (For pure Sn PSC)

Bimolecular recombination is one of the important factors, which decide the performance of the solar cell. The rate of bimolecular recombination $k$ as a function of charge carrier density $n$ is shown in figure **12d**. We observe that $k$ increases with increase in $n$ and shows a value of ~1.58 x $10^{-11}$ $cm^3s^{-1}$ for pure Pb based PSC at 100 mW/cm$^2$ illumination. The value of $k$ reported here for pure Pb based PSC is in good agreement with the literature.[34] The Pb-Sn mixed PSC shows higher value of k (~ 3.05 x $10^{-11}$ $cm^3s^{-1}$) among the three devices. However, for pure Sn the value of $k$ (~6.21 x $10^{-12}$ $cm^3s^{-1}$) is slightly higher than that of pure Pb based PSC. The value of $k$ is lower for pure Sn based PSC than the Pb-Sn mixed PSC at any value of $n$, which is again a contradictory result considering the fact that pure Sn based perovskite films have higher defect states (figure **5**).

## 8.4 Discussion

A distinct carrier dynamics based on 'B' site being Pb *vs* Sn *vs* Pb-Sn mixed requires a deeper insight to connect with PSC performance. The importance of long-lived hot carriers to the PL contribution in Sn based perovskite (specially, FASnI$_3$) has been discussed in literature.[35] However, the origin behind the long lifetime of charge carriers in halide perovskite is still a topic of debate in the community. There are many speculations regarding the long carrier lifetime such as Rashba effect, ferroelectric property, formation of polarons which prevent the recombination of free electron-hole, photon recycling, etc. Since the atomic number of Sn is lower than that of Pb, the spin-orbit coupling effect for Sn counterparts is almost three times lower than that of Pb compounds.[36] There are reports which suggest that Rashba effect causes fluctuation in conduction band structure and thus it can lead to formation of either indirect band gap or combination of both direct and indirect band gap, which can be responsible for slow decay rate of hot carriers.[37,38] Sn based perovskite materials are known to oxidized in their higher oxidation state (+4) which increase the hole density near the top of the valance band. This leads to absorption of photons higher than the fundamental band gap and is responsible for Burstein-Moss shift in Sn based perovskites.[39] However, addition of $SnF_2$ reduces the oxidation of $Sn^{2+}$ to $Sn^{+4}$, and thus reducing the formation of Sn vacancies. Therefore, $SnF_2$ helps to reduce the relaxation path in the valance band and it can be responsible for slow carrier relaxation.[40] Here, we used small amount of Cs in $MA_{0.25}FA_{0.75}SnI_3$ based perovskite to make it more stable. Since, Cs is smaller in size than MA and FA, hence it will create a chemical pressure on the inorganic lattice ($SnI_6$).[35,41] Sn based perovskite semiconductor already have defect states in the sub gap near to the valance band. As the additional Cs creates chemical pressure, the defect states become shallower and it can be also responsible for slow carrier relaxation in pure Sn based semiconductors. Mahata *et al.* reported that pure Sn based perovskite shows double polaron stabilization energy in comparison to the pure Pb based perovskites.[42] The addition of small amount of Cs along with MA and FA cations results into strong structural distortion, which shortens the Pb-Sn bond length. However, such a small amount of Cs does not change the perovskite phase (figure **4b**). Degree of hysteresis and ideality factor (intensity dependent $V_{OC}$) further indicates the relatively higher defect states induced characteristics in pure Sn based PSC. Hence, we hypothesize the reason for longer carrier lifetime of perturbed charge carrier for pure Sn based PSC despite of lower $V_{OC}$, might be related to defects.

We mapped the recombination rate constant on the basis of carrier lifetime of perturbed charge carriers *via* TPV and steady state carrier density extracted through TPC method.[29,33] However, *n* calculated from charge extracted at lower intensity is not very accurate.[43] In Sn based PSC, the FF decreases drastically below 0.5 Sun which indicates the presence trap states at lower intensity (figure **7**). Hence, the charge extraction becomes very poor at lower intensity, which leads to inaccuracy in *n* value. However, in the case of pure Pb based PSC, the FF is quite constant for all intensities which reflects a high quality perovskite film with lower number of trap states (figure **7**). Figure **12d** represents that pure Pb based PSC shows a lower value of *k* for all *n*, which suggest that the bimolecular recombination is more dominating in comparison to pure Sn and Pb-Sn mixed PSCs. The slight non-linearity of *k* with *n* leads to $\acute{\alpha}$ higher than 2 and thus, the rate law of carrier decay dynamics exhibits third order dependence on the charge carrier density (*dn/dt* α $n^{(\acute{\alpha}+1)}$).[29]

## 8.5 Conclusion

In conclusion, we present a comparative study on the charge carrier recombination dynamics in pure Pb, pure Sn and Pb-Sn mixed based PSCs. The pure Sn based PSC shows higher carrier lifetime despite having lower $V_{OC}$ than pure Pb and Pb-Sn mixed PSCs. Long lived photo-induced carrier lifetime *vs* $V_{OC}$ discrepancy suggest the role of defects for such observation. We also demonstrate that the rate law of carrier decay dynamics depends upon relative defect state density. Pure Pb based PSC have lower number of trap states. The non-linearity of $k$ with $n$ at lower intensity leads to third order dependence on $n$. Hence, this study helps to understand the recombination dynamics of photo-generated charge carriers in different perovskite based semiconductors through TPV and correlate it to trap states and long carrier recombination lifetime.

## 8.6    Post Script

This chapter presents a study over charge carrier recombination dynamics of pure Sn, pure Pb and Pb-Sn mixed based PSCs *via* TPV measurement and correlate it with defect density in the bulk of the semiconductor. Till now, we have understanding about defect and disorder in the perovskite, additive engineering to passivate the defects and study of charge transport and recombination dynamics of perovskite based semiconductors. Chapter **9** will provide summary of the thesis and a brief description of the future plans with the use of results mentioned in this thesis.

# Chapter 9

# Summary and Future Outlooks

# CHAPTER 9

# Summary and Future Outlooks

## 9.1    Summary

The thesis is focused on the study of imperfection such as disorder and defect states in solution processed perovskite thin film based semiconductor through temperature dependent optical studies and passivation of the defect states via solvent or small organic molecule additives to enhance the performance of perovskite solar cells (PSCs) in terms of efficiency and stability. The thesis also aims towards understanding of charge transport length scale in PSCs and connect it with the defect density in the bulk of the semiconductor. In addition, Pb free based PSCs are fabricated and charge carrier decay dynamics of Pb versus Sn halide based perovskite is studied through transient photovoltage measurement (TPV). Here, we summaries the thesis in detail:

1) First, we establish a methodology using optical techniques for the concise characterization of disorder in the perovskite materials. It correlates fundamental aspects such as lattice dilatation and static & dynamic disorder to the optical properties of perovskite semiconductors. This provides important insights into the electronic and structural properties of MAPbI$_3$ based perovskites which should be transferable to many related organic metal-halide perovskites. In addition, we proposed an easy way of processing perovskite nanocrystals by blending the bulk perovskite with wide-band gap organic semiconductors (BCP and/or CBP) forming type-1 heterostructures. This, in turn, largely tunes the band gap over a wider range of electromagnetic spectra due to weak confinement. We also demonstrate that a wide range of band gap tuning can be explained broadly by Bohr radius versus average size of crystallites in the efficient type-1 heterostructure.

2) For defect passivation in PSCs, we have used solvent additives (DIO and phosphinic acid) and small organic molecule (BCP) in the perovskite precursor solution and study the role of additives in three kinds of perovskite materials: $MAPbI_3$, $MAPbI_{3-x}Cl_x$ and $MAPbBr_3$. A comparative study between the pristine and additive based PSCs has been carried out. The improved performance of additive based PSC is attributed to the suppressed non radiative recombination, retarded rate of crystallization which provides a smoother & pin hole free perovskite film with bigger domains as compared to the pristine film and enhanced charge transfer rate by choosing double electron and double hole extraction layers. BCP being a hydrophobic organic material provides moisture stability to the PSC. The thesis also shows that the 70% of the efficiency retains after 3 months for passivated PSCs.

3)  The thesis also present a comparative study of charge transport length scale ($L$) for pristine and passivated $MAPbI_3$ based perovskite solar cells through scanning photocurrent microscopy (SPM). The SPM study suggested an improved $L$ and degree of ambipolarity of photo-generated charge carriers (electron and hole) in a passivated as compared to pristine $MAPbI_3$ based PSCs. These results found to be in correlation with frequency dependent photocurrent measurement, which shows that relaxation time of charge carrier is relatively lower in passivated $MAPbI_3$ based PSCs. We explained the mechanism with trap-assisted recombination, where trap states are induced by ion migration in halide perovskite films.

4)  Finally, we fabricated $FA_{0.95}Cs_{0.05}PbI_3$ (pure Pb), $MA_{0.20}FA_{0.75}Cs_{0.05}SnI_3$ (pure Sn) and $(MAPbI_3)_{0.4}$ $(FASnI_3)_{0.6}$ (Pb–Sn mixed) based PSCs and compare the charge carrier recombination dynamics of all the three PSCs through TPV measurement. Sn based PSC shows higher lifetime of perturbed charge carrier in comparison to pure Pb based PSC which will be correlated to slow relaxation lifetime of the charge carrier and recombination through the defect states in the perovskite thin films. This study also demonstrates the rate law of charge carrier decay in all the three devices and also reveals the nonlinear dependence of $k$ on $n$. Although, we agree that for better understanding of such complex results, more experimental evidences are needed.

Overall, this thesis provide a detail study of photo-physics and device physics of defect induced Pb halide based perovskite semiconductor and it will be useful to apply the fabrication and characterization techniques used in the thesis for Pb free PSCs.

## 9.2    Future Outlooks

Photo-physics of lead based perovskites are well understood in last few years but it surprises every day. Few of them are; double emission peak of perovskite materials at low temperature, contrast behaviour of enhanced steady state photoluminescence (PL) intensity *vs* decreased lifetime at low temperature with respect to room temperature in perovskite nano-crystals, origin of non-radiative recombination: photon recycling *vs* defect mediated recombination, role of ions *vs* ferroelectricity, etc. The origin of existence of dual emission is still a controversy in the perovskite community. However, it is believed that inclusion of tetragonal phase (low energy peak) in the orthorhombic phase (high energy peak) at low temperature is one of the possible reasons for dual emission feature. We have studied the temperature dependent steady state PL of pristine $MAPbI_3$ based perovskite and perovskite with BCP additive. Surprisingly, we find that dual emission peak of $MAPbI_3$ disappear at lower temperature (10 K) with addition of BCP. The steady state PL of $MAPbI_3$ with and without BCP additive films at 10 K is shown in figure **9.1**. However, a detailed optical study (temperature dependent delayed emission spectroscopy) with different weight % of BCP is required to verify and conclude this aspect and is one of the future plans.

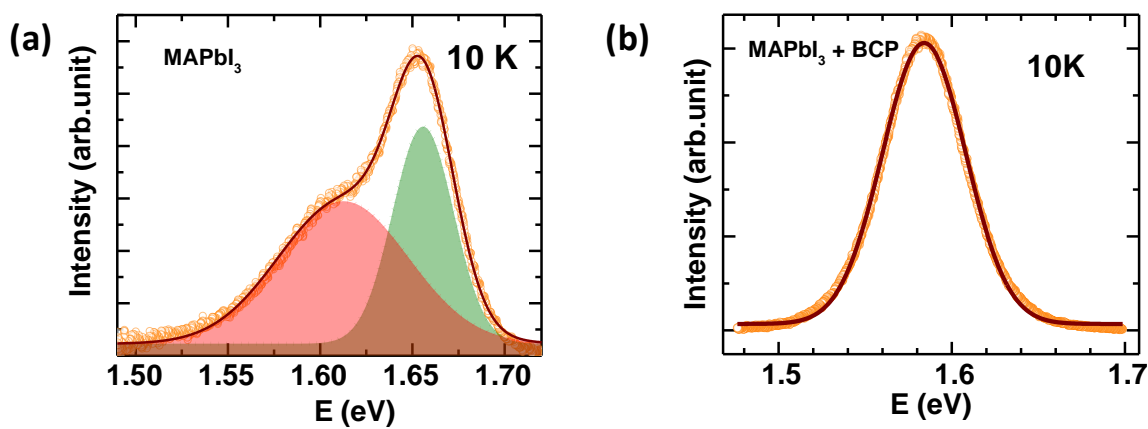

***Figure 9.1:*** *Low temperature (10 K) steady state PL spectra of $MAPbI_3$ films with and without BCP additive.*

Along with lead based perovskite, there are also reports on the optical properties of Sn based perovskite which reveals some interesting phenomenon about the material such as red shift in PL with time, stoke shift in absorption *vs* PL peak position etc. Defect mediated slow recombination of charge carriers in Sn based perovskites makes it more interesting for in-depth optical studies such as time dependent PL imaging to understand slow relaxation of charge carriers. Time delayed collection field technique can be also used to study the bimolecular recombination constant and correlate it with TPV results (Chapter **8**). Considering the demand of PSCs as an alternative for energy dependence and security purpose, another useful research is all perovskite tandem solar cells. We successfully made *p-i-n* configuration based low band gap (1.25 eV) Pb-Sn mixed PSC and wide band gap (1.6 eV) Pb based PSC. Addition of small amount of Br will significantly increase the $V_{OC}$ without compromising on the $J_{SC}$. These two cells can be use as top and bottom cell of all perovskite tandem solar cell. We have also fabricated perovskite-organic tandem solar cell with active area of 0.36 cm$^2$ and able to add $V_{OC}$ successfully. The schematic diagram of spincoating technique indicating all the layers used in the perovskite-organic tandem solar cell and primarily results are shown in figure 9.**2** and **9.3**, respectively.

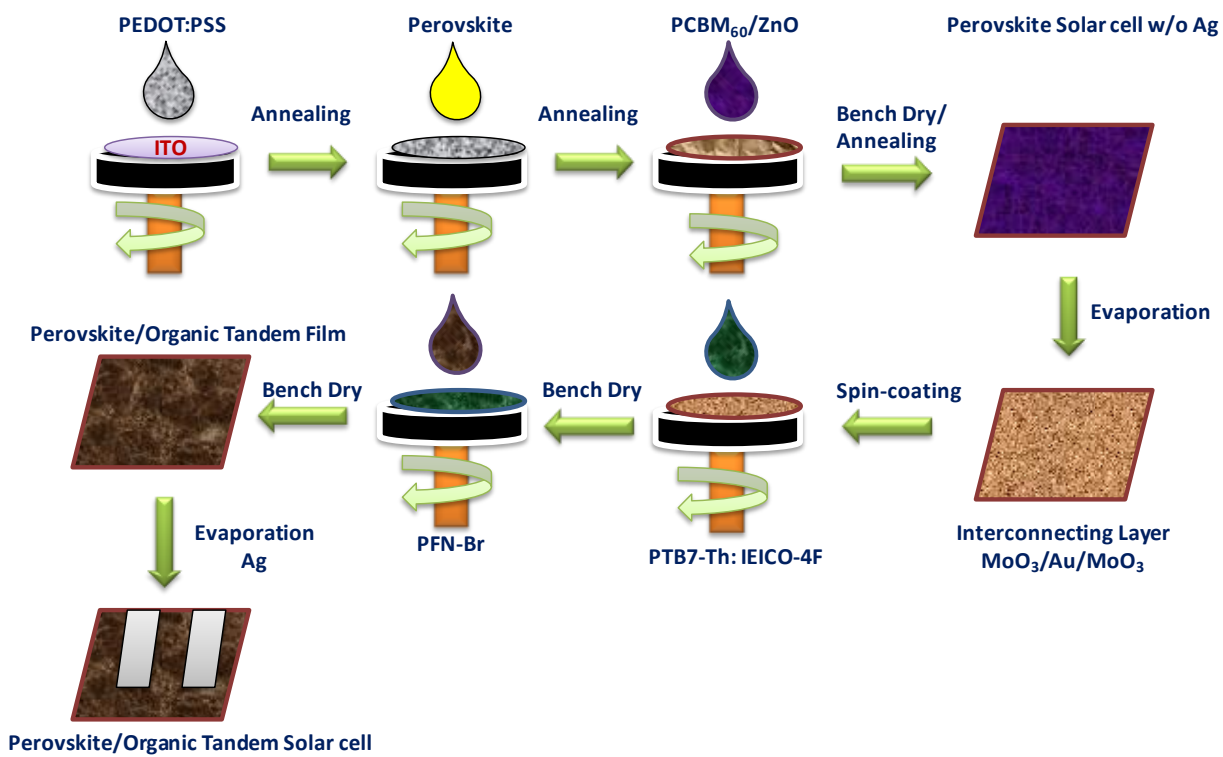

***Figure 9.2:*** *Schematic diagram of spincoating technique indicating all the layers used in the perovskite-organic tandem solar cell.*

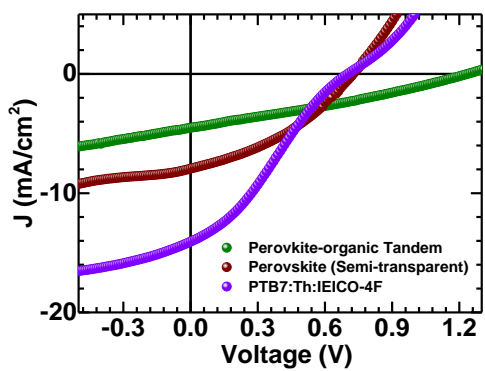

| Solar cells | $J_{SC}$ (mA/cm2) | $V_{OC}$ (V) | PCE (%) |
|---|---|---|---|
| MAPbI3 | 7.94 | 0.73 | 1.65 |
| PTB7Th:IEICO-4F | 14.07 | 0.71 | 2.12 |
| Tandem | 4.51 | 1.24 | 2.8 |

***Figure 9.3:*** *J-V characteristics of MAPbI3 based semitransparent solar cell (8 nm Ag by thermal evaporation), opaque PTB7TH:IEICO-4F bulk hetro-junction and perovskite organic tandem solar cells under 1.5AM solar simulator and corresponding photovoltaic parameter and PCE is also listed.*

*The device structure for tandem solar cell is:*

*ITO/ PEDOT:PSS/ MAPbI$_3$/ PCBM/ ZnO/ MoO$_3$ (5 nm)/ Au (5 nm)/ MoO$_3$ (5 nm)/ PTB7Th:IEICO-4F/ PFN-Br/ Ag.*